\documentclass[11pt,reqno,letter]{amsart}
\usepackage{amsmath}

\usepackage{amsfonts}
\usepackage{amssymb}
\usepackage{amsthm}

\usepackage{graphicx}
\usepackage{hyperref}
\usepackage[left=1.25in,right=1.25in,top=1.5in,bottom=1.25in]{geometry}
\usepackage{multirow}
\usepackage{verbatim}
\usepackage{float}
\usepackage{rotating}
\usepackage[dvipsnames]{xcolor}
\usepackage[normalem]{ulem}
\usepackage{booktabs}

\usepackage{tabu}

\usepackage{float}
\usepackage{booktabs}

\usepackage{subcaption}

\theoremstyle{definition}

\newtheorem*{example*}{Example}

\usepackage{calc,tikz,nicefrac}

\usepackage[section]{placeins}

\usetikzlibrary{positioning}
\usepackage{mathtools}

\usepackage{bm}

\usepackage[authoryear]{natbib}
\usepackage{pdflscape}
\usepackage{afterpage}
\usepackage{capt-of}

\renewcommand{\to}{{\rightarrow}}
\providecommand{\Infected}{{\mathcal{I}}}
\providecommand{\Recovered}{{R}}

\providecommand{\Var}{{\mathrm{Var}}}

\newcommand\indep{\protect\mathpalette{\protect\independenT}{\perp}}
\def\independenT#1#2{\mathrel{\setbox0\hbox{$#1#2$}%
    \copy0\kern-\wd0\mkern4mu\box0}}

\newcommand{\Ep}{{\mathrm{E}}}

\usepackage{neuralnetwork}

\def\bcolor{\color{ForestGreen}}
\def\pcolor{\color{blue}}
\def\icolor{\color{magenta}}
\def\wcolor{\color{gray}}
\def\ycolor{\color{red}}

\begin{document}

\title[Opening K-12 Schools and the Spread of COVID-19]
{The Association of Opening K-12 Schools with the Spread of COVID-19 in the United States: County-Level Panel Data Analysis}\thanks{We are very grateful to Co-editor and two anonymous referees whose comments results in many improvements relative to the original version.
We also thank Emily Oster for helpful comments.}

\author{Victor Chernozhukov}
\address{Department of Economics and Center for Statistics and Data Science, MIT,  MA 02139}
\email{vchern@mit.edu}
\author{Hiroyuki Kasahara}
\address{ Vancouver School of Economics, UBC, 6000 Iona Drive, Vancouver, BC.}
\email{hkasahar@mail.ubc.ca}

\author{Paul Schrimpf}
\address{ Vancouver School of Economics, UBC, 6000 Iona Drive, Vancouver, BC.}
\email{schrimpf@mail.ubc.ca}

\date{\today} 

\begin{abstract}
This paper empirically examines how the opening of K-12 schools and colleges is associated with the spread of COVID-19 using county-level panel data in the United States.  Using data on foot traffic and K-12 school opening plans, we analyze how an increase in visits to schools and opening schools with different teaching methods (in-person, hybrid, and remote) is related to the 2-weeks forward growth rate of confirmed COVID-19 cases. Our debiased panel data regression analysis with a set of county dummies,  interactions of state and week dummies, and other controls shows that an increase in visits to both K-12 schools and colleges is associated with a subsequent increase in case growth rates. The estimates indicate that fully opening K-12 schools with in-person learning is associated with a 5 (SE = 2) percentage points increase in the growth rate of cases.  We also find that the positive association of K-12 school visits or in-person school openings with case growth is stronger for counties that do not require staff to wear masks at schools. These results have a causal interpretation in a structural model with unobserved county and time confounders. Sensitivity analysis shows that the baseline results are robust to timing assumptions and alternative specifications.\end{abstract}



\maketitle



 \section{Introduction}
Does opening K-12 schools lead to the spread of COVID-19? Do mitigation strategies such as mask-wearing requirements help reduce the transmission of SARS-CoV-2  at school?  These are important policy relevant questions  in countries with low vaccination rates, especially given the emerging variants of concerns with higher transmission rates. If in-person school openings substantially increase  COVID-19 cases, then local governments could promote enforcing mitigation measures at schools (universal and proper masking, social distancing, and handwashing) to lower the risk of COVID-19 spread. 
Furthermore, the governments could prioritize vaccines for education workers and elderly parents in case of in-person school openings.  This paper uses county-level panel data on K-12 school opening plans and mitigation strategies together with foot traffic data to investigate how an increase in the visits to K-12 schools is associated with a subsequent increase in COVID-19 cases in the United States.

\section*{Data} We begin with describing our data and provide descriptive evidence. Our sample period is from  April 1, 2020, to December 2, 2020.
Our analysis uses county-level panel data in the United States. As outcome variables, we use the weekly cases and deaths as well as their growth rates. The main explanatory variables of interest are the variable for school openings with different teaching methods and mitigation measures from MCH Strategic Data and per-device visits to K-12 schools from SafeGraph foot traffic data. We also use the foot traffic data on stay-at-home devices and visits to full-time/part-time workplaces, colleges/universities, restaurants, bars, recreational facilities,  and churches. Our panel regression analysis uses the additional data on non-pharmaceutical policy interventions (NPIs) and the number of tests.

The data on cases and deaths for each county are from the \href{https://raw.githubusercontent.com/nytimes/covid-19-data/master/us-counties.csv}{New York Times}.  \href{https://www.safegraph.com}{SafeGraph} provides foot traffic data based on a panel of GPS pings from anonymous mobile devices. Per-device visits to K-12 schools, colleges/universities, restaurants, bars, recreational places, and churches are constructed from the ratio of
daily device visits to these point-of-interest locations to the number of devices residing in each county. Full-time and part-time workplace visits are the ratio of the number of devices that spent more than 6 hours and between 3 to 6 hours, respectively, at one location other than one's home location to the total number of device counts. The staying home device variable is the ratio of the number of devices that do not leave home locations to the total number of device counts.

 \href{https://www.mchdata.com}{MCH Strategy Data} provides information on the date of school openings with different teaching methods (in-person, hybrid, and remote) as well as mitigation strategies at 14703 school districts. We link school district-level MCH data to county-level data from NYT and SafeGraph using the \href{https://www2.census.gov/programs-surveys/saipe/guidance-geographies/districts-counties/sdlist-20.xls}{file} for School Districts and Associated Counties at US Census Bureau. School district data is aggregated up to county-level using the enrollment of students at each district. Specifically, we construct the proportion of students with different teaching methods for each county-day observation using the district-level information on school opening dates and teaching methods. We define each teaching method's county-level school opening date by the weighted mean of district-level school opening dates of the corresponding teaching method with enrollment weights. We also construct a county-level dummy variable of ``No mask requirement for staff,'' which takes a value of 1 if there exists at least one school district. The measure of mask requirements for staff is highly correlated with other mitigation measures, including mask requirements for students, prohibiting sports activities, and online instruction increases as shown in SI Appendix, Table S3.\footnote{MCH Strategic Data provides the school district level data on whether each school district adopts following mitigation strategies: (i) mask requirements for staff, (ii) mask requirements for students, (iii) prohibiting sports activities, and (iv) online instruction increases, among other measures.  We decided to use mask requirements for staff as the primary variable for school mitigation strategy because it has a relatively lower number of missing values than other mitigation measures.}
 A substantial fraction of school districts report ``unknown'' or ``pending''  for teaching methods and mask requirements. We drop county observations from the sample if more than 50 percent of students attend school districts that report unknown or pending teaching methods or mask requirements for panel regression analysis with teaching methods or mask requirements.


Our empirical analysis uses 7-day moving averages of daily variables to deal with periodic fluctuations within a week. SI Appendix, Fig. S4  shows the evolution of percentiles of these variables over time, while  Tables S1-S2 present descriptive statistics and correlation matrix across variables we use for our regression analysis.
The data set contains the maximum of 3144 counties for regression analysis using foot traffic data but some observations are dropped out of samples due to missing values for school opening teaching methods and staff mask requirements in some specifications.\footnote{Our regression analysis uses 2788 counties for specification with K-12 school opening with different teaching modes, while the sample contains 2204 counties for specification with mask requirements for staff.}  The analysis was conducted using R software (version 4.0.3).

\begin{table}[ht]  
\caption{Summary Statistics \label{table:summary}}\vspace{-0.1cm}
\hspace{-4cm} \resizebox{0.7\columnwidth}{!}{
\begin{minipage}{\linewidth} 
\begin{tabular}{@{\extracolsep{5pt}}lccccccc}
\\[-1.8ex]\hline
\hline \\
  & Wkly Case  & Wkly Death  & Wkly Cases & Wkly Deaths & K-12 Sch. & Workplace   & Restaurant  \\
  &  Growth &  Growth & per 1000 & per 1000 &  Visits &  Visits & Visits \\ \hline
\textbf{In-person} \\
 \qquad Before Opening \\
\qquad\quad Mean & 0.091 & 0.013 & 0.571 & 0.060 & 0.045 & 0.047 & 0.185 \\
   & (0.011) & (0.003) & (0.031) & (0.003) & (0.002) & (0.001) & (0.007) \\
 \qquad\quad N & 52258 & 52258 & 54995 & 54995 & 67070 & 67070 & 67070 \\
\qquad After Opening \\
\qquad\quad Mean & 0.143 & 0.034 & 3.038 & 0.104 & 0.161 & 0.073 & 0.188 \\
   & (0.014) & (0.006) & (0.200) & (0.005) & (0.005) & (0.001) & (0.006) \\
 \qquad\quad N & 45749 & 45749 & 45827 & 45827 & 46030 & 46030 & 46030 \\ \hline
\quad Difference in Means & 0.052 & 0.021 & 2.467 & 0.044 & 0.116 & 0.026 & 0.003 \\
   & (0.018) & (0.007) & (0.203) & (0.004) & (0.004) & (0.001) & (0.004) \\ \hline\hline
 \textbf{Hybrid} \\
 \qquad Before Opening \\
\qquad\quad Mean & 0.096 & 0.024 & 0.664 & 0.035 & 0.036 & 0.045 & 0.242 \\
   & (0.012) & (0.006) & (0.031) & (0.001) & (0.001) & (0.0004) & (0.005) \\
 \qquad\quad N & 234820 & 234820 & 243321 & 243321 & 260573 & 260551 & 260573 \\
\qquad After Opening \\
\qquad\quad Mean & 0.121 & 0.042 & 2.368 & 0.057 & 0.126 & 0.064 & 0.249 \\
   & (0.012) & (0.006) & (0.132) & (0.002) & (0.003) & (0.001) & (0.004) \\
 \qquad\quad N & 166605 & 166605 & 166660 & 166660 & 167206 & 167206 & 167206 \\\hline
 \quad Difference in Means & 0.025 & 0.019 & 1.703 & 0.022 & 0.090 & 0.019 & 0.007 \\
   & (0.016) & (0.008) & (0.136) & (0.002) & (0.003) & (0.001) & (0.004) \\ \hline\hline
 \textbf{Remote} \\
 \qquad Before Opening \\
\qquad\quad Mean & 0.099 & 0.035 & 0.742 & 0.032 & 0.032 & 0.045 & 0.278 \\
   & (0.012) & (0.008) & (0.035) & (0.002) & (0.001) & (0.0004) & (0.008) \\
 \qquad\quad N & 76796 & 76796 & 78581 & 78581 & 82165 & 82165 & 82165 \\
\qquad After Opening \\
\qquad\quad Mean & 0.103 & 0.033 & 1.944 & 0.047 & 0.088 & 0.058 & 0.287 \\
   & (0.012) & (0.009) & (0.115) & (0.003) & (0.003) & (0.001) & (0.007) \\
 \qquad\quad N & 50127 & 50127 & 50048 & 50048 & 50183 & 50183 & 50183 \\\hline
\quad \ Difference in Means & 0.004 & -0.002 & 1.202 & 0.015 & 0.056 & 0.013 & 0.009 \\
   & (0.017) & (0.012) & (0.120) & (0.002) & (0.003) & (0.001) & (0.006) \\  \hline\hline \\[-1.8ex]
\end{tabular}
 \end{minipage}} 
   {\scriptsize
\begin{flushleft}
Notes:  Based on observations from April 15, 2020 to December 2, 2020. Standard errors that are two-way clustered on county and date are reported in parentheses. 
 \end{flushleft}} 
\end{table}

\begin{figure}[ht]
  \caption{The evolution of  cases, deaths, and visits to K-12 schools and restaurants before and after the opening of K-12 schools\label{fig:case-growth}}\smallskip
\resizebox{\columnwidth}{!}{
\hspace{-0.4cm}\begin{minipage}{\linewidth}
        \begin{tabular}{cc}
 \small \textbf{(a) Wkly Cases by K-12 Opening Modes }  &   \small\textbf{(b) Wkly Deaths by  K-12 Opening Modes} \\
    \includegraphics[width=0.45\textwidth]{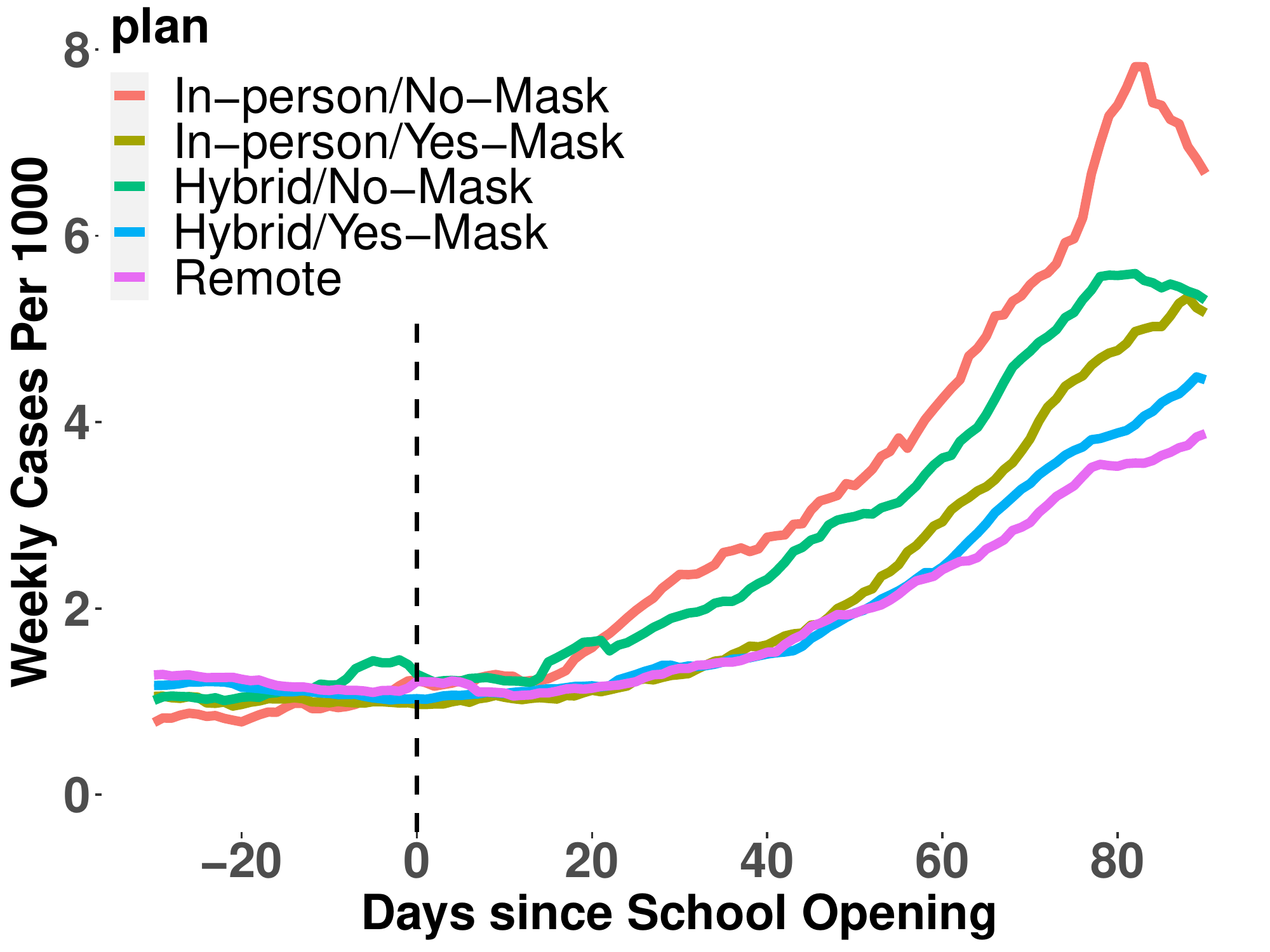}&
  \includegraphics[width=0.45\textwidth]{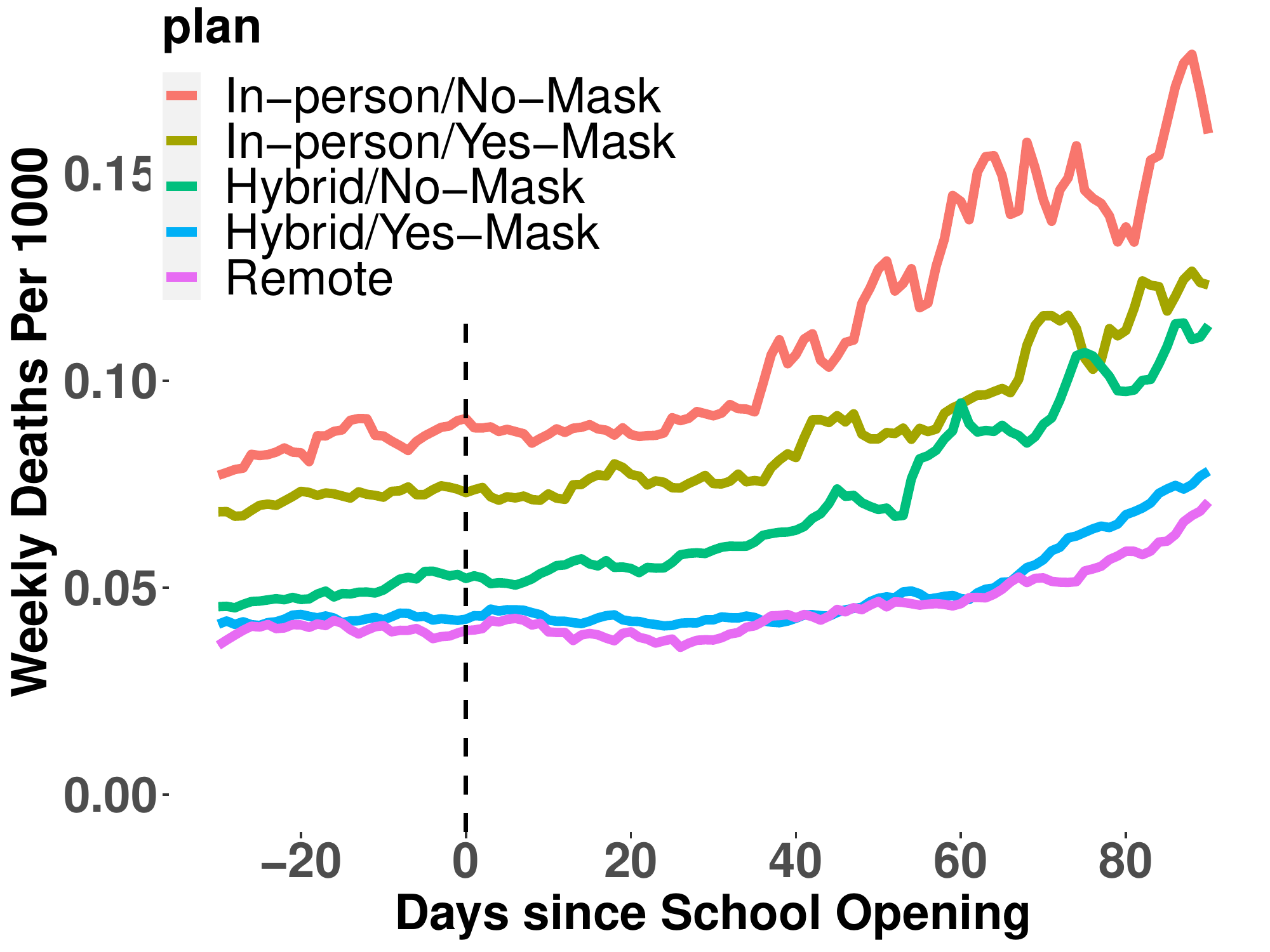}\\
  & \\
 \small    \textbf{(c) Wkly K-12 School Visits } &     \small  \textbf{(d) Wkly Full-Time Workplace Visits}\\
 \includegraphics[width=0.45\textwidth]{tables_and_figures/schoolmode-event-school}&
 \includegraphics[width=0.45\textwidth]{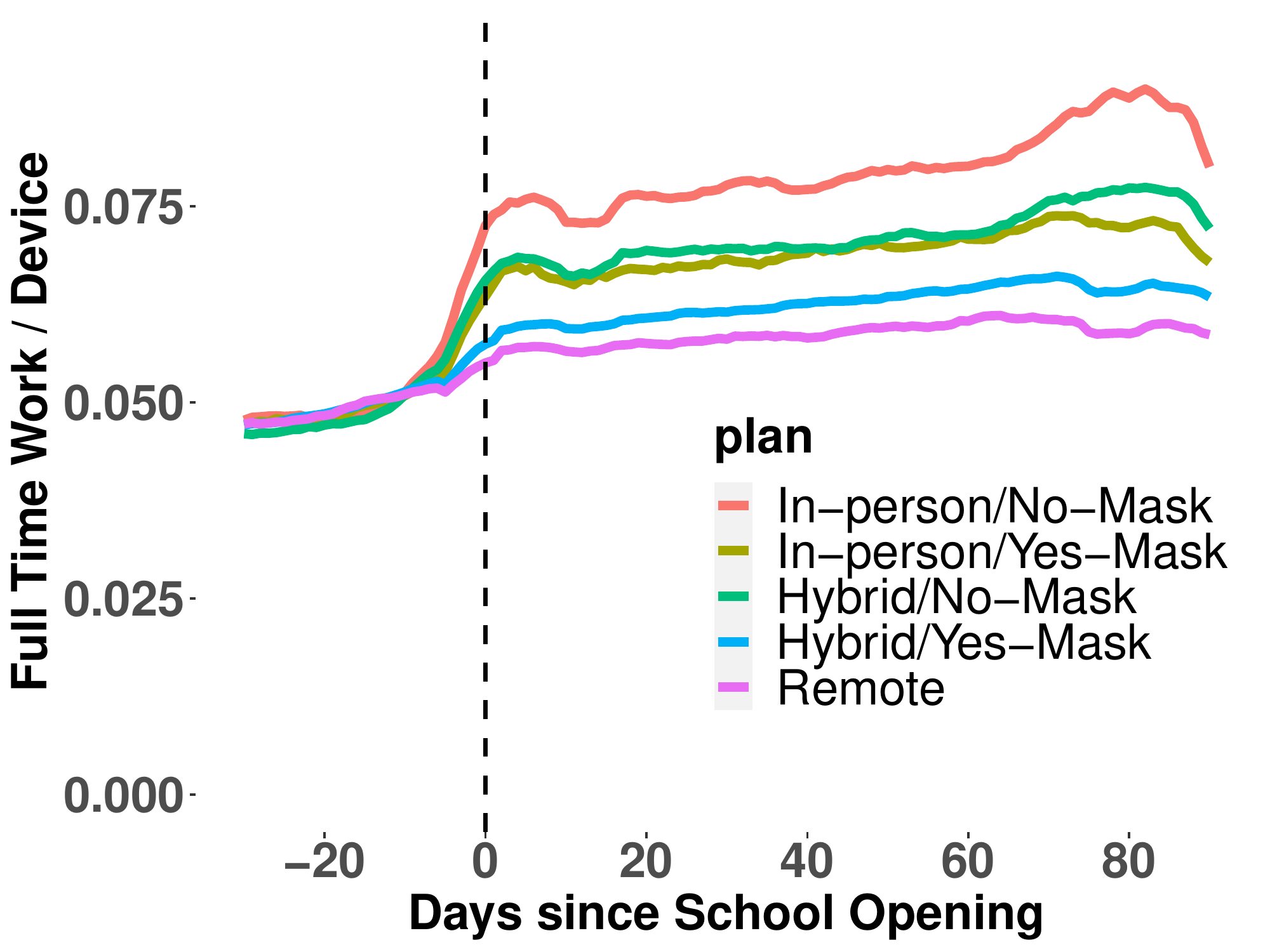}
    \end{tabular}
  \end{minipage}}
\vspace{-0.2cm}  {\scriptsize
\begin{flushleft}
Notes:  (a)-(b) plot the evolution of weekly cases or deaths per 1000 persons averaged across counties within each group of counties classified by K-12 school teaching methods and mitigation strategy of mask requirements against the days since K-12 school opening. We classify counties that implement in-person teaching as their dominant teaching method into ``In-person/Yes-Mask'' and  ``In-person/No-Mask'' based on whether at least one school district requires staff to wear masks or not. Similarly, we classify counties that implement hybrid teaching into ``Hybrid/Yes-Mask'' and  ``Hybrid/No-Mask'' based on whether mask-wearing is required for staff. We classify counties that implement remote teaching as ``Remote.''   (c) and (d) plot the evolution of the 7-day average of per-device visits to K-12 schools and full-time workplaces, respectively, against the days since K-12 school opening using the same classification as (a) and (b).
 \end{flushleft}}
\end{figure}

Table \ref{table:summary} reports the means for the growth rate of weekly confirmed cases and deaths measured by the log-difference over 7 days in reported weekly cases/deaths, where the log of weekly cases and deaths is set to be $-1$ when we observe zero weekly cases and deaths, weekly cases and deaths per 1000, and per-device visits to K-12 schools, workplaces, and restaurants by teaching methods and separately for periods before and after K-12 school openings. Standard errors for the means that are two-way clustered on county and date are reported in parentheses. Here and in the event-study analysis, we classify counties into three groups (in-person, hybrid, and remote) by the dominant teaching method under which the highest proportion of students are learning within a county.

The school opening dates spread from the beginning of August to late September across counties, where
hybrid teaching  is more common than remote or in-person teaching (SI Appendix, Fig. S4(k)).  Reflecting the steady increase in cases from late September to November of 2020 in the U.S. (SI Appendix, Fig. S4(b)),  the growth rates of cases and deaths, as well as the number of weekly confirmed cases and deaths, are higher in the period after the school opening compared to the period before. As shown in Table \ref{table:summary}, this rise in cases and deaths after the school opening is more pronounced in the counties with in-person or hybrid teaching than those with remote teaching.  The K-12 school and workplace visits are also higher after the school openings than before, especially for counties with in-person and hybrid school openings. On the other hand, the mean per-device restaurant visits do not change much before and after the school opening, regardless of teaching methods.

 Fig.  \ref{fig:case-growth} provides visual evidence for the association of opening K-12 schools with the spread of COVID-19 as well as the role of school mitigation strategies. Fig.  \ref{fig:case-growth}(a) and (b) plot the evolution of average weekly cases and deaths per 1000 persons, respectively, against days since school opening for different teaching methods and mask requirements for staff. In Fig.  \ref{fig:case-growth}(a), the average number of weekly cases starts increasing after two weeks of opening schools in-person or hybrid for counties with no mask mandates for staff, possibly suggesting that mask mandates at school reduce the transmissions of SARS-CoV-2. In Fig.  \ref{fig:case-growth}(b), the number of deaths starts rapidly increasing after 3 to 5 weeks of opening schools for counties that adopt in-person/hybrid teaching methods with no mask mandates.  Alternative mitigation strategies of requiring mask-wearing to the student, prohibiting sports activities, and promoting online instruction also appear to help reduce the number of cases after school openings (see SI Appendix, Fig.  S5(i)-(p)).

Fig.  \ref{fig:case-growth}(c) shows that opening K-12 schools in-person or hybrid increases the 7-day averages of per-device visits to K-12 schools more than opening remotely, especially when no mask mandates are in place.  Fig.  \ref{fig:case-growth}(d) and SI Appendix, Fig.  S5(e)-(f) show that visits to full-time and part-time workplaces increase after school openings with in-person teaching, suggesting that the opening of schools allow parents to return to work.\footnote{Although the workplace visits appear to start increasing before school opens in Fig.  \ref{fig:case-growth}(d), this reflects a measurement error about school opening dates as well as the use of the 7-day average visits. SI Appendix, Fig. S6(d)  shows that a sharp increase in the daily visits to workplaces only happens on the day of school openings without any increase before for a subset of counties such that the school opening date is the same across all school districts within a county. } On the other hand, we observe no drastic changes in per-device visits to restaurants, recreational facilities, and churches after school openings (SI Appendix, Fig.  S5(b)-(d)).

\section{Event-Study Analysis}

We further analyze how cases and deaths change over time after school openings by an ``event-study'' analysis \citep[e.g.,][]{Callaway2020,Goodman-Bacon2018,Sun2020}. We divide the sample into three subsamples, where each subsample contains the observation with similar school opening dates. Then,  for each of the subsamples, we run the following regression with weekly dummies of leads and lags for three school opening modes (i.e., in-person, hybrid, and remote) with county fixed effects but without time fixed effects:
\begin{equation}\label{eq:event}
Y_{it} = \sum_{p\in\mathcal{P}}\sum_{w=-8}^{22} \gamma_w^p D_{\tau,it}^p   + \alpha_i +\epsilon_{it},
\end{equation}
where   $Y_{it}$ is the number of  weekly confirmed cases/deaths per 1000, $D_{\tau,it}^p$ takes the value equal to 1 if school has been opened  for $\tau$ weeks (or will be opened after $-\tau$ weeks if $\tau<0$) with teaching method $p\in \mathcal{P}:= \{\text{in-person},\text{hybrid},\text{remote}\}$  in county $i$ at day $t$. The  $\alpha_i$ represents a county-county-specific baseline mean $i$. We consider the event window of 8 weeks before the school opening and the maximum of 22 weeks after the school opening, where the lag windows are different across subsamples given that our sample ends on December 2, 2020.

\begin{figure}[!ht]
  \caption{The event-study regression estimates before and after the school opening\label{fig:event}}\smallskip
\hspace*{-3.5cm}  \resizebox{0.7\columnwidth}{!}{
\begin{minipage}{\linewidth}  \centering
        \begin{tabular}{cc}
\textbf{\Large (a)  Cases by Opening Modes}&\textbf{\Large (b)  Deaths by Opening Modes}\smallskip\\ 
 \includegraphics[width=0.66\textwidth]{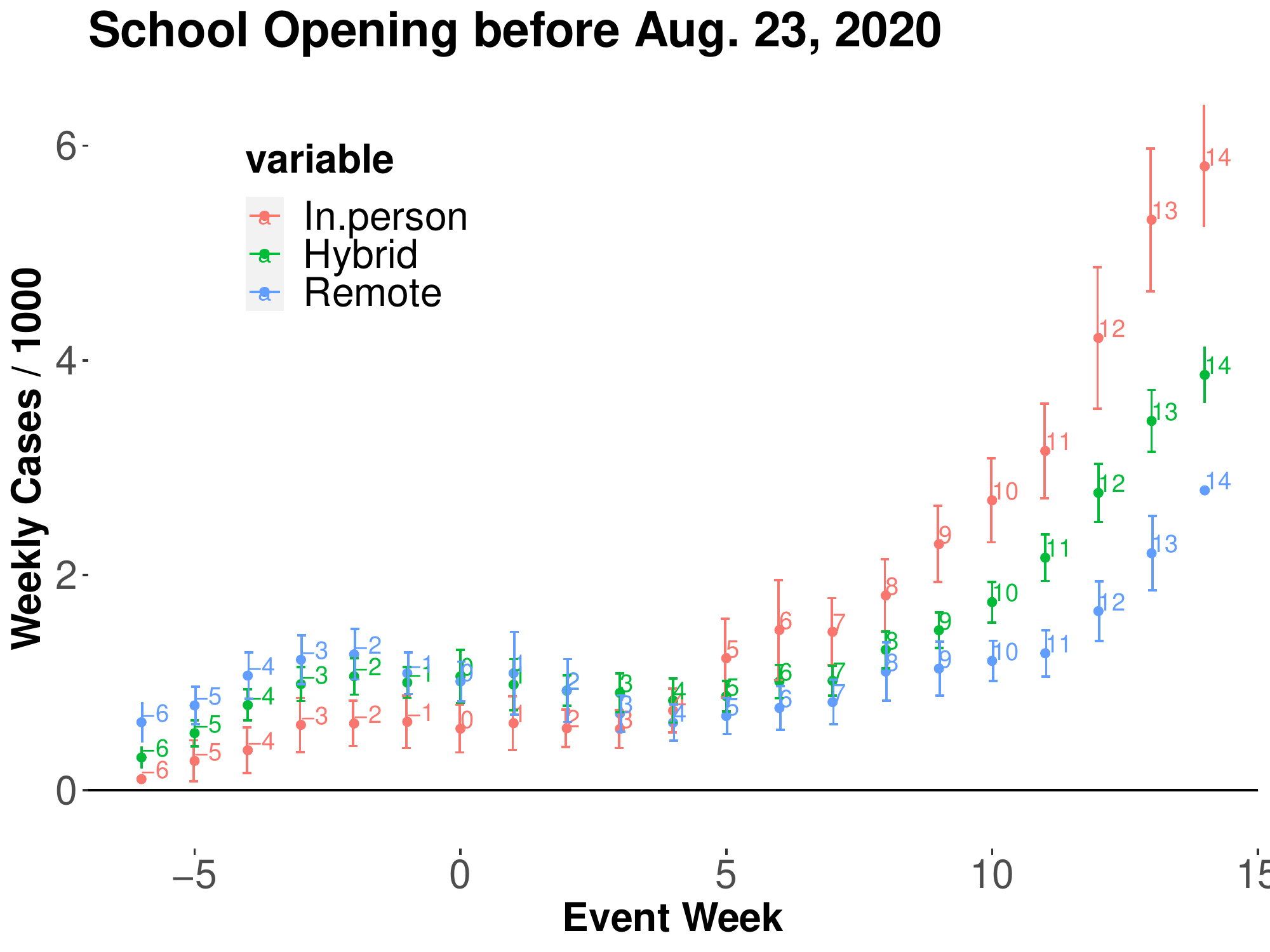}&
 \includegraphics[width=0.66\textwidth]{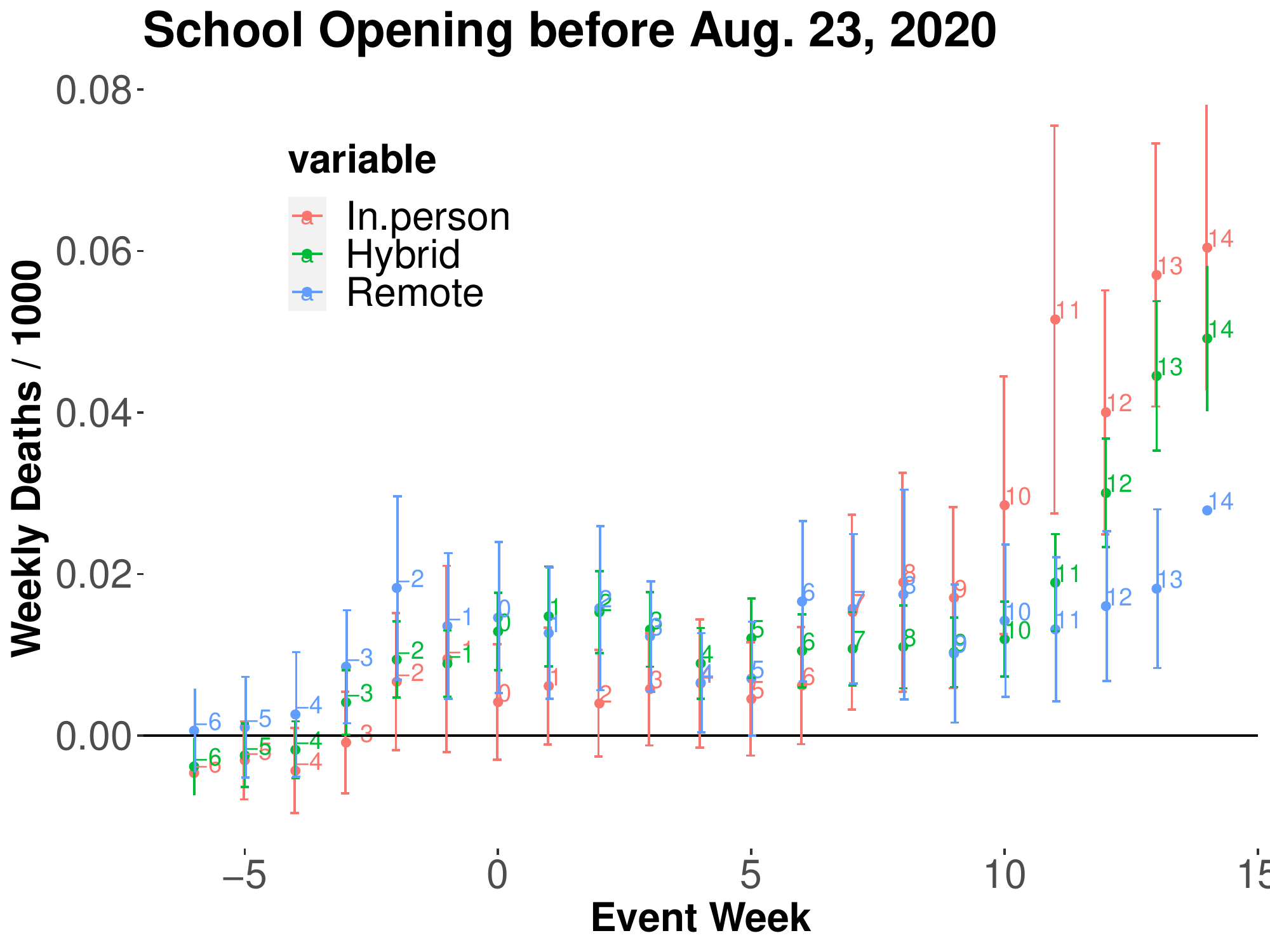}\smallskip\\
  \includegraphics[width=0.66\textwidth]{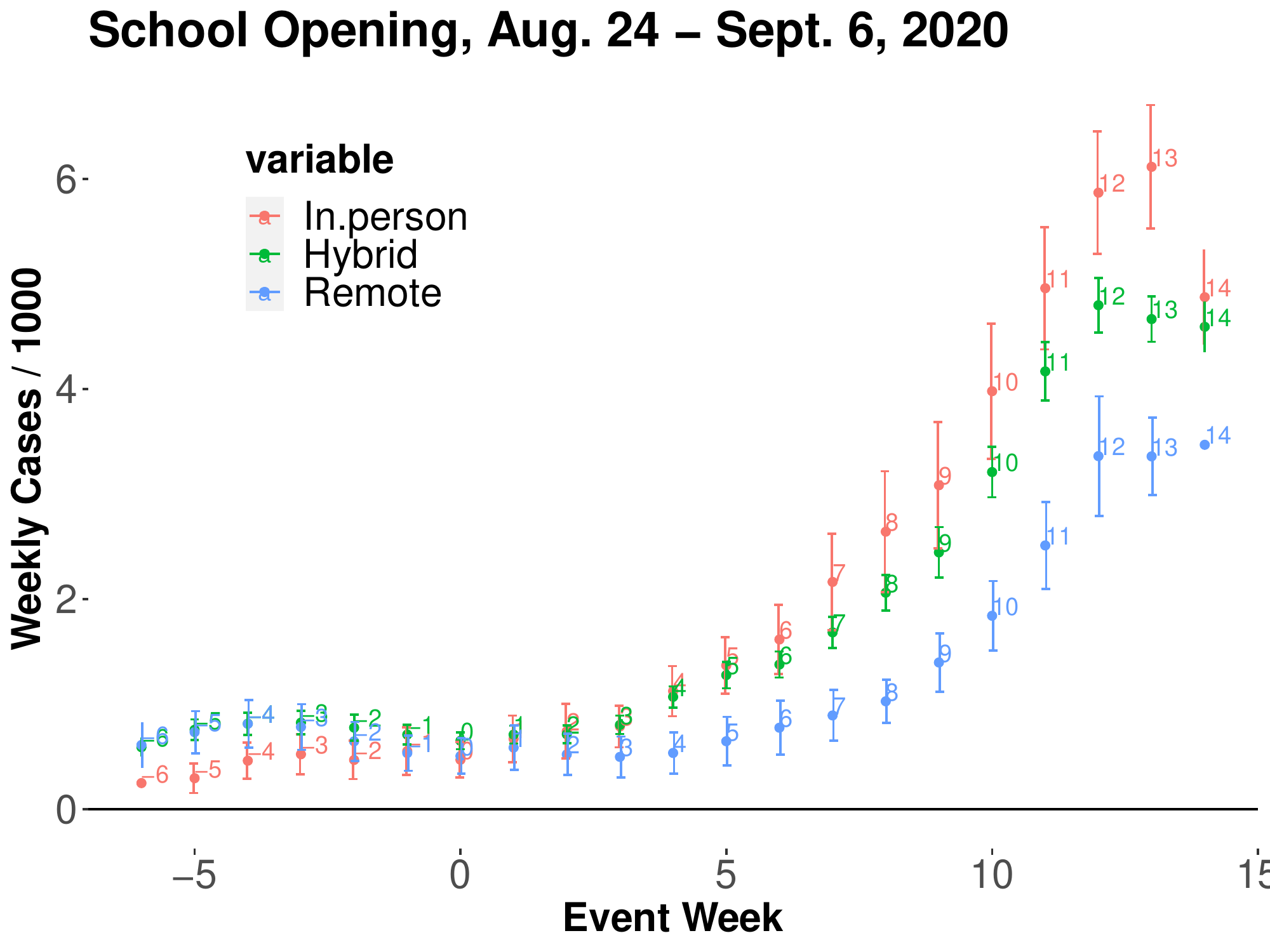}&
  \includegraphics[width=0.66\textwidth]{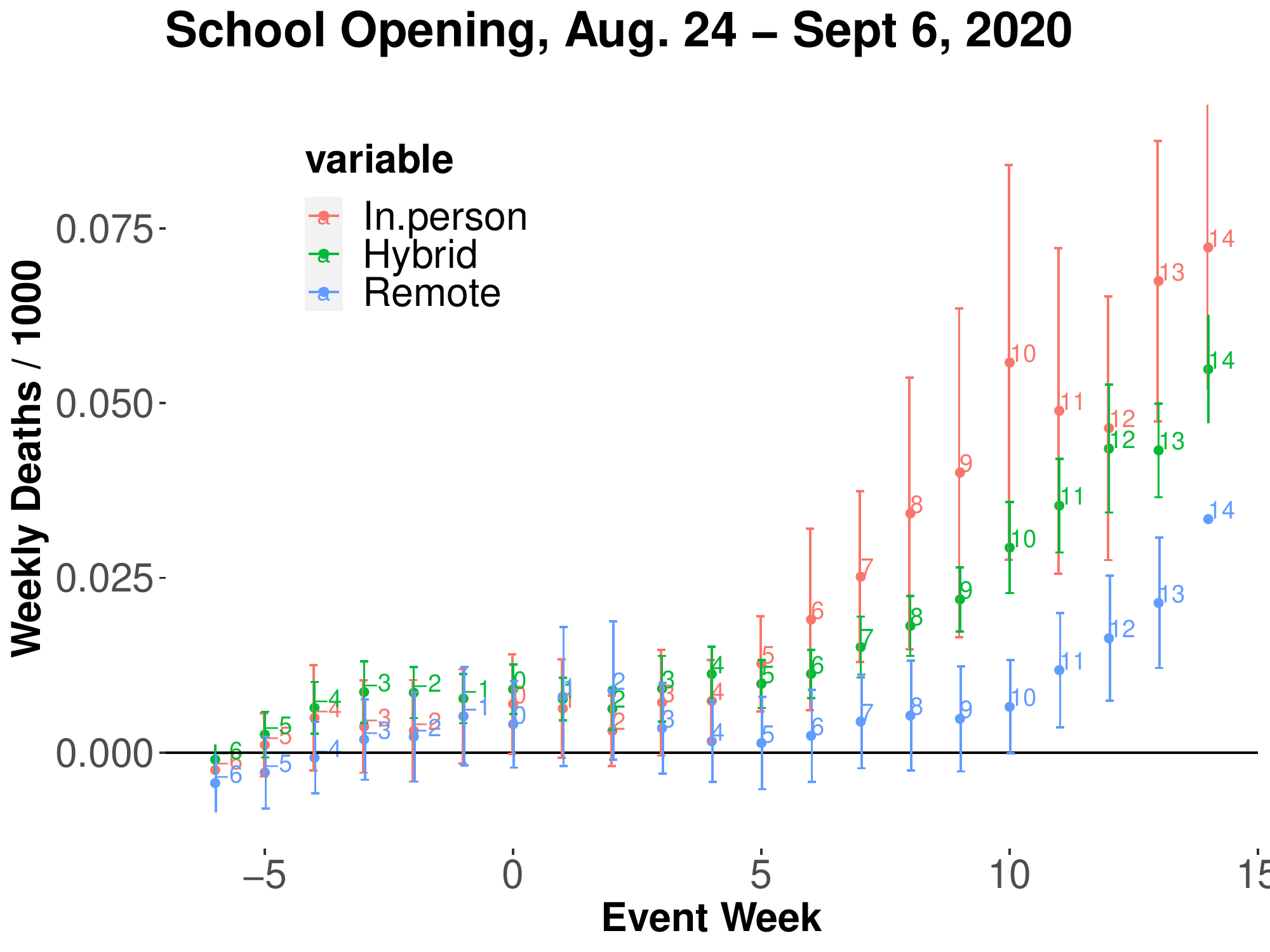}\smallskip\\
 \includegraphics[width=0.66\textwidth]{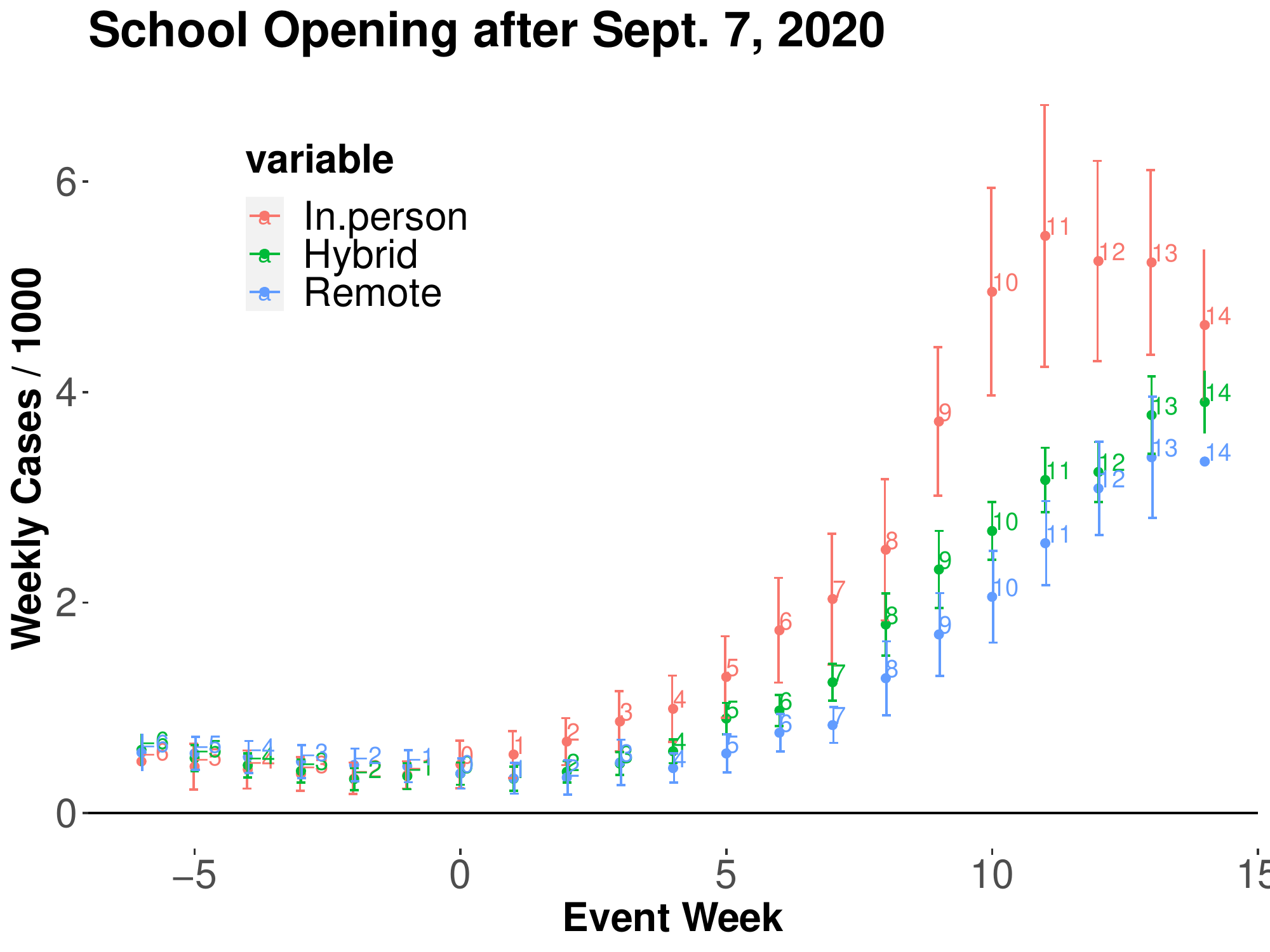}
&\includegraphics[width=0.66\textwidth]{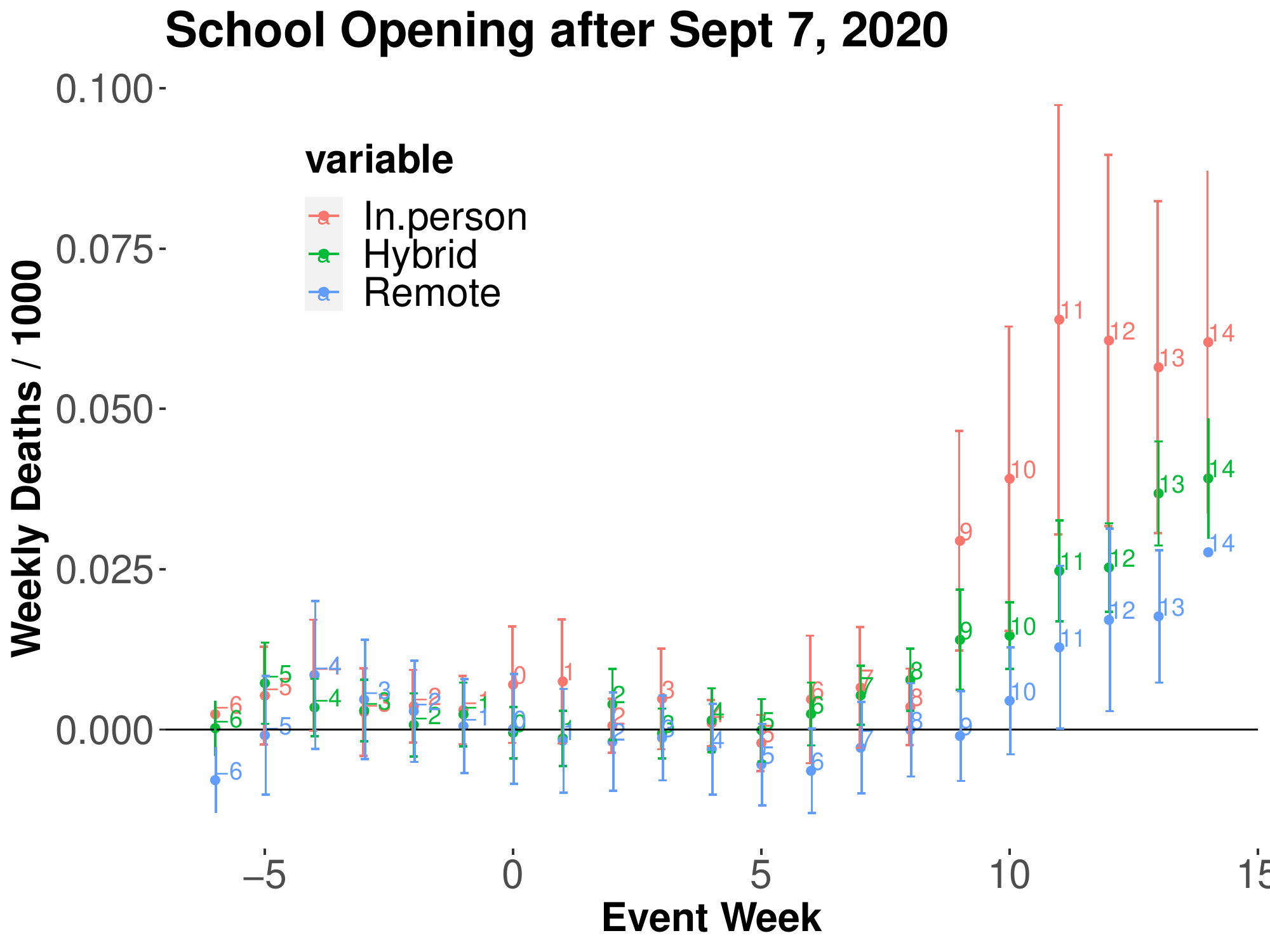}\smallskip\\
        \end{tabular}
  \end{minipage}}
 {\scriptsize
\begin{flushleft}
Notes: The figures plot the estimated coefficients for weekly dummies of leads and lags in regression specification [\ref{eq:event}] with 95 percent confidence intervals for three subsample periods. 
 \end{flushleft}} \vspace{-0.1cm}
\end{figure}

Figure \ref{fig:event} graphs the estimated coefficients over time with 95 percent confidence intervals with standard errors clustered by counties:  the first subsample uses the county observations which opened schools before  Aug 23, 2020, the second consists of the counties that opened schools between Aug 24 and Sept 6, and the third uses the counties with the school opening dates after Sept 7, 2020.   The results illustrate that the gap in weekly cases/deaths per 1000 between remote opening and full/hybrid opening grow overtime after the school opening date for all three subsamples. 

 We also estimate the dynamic treatment  effect of in-person  or hybrid school openings, as well as that of no mask mandates for staffs, relative to remote openings by the estimation method of \cite{Callaway2020} using their \href{https://cran.r-project.org/web/packages/did/vignettes/did-basics.html}{did R package}.
We define a group by a set of counties with the same school opening date and then estimate the group-time specific average treatment effect of in-person or hybrid opening using the``never-treated-units'' and ``not-yet-treated-units'' as the controls while excluding the ``already-treated-units''  from the control group. Here, we take the counties with the remote opening plan as the ``never-treated-units.''\footnote{As discussed in \cite{Goodman-Bacon2018} and \cite{Sun2020}, two-way fixed effects (TWFE) regressions with both time fixed effects and unit fixed effects may lead to a biased estimator for the treatment effect when treatment timings differ across units and, at the same time, the treatment effects are dynamic and evolve over time. The bias arises because the treatment estimate under TWFE regression partly relies on the wrong control units that have already been exposed to treatment under the staggered treatment design.}  Because almost all counties opened their schools by late September, the estimated treatment effect for the lags primarily reflects the treatment effect of in-person or hybrid openings relative to the counties with remote openings rather than the counties that had not opened schools yet. We report the estimates for the average of the group-time specific average treatment effects of in-person or hybrid opening against remote opening across groups with different school opening dates. The result can be causally interpretable under the group-by-group parallel trend assumption \cite{Callaway2020}.

\begin{figure}[!ht]
  \caption{The average treatment estimates  obtained using the DID method from Callaway and Santana (2020) \label{fig:ca}}  \vspace{-0.1cm}
\hspace{-2cm}\resizebox{0.9\columnwidth}{!}{
\begin{minipage}{\linewidth}\medskip
\centering
        \begin{tabular}{cccc}
 {\tiny (a) Cases: In-person   }& {\tiny (b) Cases:  Hybrid   }& {\tiny (c) Cases: In-person/No-Mask  }& {\tiny (d) Cases:  Hybrid/No-Mask } \\
 \includegraphics[width=0.25\textwidth]{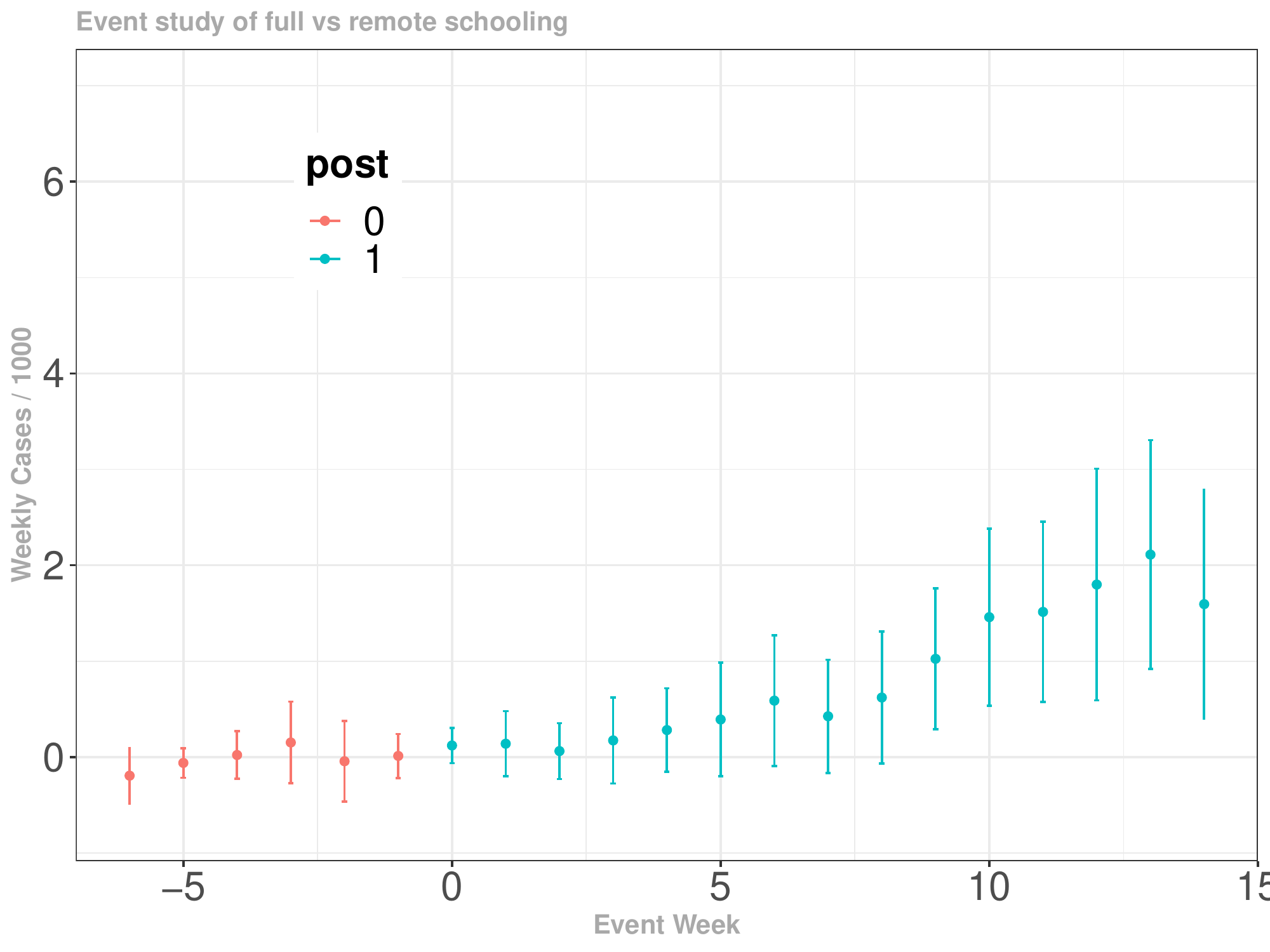}& \includegraphics[width=0.25\textwidth]{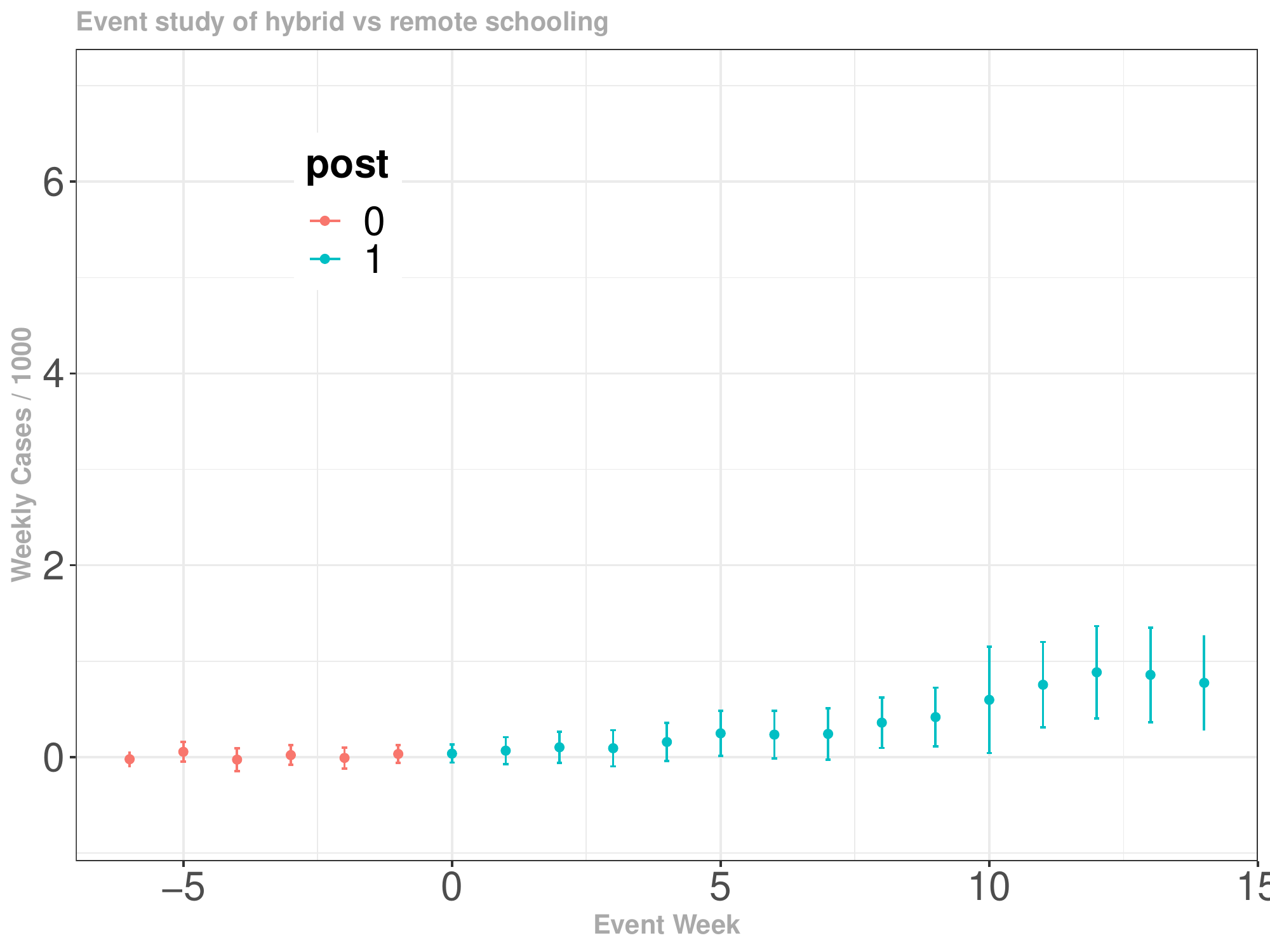} & \includegraphics[width=0.25\textwidth]{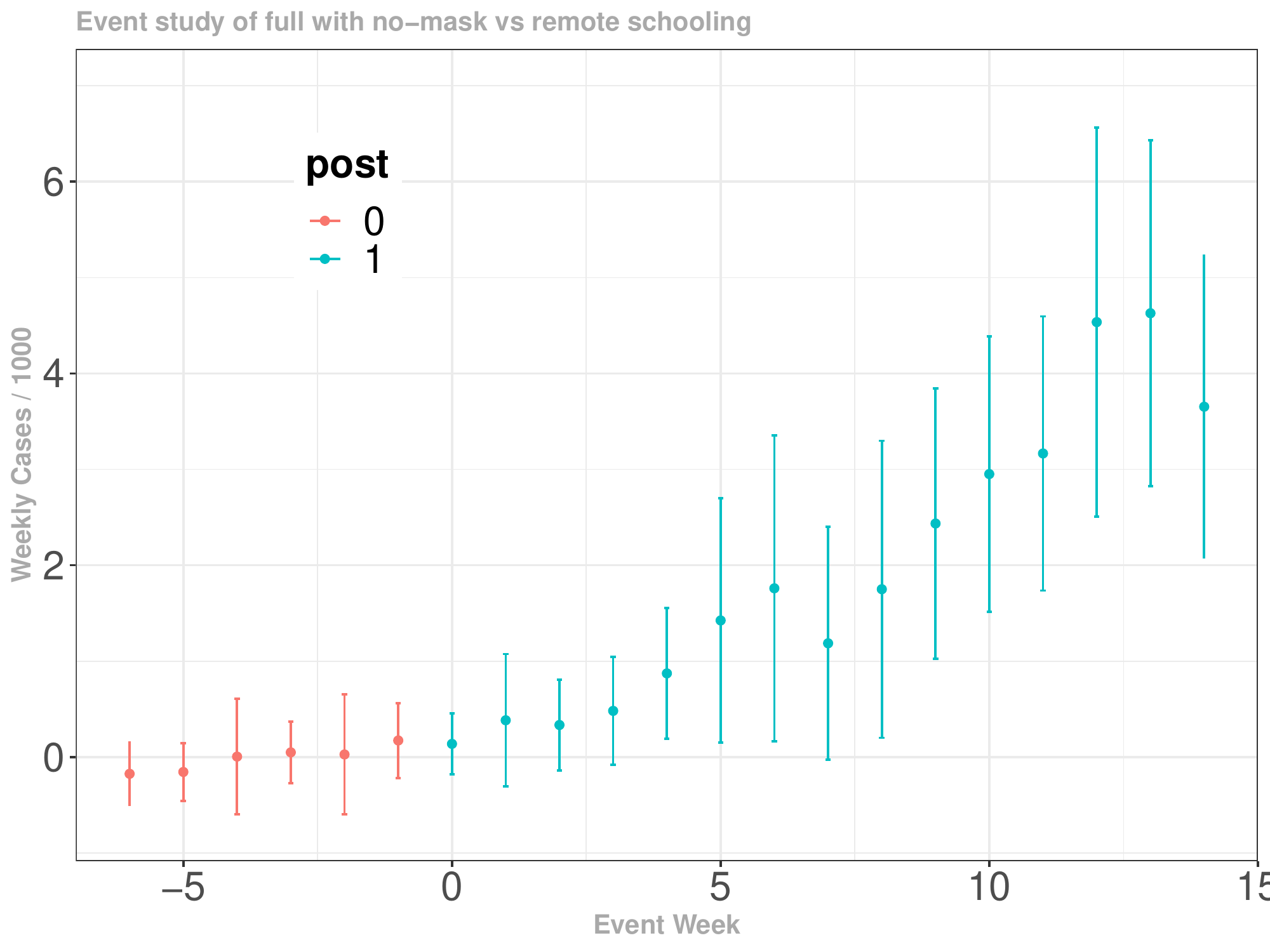}& \includegraphics[width=0.25\textwidth]{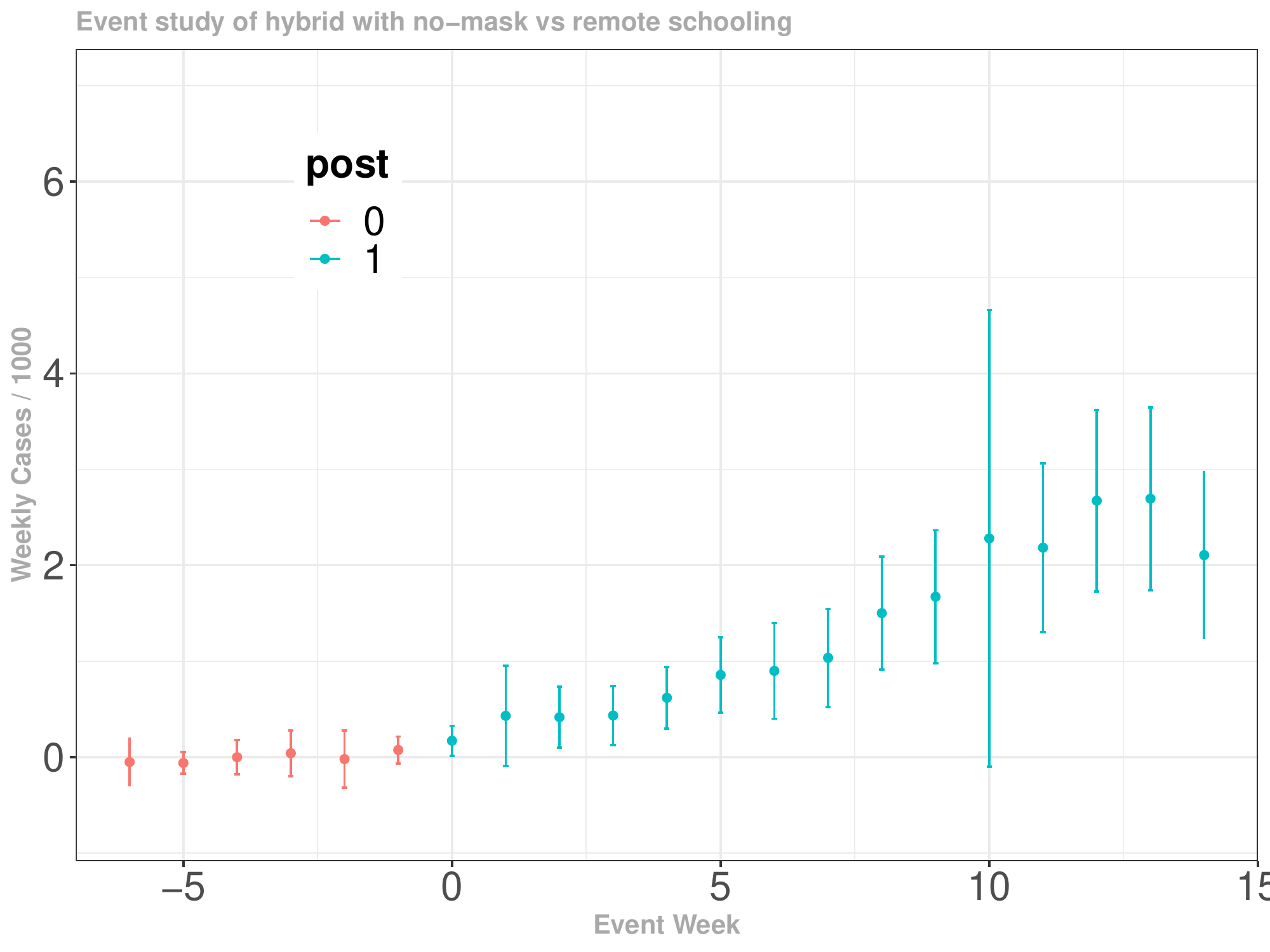}\smallskip\\
 {\tiny (e) Deaths: In-person   }& {\tiny (f) Deaths:  Hybrid  }& {\tiny (g) Deaths: In-person/No-Mask  }& {\tiny (h) Deaths:  Hybrid/No-Mask   } \\

 \includegraphics[width=0.25\textwidth]{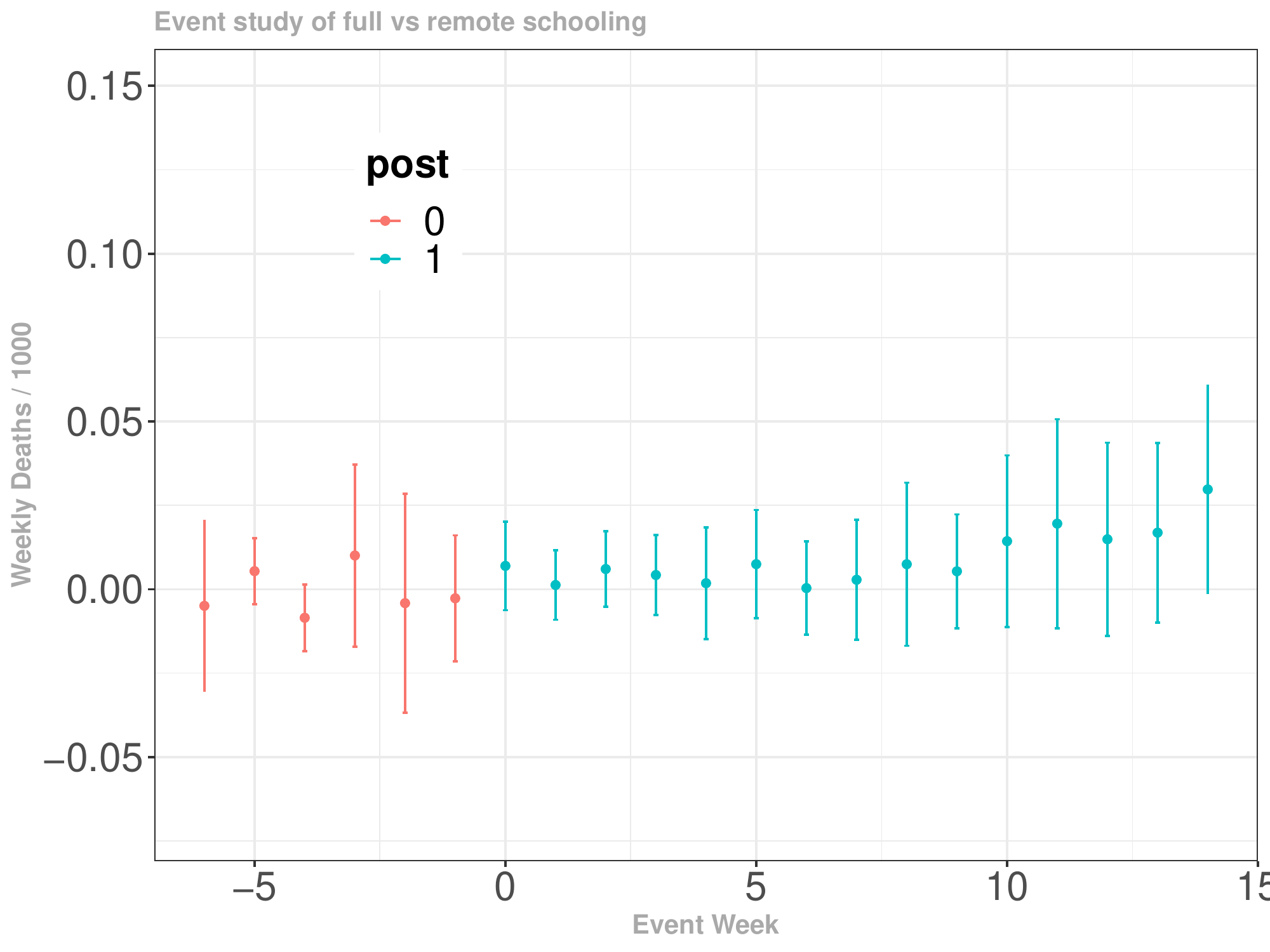}& \includegraphics[width=0.25\textwidth]{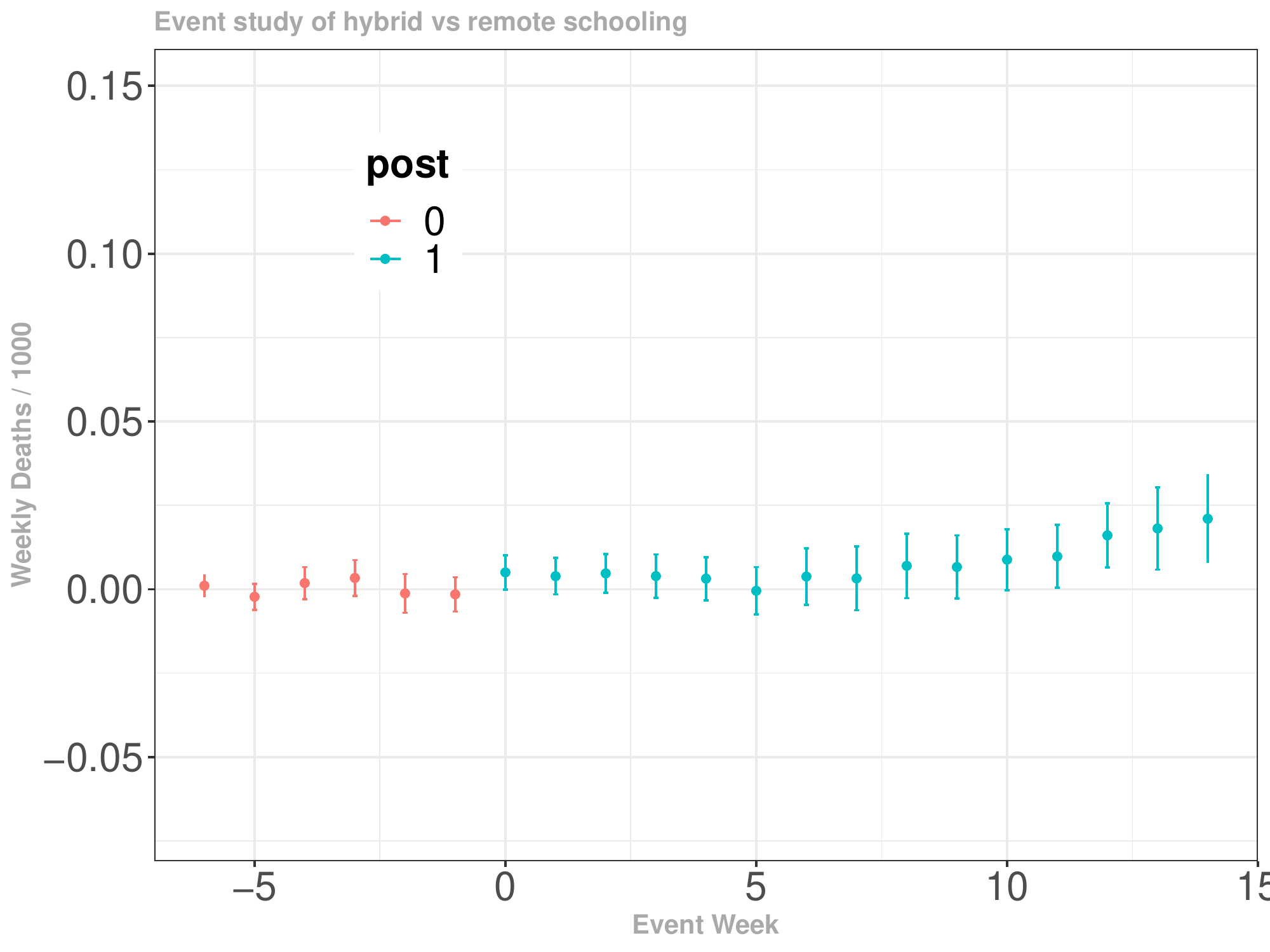}  &
  \includegraphics[width=0.25\textwidth]{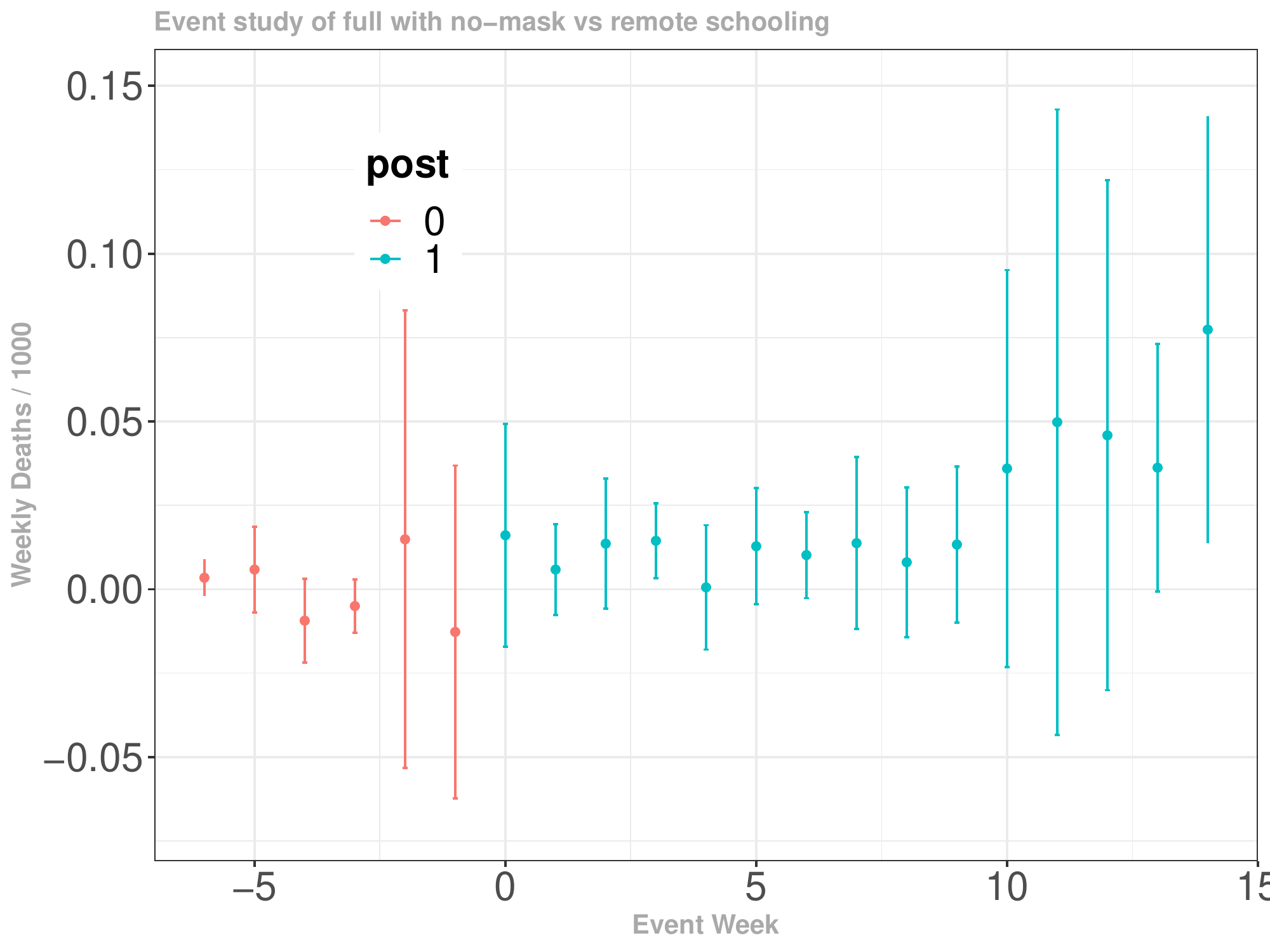}& \includegraphics[width=0.25\textwidth]{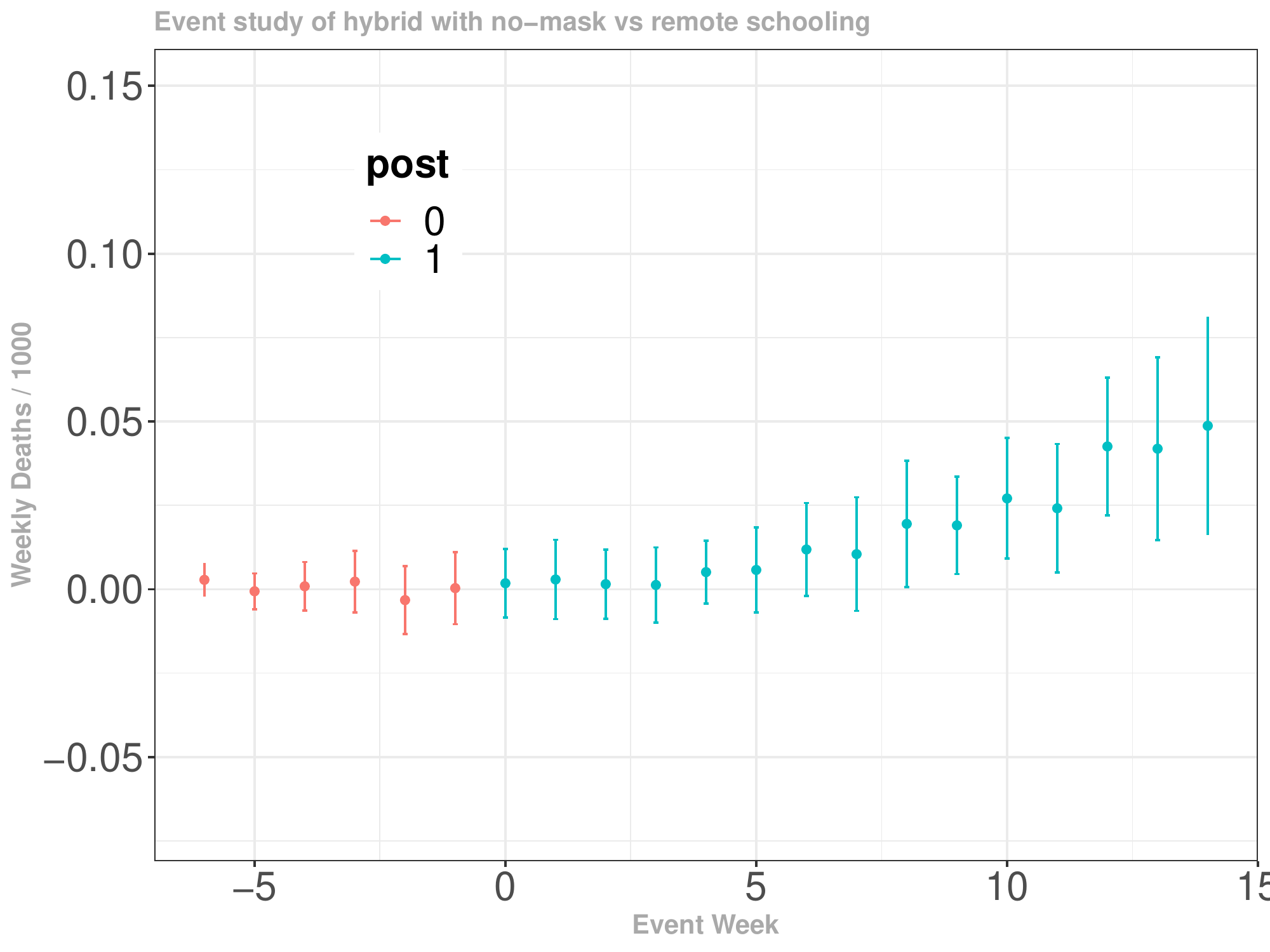} \smallskip\\
 {\tiny (i) log(Cases): In-person  }& {\tiny (j) log(Cases):  Hybrid  }& {\tiny (k) log(Cases): In-person/No-Mask }& {\tiny (l) log(Cases):  Hybrid/No-Mask  } \\

 \includegraphics[width=0.25\textwidth]{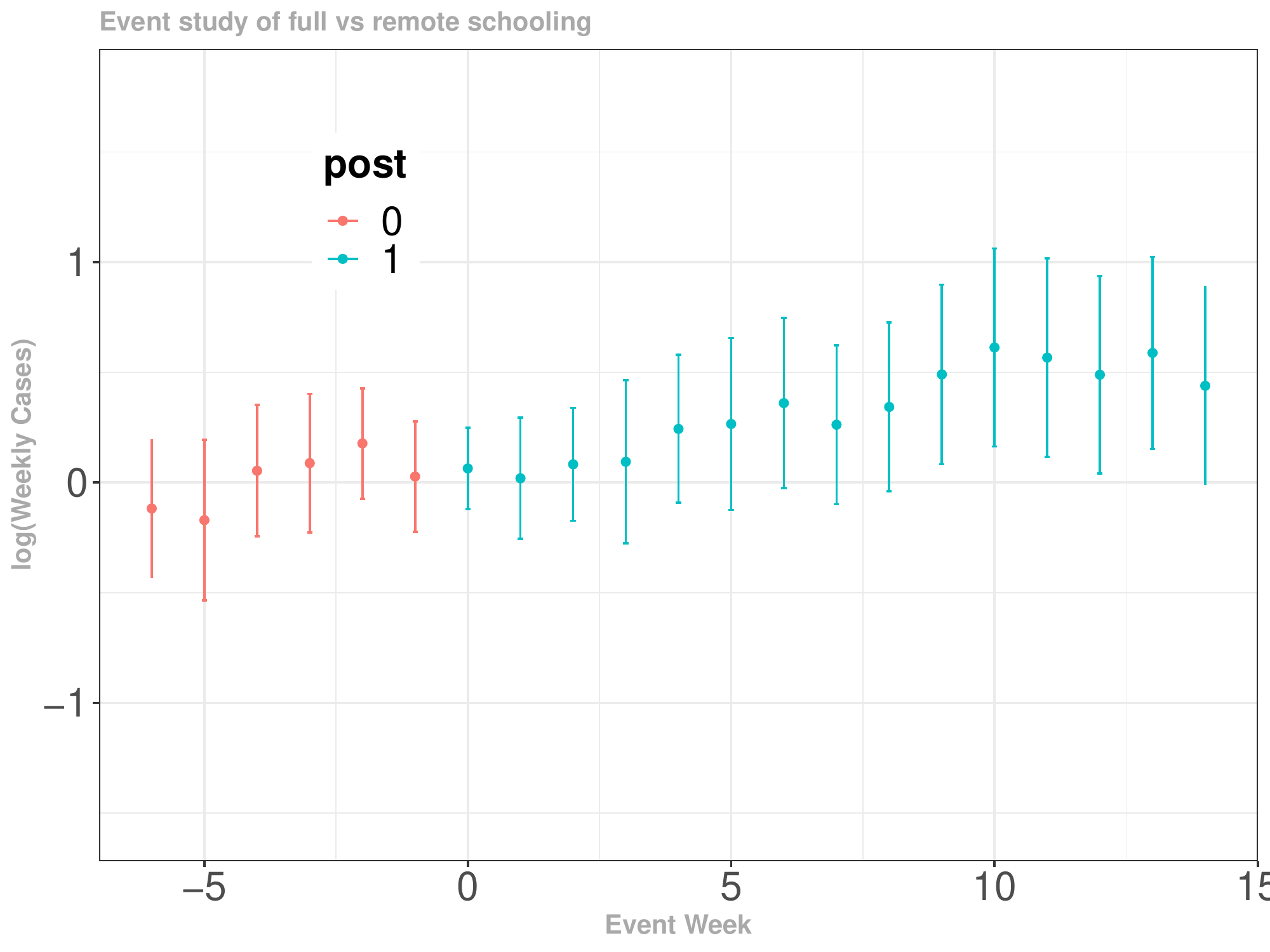}& \includegraphics[width=0.25\textwidth]{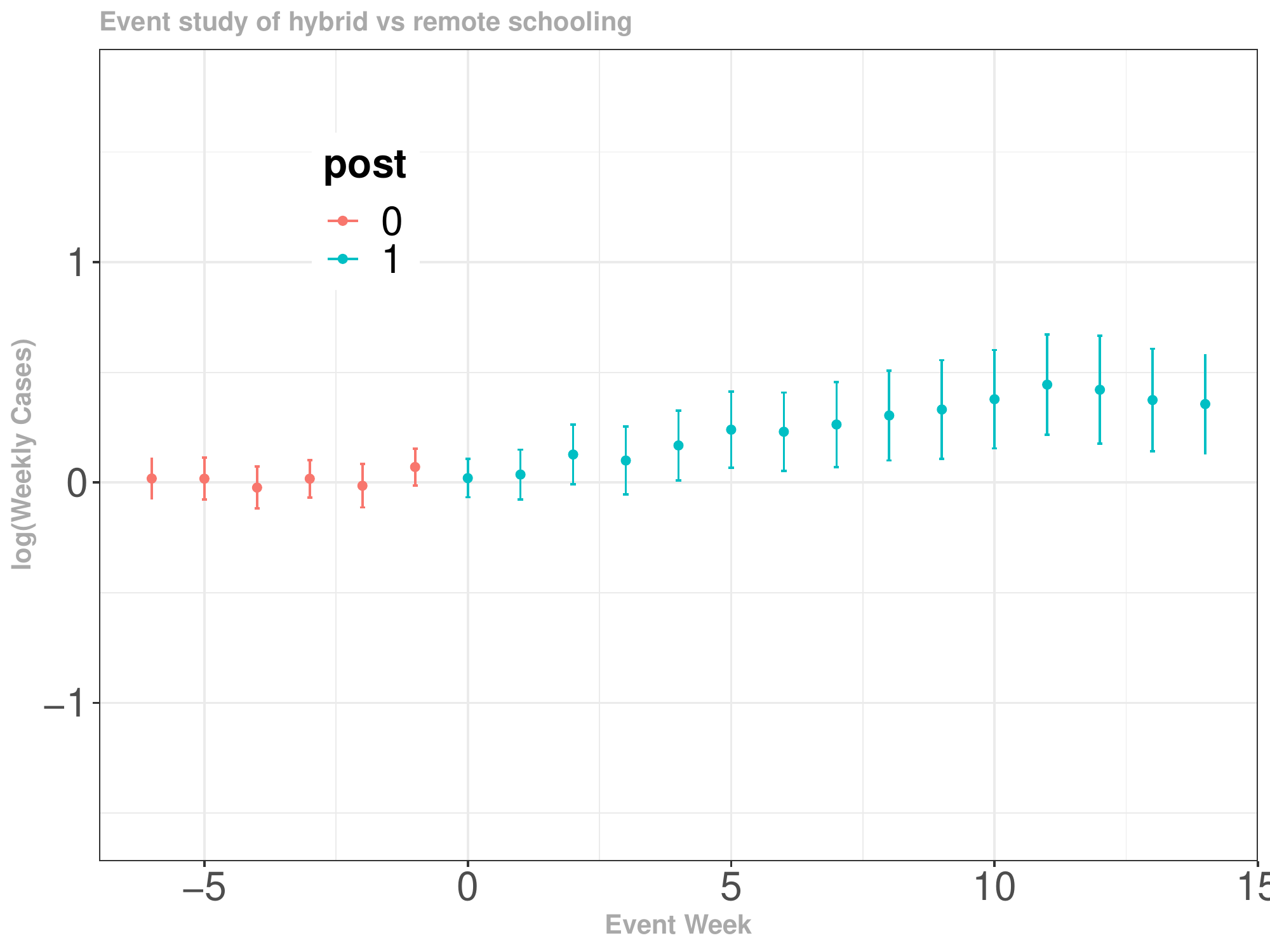} & \includegraphics[width=0.25\textwidth]{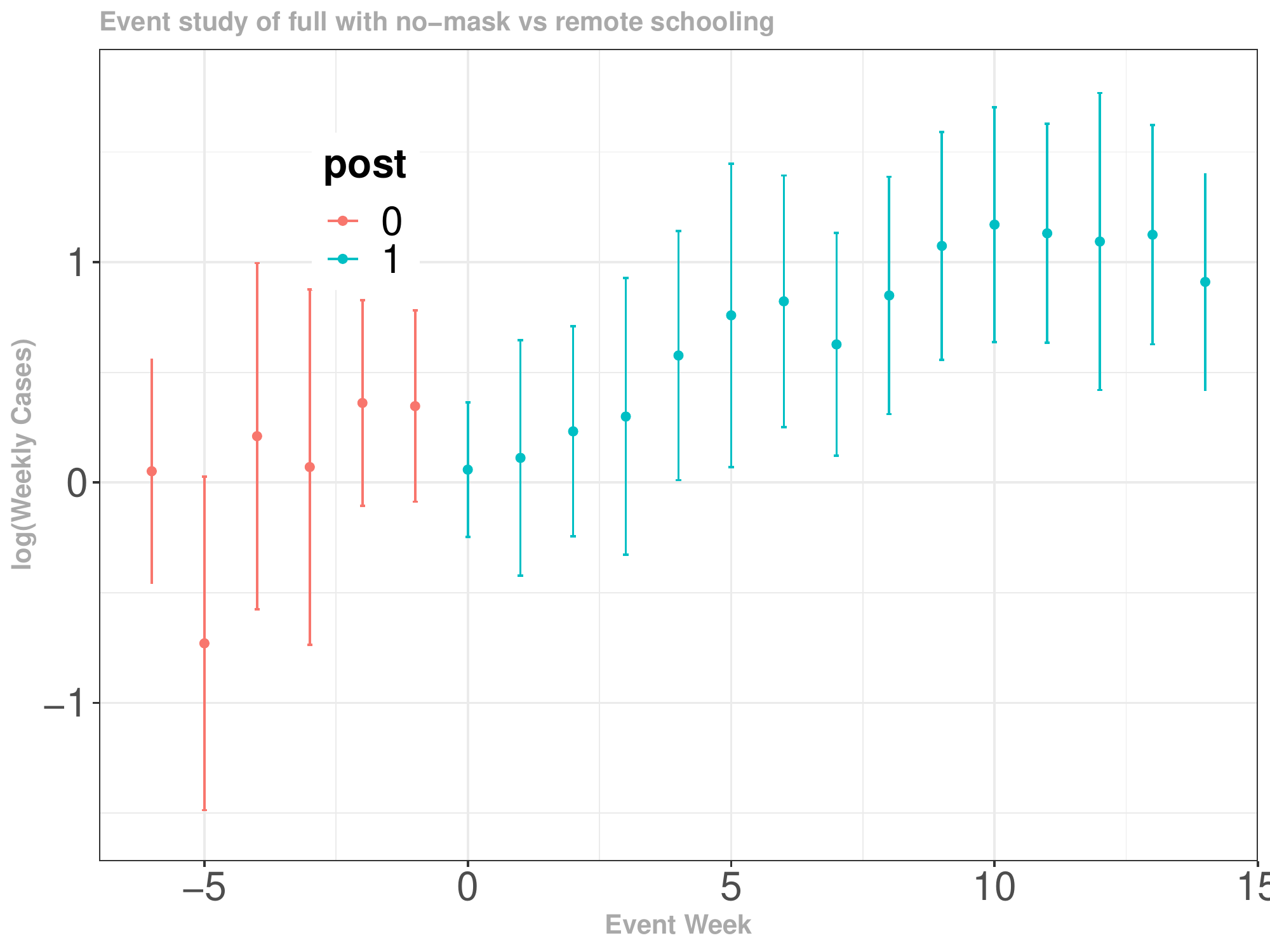}& \includegraphics[width=0.25\textwidth]{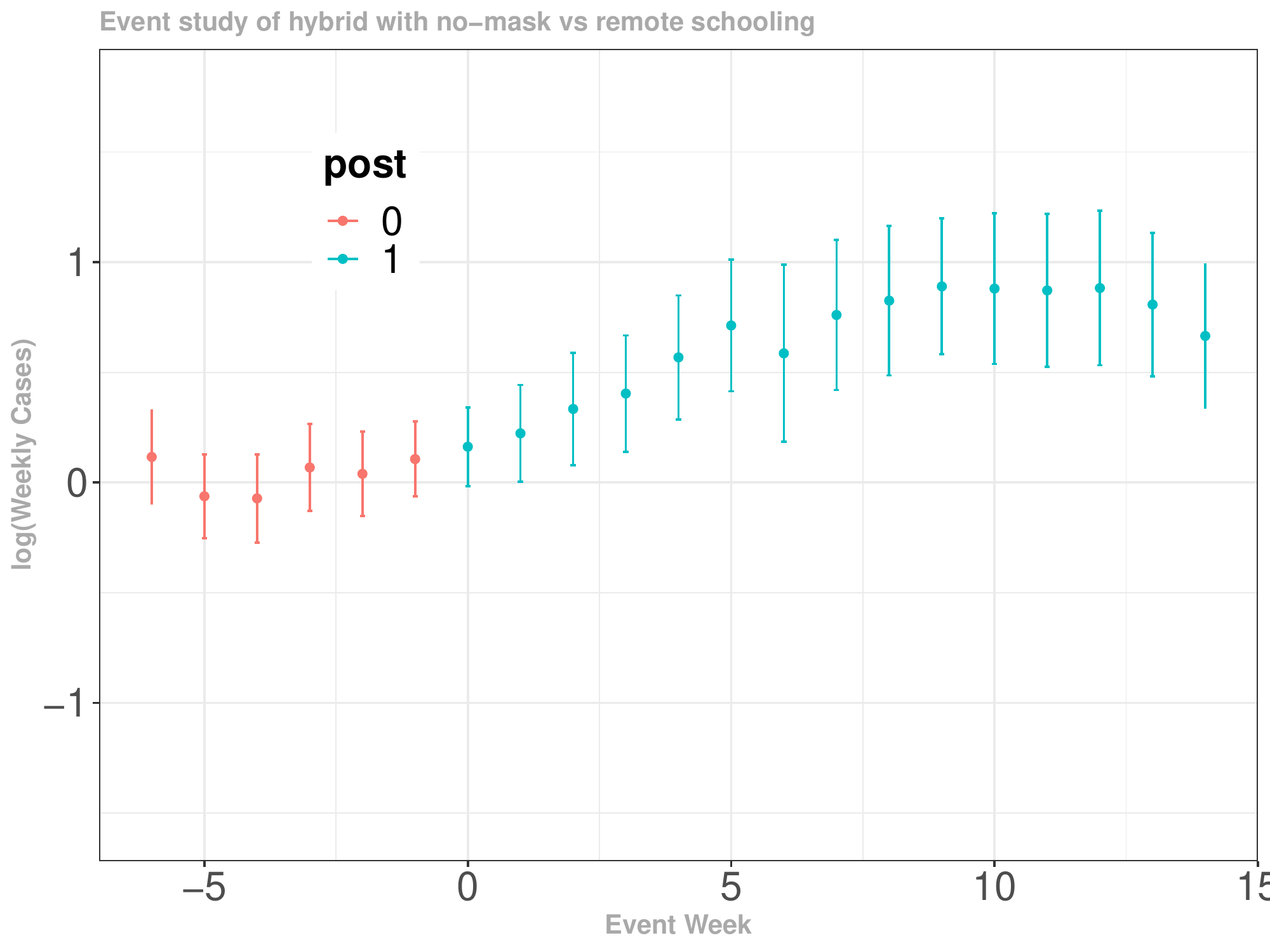}\smallskip\\
 {\tiny (m) log(Deaths): In-person  }& {\tiny (n) log(Deaths):  Hybrid  }& {\tiny (o) log(Deaths): In-person/No-Mask }& {\tiny (p) log(Deaths):  Hybrid/No-Mask  } \\

 \includegraphics[width=0.25\textwidth]{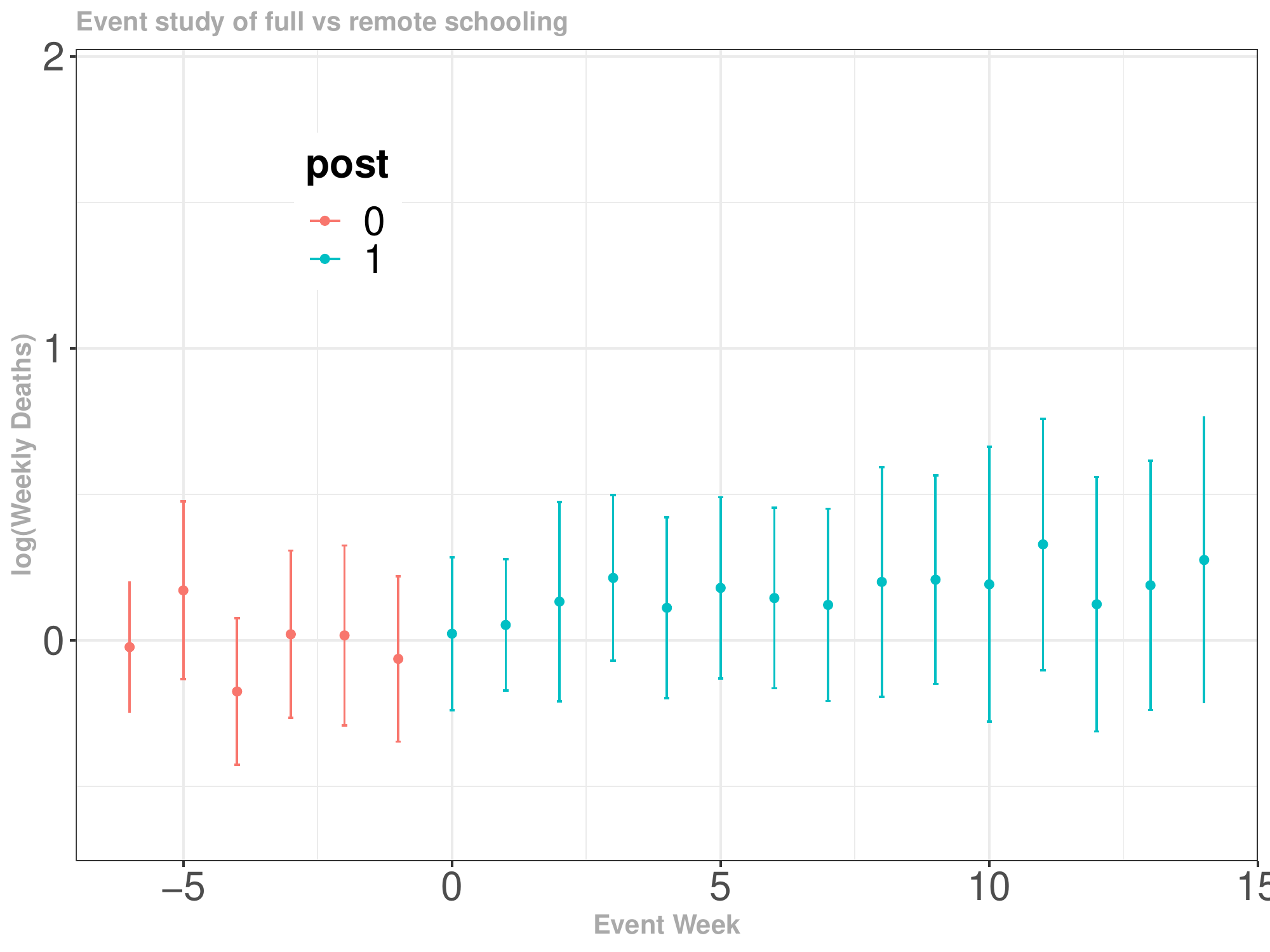}& \includegraphics[width=0.25\textwidth]{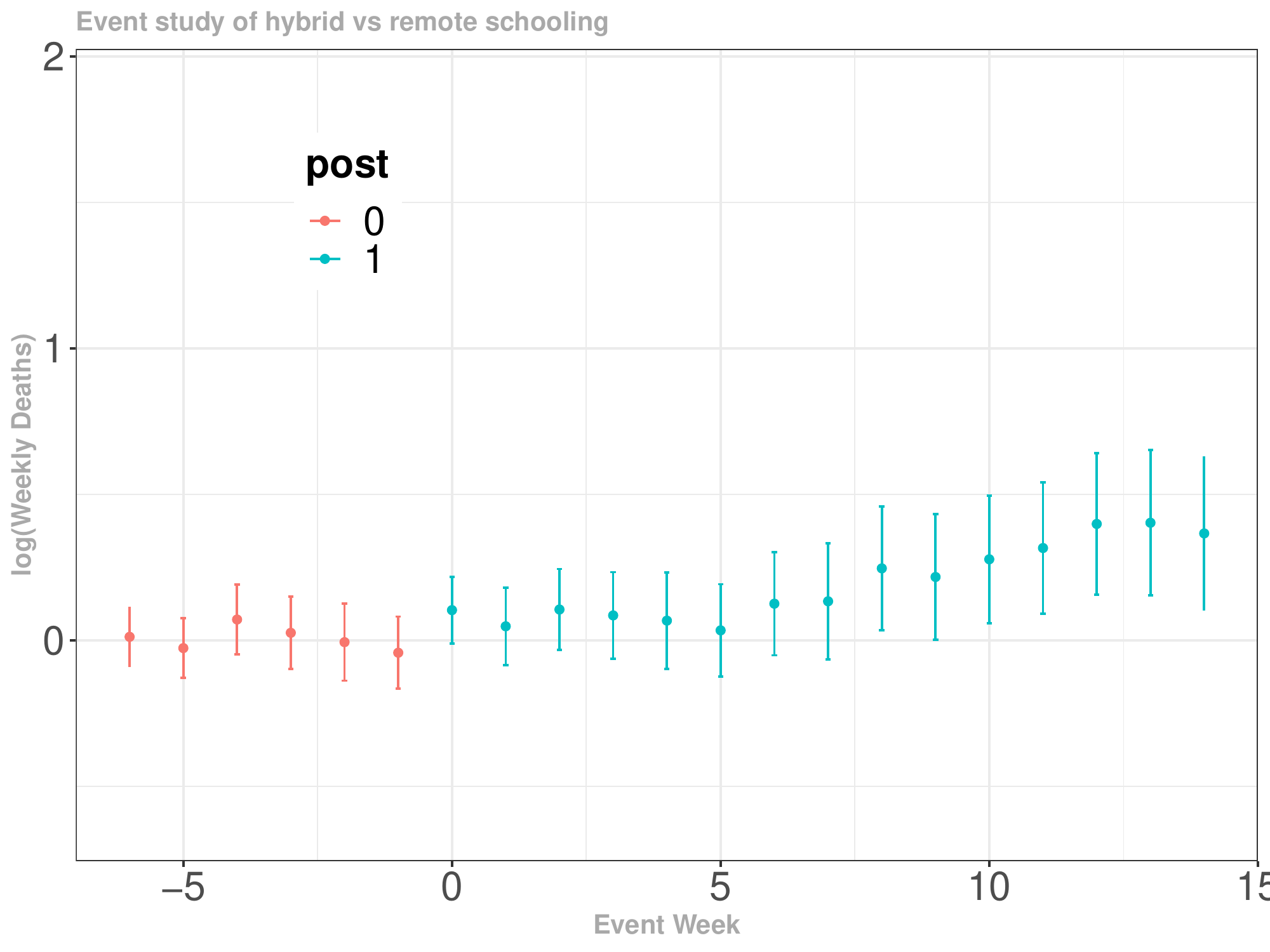} & \includegraphics[width=0.25\textwidth]{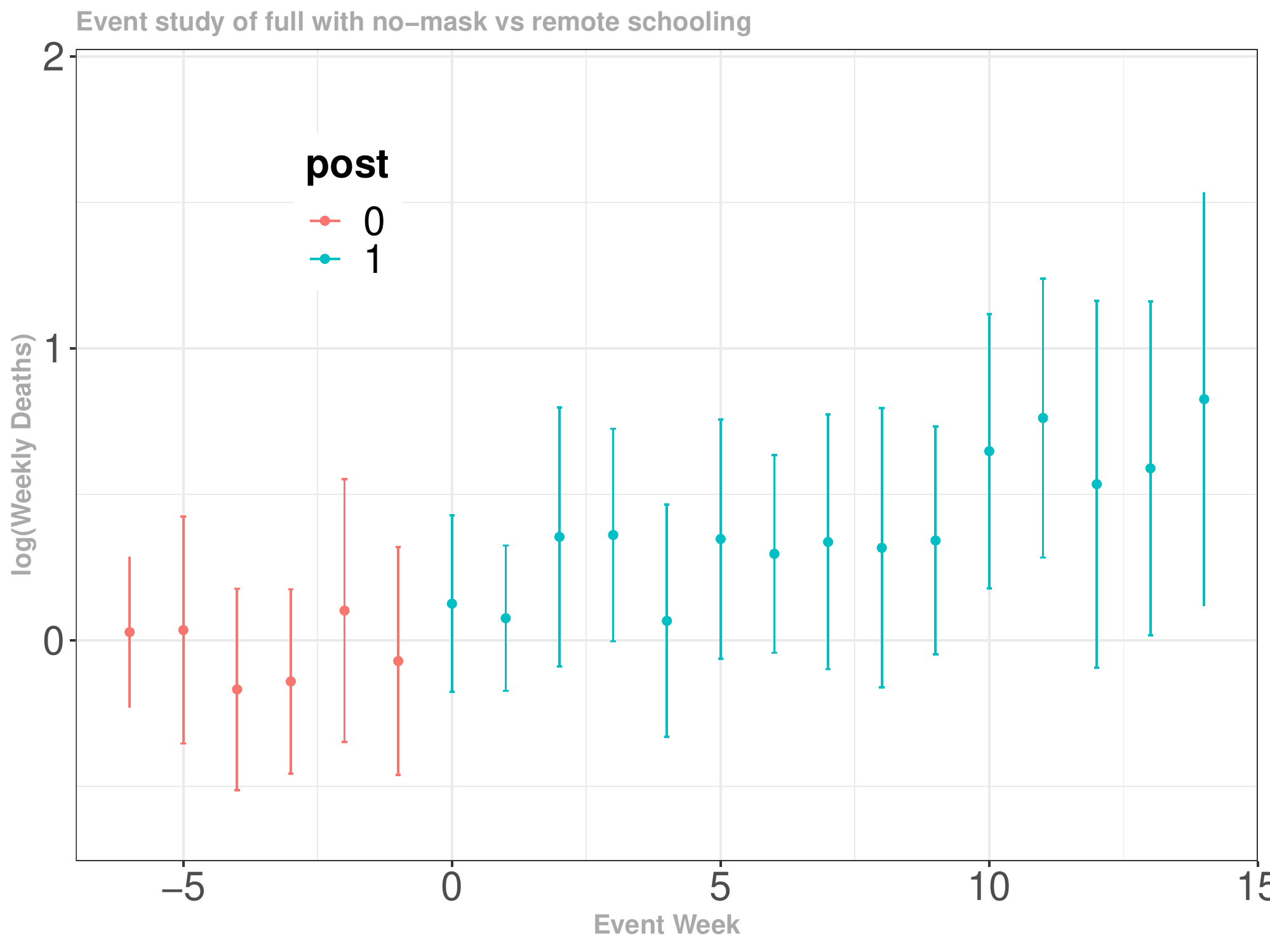}& \includegraphics[width=0.25\textwidth]{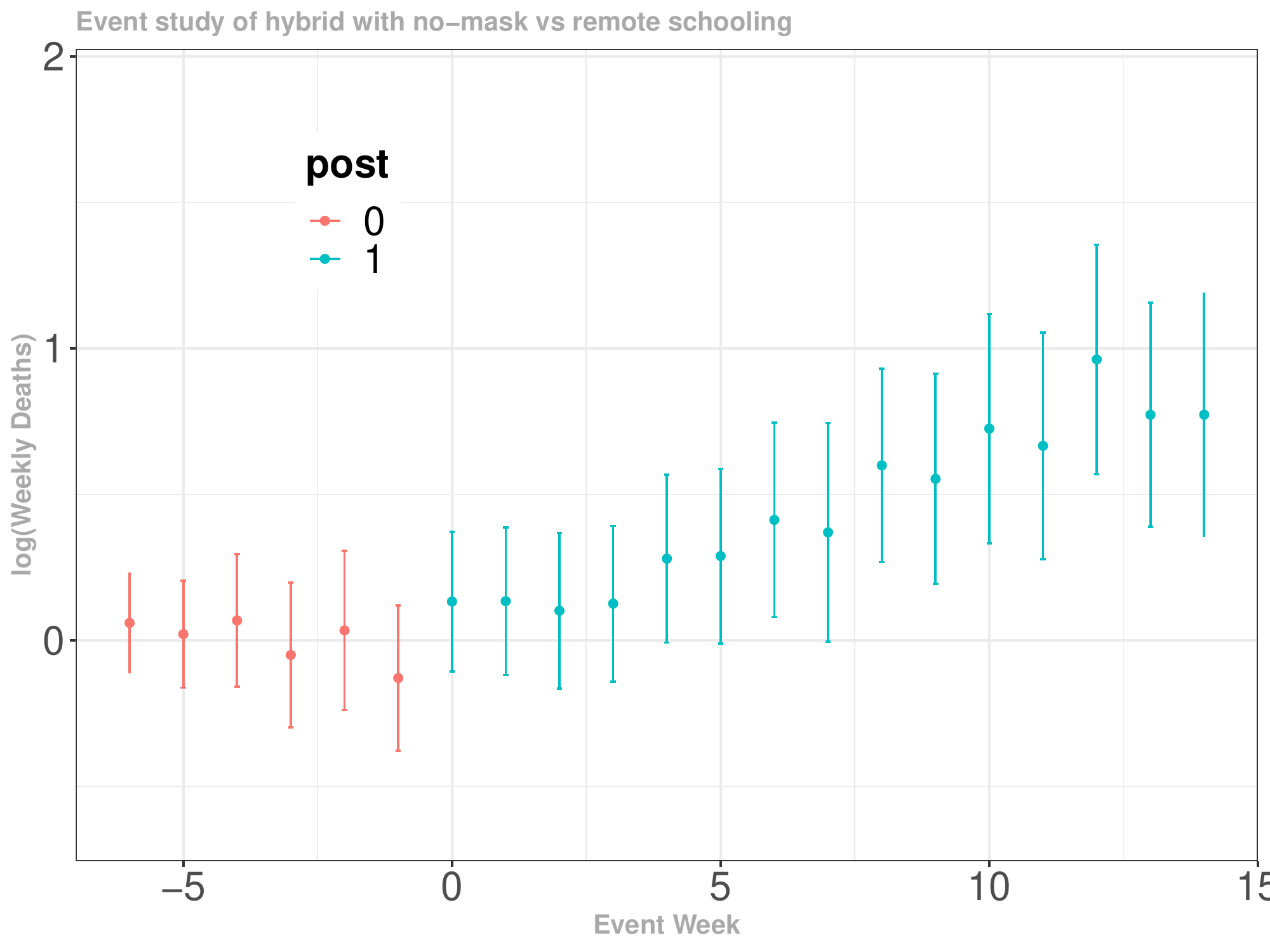}\smallskip\\
 {\tiny (q) School: In-person  }& {\tiny (r) School:  Hybrid  }& {\tiny (s) School: In-person/No-Mask }& {\tiny (t) School:  Hybrid/No-Mask  } \\

 \includegraphics[width=0.25\textwidth]{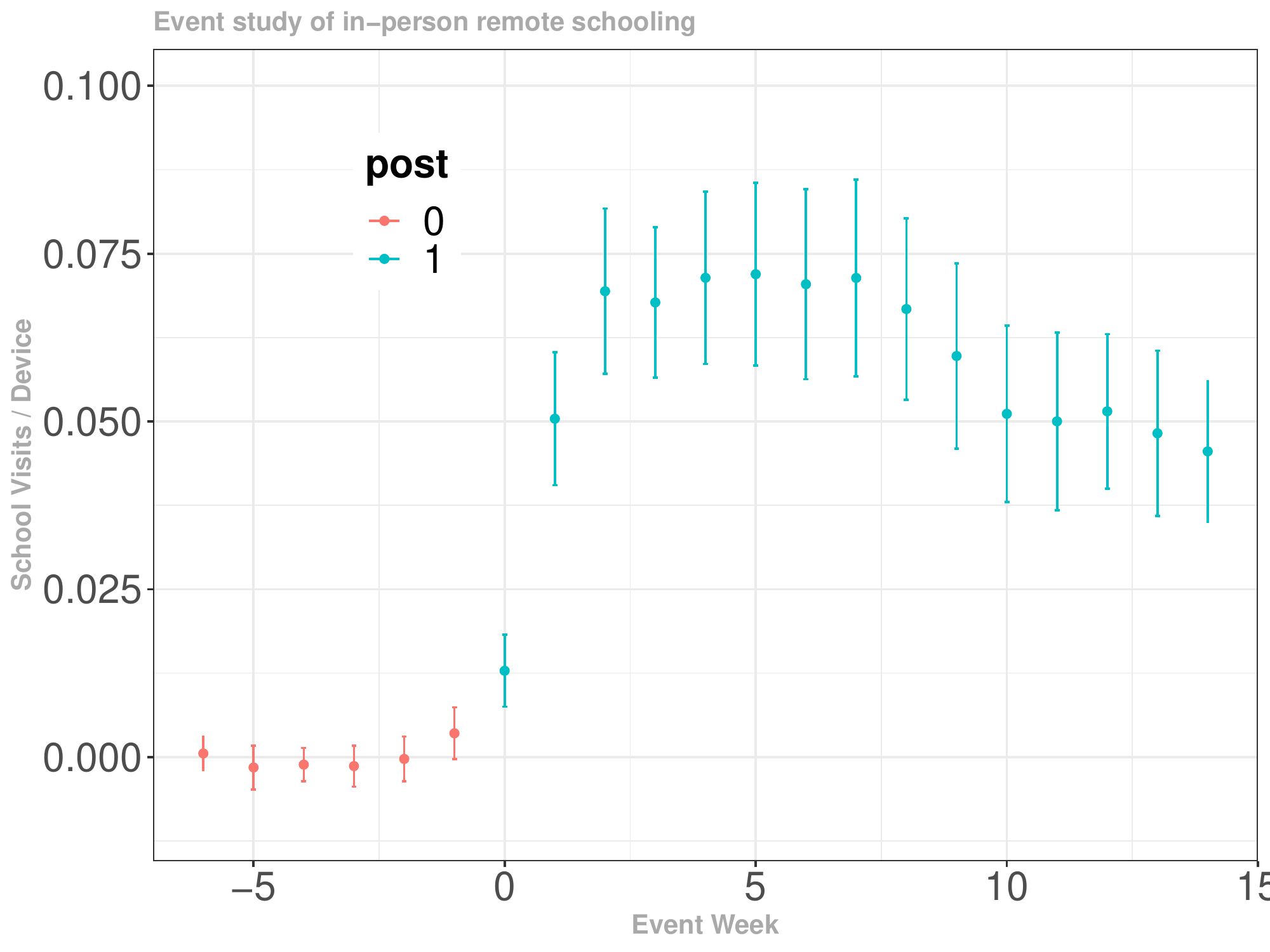}& \includegraphics[width=0.25\textwidth]{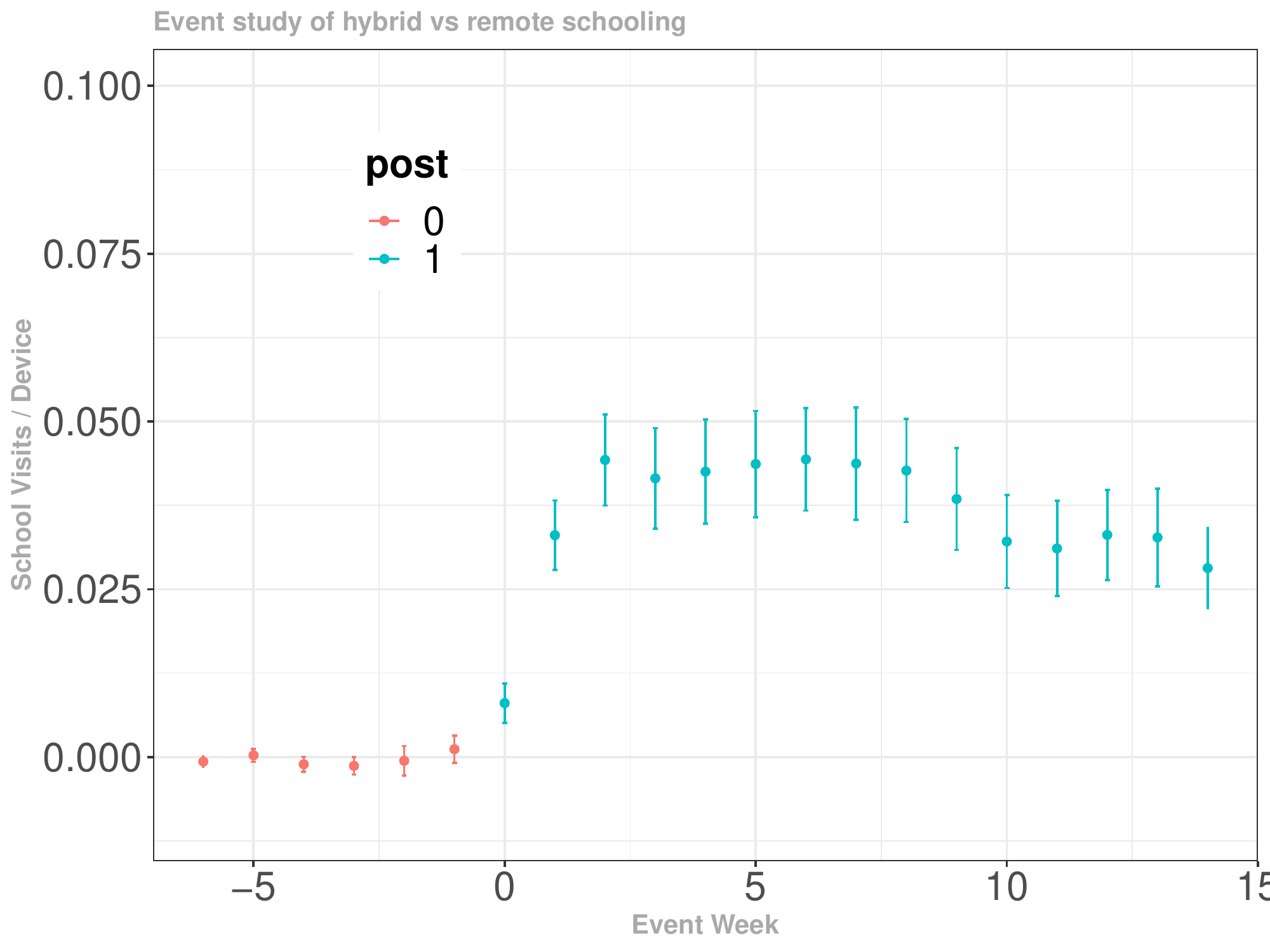}
 &
 \includegraphics[width=0.25\textwidth]{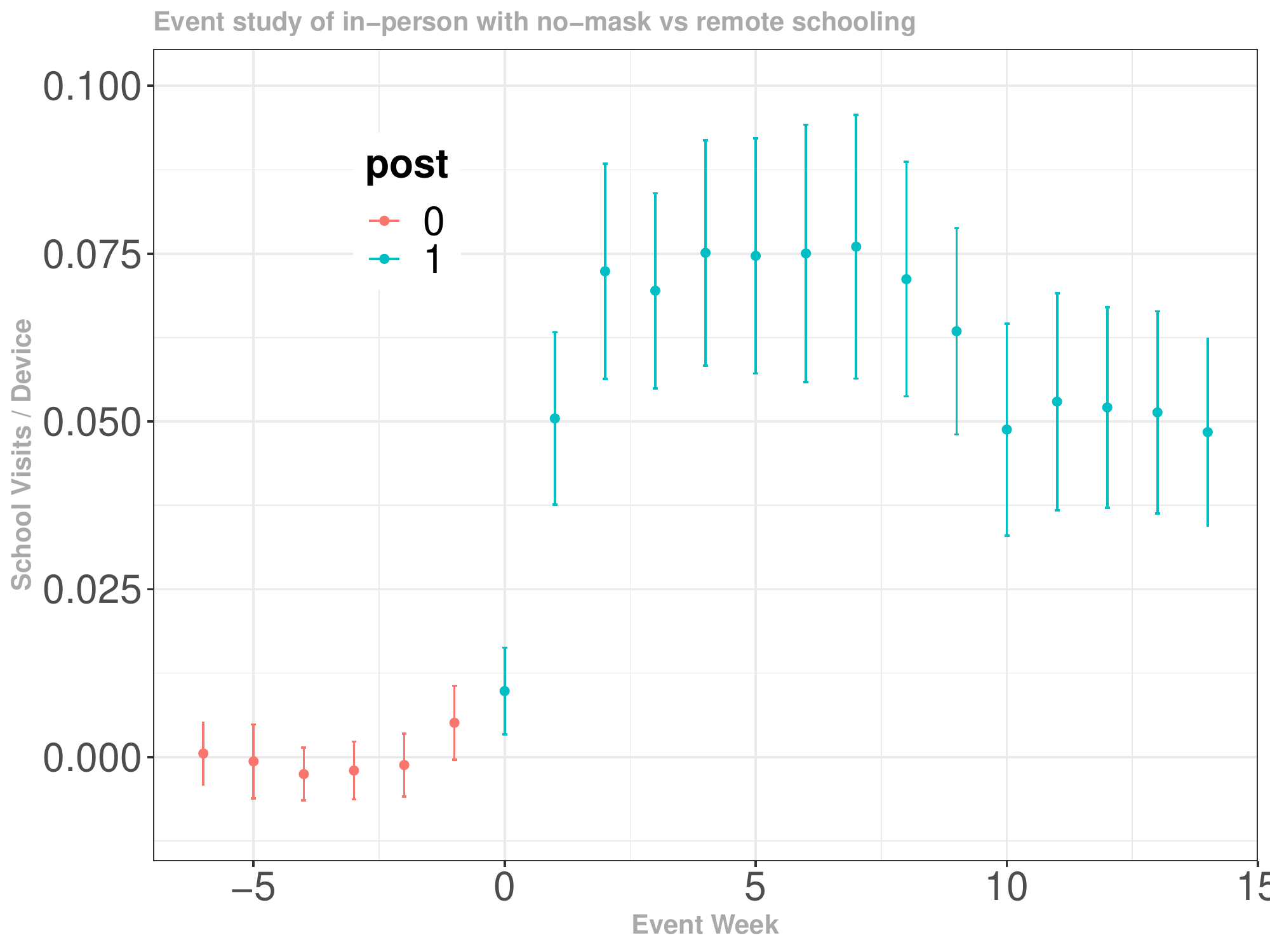}& \includegraphics[width=0.25\textwidth]{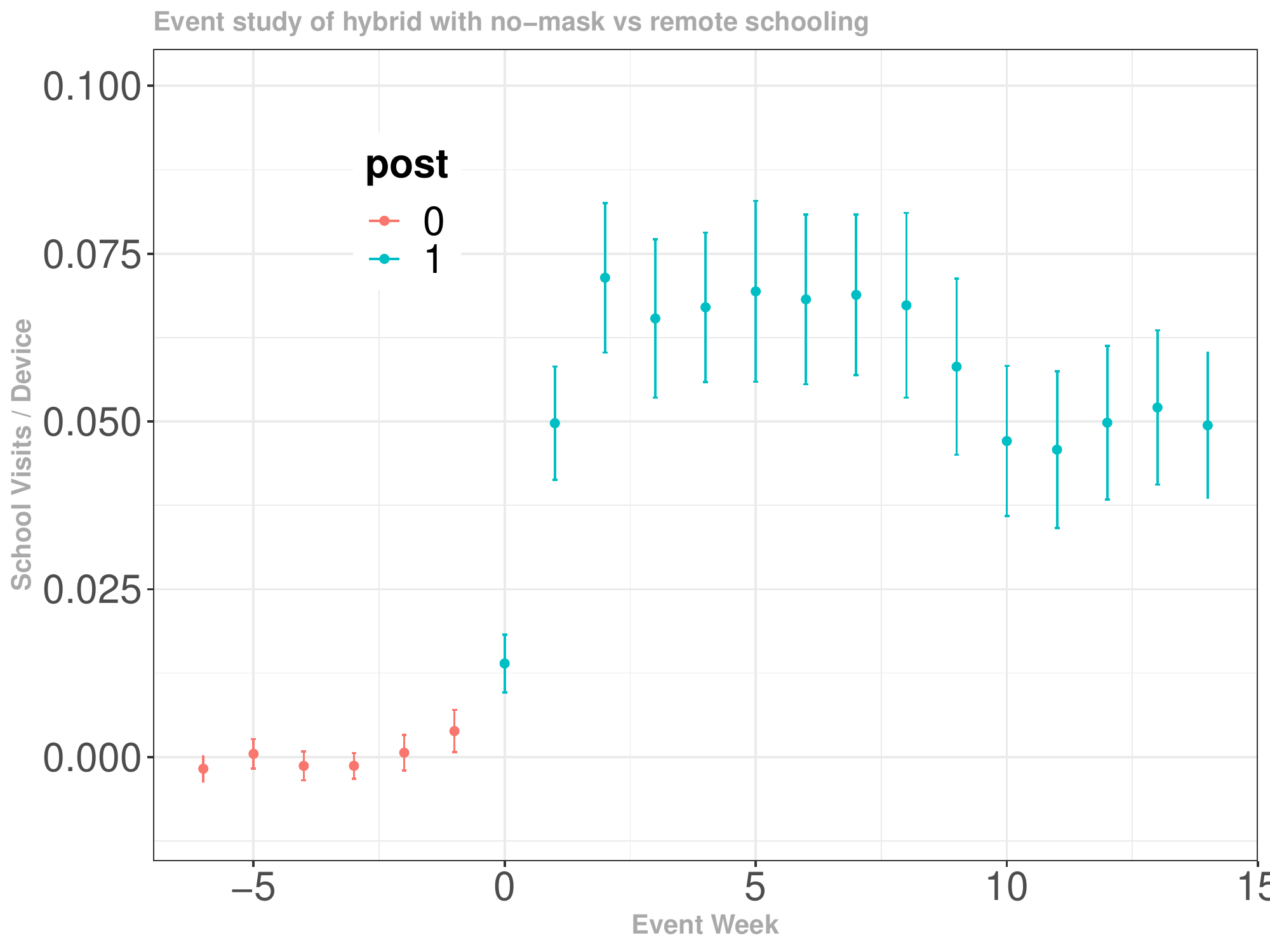}  \smallskip\\
  {\tiny (u) Full-time: In-person  }& {\tiny (v) Full-time:  Hybrid  }& {\tiny (w) Full-time: In-person/No-Mask }& {\tiny (x) Full-time:  Hybrid/No-Mask  }\\
 \includegraphics[width=0.25\textwidth]{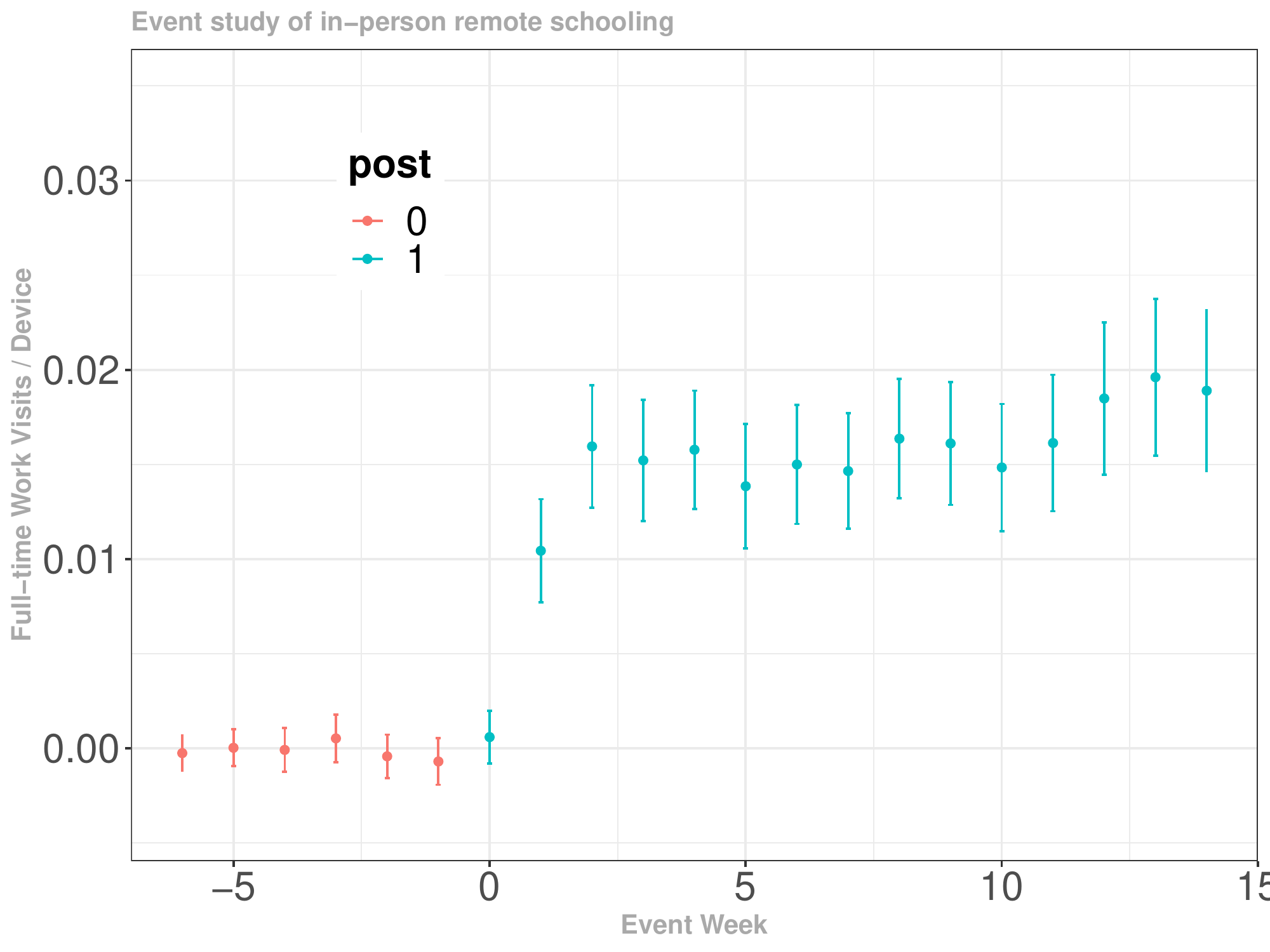}& \includegraphics[width=0.25\textwidth]{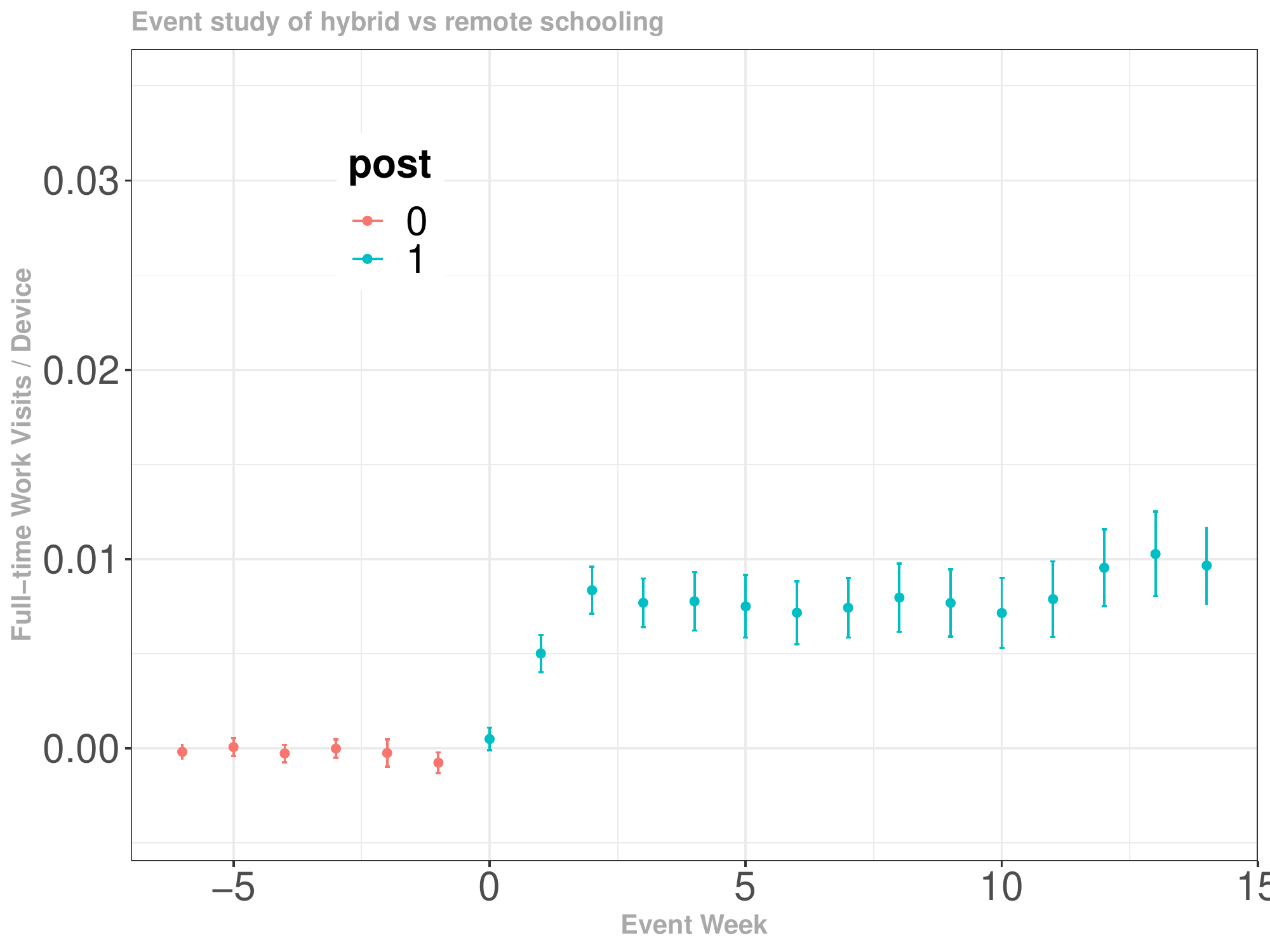}
 &
 \includegraphics[width=0.25\textwidth]{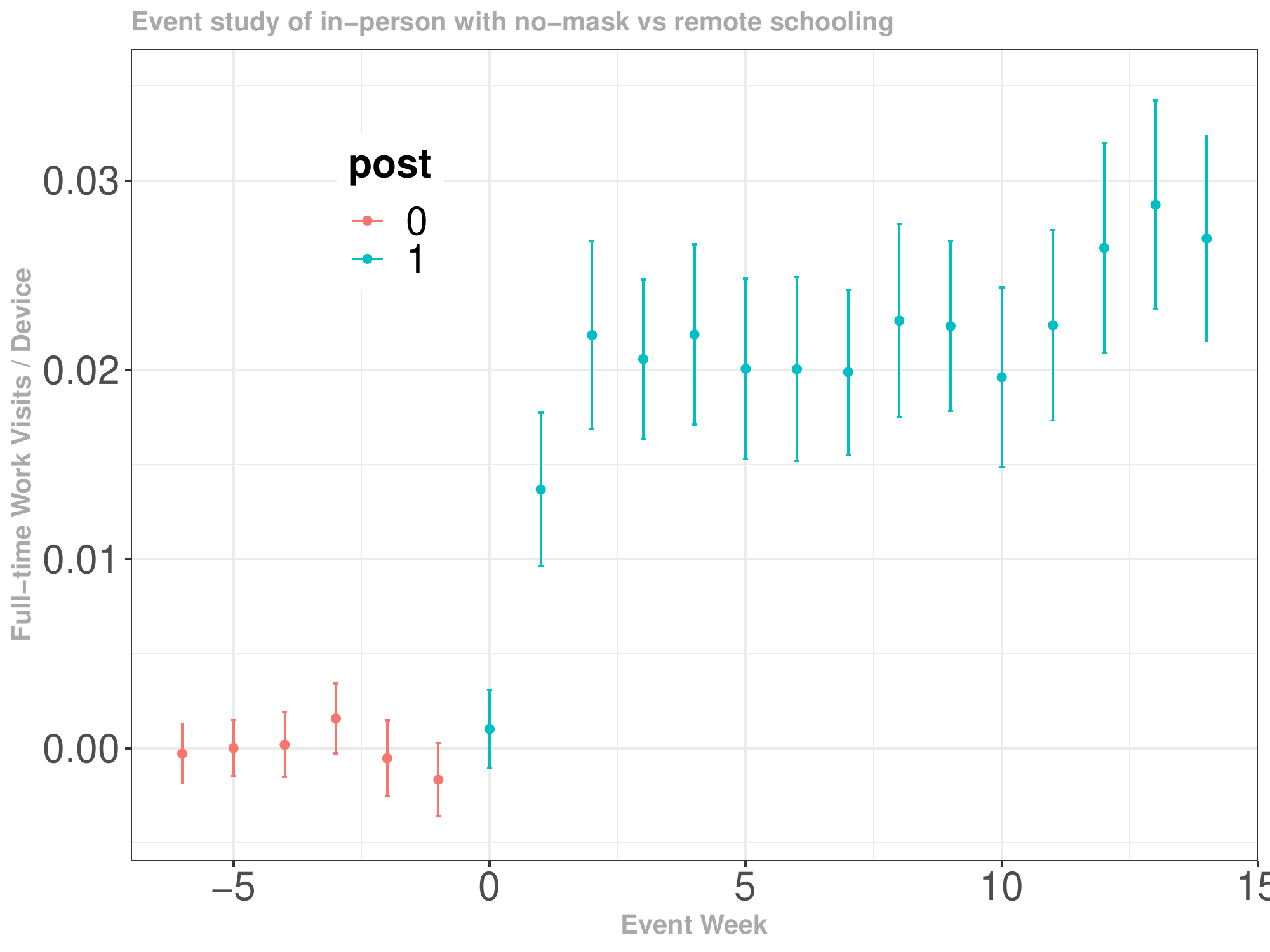}& \includegraphics[width=0.25\textwidth]{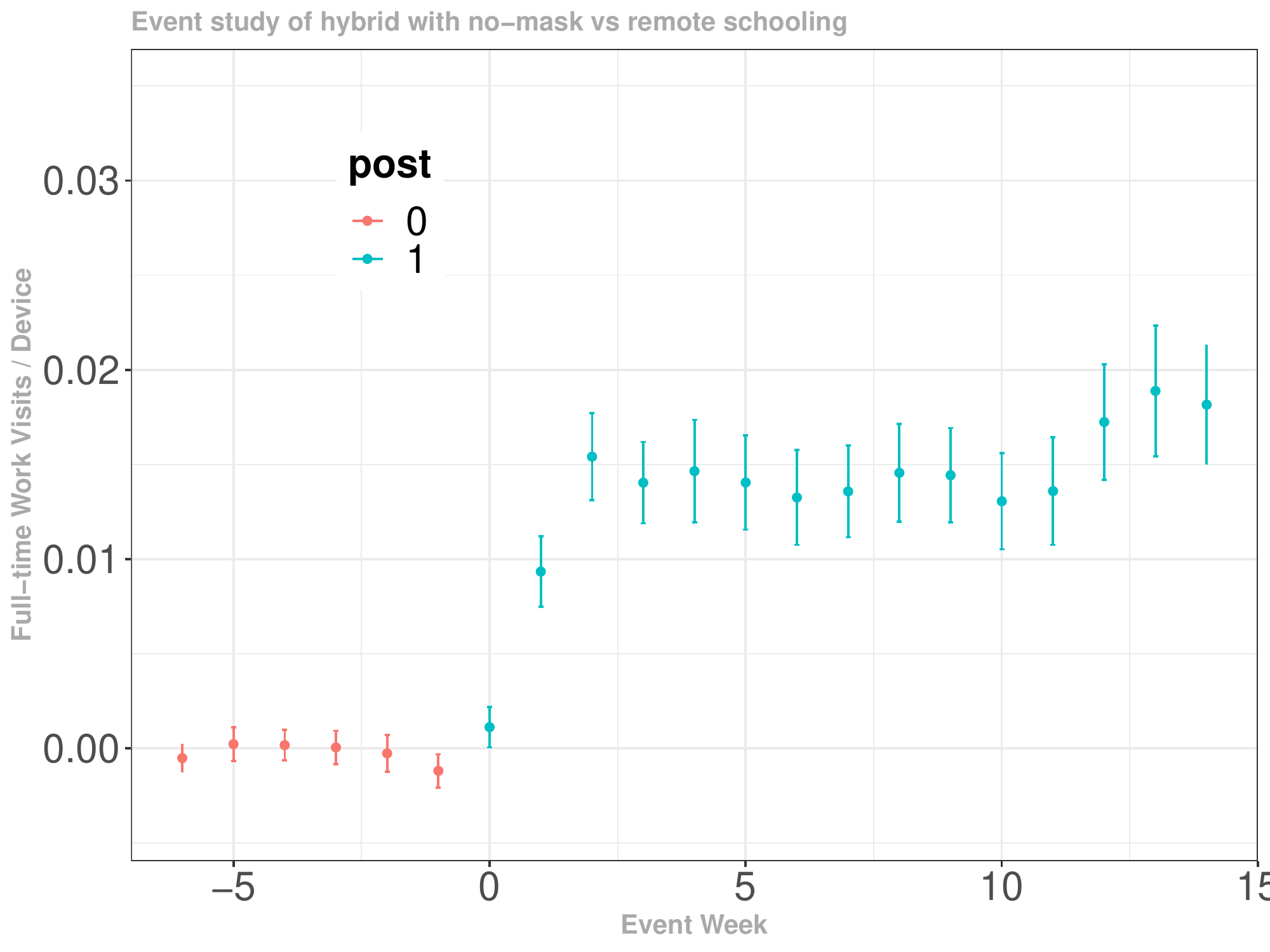}  \smallskip\\
\end{tabular}
  \end{minipage}}
 {\scriptsize
\begin{flushleft}
Notes: (a) plots the estimates and 95\% simultaneous confidence intervals of  the average dynamic treatment effect of in-person openings relative to the counties with remote openings as well as the counties that have not opened yet on cases per 1000 using a subset of counties with either in-person opening or remote-opening, where we use the estimation method of \cite{Callaway2020} implemented by their \href{https://cran.r-project.org/web/packages/did/vignettes/did-basics.html}{did R package}.  Similarly, (b), (c), and (d) plots the estimates of  the average dynamic treatment effect of school opening with hybrid, in-persion/mask mandates, and hybrid/mask mandates teaching methods, respectively, using a subset of counties with the corresponding teaching method as well as remote-opening. (e)-(h), (i)-(l), (m)-(p), (q)-(t), and (u)-(x) report the estimates of the average dynamic treatment effect on deaths per 1000, log(cases), log(deths), per-devise visits to K-12 schools, and per-devise visits to full-time workplaces, respectively.
 \end{flushleft}} \vspace{-0.5cm}
\end{figure}

Fig. \ref{fig:ca} presents the estimated group-time average treatment effects with 95 percent simultaneous confidence intervals.  Fig.  \ref{fig:ca}(a)-(h) show that  cases and deaths in counties with in-person or hybrid openings relative to those with remote openings increase after school openings; furthermore, these increases are more pronounced for counties without any mask mandate for staffs in Fig.  \ref{fig:ca}(c)(d)(g)(h). Fig. \ref{fig:ca}(i)-(p) similarly indicate that the log of weekly cases and deaths in counties with in-person or hybrid openings gradually increase after the school opening date, especially for counties with no mask mandates. In the figures, the estimates in the pre-treatment period are flat and not statistically different from zero, consistent with parallel trend assumption.

Consistent with the findings in Fig. \ref{fig:case-growth}(c)(d),    Fig. \ref{fig:ca}(q)-(x) show that visits to K-12 schools and full-time workplaces increase after the opening of schools in counties with in-person/hybrid teaching relative to those with remote teaching. In contrast, SI Appendix, Fig. S5 indicates no evidence for the association of the school opening date with the visits to restaurants, bars, recreational facilities, and churches, suggesting that other unobserved county-level confounders that affect people's mobilities to these places (e.g., lockdown policies) may not be systematically related to the timing of school opening, teaching modes, and mitigation measures.

Our finding is consistent with  \cite{Lessler2021} who examine the data from a massive online survey and find the association between in-person schooling and COVID-19-related outcomes across counties and the importance of school-based mitigation measures for reducing transmission risks in the United States.  In contrast, using an event-study design, \cite{ISPHORDING2021} and \cite{Bismarck-Osten2020} find no evidence that fully opening schools increased case number within the 3-4 weeks of school openings in Germany.  One possible reason for these contradictory findings is that the mitigation measures in German
schools may have been more effective in containing in-school transmissions than the measures adopted by the US schools with in-person openings. Another important source of the difference is that the event window length of  3-4 weeks in \cite{ISPHORDING2021} and \cite{Bismarck-Osten2020} may be too short to identify the effect of school openings on the confirmed cases because asymptomatic, undetected cases are prevalent among children \citep{Leidman2021}.

\section{Dynamic Panel Regression Analysis}
While our event-study analysis provides important evidence,  the parallel trend assumption might be at odds with the implications of the main epidemiological models that  the spread of cases is highly non-linear and that  the current number of cases dynamically depends on the number of past infected individuals through the virus transmission, which in turn is influenced by school openings, containment policies, and importantly people's voluntary behavioral changes in response to information \citep{Callaway2021,chernozhukov2021}. Furthermore, the validity of the parallel trend assumption is sensitive to the transformation of outcome variable \cite{roth2021parallel}. Therefore, as an alternative approach, we analyze the effect of  opening K-12 schools on case growth by panel data regression under the uncounfoundedness assumption, where a specification is motivated by  the Susceptible-Infectious-Recovered-Deceased (SIRD) model.    

\subsection{Methods} 
\begin{figure}[!ht]
  \caption{The causal path diagram for our model \label{fig:dag}}\bigskip
  
 \resizebox{\columnwidth}{!}{
   \hspace{5cm}    \includegraphics[clip, trim=5.5cm 18.7cm 2cm 3.8cm, width=1.00\textwidth]{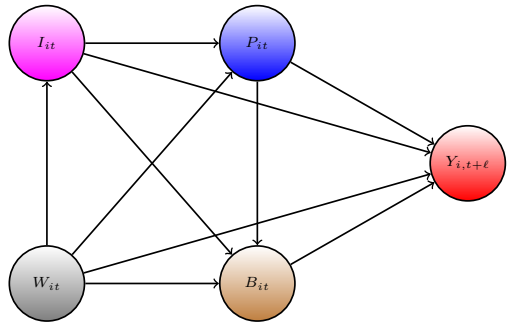}
      }
\end{figure}

SI Appendix provides the details for  our research design which closely follows \cite{chernozhukov2021}.
Fig. \ref{fig:dag} is a causal path diagram \citep[see][]{pearl:causality} for our model that describes how policies, behavior, and information interact together:
\begin{itemize}
\item The \textit{forward} health outcome,
$Y_{i,t+\ell}$, is determined last after all other variables have been determined;
\item The policies, $P_{it}$,  affect health outcome $Y_{i,t+\ell}$ either directly, or indirectly through mediators, human behavior $B_{it}$, which may only be partially observed.
\item  Information variables, $I_{it}$, such as lagged values of outcomes can affect human behavior and policies, as well as  outcomes directly;
\item The confounders $W_{it}$, which vary across counties and time, affect all other variables; these include testing rates and
unobserved but estimable county and state-week effects.
\end{itemize}

The index $i$ denotes the county $i$, and $t$ and $t+\ell$ denotes the time, where $\ell$ represents the time lag between infection and case confirmation or death. Our health outcomes are the growth rates in Covid-19 cases and deaths. Policy variables include school reopening in various modes, mask mandates, ban gathering, and stay-at-home orders, while information variables include lagged values of cases or deaths.

The causal structure allows for the effect of the policy to be either direct or indirect. For example, school openings not only directly affect case growth through the within-school transmission but also indirectly affect case growth by increasing parents' mobility. The structure also allows for changes in behavior to be brought by the change in policies and information. The information variables, such as the actual recorded number of past  confirmed cases, can cause people to spend more time at home, regardless of adopted policies; these changes in behavior, in turn, affect the transmission of SARS-CoV-2.

To further motivate our panel regression specification, we consider the SIRD
model:
\begin{align}
  \dot{S}(t) & = -\frac{S(t)}{N} \beta(t) \Infected(t),\quad
  \dot{\Infected}(t)   = \frac{S(t)}{N} \beta(t) \Infected(t) - \gamma  \Infected(t), \label{eq:i}\\
  \dot{\Recovered}(t) & = (1-\kappa) \gamma  \Infected(t),\quad   \dot{D}(t)  = \kappa \gamma \Infected(t), \nonumber 
  \label{eq:d}
\end{align}
where $S$, $\Infected$, $\Recovered$, and $D$ denote the number of susceptible,
infected, recovered, and deceased individuals in a given state.  Each of these variables is a function of time and dots indicate time derivative. $N$ is the population, $\beta(t)$ is the rate of infection
spread, $\gamma$ is the rate of recovery or death, and $\kappa$ is the
probability of death conditional on infection. Confirmed cases, $C(t)$, evolve as
\begin{equation}
  \dot{C}(t) = \tau(t) \Infected(t), \label{eq:c}
\end{equation}
where $\tau(t)$ is the testing proportion (detection rate). Differentiating the logarithm of [\ref{eq:c}] and substituting [\ref{eq:i}], we have
\begin{align}
  \frac{\ddot{C}(t)}{\dot{C}(t)}
              & =
                \frac{S(t)}{N} \beta(t) -\gamma  + \frac{\dot{\tau}(t)}{\tau(t)}, \label{eq:c2}
\end{align}
which indicates that the growth rate of cases depends on the rate of infection spread and the change in detection rates.

Our empirical specification is a discrete-time analog of equation [\ref{eq:c2}] by approximating the case growth rate with the log difference in weekly confirmed cases and specifying $\frac{S(t)}{N}\beta(t)$ as a linear function of the variable for K-12 school visits, policies,  past cases,
 county fixed effects, and state-week fixed effects:
\begin{align}
&\Delta_7 \log \text{\textit{Case}}_{it}  =  \beta' \text{\textit{Visit}}_{i,t-14}  +   \sum_{\tau=14,21,28} \beta_{y,\tau} \log \text{\textit{Case}}_{i,t-\tau}  \nonumber\\
&\quad\qquad + \gamma' \text{\textit{NPI}}_{i,t-14}  + \theta \text{\textit{Test}}_{it}+\alpha_i +\delta_{s(i),w(t)} + \epsilon_{it},  \label{eq:panel-1}
\end{align}
where $i$ is county, $t$ is day. The outcome variable $\Delta_7 \log \text{\textit{Case}}_{it}:=\log \text{\textit{Case}}_{it}-\log \text{\textit{Case}}_{i,t-7}$ is the log-difference  over 7 days in reported weekly cases with $\text{\textit{Case}}_{it}$ denoting the number of confirmed cases from day $t-6$ to $t$. For the observation with zero weekly cases,  we set the value of the log of weekly cases, $\log \text{\textit{Case}}_{it}$, to be $-1$.

$\text{\textit{Visit}}_{i,t-14}$ includes per-device  K-12 school visits and college visits lagged by 14 days from the SafeGraph foot traffic data (SI Appendix, Fig. S4(c)(f)). The direct measure of K-12 school visits has advantages over that of school opening modes from the MCH data because the latter is prone to measurement errors caused by unrecorded changes in teaching methods and school closures beyond the information recorded in the MCH data.   The measure of K-12 school visits may also capture student density heterogeneity within the same opening mode, especially for the hybrid teaching method.

As confounders, we consider a set of county fixed effects, $\alpha_i$, as well as state-week fixed effects, $\delta_{s(i),w(t)}$. County fixed effects $\alpha_i$ control permanent differences across counties in unobserved personal risk-aversion and attitude toward mask-wearing, hand washings, and social distancing. The coefficient $\delta_{s(i),w(t)}$  on the interaction terms between statey dummy variables and week dummy variables capture any change over time in people's behaviors and non-pharmaceutical policy interventions (NPIs) that are common within a state; they also control for changes in weather, temperature, and humidity within a state.  $\text{\textit{NPI}}_{i,t-14}$ includes
county-level NPIs (mask mandates, banning gathering of more than 50 persons, stay-at-home orders) lagged by two weeks that control for the effect of people's behavioral changes driven by county-level policies on case growths. NPI  data on stay-at-home orders and gathering bans is from \href{https://github.com/JieYingWu/COVID-19_US_County-level_Summaries}{Jie Ying Wu} \cite{killeen2020} while  \href{https://drive.google.com/uc?export=download&id=1qVIhPaBQ-apdDjOaKV2eA9SgZNkLMLAm}{the data on mask policies} is from  \cite{Wright2020}. These NPI data contain information up to the end of July; in our regression analysis, we set the value of these policy variables after August to be the same as the last day of observations.

$\text{\textit{Test}}_{it}$ is the growth rate of the number of tests recorded at the daily frequency for each state to capture the changes in detection rates ${\dot{\tau}(t)}/{\tau(t)}$ in [\ref{eq:c2}].  This variable is important to fill the gap between confirmed cases and actual infections.

Finally, the logarithm of past weekly confirmed cases denoted by $\log \text{\textit{Case}}_{i,t-\tau}$ for $\tau=14, 21$, and $28$  are included  in [\ref{eq:c2}] as important confounders representing information variables. First, because the timing and the mode of school openings are likely to be affected by the number of lagged cases or deaths (e.g., the decision to reopen schools in California and Oregon depended on trends in local case counts \citep{Goldhaber-Fiebert2020}), controlling for the past weekly cases is critical for the unconfoundedness assumption.\footnote{Referring to the causal path diagram in Fig. \ref{fig:dag}, the information variables affect both policies (e.g., school openings) and  the people's behavior (see \cite{chernozhukov2021} and Table \ref{tab:PItoB} for empirical evidence). Even if information variables do not affect outcome directly, they are important confounders because in the terminology of Pearl \cite{pearl:causality}, they open a ``backdoor'' path from outcome to policy, which creates a non-causal association between policies and outcomes. By controlling for information variables (and other confounders), we block non-causal associations, revealing the association generated by the causal effect of policies on outcomes. See our discussion in SI Appendix as well as \cite{pearl:causality} for more details.} Second, controlling for past confirmed cases is important because people may voluntarily change their risk-taking behavior in response to the new information provided by the \textit{confirmed} cases rather than the \textit{actual}, but unknown, number of infected individuals.

We also consider an alternative specification  using  the proportion of K-12 students attending schools with teaching method $p\in \{\text{in-person},\text{hybrid},\text{remote}\}$ constructed from the MCH data in place of the visits to K-12 schools from the foot traffic data. Furthermore, we investigate the role of mitigation strategies at school on the transmission of SARS-CoV-2 by examining how the coefficients of K-12 school visits and K-12 school openings depend on the mask-wearing requirement for staff with an interaction term between school opening variables and a dummy variable for staff mask-wearing requirements.


For parameter identification, we assume that the error $\epsilon_{it}$ in [\ref{eq:panel-1}] is orthogonal to the observed explanatory variables of school visits/openings, NPIs, test rates, and the past log cases, county fixed effects, and state-week fixed effects. The estimated parameters for school openings can be causally interpreted under the unconfoundedness assumption that the variables related to school openings (school visits, opening dates, teaching methods) are as good as randomly assigned after conditioning on other controls, county fixed effects, and state-week fixed effects.

Because the fixed effects estimator with a set of county dummies for dynamic panel regression could be severely biased when the time dimension is short compared to cross-sectional dimension \citep{Nickell1981}, we employ the debiased estimator by implementing bias correction  \citep[e.g.,][]{chen2019mastering} although the fixed effects estimator without bias correction gives qualitatively similar results. See SI Appendix, Table S5 and  Fig. S8 for the results from the fixed effects estimator without bias correction.

 \subsection*{Result}
Table \ref{tab:PItoY}  reports the debiased estimates of panel data regression.  Clustered standard errors at the state level are reported in the bracket to provide valid inference under possible dependency over time and across counties within each state.
The results suggest that an increase in the visits to K-12 schools and opening K-12 schools with in-person learning mode is associated with an increase in the growth rates of cases with two weeks lag when schools implement no mask mandate for staff.

In column (1),  that of per-device visits to K-12 schools is 0.47 (SE = 0.07).  The change in top 5 percentile values of per-device visits to  K-12 schools between June and September among counties is around 0.15 in SI Appendix, Fig. S4(c). Taking this value as a benchmark for full openings,  fully opening K-12 schools may have contributed to (0.47$\times$0.15=) 7 percentage points increase in case growth rates. Column (3) indicates that openings of K-12 schools with the in-person mode are associated with 5 (SE = 2) percentage point increases in weekly case growth rates. It also provides evidence that openings of K-12 schools with remote learning mode are associated with lower case growth, perhaps because remote school opening induces more precautionary behavior to reduce transmission risk.

In column (2), the estimated coefficient of the interaction between K-12 school visits and no mask-wearing requirements for staff is 0.24 (SE=0.07),   providing evidence that mask-wearing requirements for staff may have reduced the transmission of SARS-CoV-2  at schools. Similarly, in column (4), the coefficients on the interaction of in-person and hybrid school openings with no mask mandates are positively estimated as 0.04 (SE=0.02) and 0.05 (SE=0.02), respectively.   These estimates likely reflect not only the effect of mask-wearing requirements for staff but also that of other mitigation measures. For example, school districts with staff mask-wearing requirements frequently require students to wear masks and often increase online instructions.

Other studies on COVID-19 spread in schools have also pointed to the importance of mitigation measures. In contact tracing studies of cases in schools, \cite{gillespie2021} found that 6 out of 7 traceable case clusters were related to clear noncompliance with mitigation protocols, and  \cite{zimmerman2020} found that most secondary transmissions were related to absent face coverings.
\cite{hobbs2020} find that children who tested positive for COVID-19 are considerably less likely to have had reported consistent mask use by students and staff inside their school.

\begin{figure}[!ht]
  \caption{Sensitivity analysis for the estimated coefficients of K-12 visits and college visits of case growth regressions: Debiased Estimator \label{fig:sensitivity}}\smallskip
  
\resizebox{0.7\columnwidth}{!}{
        \begin{tabular}{c}
      (a) Case Growth Estimates   \\
      \includegraphics[width=0.48\textwidth]{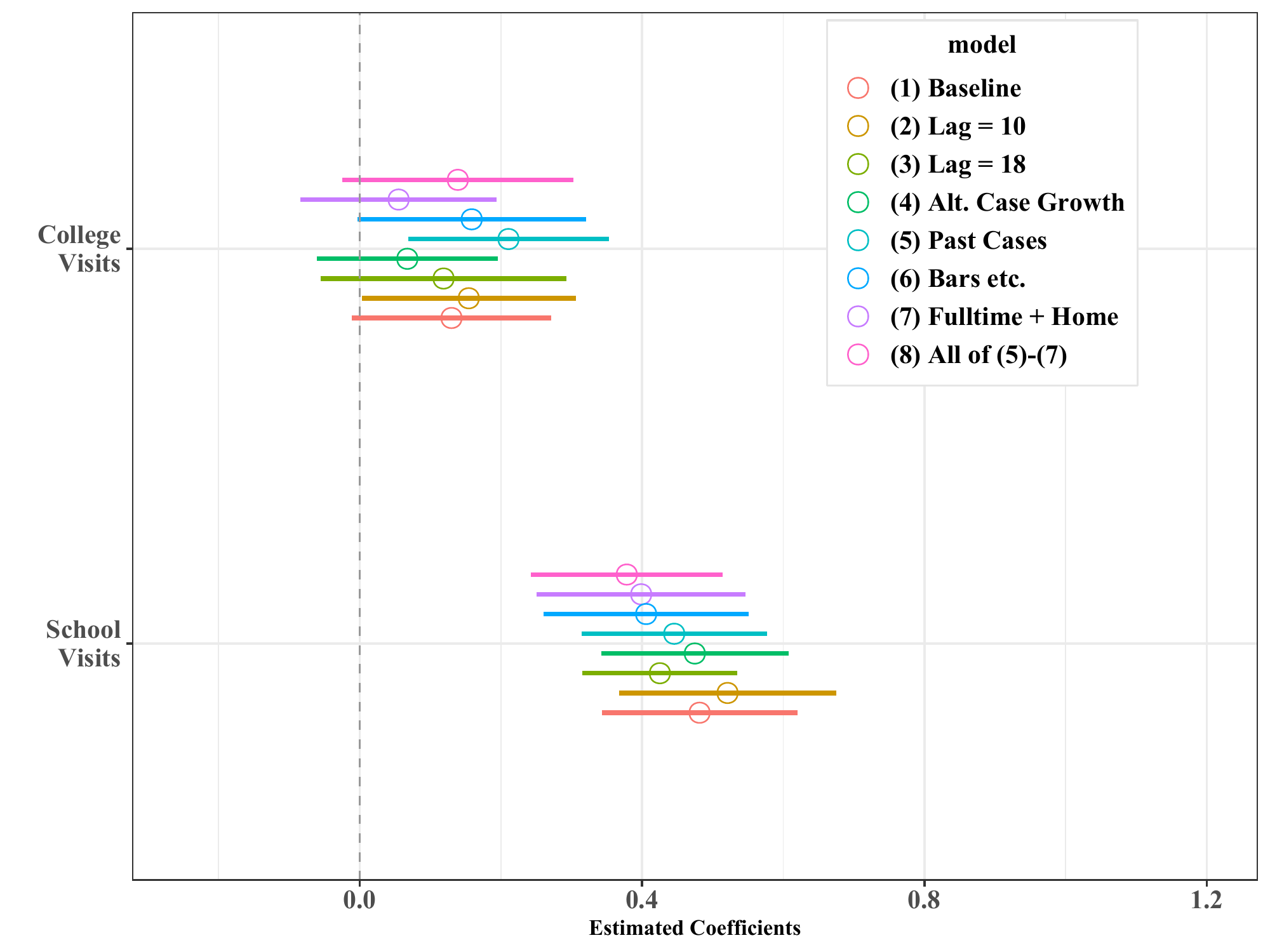} \\
       (b) Case Growth Estimates with School Visits $\times$ No Mask   \\
     \includegraphics[width=0.48\textwidth]{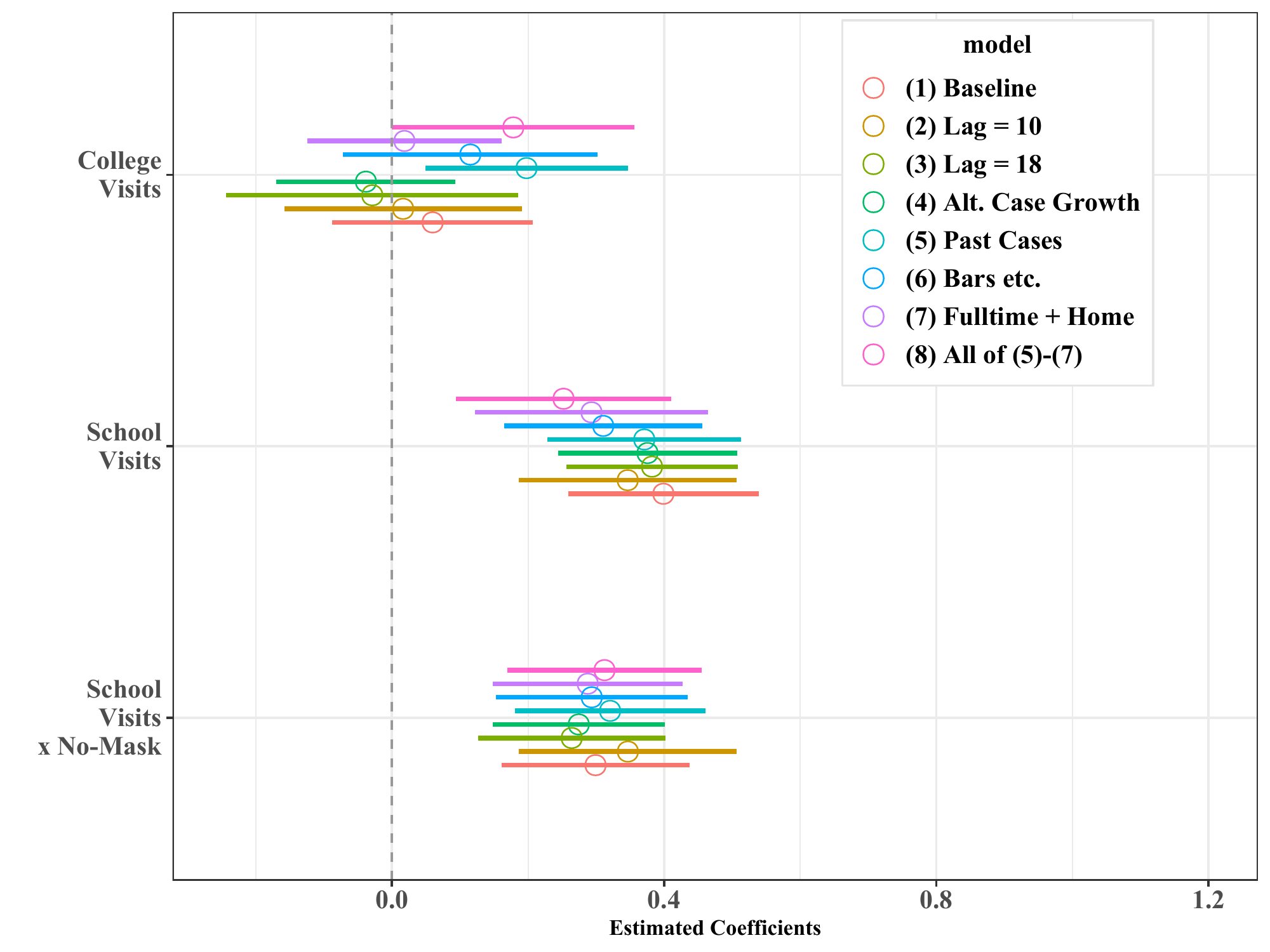}  \\
       \end{tabular}}
\vspace{-0.2cm}   {\scriptsize
\begin{flushleft}
Notes:  (a) presents the estimated of college visits and K-12 school visits with the 90 percent confidence intervals across different specifications taking the column (1) of Table  \ref{tab:PItoY}  as baseline. (b) presents the estimates of  college visits , K-12 school visits, and the interaction between K-12 school visits and no mask wearing requirement for staff  taking column (2) of Table  \ref{tab:PItoY}  as baseline. The results are based on the debiased estimator.  SI Appendix, Fig. S8 presents the results based on the  estimator without bias correction.
\end{flushleft}  }
\end{figure}

The estimated coefficient of per-device visits to colleges is 0.14 (SE = 0.07) in column (1) of  Table \ref{tab:PItoY}.  With the change in top 5 percentile values of college visits between June and September as a benchmark for full openings,  which is about 0.1,  fully opening colleges may be associated with (0.14$\times$0.1=) 1.4 percentage points increase in case growth. Therefore, the estimated association of opening colleges with case growth is much smaller than that of opening K-12 schools. Furthermore, alternative specifications in columns (2) and (4) of Table \ref{tab:PItoY} and those in Fig. \ref{fig:sensitivity}  yield
smaller, and sometimes insignificant, estimates for college visits. This sensitivity may be due to the limited variation in changes in college visits over time across counties in the data, where the 75 percentile value of college visits is consistently very low (SI Appendix, Fig. S4(f)). SI Appendix presents more discussions on the association of opening colleges with the spread of COVID-19, where Fig. S2 and S3 provide descriptive evidence that opening colleges and universities may be associated with the spread of COVID-19 in counties where a large public universities are located.

Consistent with evidence from US state-level panel data analysis in \cite{chernozhukov2021}, the estimated coefficients of county-wide mask mandate policy are negative and significant in columns (1)-(4), suggesting that mandating masks reduces case growth. The estimated coefficients of ban gatherings and stay-at-home orders are also negative. The negative estimates of the log of past weekly cases are consistent with a hypothesis that the information on higher transmission risk induces people to take precautionary actions voluntarily to reduce case growth.  The table also highlights the importance of controlling for the test growth rates as a confounder.

Evidence on the role of schools in the spread of COVID-19 from other studies is mixed. Papers that focus on contract tracing of cases among students find limited spread from student infections \citep{zimmerman2020,brandal2021,ismail2020,gillespie2021,falk2021,willeit2021}.  There is also some evidence that school openings are associated with increased cases in the surrounding community. \cite{bignami2021} provides suggestive evidence that school openings are associated with increased cases in Montreal neighborhoods. \cite{auger2020} use US state-level data to argue that school closures at the start of the pandemic substantially reduced the infection.

Three closely related papers also examine the relationship between schools and county-level COVID-19 outcomes in the US. \cite{goldhaber2021} examine the relationship between schooling and cases in counties in Washington and Michigan. They find that in-person schooling is only associated with increased cases in areas with high pre-existing COVID-19 cases. Similarly, \cite{harris2021} analyze US county-level data on COVID-19 hospitalizations and find that in-person schooling is not associated with increased hospitalizations in counties with low pre-existing COVID-19 hospitalization rates. The outcome variable of our regression analysis is case growth rates instead of new cases or hospitalizations. Consistent with \cite{goldhaber2021} and \cite{harris2021}, our finding of a constant increase in growth rates implies a greater increase in cases in counties with more pre-existing cases. \cite{reinbold2021} finds that counties with hybrid or remote openings had fewer cases than those with in-person openings but finds no association of teaching modes with deaths during the    first three weeks of the school year in Illinois,. Our finding on death rates does not necessarily contradict that of  \cite{reinbold2021} because the three weeks period is too short to identify the effect on deaths, whereas we examine the effect on deaths after 3 to 5 weeks of openings.

\begin{table}[!htbp] \centering
 \caption{The Association of School/College Openings and NPIs with Case Growth in the United States: Debiased Estimator}\vspace{-0.3cm}
 \label{tab:PItoY}
\resizebox{0.8\columnwidth}{!}{
\begin{tabular}{@{\extracolsep{1pt}}lcc|cc}
\\[-1.8ex]\hline
\hline \\ [-1.8ex]
 & \multicolumn{4}{c}{\textit{Dependent variable:  Case Growth Rates}} \\
\cline{2-5}
& (1) & (2) & (3) & (4)\\
\hline 
K-12 Visits, 14d  lag  & 0.467$^{***}$ & 0.386$^{***}$ &  &  \\
  & (0.070) & (0.070) &  &  \\
 K-12 Visits $\times$ No-Mask, 14d  lag &  & 0.297$^{***}$ &  &  \\
  &  & (0.070) &  &  \\
K-12  In-person, 14d  lag  &  &  & 0.047$^{***}$ & 0.023 \\
  &  &  & (0.017) & (0.021) \\
K-12 Hybrid, 14d  lag  &  &  & $-$0.008 & $-$0.037$^{***}$ \\
  &  &  & (0.014) & (0.013) \\
K-12  Remote, 14d  lag  &  &  & $-$0.082$^{***}$ & $-$0.102$^{***}$ \\
  &  &  & (0.016) & (0.015) \\
 K-12   In-person $\times$ No-Mask, 14d  lag &  &  &  & 0.041$^{**}$ \\
  &  &  &  & (0.019) \\
 K-12  Hybrid $\times$ No-Mask, 14d  lag &  &  &  & 0.049$^{***}$ \\
  &  &  &  & (0.017) \\ \hline
 College Visits, 14d  lag  & 0.139$^{*}$ & 0.070 & 0.132$^{**}$ & 0.010 \\
  & (0.071) & (0.073) & (0.064) & (0.076) \\
Mandatory mask, 14d  lag & $-$0.113$^{***}$ & $-$0.123$^{***}$ & $-$0.128$^{***}$ & $-$0.128$^{***}$ \\
  & (0.018) & (0.017) & (0.020) & (0.019) \\
Ban gatherings, 14d  lag & $-$0.124$^{***}$ & $-$0.136$^{***}$ & $-$0.135$^{***}$ & $-$0.137$^{***}$ \\
  & (0.033) & (0.044) & (0.033) & (0.042) \\
Stay at home, 14d  lag & $-$0.264$^{***}$ & $-$0.260$^{***}$ & $-$0.261$^{***}$ & $-$0.268$^{***}$ \\
  & (0.031) & (0.039) & (0.034) & (0.040) \\\hline
 log(Cases), 14d  lag  & $-$0.101$^{***}$ & $-$0.101$^{***}$ & $-$0.098$^{***}$ & $-$0.099$^{***}$ \\
  & (0.009) & (0.010) & (0.010) & (0.010) \\
 log(Cases), 21d  lag & $-$0.061$^{***}$ & $-$0.060$^{***}$ & $-$0.060$^{***}$ & $-$0.059$^{***}$ \\
  & (0.005) & (0.005) & (0.005) & (0.005) \\
 log(Cases), 28d  lag  & $-$0.030$^{***}$ & $-$0.033$^{***}$ & $-$0.031$^{***}$ & $-$0.034$^{***}$ \\
  & (0.003) & (0.003) & (0.004) & (0.004) \\
  Test Growth Rates & 0.009$^{**}$ & 0.008$^{*}$ & 0.009$^{**}$ & 0.009$^{**}$ \\
  & (0.004) & (0.004) & (0.004) & (0.004)\\
 \hline 
County Dummies & Yes & Yes &  Yes  &  Yes  \\
State$\times$ Week Dummies&Yes & Yes &  Yes  &  Yes  \\
\hline \\[-1.8ex] Observations & 690,297 & 545,131 & 612,963 & 528,941 \\
R$^{2}$ & 0.092 & 0.093 & 0.092 & 0.094 \\ \hline
\hline 
\end{tabular}}
  {\scriptsize
\begin{flushleft}
Notes: Dependent variable is the log difference over 7 days in weekly positive cases. Regressors are 7-days moving averages of corresponding daily variables  and lagged by 2 weeks to reflect the time between infection and case reporting except for test growth rates. All regression specifications include county fixed effects and state-week fixed effects.
The debiased fixed effects estimator is applied.  The results from the estimator without bias correction is presented in  SI Appendix, Table S5.
Asymptotic clustered standard errors at the state level are reported in bracket.   {$^{*}$p$<$0.1; $^{**}$p$<$0.05; $^{***}$p$<$0.01}
\end{flushleft}}
\end{table}

We next provide sensitivity analysis by changing our regression specifications and assumptions about delays between infection and reporting cases as follows:
\begin{itemize}
\item[(1)] Baseline specifications in columns (1) and (2) of Table \ref{tab:PItoY}.
\item[(2)]\hspace{-0.15cm},(3) Alternative time lags of 10 and 18 days for visits to colleges and K-12 schools as well as NPIs.
\item[(4)] Setting the log of weekly cases to $0$ when we observe zero weekly cases to compute the log-difference in weekly cases for the outcome variable.
\item[(5)] Add the log of weekly cases lagged by 5 weeks and per-capital \textit{cumulative} number of cases lagged by 2 weeks as controls.
\item[(6)] Add per-device visits to restaurants, bars, recreational places, and churches  lagged by 2 and 4 weeks as controls.
\item[(7)] Add per-device visits to full-time and part-time workplaces and a proportion of devices staying at home lagged by 2 weeks as controls.
\item[(8)] All of (5)-(7).
\end{itemize}

Because the actual time lag between infection and reporting cases may be shorter or longer than 14 days, we consider the alternative time lags in (2) and (3). Specification (4) checks the sensitivity of handling zero weekly cases to construct the outcome variable of the log difference in weekly cases.

A major concern for interpreting our estimate in Table \ref{tab:PItoY}  as the causal effect is that a choice of opening timing, teaching methods, and mask requirements may be endogenous. Our baseline specification mitigates this concern by controlling for county-fixed effects, state-week fixed effects, and the log of past cases. However, a choice of school openings may still be correlated with time-varying unobserved factors at the county level.  Therefore, we estimate a specification with additional time-varying county-level controls in (5)-(8).

Fig. \ref{fig:sensitivity}(a) takes column (1) of Table \ref{tab:PItoY} as a baseline specification and plots the estimated coefficients for visits to colleges and K12 schools with the 90 percent confidence intervals across different specifications using the debiased estimator; the estimates using the standard estimator \textit{without} bias correction are qualitatively \textit{similar} and reported in SI Appendix, Fig. S8. The estimated coefficients of K-12 school visits and college visits are all positive across different specifications, suggesting that an increase in visits to K-12 schools and colleges is robustly associated with higher case growth. On the other hand, the estimated coefficients often become smaller when we add more controls. In particular, relative to the baseline, adding full-time/part-time workplace visits and staying home devices leads to somewhat smaller estimated coefficients for K-12 school and college visits, suggesting that opening K-12 schools and colleges is associated with people returning to work and/or going outside more frequently.

In Fig. \ref{fig:sensitivity}(b), the estimated interaction terms of K-12 school visits and no mask-wearing requirements for staff in column (2) of Table \ref{tab:PItoY} are all positive and significant, robustly indicating a possibility that mask-wearing requirement for staff may have helped to reduce the transmission of SARS-CoV-2 at schools when K-12 schools opened with the in-person teaching method.

SI Appendix, Tables S6-S7 provide further robustness checks, showing that the results are similar even when we use the log of weekly cases in place of the log difference of weekly cases as the outcome variable.

\subsection{Association between School Openings and Mobility}

As highlighted by a modeling study for the United Kingdom
\citep{panovska2020lancet}, there are at least two reasons why opening  K-12 schools in-person may increase the spread of COVID-19. First,  opening K-12 schools increases the number of contacts within schools, which may increase the risk of transmission among children, parents,  education workers, and communities at large. Second, reopening K-12 schools allow parents to return to work and increase their mobility in general,  which may contribute to the transmission of  COVID-19 at schools and workplaces.

To give insight on the role of reopening K-12 schools for parents to return to work and to increase their mobility, we conduct panel data regression analysis by taking visits to full-time workplaces and a measure of staying home devices as outcome variables and use a similar set of regressors as in Table \ref{tab:PItoY} but without taking 2 weeks time lags.

Table \ref{tab:PItoB} shows how the proportion of devices at full-time workplaces and that of staying home devices are associated with visits to  K-12 schools as well as their in-person openings. In columns (1) and (2), the estimated coefficients of per-device K-12 school visits and opening K-12 schools for full-time work outcome variables are positive and especially large for in-person K-12 school opening. Similarly, the estimates in columns (3) and (4) suggest the negative association of per-device K-12 school visits and opening K-12 schools with the proportion of devices that do not leave their home. SI Appendix, Table S4 also provides evidence that the visits to restaurants and bars are positively associated with the visits to  K-12 schools and colleges. This is consistent with a hypothesis that opening K-12 school allows parents to return to work and spend more time outside. This result may also reflect education workers returning to work. In columns (3) and (4), the positive coefficient of current and past log cases suggests that people voluntarily choose to stay home when the transmission risk is high.

Table \ref{tab:PBItoY}  presents regression analysis similar to that in Table \ref{tab:PItoY} but including the proportion of devices at full-time/part-time workplaces and those at home as additional regressors, which corresponds to specification (7) in Fig. \ref{fig:sensitivity}. The estimates indicate that the proportion of staying home devices is negatively associated with the subsequent case growth, while the proportion of devices at full-time workplaces is positively associated with the case growth. Combined with the estimates in Table \ref{tab:PItoB}, these results suggest that school openings may have increased the transmission of SARS-CoV-2 by encouraging parents to return to work and to spend more time outside. This mechanism can partially explain the discrepancy between our findings and various studies that focus on cases among students. Contract tracing of cases in schools, such as \cite{falk2021}, \cite{zimmerman2020}, \cite{willeit2021}, \cite{brandal2021}, and \cite{ismail2020}, often finds limited direct spread among students. On the other hand, \cite{vlachos2021} finds that parents and teachers of students in open schools experience increases in infection rates.

In columns (1)-(2) of Table \ref{tab:PBItoY}, the estimated coefficients on K-12 school visits remain positive and large even after controlling for the mobility measures of returning to work and being outside home, which are mediator variables to capture the indirect effect of school openings on case growth through its effect on mobility. The coefficient on K-12 school visits is approximately 75\% as large in Table \ref{tab:PBItoY} as in Table \ref{tab:PItoY}, suggesting that that within-school transmission may be the primary channel through which school openings affect the spread of COVID-19.


\subsection*{Death Growth Regression} We also analyze the effect of school openings on the death growth by estimating:
\begin{align}
&\Delta_{21} \log \text{\textit{Death}}_{it}  =  \beta' \text{\textit{Visit}}_{i,t-35}  +   \sum_{\tau=35,42,49} \beta_{y,\tau} \log \text{\textit{Death}}_{i,t-\tau}  \nonumber\\
&\quad\qquad+ \gamma' \text{\textit{NPI}}_{i,t-35} +\alpha_i + \delta_{s(i),w(t)} + \epsilon_{it},  \label{eq:panel-3}
\end{align}
where the outcome variable $\Delta_{21} \log \text{\textit{Death}}_{it} :=\log \text{\textit{Death}}_{it}-\log \text{\textit{Death}}_{i,t-21}$ is  the log difference  over 21 days in reported weekly deaths with $\text{\textit{Death}}_{it}$ denoting  the number of reported deaths from day $t-6$ to $t$. The log of weekly deaths, $\log \text{\textit{Death}}_{it}$, is set to be $-1$ when we observe zero weekly deaths. We take the log difference over 21 days rather than 7 days for measuring death growth because the time lag between infection and death reporting is stochastic and spreads over at least 2 weeks.\footnote{ For deceased persons above 65 years old between March 1, 2020 and January 31, 2021, CDC estimates that the interquartile range for the number of days from symptom onset to death is (9,25) days while the interquartile range from death to reporting is (4,59) days. See  Table 2 of \url{https://www.cdc.gov/coronavirus/2019-ncov/hcp/planning-scenarios.html}.  }

The explanatory variables in [\ref{eq:panel-3}]  are lagged by 35 days to capture the time lag of infection and death reporting. Taking a longer time lag than 35 days may capture the effect of school openings on deaths through the secondary infection better (e.g., an infection from children to parents/grandparents); therefore, we also consider a lag length of 42 and 49 days.

\begin{figure}[!ht]
  \caption{Sensitivity analysis for the estimated coefficients of K-12 visits and college visits of death growth regressions: Debiased Estimator \label{fig:sensitivity-death}}\smallskip
  
\resizebox{0.7\columnwidth}{!}{
        \begin{tabular}{c}
      (a) Death Growth Estimates   \\
      \includegraphics[width=0.49\textwidth]{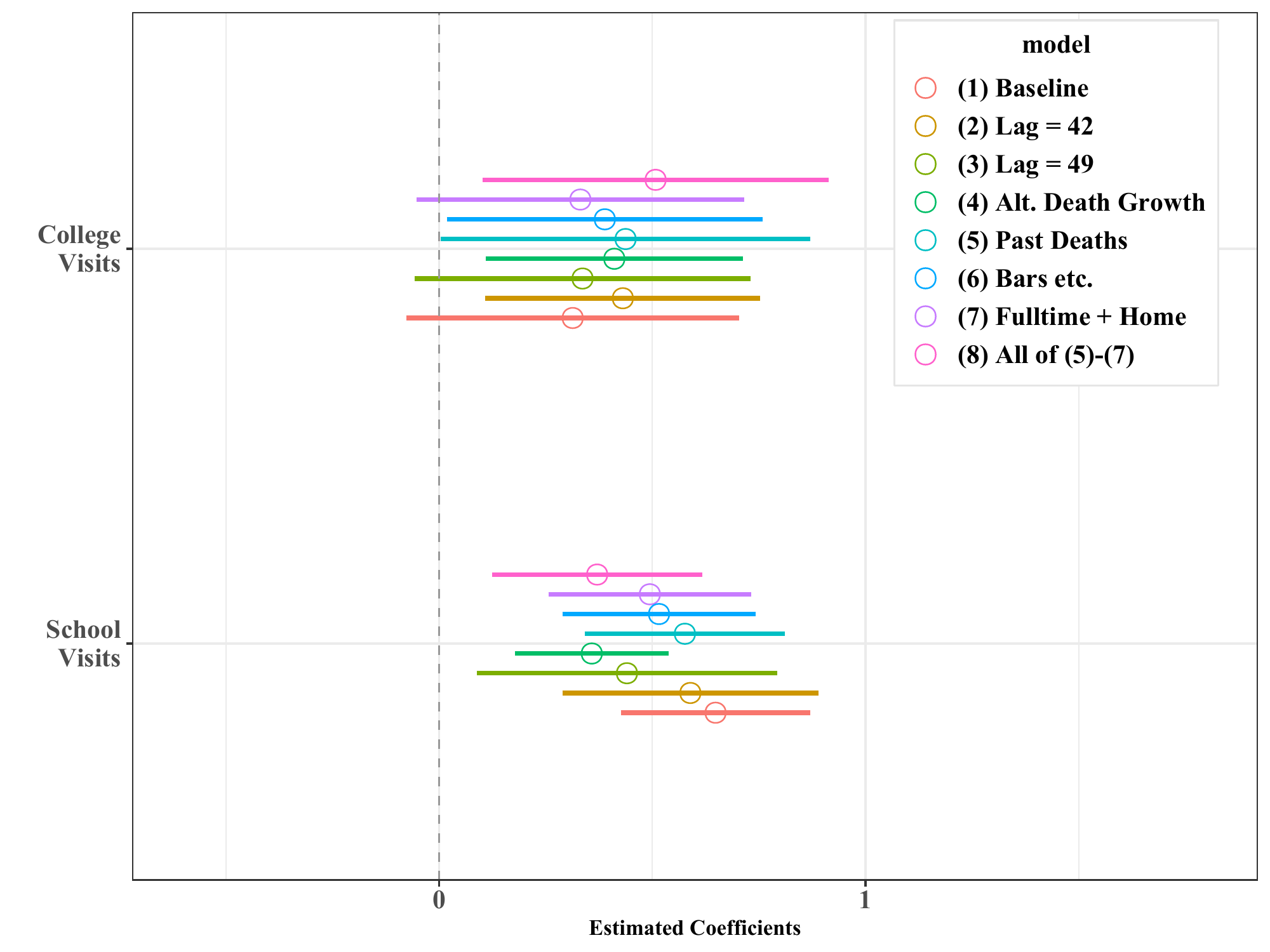} \\
       (b) Death Growth Estimates with School Visits $\times$ No Mask   \\
       \includegraphics[width=0.49\textwidth]{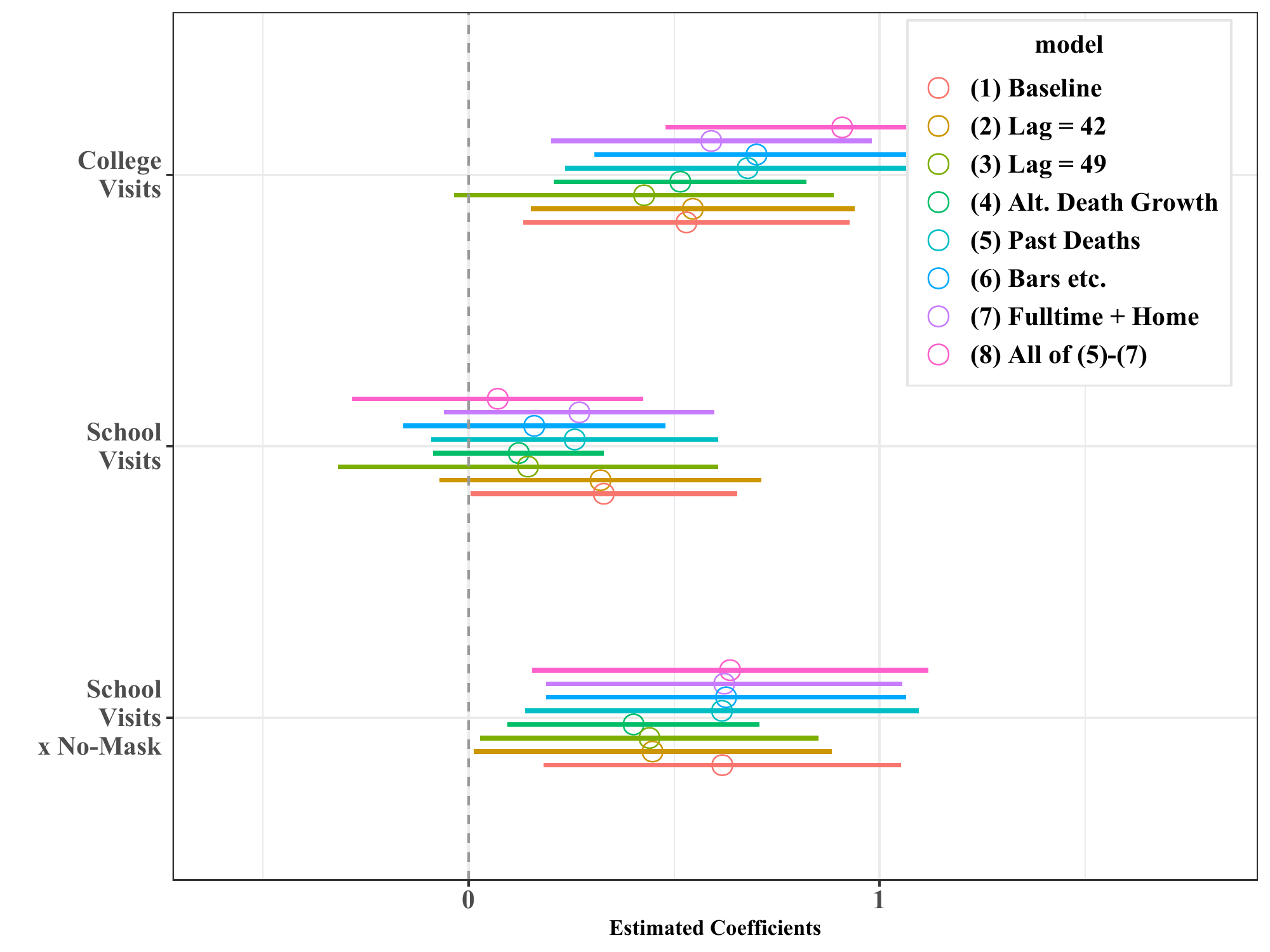}
       \end{tabular}}
\vspace{-0.2cm}   {\scriptsize
\begin{flushleft}
Notes:  (a) presents the estimated of college visits and K-12 school visits with the 90 percent confidence intervals across different specifications taking the column (1) of SI Appendix, Table S9 as baseline. (b) presents the estimates of  college visits , K-12 school visits, and the interaction between K-12 school visits and no mask wearing requirement for staff  taking column (2) of SI Appendix, Table S9  as baseline.
\end{flushleft}  }
\end{figure}

Fig. \ref{fig:sensitivity-death} illustrates the estimated coefficients of visits to colleges and K-12 schools across different specifications for death growth regressions. Fig. \ref{fig:sensitivity-death}(a) shows that the coefficient of visits to colleges and K-12 schools are positively estimated for (1) baseline, (2)-(3) an alternative time lag of  42 and 49 days, (4)  setting the log of weekly deaths to $0$ when we observe zero weekly deaths to compute death growth over 3 weeks,  and adding more controls in (5)-(8), providing robust evidence that an increase in visits to colleges and K-12 schools is positively associated with the subsequent increase in weekly death growth rates.   Fig. \ref{fig:sensitivity-death}(b) corresponds to  Fig. \ref{fig:sensitivity}(b), showing that the association of K-12 school visits with death growth is stronger when no mask mandate for staff is in place.

SI Appendix, Table S9 reports the baseline estimates while Table S10 shows those for the specification with staying home devices and workplace visits. The estimated coefficients on K-12 school visits remain positive and large even after controlling for the variables of returning to work and being outside the home in Table S10, suggesting the possible role of within-school transmission for a rise in deaths after school openings. These results are consistent with our findings in case growth regressions.


\section{Limitations}
Our study has the following limitations. First, our study is observational and, therefore, should be interpreted with great caution. It only has a causal interpretation in a structural model under the unconfoundedness assumptions that might not hold in reality. While we present sensitivity analysis with various controls, including county dummies and interactions of state dummies and week dummies,  the decisions to open K-12 schools may be endogenous and correlated with other unobserved time-varying county-level factors that affect the spread of COVID-19. For example, people's attitudes toward social distancing, hand-washing, and mask-wearing may change over time (which we cannot observe in the data). Their changes may be correlated with school opening decisions beyond the controls we added to our regression specifications.

Our analysis is also limited by the quality and the availability of the data as follows. The reported number of cases is likely to understate true COVID-19 incidence, especially among children and adolescents because they are less likely to be tested than adults given that children exhibit milder or no  symptoms. This is consistent with CDC data which shows the lower testing volume and the higher rate of positive test among children and adolescents than adults \citep{Leidman2021}.  County-level testing data is not used because of a lack of data, although state-week fixed effects control for the weekly difference across counties within the same state and we also control daily state-level test growth rates.

Because foot traffic data is constructed from mobile phone location data, the data on K-12 school visits likely reflects the movements of parents and older children who are allowed to carry mobile phones to schools and excludes those of younger children who do not own mobile phones.\footnote{We also focus on limited Points-Of-Interest: K-12 schools, colleges and universities, restaurants, drinking places, other recreational places including recreational facilities, and churches. We check the robustness by including visits to assisted living facilities for the elderly and nursing care facilities as additional controls, but the results are not sensitive to their inclusion.}

Because COVID-infected children and adolescents are known to be less likely to be hospitalized or die from COVID, the consequence of transmission among children and adolescents driven by school openings crucially depends on whether the transmission of SARS-CoV-2 from infected children and adolescents to the older population can be prevented.\footnote{In the meta-analysis of 54 studies
on the household transmission of SARS-CoV-2 \cite{zachary2020}, estimated household secondary attack rate \textit{to} child contacts was 16.8\%. \cite{miyahara2021} reports that household secondary attack rate \textit{from} children and adolescence to other family members was 23.8\% and higher than other age groups in Japan.}  Our analysis does not provide any empirical analysis on how school opening is associated with the transmission across different age groups due to data limitations.\footnote{CDC collects the data on the number of reported cases by age groups from each state whenever such data is available. However, for many counties, the reported cases by age groups are missing or there exists a substantial gap between the sum of cases across different age groups reported by CDC and the total number of  cases reported in NYT case data (see, for example, the case of Ingham, MI, in  SI Appendix,  Fig. S3).} \cite{vlachos2021} show that teachers in open schools experience higher COVID-19 infection rates compared to teachers in closed schools. They also show that this increase in infection rate also occurs in partners of teachers and parents of students in open schools.

The impact of school openings on the spread of COVID-19 on case growth may be different across counties and over time because it may depend not only on in-school mitigation measures but also on contact tracing,  testing strategies, and the prevalence of community transmissions \citep{Goldhaber-Fiebert2020,ziauddeen2020}.  We do not investigate how the association between school openings and case growths depends on contact tracing and testing strategies at the county level due to data limitation.

The result on the association between school opening and death growth in Fig. \ref{fig:sensitivity-death} is suggestive but must be viewed with caution. The time lag between infection and death is stochastic and spreads over time, making it challenging to uncover the relationship between the timing of school openings and subsequent deaths.  Furthermore, while we provide sensitivity analysis for handling zero weekly deaths to approximate death growth, our construction of the death growth outcome variable remains somewhat arbitrary. %

Finally, our result \textit{does not} imply that K-12 schools should be closed. Closing schools have negative impacts on children's learning \citep{Engzelle2021} and may cause declining physical and mental healths among children and their parents \citep{TAKAKU2021,Fordn614,Gadermanne042871}. On the other hand,
there is emerging evidence of long-term harm on children's health induced by COVID  \citep{parcha2021}.
The decision to open or close K-12 schools requires careful assessments of the cost and the benefit by policymakers. However,  given their relatively low implementation costs,  our findings strongly support policies that enforce masking and other precautionary actions at school and prioritizing vaccines for education workers and elderly parents/grandparents.

\begin{table}[!htbp] \centering
 \caption{The Association of School/College Openings  with Full-time Workplace Visits and Staying Home   in the United States: Standard Fixed Effects Estimator without Bias Correction}\vspace{-0.3cm}
 \label{tab:PItoB}
 \smallskip
\resizebox{0.7\columnwidth}{!}{
 \begin{tabular}{@{\extracolsep{1pt}}lcc|cc}
\\[-1.8ex]\hline
\hline
 & \multicolumn{4}{c}{\textit{Dependent variable}} \\
\cline{2-5}
 & Full Time  & Full Time  &Stay Home& Stay Home \\
  & (1) & (2) & (3) & (4)\\
\hline \\[-1.8ex]
  K-12 School Visits & 0.085$^{***}$ &  & $-$0.019 &  \\
  & (0.006) &  & (0.026) &  \\
  Open K-12 In-person &  & 0.999$^{***}$ &  & $-$0.924$^{**}$ \\
  &  & (0.125) &  & (0.382) \\
  Open K-12 Hybrid &  & 0.490$^{***}$ &  & $-$0.127 \\
  &  & (0.051) &  & (0.186) \\
  Open K-12 Remote &  & 0.236$^{***}$ &  & $-$0.271 \\
  &  & (0.048) &  & (0.307) \\ \hline
  College Visits & $-$0.040$^{***}$ & $-$0.046$^{***}$ & $-$0.144$^{***}$ & $-$0.148$^{***}$ \\
  & (0.004) & (0.006) & (0.024) & (0.026) \\
 Mandatory mask & $-$0.057 & $-$0.141$^{**}$ & 0.016 & 0.036 \\
  & (0.042) & (0.053) & (0.259) & (0.250) \\
  Ban gatherings & 0.060 & 0.075 & 0.436 & 0.351 \\
  & (0.047) & (0.051) & (0.560) & (0.520) \\
  Stay at Home & $-$0.060$^{*}$ & $-$0.066$^{*}$ & 2.793$^{***}$ & 2.809$^{***}$ \\
  & (0.031) & (0.033) & (0.329) & (0.340) \\ \hline
  log(Cases) & 0.005 & 0.003 & 0.288$^{***}$ & 0.282$^{***}$ \\
  & (0.004) & (0.005) & (0.028) & (0.028) \\
  llog(Cases), 7d lag & $-$0.001 & $-$0.005$^{*}$ & 0.209$^{***}$ & 0.207$^{***}$ \\
  & (0.002) & (0.003) & (0.019) & (0.017) \\
log(Cases), 14d lag & $-$0.0005 & $-$0.003 & 0.098$^{***}$ & 0.097$^{***}$ \\
  & (0.002) & (0.002) & (0.023) & (0.024) \\ 
   \hline \\[-1.8ex]
Observations & 670,895 & 595,872 & 670,895 & 595,872 \\
R$^{2}$ & 0.870 & 0.853 & 0.889 & 0.888 \\
\hline
\hline \\[-1.8ex]  
\end{tabular}
}
{\scriptsize \begin{flushleft}
Notes:  Dependent variables are full-time workplace visits and staying home devices per residing device. All regression specifications include county fixed effects and state-week fixed effects,. The standard fixed effects estimator without bias correction is used. Clustered standard errors at the state level are reported in the bracket.  {$^{*}$p$<$0.1; $^{**}$p$<$0.05; $^{***}$p$<$0.01}
\end{flushleft}}
\end{table}

\begin{table}[!htbp] \centering
 \caption{The Association of School/College Openings, Full-time/Part-time Work, and Staying Home with Case Growth in the United States: Debiased  Estimator}\vspace{-0.3cm}
 \label{tab:PBItoY}
\resizebox{0.7\columnwidth}{!}{
\begin{tabular}{@{\extracolsep{1pt}}lcc|cc}
\\[-1.8ex]\hline
\hline \\ [-1.8ex]
 & \multicolumn{4}{c}{\textit{Dependent variable:  \textbf{Case Growth Rates}}} \\
\cline{2-5}
& (1) & (2) & (3) & (4)\\
\hline 
   K-12 Visits, 14d  lag & 0.393$^{***}$ & 0.283$^{***}$ &  &  \\
  & (0.075) & (0.087) &  &  \\
   K-12 Visits $\times$ No-Mask, 14d  lag &  & 0.287$^{***}$ &  &  \\
  &  & (0.071) &  &  \\
  K-12  In-person, 14d  lag&  &  & 0.015 & $-$0.007 \\
  &  &  & (0.016) & (0.020) \\
  K-12  Hybrid, 14d  lag &  &  & $-$0.028$^{**}$ & $-$0.055$^{***}$ \\
  &  &  & (0.013) & (0.013) \\
  K-12  Remote, 14d  lag &  &  & $-$0.094$^{***}$ & $-$0.115$^{***}$ \\
  &  &  & (0.015) & (0.014) \\
  K-12   In-person $\times$ No-Mask, 14d  lag &  &  &  & 0.034$^{*}$ \\
  &  &  &  & (0.020) \\
  K-12   Hybrid $\times$ No-Mask, 14d  lag &  &  &  & 0.043$^{***}$ \\
  &  &  &  & (0.017) \\ \hline
  Full-time Work Device, 14d  lag& $-$0.117 & 0.186 & 0.956$^{**}$ & 0.967$^{**}$ \\
  & (0.417) & (0.490) & (0.384) & (0.436) \\
 Part-time Work Device, 14d  lag & 0.262 & 0.466 & 0.820$^{***}$ & 0.915$^{***}$ \\
  & (0.259) & (0.305) & (0.276) & (0.309) \\
 Staying Home Device, 14d  lag& $-$0.290$^{***}$ & $-$0.283$^{***}$ & $-$0.352$^{***}$ & $-$0.332$^{***}$ \\
  & (0.057) & (0.069) & (0.061) & (0.067) \\ \hline 
Observations & 690,297 & 545,131 & 612,963 & 528,941 \\
R$^{2}$ & 0.092 & 0.093 & 0.092 & 0.094 \\
\hline
\hline 
\end{tabular}}
  {\scriptsize
\begin{flushleft}
Notes: Dependent variable is the log difference over 7 days in weekly positive cases. All regression specifications include county fixed effects and state-week fixed effects, college visits, three NPIs, and 2, 3, and 4 weeks lagged log of cases. See SI Appendix, Table S8 for the estimated coefficients for NPIs and the log of current and past cases.
The debiased fixed effects estimator is applied.  Asymptotic clustered standard errors at the state level are reported in the bracket.  {$^{*}$p$<$0.1; $^{**}$p$<$0.05; $^{***}$p$<$0.01}
\end{flushleft}}
\end{table}

\begin{footnotesize}

\bibliographystyle{jpe}
\bibliography{covid}

\end{footnotesize}

\newpage

\section{Supplementary Information Appendix}

\subsection*{The Model and Methods}\label{sec:causal-mode-SI}

\subsubsection*{The Structural Causal Model}
Our approach draws on the framework presented in our previous paper \cite{chernozhukov2021}. Here we summarize the approach 
for completeness, highlighting the main difference  (here we do not assume that all relevant social distancing behavioral variables are observed).

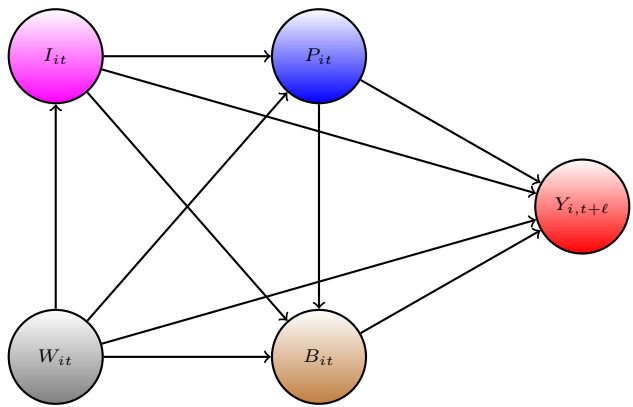
\begin{figure}[ht]
\begin{center}
\begin {tikzpicture}[-latex, auto, node distance =2cm and 3.5cm, on grid, thick,
  empty/.style ={circle, top color=white, bottom color = white, draw, white, text=white , minimum width =1.25 cm},
  policy/.style ={circle, top color=white, bottom color = blue, draw, black, text=black , minimum width =1.25 cm},
    behavior/.style ={circle, top color=white, bottom color = brown, draw, black, text=black , minimum width =1.25 cm},
   observed/.style ={circle, top color=white, bottom color = magenta, draw, black, text=black , minimum width =1.25 cm} ,
   confounder/.style ={circle, top color=white, bottom color =  gray, draw, black, text=black , minimum width =1.25 cm} ,
outcome/.style ={circle, top color=white, bottom color = red, draw, black, text=black , minimum width =1.25 cm} ]

\node[policy]   (P) {\tiny $P_{it}$};
\node[empty] (E) [below=of P] {\tiny $I_{it}$};
\node[outcome]  (Y) [right =of E] {\tiny $Y_{i,t+\ell}$};
\node[behavior] (B) [below =of E] {\tiny $B_{it}$};
\node[observed] (I) [left =of P] {\tiny $I_{it}$};
\node[confounder] (W) [left=of B] {\tiny  $W_{it}$};
\path[->] (P) edge (Y);
\path[->] (B) edge (Y);
\path[->] (P) edge (B);
\path[->] (I) edge (B);
\path[->] (I) edge (P);
\path[->] (W) edge (Y);
\path[->] (W) edge (B);
\path[->] (W) edge (P);
\path[->] (W) edge (I);
\path[->] (I) edge (Y);

\end{tikzpicture} \end{center}
\caption{The causal path diagram for our model.}\label{Wright}
\end{figure}

We begin with a qualitative description of the model via a causal path diagram shown in Figure \ref{Wright}, which describes how policies, behavior, and information interact together:
\begin{itemize}
\item The \textit{forward} health outcome,
$Y_{i,t+\ell}$, is determined last, after all other variables have been determined;
\item The  adopted vector of policies, $P_{it}$,  affect health outcome $Y_{i,t+\ell}$ either directly, or indirectly by altering  individual distancing and other precautioanry behavior  $B_{it}$, which may be only partially observed;
\item  Information variables, $I_{it}$, such as lagged values of outcomes and other lagged observable variables (see robustness checks) can affect human behavior and  policies, as well as  outcomes;
\item The confounding factors $W_{it}$, which vary across counties and time, affect all other variables; these include
unobserved though estimable county, time, state, state-week effects.
\end{itemize}
The index $i$ denotes observational unit, the county, and $t$ and $t+\ell$ denotes the time, where  $\ell$ represents the typical time lag  between infection and case confirmation or death.

Our main outcomes of interest are the growth rates in Covid-19 cases and deaths and policy variables include school reopening in various modes, mask mandates, ban gathering, and stay-at-home orders,  and the information variables include lagged values of outcome (as well as other variables described in the sensitivity checks). 

The role of behavioral variables in the model is two-fold. First, the presence of these variables in the model requires us to control for the information variables -- even when information variables affect outcomes only through policies or behavior.  In this case conditioning on the information blocks the backdoor path (see, \cite{pearl:causality}) creating confounding $${\ycolor Y_{i,t+\ell}} \longleftarrow {\bcolor B_{it}} \longleftarrow  {\icolor I_{it}} \longrightarrow {\pcolor P_{it}}.$$  Therefore conditioning on the information is important even when there is no direct effect  ${\icolor I_{it}} \longrightarrow {\ycolor Y_{i,t+\ell}}$. This observation motivates our main dynamic specification below, where information variables include lagged growth rates and new cases or new deaths per capita. Second, while not all behavioral variables may be observable, we can still study as the matter of supporting analysis, the effects of policies on observed behavioral variables (the portion of time in workplace, restaurants, and bars) and of behavioral variables on outcomes, thereby gaining insight as to whether policies have changed private behavior and to what extent this private behavior changed the outcomes  (for the analysis, of early pandemic data in this vein, see our previous paper).

The causal structure allows for the effect of the policy to be either direct or indirect. The structure also allows for changes in behavior to be brought by the change in policies and information. These are all realistic properties that we expect from the context of the problem. Policies such as closures and reopenings of schools, closures or reopening of non-essential business, and restaurants, affect the behavior in strong ways.  In contrast, policies such as mandating employees to wear masks can potentially affect the Covid-19 transmission directly.  The information variables, such as recent growth in the number of cases, can cause people to spend more time at home, regardless of adopted policies; these changes in
behavior, in turn, affect the transmission of Covid-19.

The causal ordering induced by this directed acyclical graph is determined by the following
timing sequence: 
\begin{itemize}
\item[(1)]  information and confounders get determined at $t$,
\item[(2)] policies are set in place, given information and confounders at $t$;
\item[(3)] behavior is realized, given policies, information, and confounders at $t$;
\item[(4)] outcomes get realized at $t+\ell$ given policies, behavior, information, and confounders.
\end{itemize}

The model also allows for direct dynamic effects of information variables on the outcome through autoregressive structures that capture persistence in growth patterns. We do not highlight these dynamic effects and only study the short-term effects (longer-run effects get typically amplified; see our previous paper \cite{chernozhukov2021} for more details.)

Our quantitative model for causal structure in Figure \ref{Wright} is given by the following econometric structural equation model:\begin{equation} \label{eqn:SEM1} \tag{SEM}
  \begin{aligned}
   &  {\ycolor Y_{i,t+\ell}} (b,p,\iota) &:=& {\bcolor \alpha ' b}  +  {\pcolor \pi 'p} +
    {\icolor \mu'\iota } + {\wcolor \delta_Y 'W_{it}} + \varepsilon^y_{it}, \\
   &  {\bcolor B_{it}} (p,\iota) &:= & {\pcolor \beta'p } + {\icolor \gamma'\iota} +      {\wcolor\delta_B 'W_{it} } + \varepsilon^b_{it}, \\
   & {\pcolor P_{it}}   (\iota) & := &  p ({\icolor \eta'\iota}, {\wcolor W_{it}},  \varepsilon^p_{it} ), 
         \end{aligned}
 \end{equation}
which is a collection of structural potential response functions (potential outcomes), where the stochastic schocks
are decomposed into an observable part $\delta' W$ and unobservable part $\varepsilon$.  Lower case letters
$\iota$, $b$ and $p$ denote the potential values of information, behavior, and policy variables. The restrictions on shocks are described below. 

The observed outcomes, policy, and behavior variables are generated by setting $\iota= {\icolor I_{it}}$ and propagating
the system from the last equation to the first:
\[
  \begin{aligned}
& {\ycolor Y_{i,t+\ell}}  & := & Y_{i,t+\ell} ( {\bcolor B_{it} } ,{\pcolor P_{it}}, {\icolor I_{it}}), \\
& {\bcolor B_{it} } & := &   B_{it}({\pcolor P_{it} } ,{\icolor I_{it}}), \\
& {\pcolor P_{it} }& := &  P_{it}({\icolor I_{it}}). \end{aligned}
\]

The orthogonality restrictions on the stochastic components are as follows: The stochastic shocks $\varepsilon^y_{it}$
and  $\varepsilon^p_{it}$ are centered and furthermore,
\begin{equation}\label{eqn:SEM2} \tag{O}
\begin{aligned}
   & \varepsilon^y_{it} &  \perp &  \quad (\varepsilon^b_{it}, P_{it}, {\wcolor W_{it}}, {\icolor I_{it}}), \\
&  \varepsilon^b_{it}  & \perp & \quad  (P_{it}, {\wcolor W_{it}}, {\icolor I_{it}}), \\
&   \varepsilon^p_{it} &  \indep &  \quad ({\wcolor W_{it}}, {\icolor I_{it}}),
\end{aligned}
\end{equation}
where we say that $V \perp U$ if $\Ep VU = 0$. This is a standard way of representing restrictions on errors in structural equation modeling. The last equation states that variation in policies is exogenous conditionally on confounders and information variables.


The system above together with  orthogonality restrictions (\ref{eqn:SEM2}) implies the following collection of stochastic equations for realized variables:
\begin{align}
   &  {\ycolor  Y_{i,t+\ell}}
    = {\bcolor\alpha ' B_{it}} + {\pcolor\pi 'P_{it}} + {\icolor\mu'I_{it}} + {\wcolor\delta_Y 'W_{it}}  + \varepsilon^y_{it},
    &  & \varepsilon^y_{it} \perp {\bcolor B_{it}}, {\pcolor P_{it}}, {\icolor I_{it}}, {\wcolor W_{it}} \label{eq:R1} \tag{BPI$\to$Y} \\
    &  {\bcolor B_{it}}
     =  {\pcolor \beta' P_{it}} + {\icolor \gamma'I_{it}} +  {\wcolor \delta_B' W_{it}} + \varepsilon^b_{it},
   & & \varepsilon^b_{it} \perp {\pcolor P_{it}}, {\icolor I_{it}}, {\wcolor W_{it}}  \label{eq:R2} \tag{PI$\to$B} 
         \end{align}

As discussed below, the information variable includes case growth. Therefore, the orthogonality restriction  $ \varepsilon^y_{it} \perp   {\pcolor P_{it}}$ holds if  the government does not have knowledge on future case growth beyond what is predicted by the information set and the  confounders; even when the government has some knowledge on $\varepsilon^y_{it}$, the orthogonality restriction may hold if there is a time lag for the government to implement its policies based on $\varepsilon^y_{it}$.

We stress that our  main analysis does not require all components of ${\bcolor B_{it}}$ to be observable. 

\textbf{Main Implication.} The model stated above implies the following projection equation:

\begin{align}
   {\ycolor  Y_{i,t+\ell}}
   =    a'  {\pcolor P_{it}} +b'  {\icolor I_{it} }+ c' {\wcolor W_{it}}  + {\bar \varepsilon}_{it},  \quad   {\bar \varepsilon}_{it} \perp
  {\pcolor P_{it}},  {\icolor I_{it}}, {\wcolor W_{it}},  \label{eq:R4} \tag{PI$\to$Y}
     \end{align}
where 
$$
a' := ( {\bcolor\alpha '}  {\pcolor \beta' }+{\pcolor\pi'} ), \quad b' := ( {\bcolor\alpha '}  {\icolor \gamma'} + {\icolor \mu'}),
\quad c' :=  ( {\bcolor\alpha '}  {\wcolor \delta_B' }+{\wcolor\delta_Y'} )
$$
This follows immediately from plugging equation (PI $\to$ B) to  equation (BPI $\to$ Y) and verifying
that the composite stochastic shock $\bar \varepsilon_{it}$ obeys the orthogonality condition stated 
in  (\ref{eq:R4}).

The main parameter of interest is the structural causal effect of the policy:
$$
a' = ( {\bcolor\alpha '}  {\pcolor \beta' }+{\pcolor\pi'} ).
$$
It comprises direct policy effect ${\pcolor\pi'}$ as well as the indirect effect $ {\bcolor\alpha '}  {\pcolor \beta' }$, realized by 
the policy changing observed and unobserved behavior variables ${\bcolor B_{it}}$.   This coefficient $a$ and $b$ can estimated directly using the dynamic panel data methods described in more detail below.

As additional analysis, we can estimate the determinants for the observed behavioral mobility measures-- the observed part of  ${\bcolor B_{it}}$.


\subsubsection*{Identification and Parameter Estimation}

The orthogonality equations imply that the main equation is the projection equation, and parameters $a$ and $b$ are identified if ${\pcolor P_{it}}$ and ${\icolor I_{it}}$ have sufficient variation left after partialling out the effect of controls:
\begin{equation}\label{eqn:SEM-PO}
 \begin{aligned}
   & {\ycolor { \tilde Y_{i,t+\ell}}}  & =& a '{\pcolor {\tilde P_{it}}} +c' {\icolor \tilde I_{it} } + \bar \varepsilon_{it},&\bar \varepsilon_{it} &\perp{\tilde  P_{it}}, { \tilde I_{it}},  \\
   \end{aligned}
\end{equation}
where $ \tilde V_{it} = V_{it}   -     {\wcolor W_{it}'} \Ep[{\wcolor W_{it}W_{it}'}]^{-} \Ep[{\wcolor W_{it}} V_{it}]$ denotes
the residual after removing the orthogonal projection of $V_{it}$ on ${\wcolor W_{it}}$. The residualization is a linear operator, implying that (\ref{eqn:SEM-PO}) follows immediately from the above. The parameters of (\ref{eqn:SEM-PO})  are identified as projection coefficients in these equations, provided that residualized vectors have non-singular variance 
matrix: \begin{equation}
 \ \Var ({\pcolor \tilde P_{it}'}, {\icolor \tilde I_{it}'})> 0.
 \end{equation}

Our main estimation method is the fixed effects estimator, where the county, state, state-week effects are treated as 
unobserved components of ${\wcolor W_{it}}$ and estimated directly from the panel data, so they are rendered 
(approximately) observable once the history is sufficiently long. The stochastic shocks $\{ \varepsilon_{it}\}_{t=1}^T$
are treated as independent across states and can be arbitrarily dependent across time $t$ within a state.  In other words, the standard errors will be clustered at the state level.    When histories are not long,  substantial biases emerge 
from working with the estimated version ${\wcolor \widehat W_{it}}$ of ${\wcolor W_{it}}$  (known as the Nickel bias \citep{Nickell1981}) and they need to be removed using debiasing methods. In our context, debiasing changes  the magnitudes of the original biased fixed effect estimator but does not change the qualitative conclusions reached without any debiasing. 

\subsection*{Formulating Outcome and Key Confounders via SIR model}\label{sec:sirmodel}
Letting $C_{it}$ denote the cumulative number of  confirmed cases in county $i$ at time $t$, our outcome
\begin{equation} \label{eq:y}
 \Delta_7 \log(\Delta_7 C_{it}):= \log( \Delta_7 C_{it} ) -
\log( \Delta_7 C_{i,t-7})
\end{equation}
approximates the weekly growth rate in new cases from $t-7$ to $t$.\footnote{We may show that $ \log( \Delta_7 C_{it} ) -
\log( \Delta_7 C_{i,t-7})$ approximates the average
growth rate of cases from $t-7$ to $t$.} Here $\Delta_7$ denotes the differencing operator over 7 days from $t$ to $t-7$, so that $\Delta_7 C_{it}:=C_{it}-C_{i,t-7}$ is the number of new confirmed cases from day $t$ and day $t-6$.

We chose this metric as this is the key metric for policymakers deciding when to relax Covid mitigation policies.  The U.S. government's guidelines for state reopening
recommend that states display a
``downward trajectory of documented cases within a 14-day period''
\citep{whitehouse2020}. A negative value of
$Y_{it}$ is an indication of meeting these criteria for reopening. By focusing on weekly cases rather than daily cases, we smooth idiosyncratic daily fluctuations as well as periodic fluctuations associated with the days of the week.

Our measurement equation for estimating equations (\ref{eq:R1}) and (\ref{eq:R4}) will take the form:
\begin{align}
{\ycolor \Delta_7 \log(\Delta_7 C_{it})}  =    X_{i,t-14} '   \theta  +  \delta_T \Delta_7   \log(T_{it})  + \epsilon_{it},
 \label{eq:M} \tag{M-C}
\end{align}
where $i$ is county, $t$ is day, $C_{it}$ is cumulative confirmed
cases, $T_{it}$ is the number of tests over 7 days, $\Delta$ is
a 7-days differencing operator, $\epsilon_{it}$ is an unobserved error term.
 $X_{i,t-14}$  collects other behavioral, policy, and confounding variables, depending
on whether we estimate (\ref{eq:R1}) or (\ref{eq:R4}), where the lag of $14$ days captures the time lag between infection and confirmed case (see \cite{midas2020}). 
   Here
$$\Delta_7   \log(T_{it} ):=  \log(T_{it}) - \log(T_{i,t-7})  $$ 
is the key confounding variable,
derived from considering the SIR model below. We describe other confounders in the empirical analysis section.


 Our main estimating equation (\ref{eq:M}) is motivated by a variant of SIR
model, where we add confirmed cases and infection detection via testing.
Let $S$, $\Infected$, $\Recovered$, and $D$ denote the number of susceptible,
infected, recovered, and dead individuals in a given state. Each of these variables are a function of time. We model
them as evolving as
\begin{align}
  \dot{S}(t) & = -\frac{S(t)}{N} \beta(t) \Infected(t) \label{eq:s} \\
  \dot{\Infected}(t) & = \frac{S(t)}{N} \beta(t) \Infected(t) - \gamma  \Infected(t) \label{eq:i}\\
  \dot{\Recovered}(t) & = (1-\kappa) \gamma  \Infected(t) \label{eq:r}\\ 
  \dot{D}(t) & = \kappa \gamma \Infected(t) 
  \label{eq:d}
\end{align}
where $N$ is the population, $\beta(t)$ is the rate of infection
spread, $\gamma$ is the rate of recovery or death, and $\kappa$ is the
probability of death conditional on infection.

Confirmed cases, $C(t)$, evolve as
\begin{equation}
  \dot{C}(t) = \tau(t) \Infected(t), \label{eq:c}
\end{equation}
where $\tau(t)$ is the rate that infections are detected.

Our goal is to examine how the rate of infection $\beta(t)$ varies with observed policies
and measures of social distancing behavior. A key challenge is that we only
observed $C(t)$ and $D(t)$, but not $\Infected(t)$. The unobserved $\Infected(t)$ can
be eliminated by differentiating (\ref{eq:c}) and using (\ref{eq:i})  as
\begin{align}
  \frac{\ddot{C}(t)}{\dot{C}(t)}
              & =
                \frac{S(t)}{N} \beta(t) -\gamma  + \frac{\dot{\tau}(t)}{\tau(t)}. \label{eq:c2}
\end{align}
We consider a discrete-time analogue of equation (\ref{eq:c2}) to motivate our empirical
specification by relating the detection rate $\tau(t)$  to the number of tests $T_{it}$ while specifying $\frac{S(t)}{N}\beta(t)$ as a linear function of variables $X_{i,t-14}$.
This results in
\begin{align}
  \underbracket{\Delta_7 \log(\Delta_7 C_{it})}_{\frac{\ddot{C}(t)}{\dot{C}(t)}}
  =
      \underbracket{X_{i,t-14}' \theta + \epsilon_{it}}_{\frac{S(t)}{N}\beta(t) -\gamma}
       +
       & \underbracket{\delta_T \Delta
      \log(T)_{it}}_{\frac{\dot{\tau}(t)}{\tau(t)} } \nonumber
\end{align}
which is equation (\ref{eq:M}), where $X_{i,t-14}$ captures a vector of variables related to $\beta(t)$.

\begin{quote}
\textsc{Structural Interpretation}. The component $X_{i,t-14}' \theta$
is the projection of $\beta_i(t)S_{i}(t)/N_{i}(t)  - \gamma$ on   $X_{i,t-14}$ (including
testing variable).
\end{quote}

\textbf{Growth Rate in Deaths as Outcome}. By differentiating (\ref{eq:d}) and (\ref{eq:c}) with respect to $t$ and using (\ref{eq:c2}), we obtain
\begin{align}
\frac{\ddot{D}(t) }{\dot D(t)}& = \frac{\ddot{C}(t) }{\dot C(t)}  - \frac{\dot{\tau}(t) }{ \tau(t)}    =  \frac{S(t)}{N}\beta(t)  -   \gamma.\label{eq:d2}
\end{align}
Our measurement equation for the growth rate of deaths is based on equation (\ref{eq:d2}) but   account for a $35$ day lag between infection and death as
\begin{align}
{\ycolor \Delta_{21} \log(\Delta_7 D_{it})}  = X_{i,t-35}' \theta + \epsilon_{it},\label{eq:M-D} \tag{M-D}
\end{align}
where
\begin{equation} \label{eq:y-d}
 \Delta_{21} \log(\Delta_7 D_{it}):= \log( \Delta_7 D_{it} ) -
\log( \Delta_7 D_{i,t-21})
\end{equation}
approximates the weekly growth rate in deaths from $t-7$ to $t$ in state $i$.  Sensitivity analysis also provides
results for the case of $28$ and $35$ lag.

\subsection*{Debiased Fixed Effects Dynamic Panel Data Estimator}
We apply Jackknife bias corrections;  see \cite{chen2020}  and  \cite{HahnNewey2004} for more details. Here, we briefly describe the debiased fixed effects estimator we use.

Given our panel data with sample size $(N,T)$, denote a set of counties by $\mathcal{N}=\{1,2,...,N\}$.  We randomly and repeatedly partition $\mathcal{N}$ into two sets as $\mathcal{N}_1^j$ and $\mathcal{N}_2^j=\mathcal{N}\setminus \mathcal{N}_1^j$ for $j=1,2,...,J$, where $\mathcal{N}_1^j$ and $\mathcal{N}_2^j$ (approximately) contain the same number of counties.  For each of $j=1,...,J$, consider two sub-panels 
(where $i$ stands for county and $t$ stands for the day)  defined by
$${\bf S}_1^j = {\bf S}_{11}^j  \cup {\bf S}_{22}^j\quad \text{ and }\quad {\bf S}_2^j  = {\bf S}_{12} ^j \cup {\bf S}_{21}^j $$ with ${\bf S}_{1k} ^j=  \{(i,t) :  i \in \mathcal{N}_k,  t \leq \lceil T/2 \rceil \}$ and 
${\bf S}_{2k} ^j=  \{(i,t) :  i \in \mathcal{N}_k,  t  \geq \lfloor T/2 + 1\rfloor\}$ for $k=1,2$, where $\lceil . \rceil$  and  $\lfloor . \rfloor$ are the ceiling and floor functions.  Each of these two subpanels, ${\bf S}_1^j$ and ${\bf S}_2^j$,  includes observations for all cross-sectional units and time periods. 
 
 We form the estimator with bias-correction as
$$
\widehat \beta_{\rm BC}  :=  2 \widehat \beta -  \widetilde \beta\quad\text{with}\quad
 \widetilde \beta := \frac{1}{J}\sum_{j=1}^J\widetilde \beta_{{\bf S}^j_1\cup{\bf S}_2^j},
$$
where $\widehat \beta$ is the standard estimator with  a set of $N$ county dummies while $\widetilde \beta_{{\bf S}_1^j\cup{\bf S}_2^j}$ denotes the estimator using the data set  ${\bf S}^j_1\cup{\bf S}_2^j$ but treats the counties in  ${\bf S}^j_1$ differently from those in  ${\bf S}^j_2$ to form the estimator--- namely, we include approximately $2N$ county dummies to compute $\widetilde \beta_{{\bf S}^j_1\cup{\bf S}^j_2}$. Thus, $(\widehat \beta -  \widetilde \beta)$ is the approximation to the bias of $\widehat \beta$, subtracting which from $\widehat \beta$ gives the formula given above. 
We set $J=2$ in our empirical analysis. When we choose $J=5$ for some specifications, we  obtained similar results.
 
An alternative  jacknife bias-corrected estimator is $\widehat \beta_{\rm CBC}  =  2 \widehat \beta -  \frac{1}{J}\sum_{j=1}^J(\widetilde \beta_{{\bf S}_1^j}+\widetilde \beta_{{\bf S}_2^j})/2$, where $\widetilde \beta_{{\bf S}_k^j}$  denotes the fixed effect estimator using the subpanel  ${\bf S}_k^j$ for $k=1,2$. In our empirical analysis, these two cross-over jackknife bias corrected estimators give  similar result; in simulation experiments, the first form performed somewhat better, so we settled out choice on it.
  
  We report asymptotic standard errors with state-level clustering, justified by the standard asymptotic theory of bias corrected estimators. The rationale for state-level clustering is that the stochastic shocks in the model can be correlated across counties, especially within the state.  A simple way to model this is to allow for the arbitrary within-state correlation and adjust the standard errors to account for this (state-level clustering).

\subsection*{Discussion on the Effect of College Visits with Cases and Deaths}

\begin{figure}[!ht]
  \caption{The number of cases by age groups and the number of visits to colleges/universities and bars in Dane county, WI, and  Lane county, OR\label{fig:dane}}
  \resizebox{0.9\columnwidth}{!}{
\begin{minipage}{\linewidth}
        \begin{tabular}{c|c}
         \textbf{  Dane County, WI } & \textbf{Lane County, OR }\\
  Cases by Age Groups &  Cases by Age Groups  \\
      \includegraphics[width=0.50\textwidth]{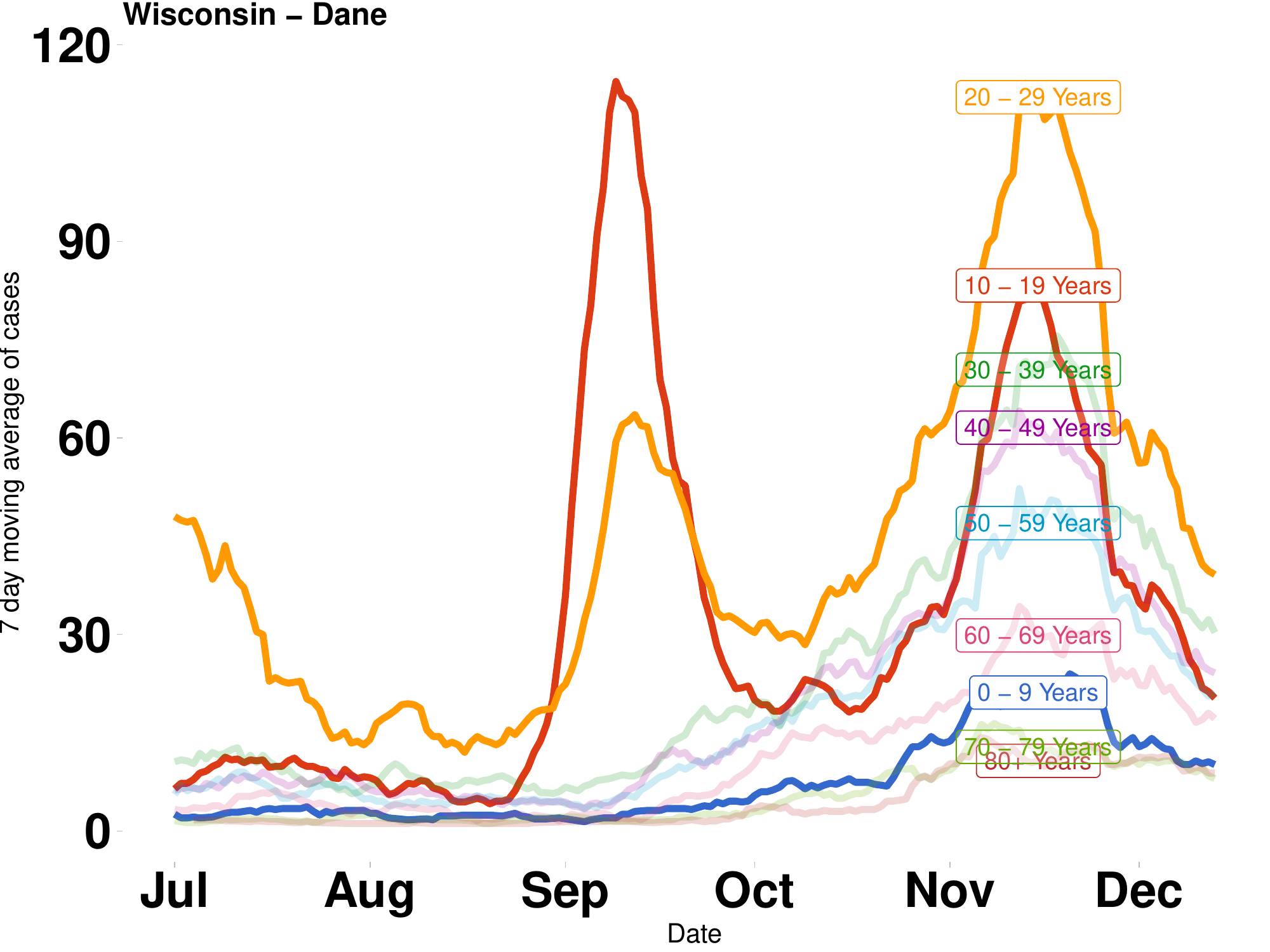}&
      \includegraphics[width=0.50\textwidth]{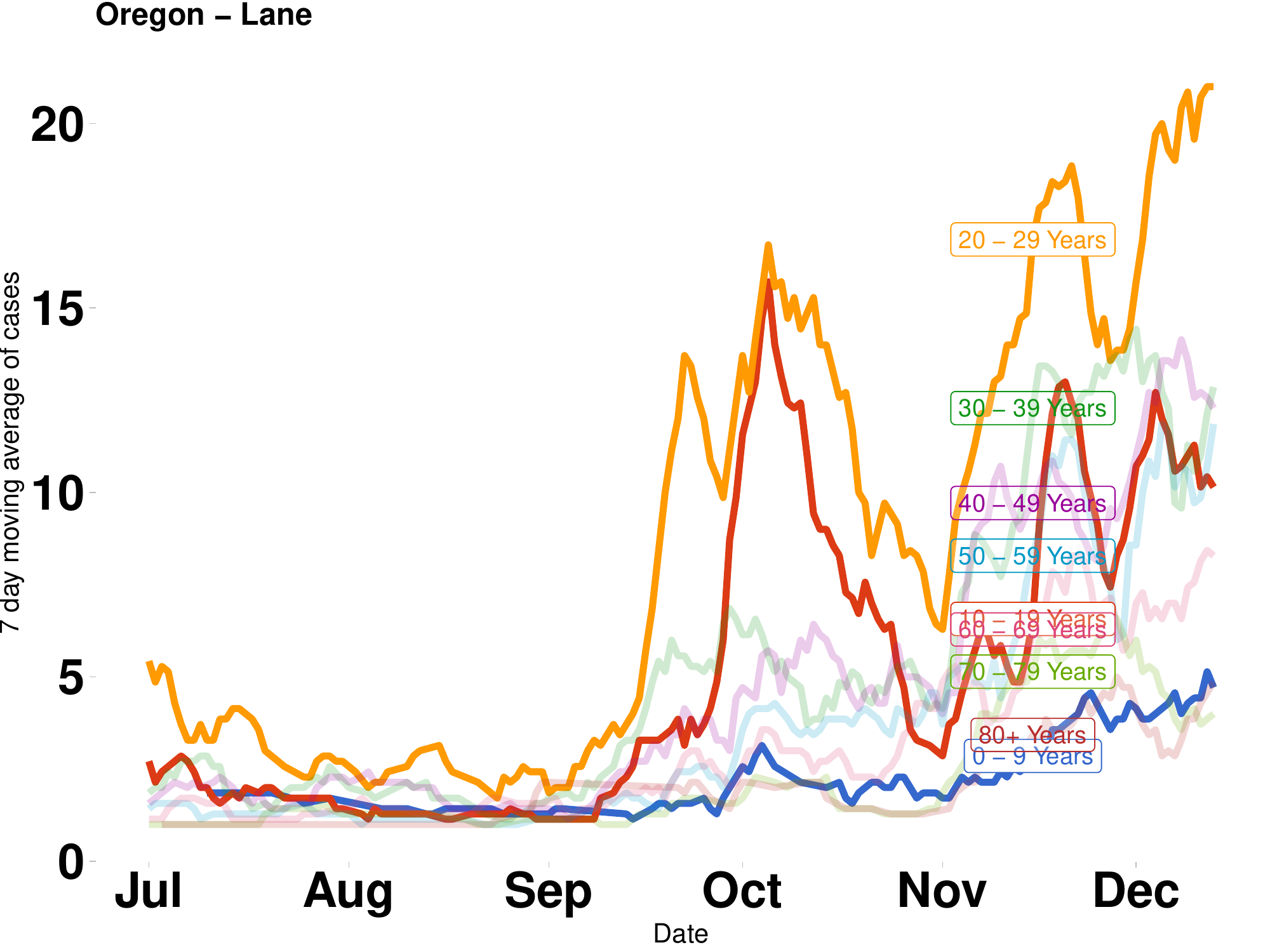}\\
    College Visits  &  College Visits  \\
      \includegraphics[width=0.50\textwidth,height=0.2\textwidth]{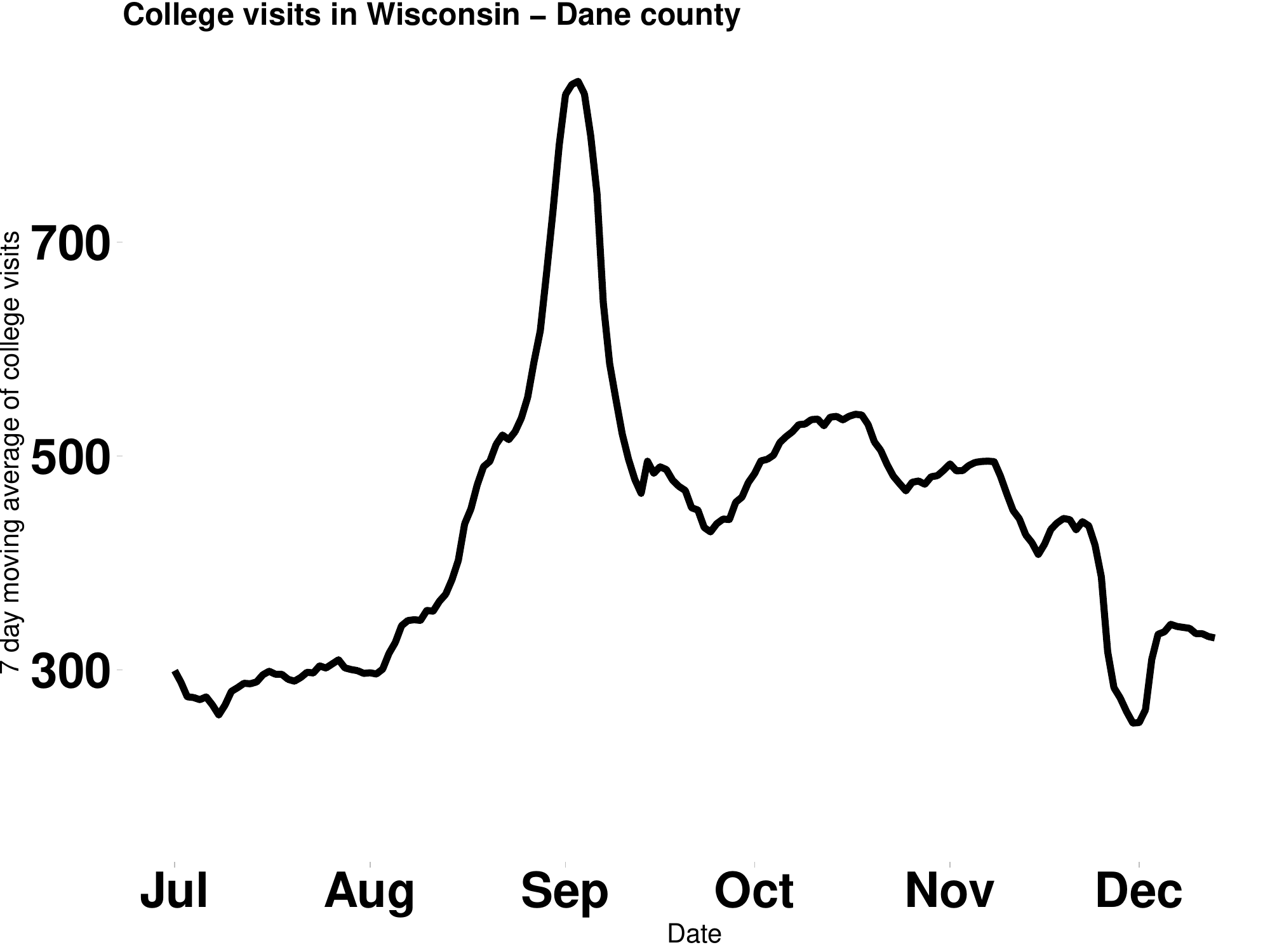}&
      \includegraphics[width=0.50\textwidth,height=0.2\textwidth]{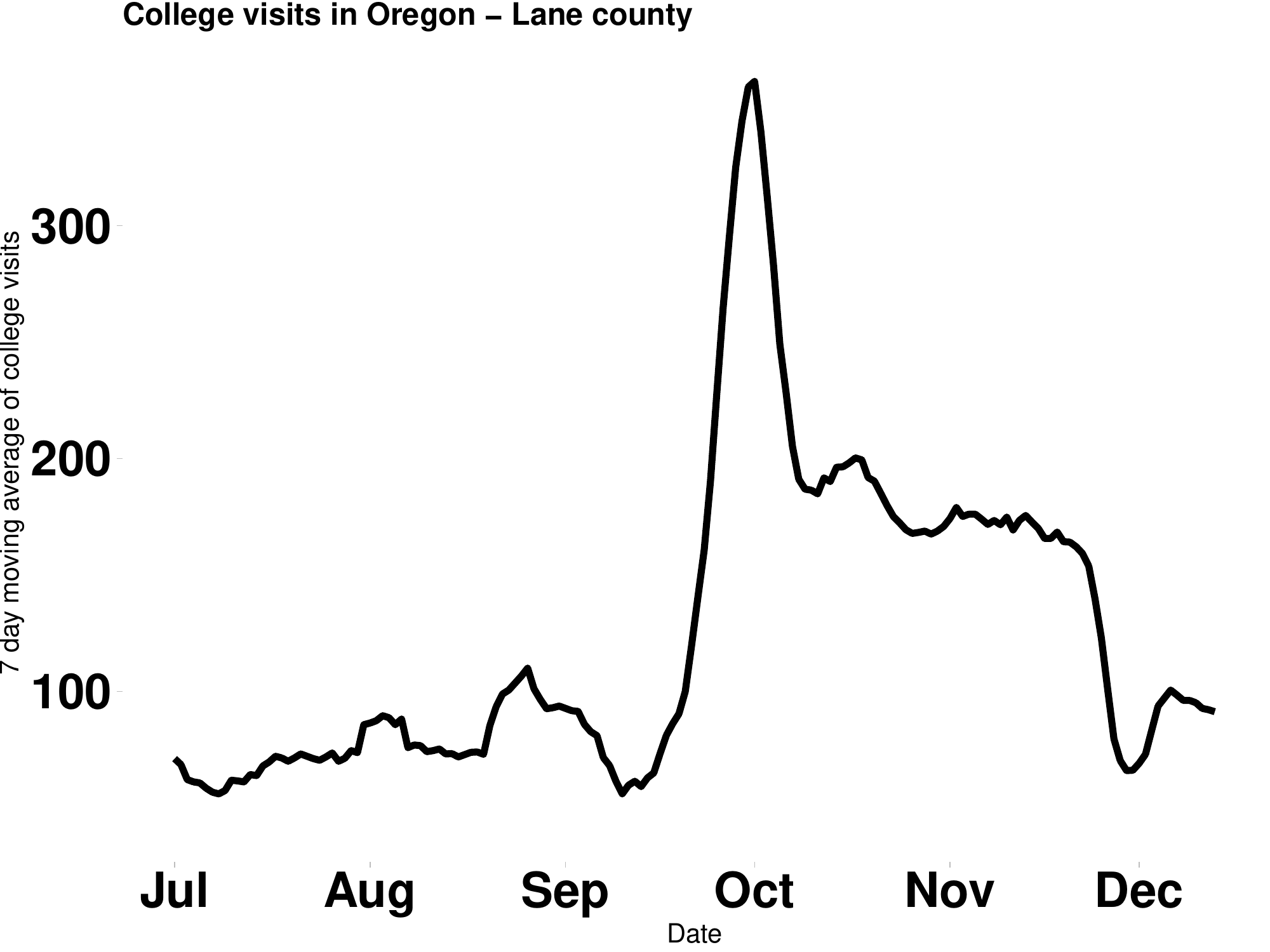}\\
  Bar Visits  &  Bar Visits   \\
      \includegraphics[width=0.50\textwidth,height=0.2\textwidth]{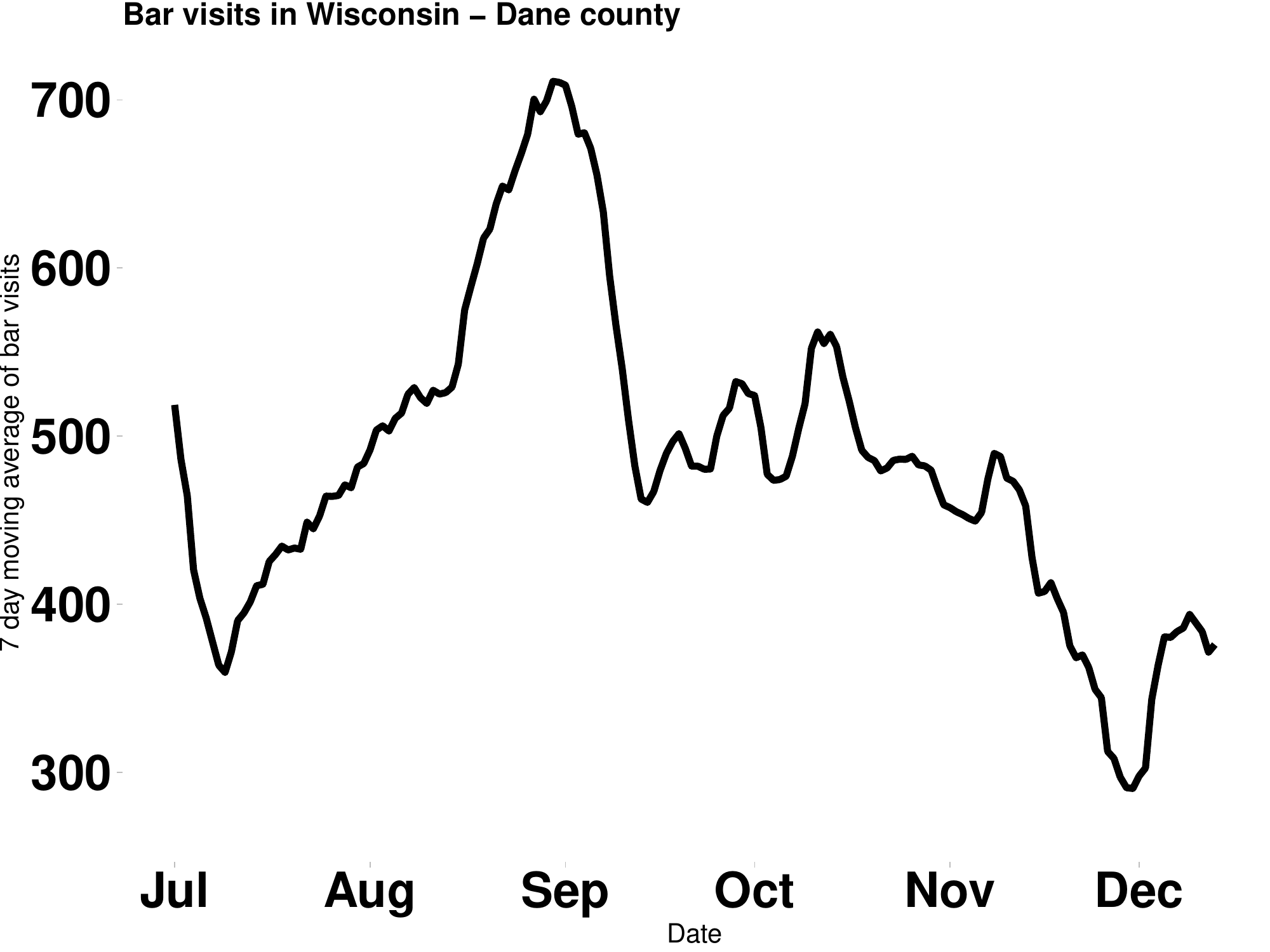}&
      \includegraphics[width=0.50\textwidth,height=0.2\textwidth]{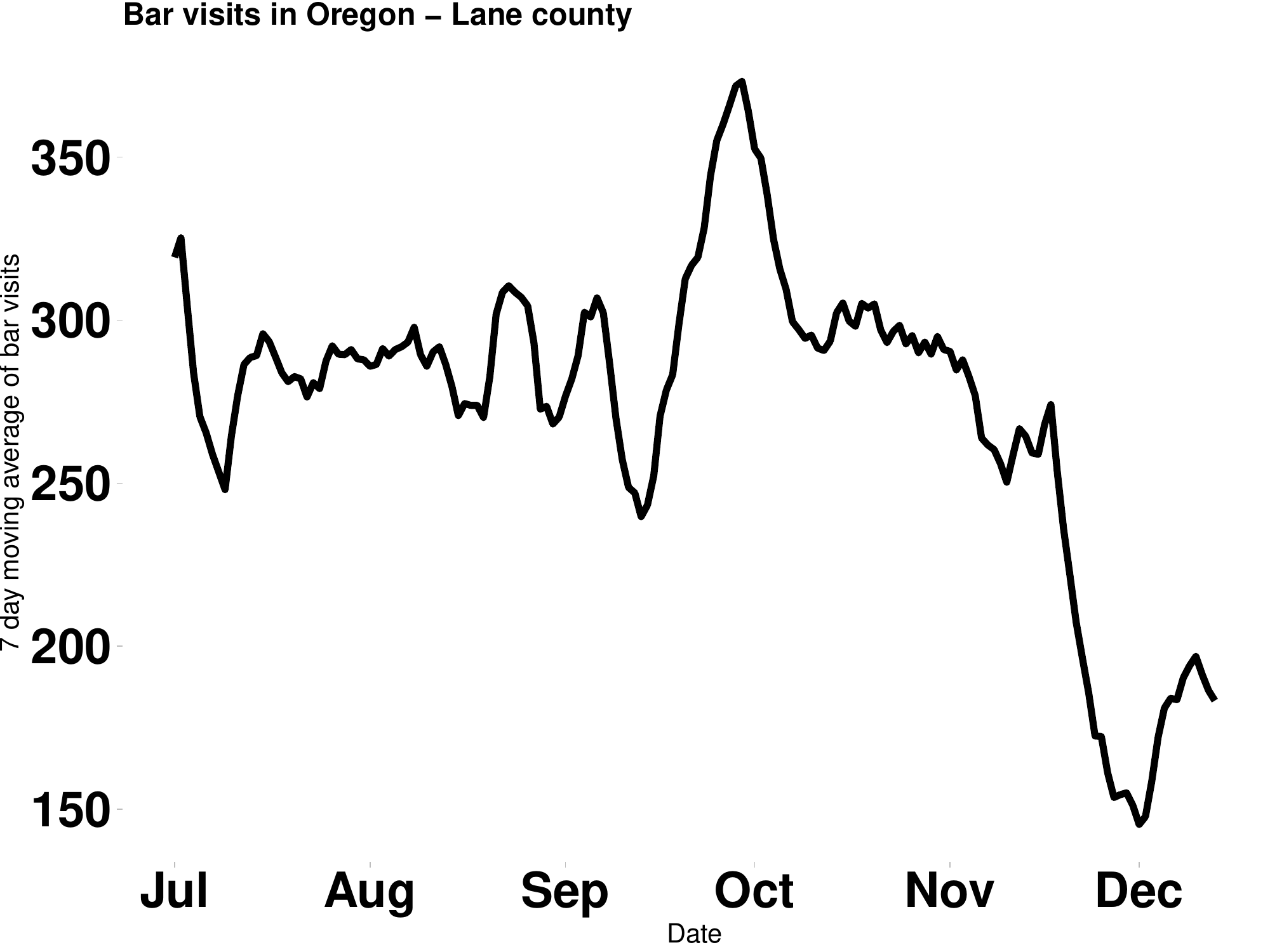}\\
   Restaurant Visits  &  Restaurant Visits   \\
      \includegraphics[width=0.50\textwidth,height=0.2\textwidth]{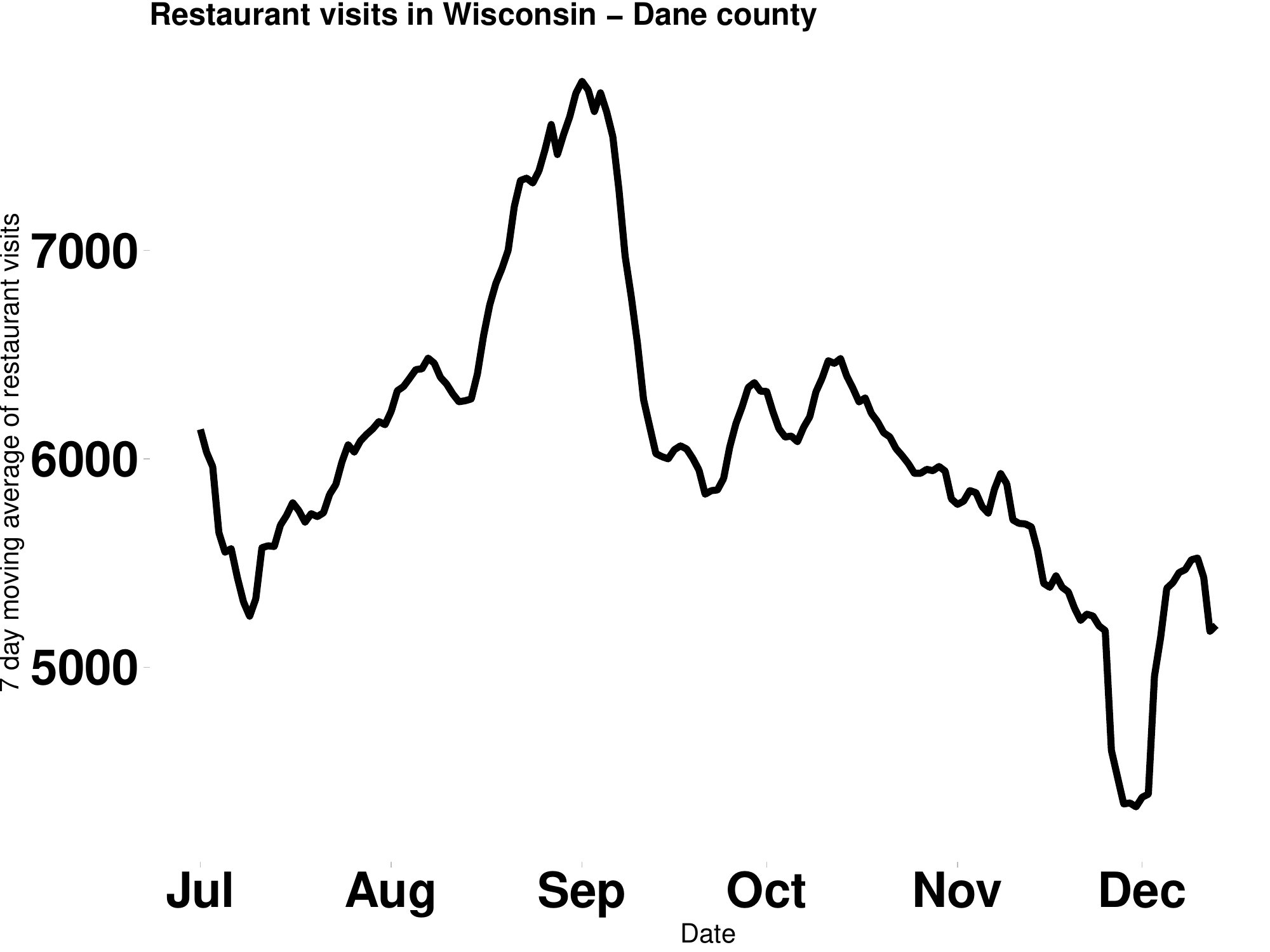}&
      \includegraphics[width=0.50\textwidth,height=0.2\textwidth]{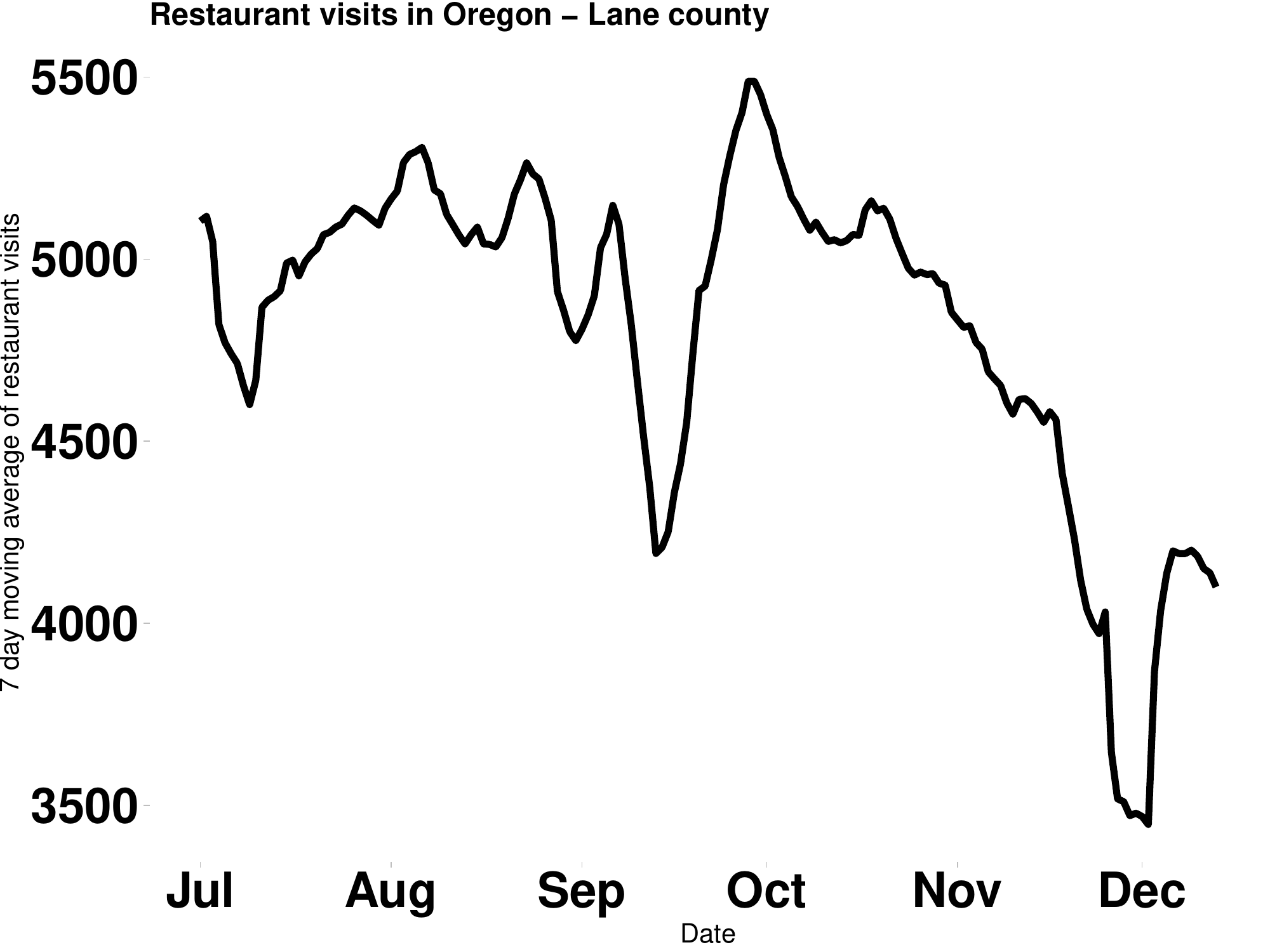}\\
    \end{tabular}
  \end{minipage}}
\vspace{-0.2cm}  {\scriptsize
\begin{flushleft}
Notes:  The first, the second, and the third figures in the left panel show the evolution of the number of cases by age groups, the number of visits to colleges/universities, and bars, respectively, in Dane County, WI. The right panel shows the corresponding figures for Lane County, OR.
 \end{flushleft}    }
 \end{figure}

\begin{figure}[!ht] 
  \caption{The number of cases by age groups and the number of visits to colleges/universities,  bars, restaurants, recreation facilities,   K-12 schools,  and a comparison of reported cases between CDC and NYT data \label{fig:dane-SI}} \medskip
  \hspace{-4cm}\resizebox{0.9\columnwidth}{!}{
\begin{minipage}{\linewidth} 
        \begin{tabular}{c|c|c|c|c}  
         \large{ \textbf{ Pima, AZ} }&   \large {  \textbf{Ingham, MI}}&  \large  {  \textbf{Centre, PA }}& 
          \large  { \textbf{Story, IA}}&   \large {  \textbf{Champaign, IL}}\\ \hline
   \small Cases by Age Groups &  \small   Cases by Age Groups &  \small   Cases by Age Groups &   \small  Cases by Age Groups&  \small  Cases by Age Groups  \\ 
      \includegraphics[width=0.20\textwidth]{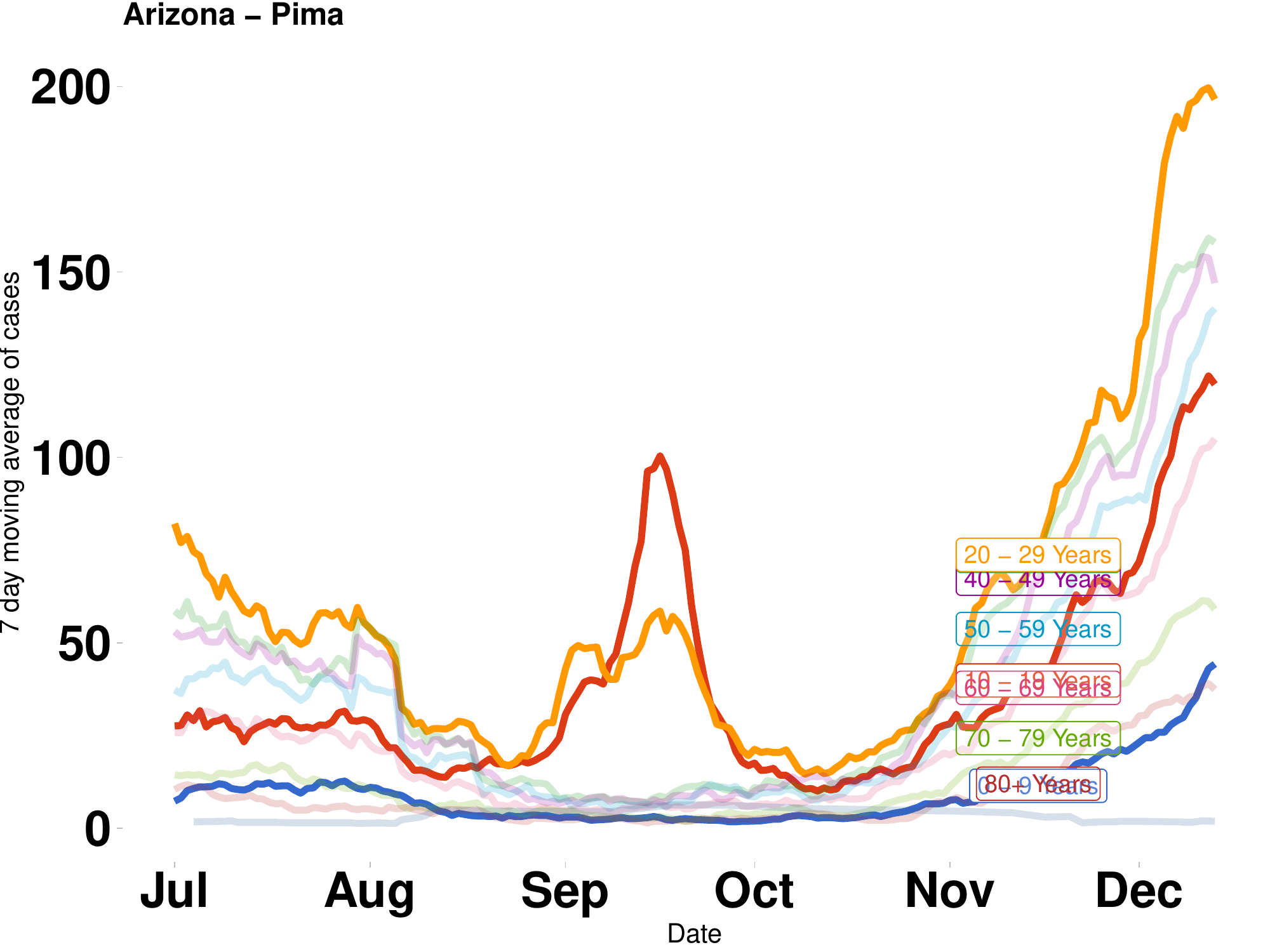}&
      \includegraphics[width=0.20\textwidth]{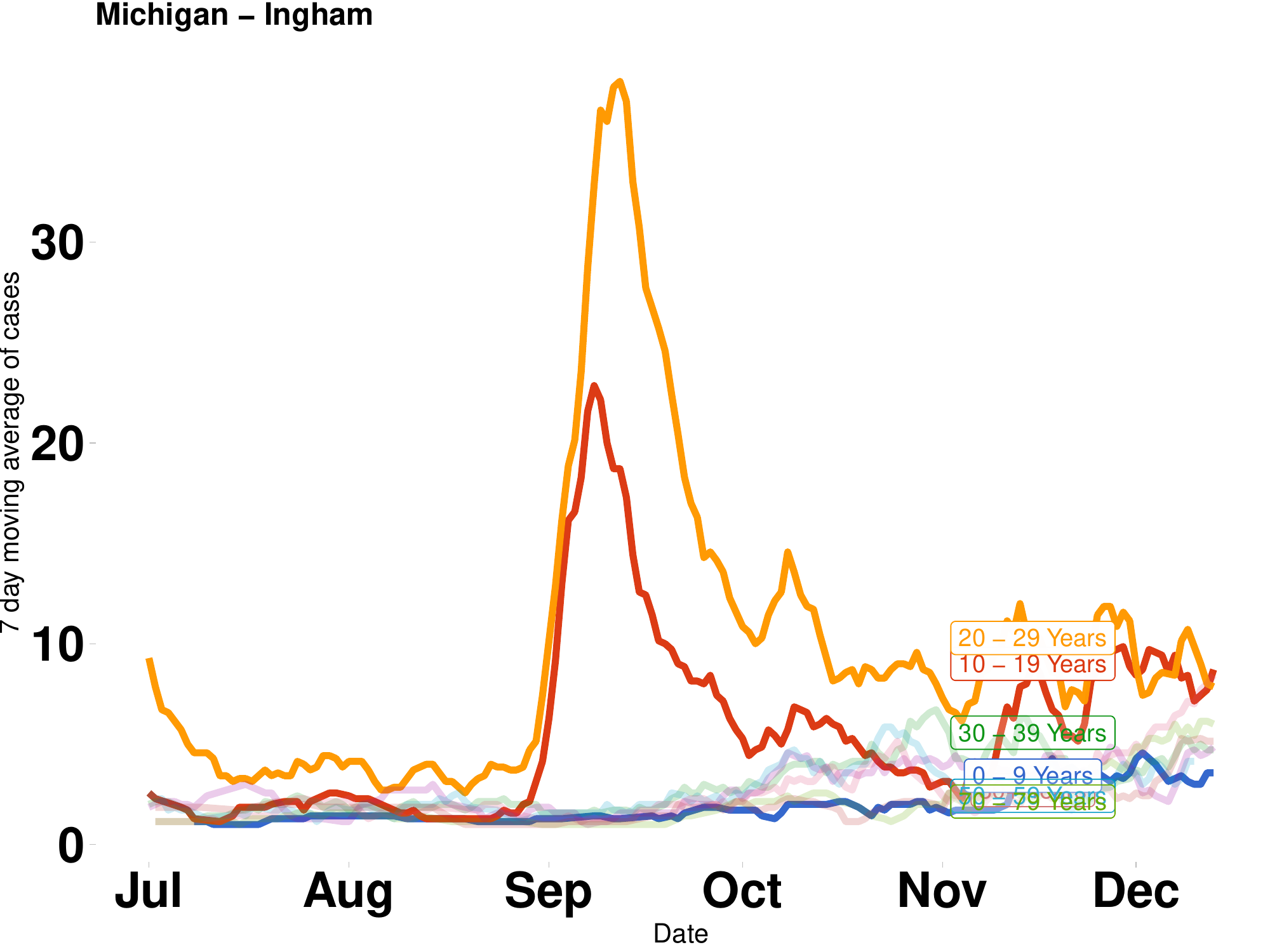}&
      \includegraphics[width=0.20\textwidth]{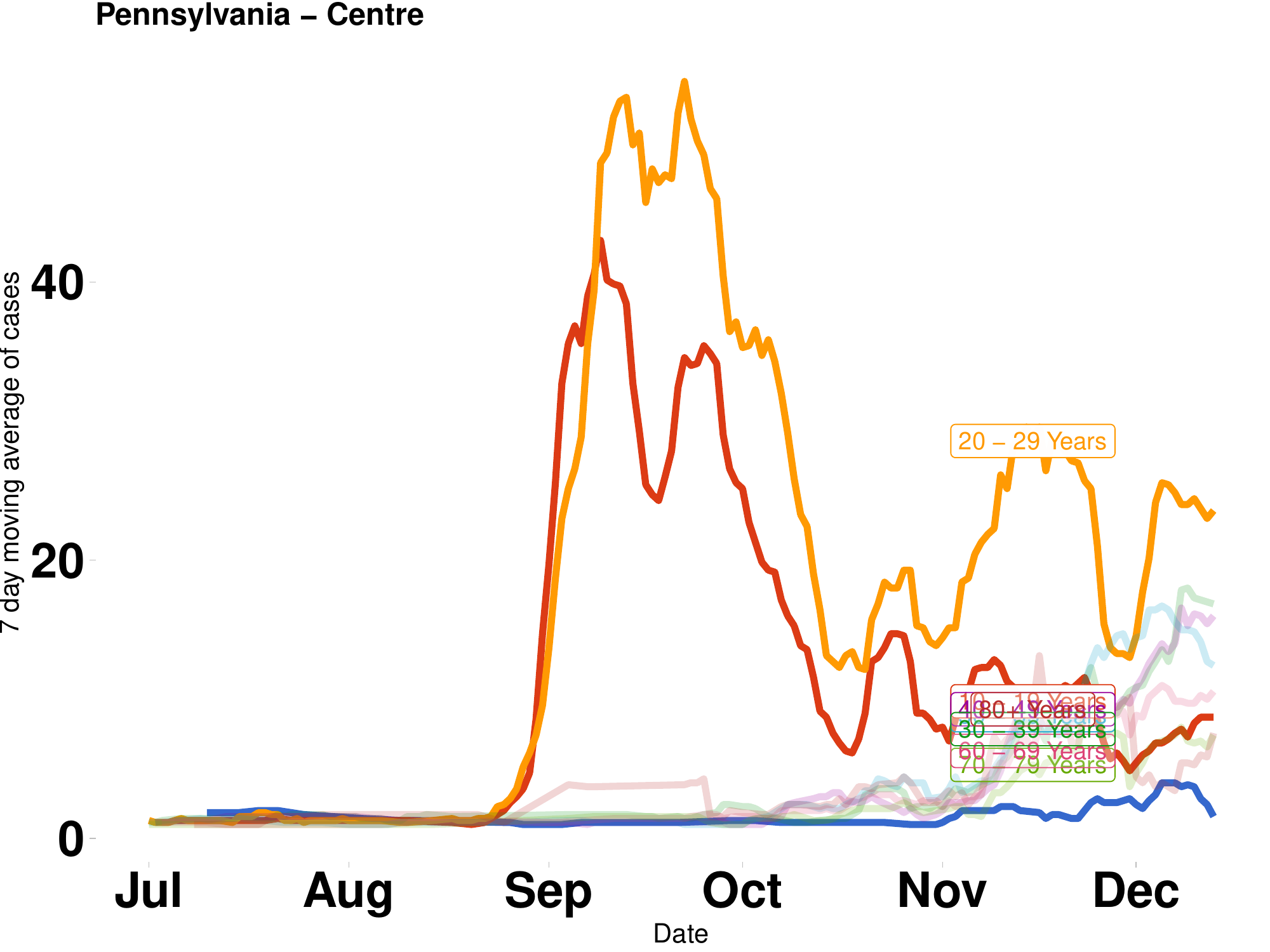}&
      \includegraphics[width=0.20\textwidth]{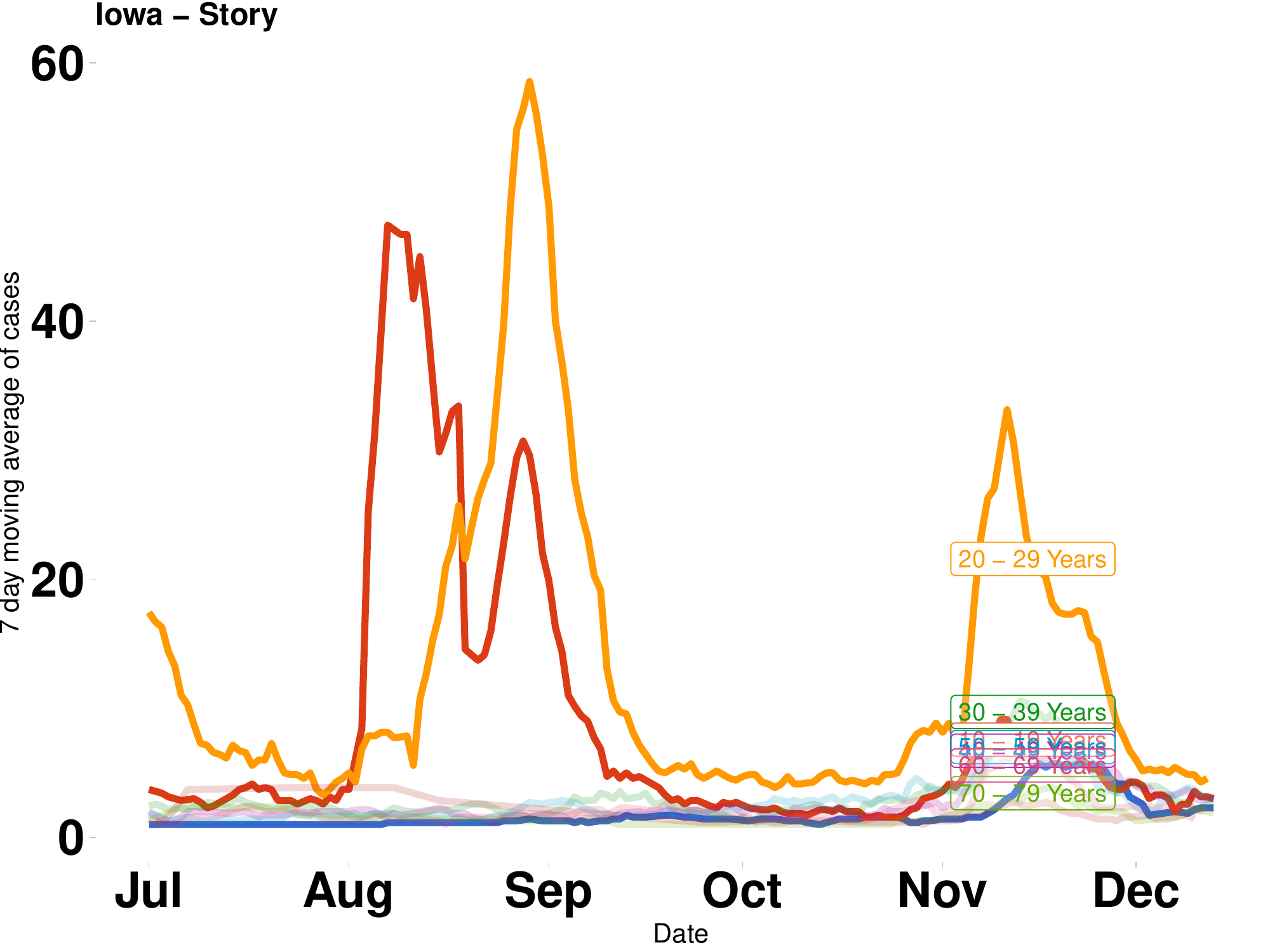}& 
      \includegraphics[width=0.20\textwidth]{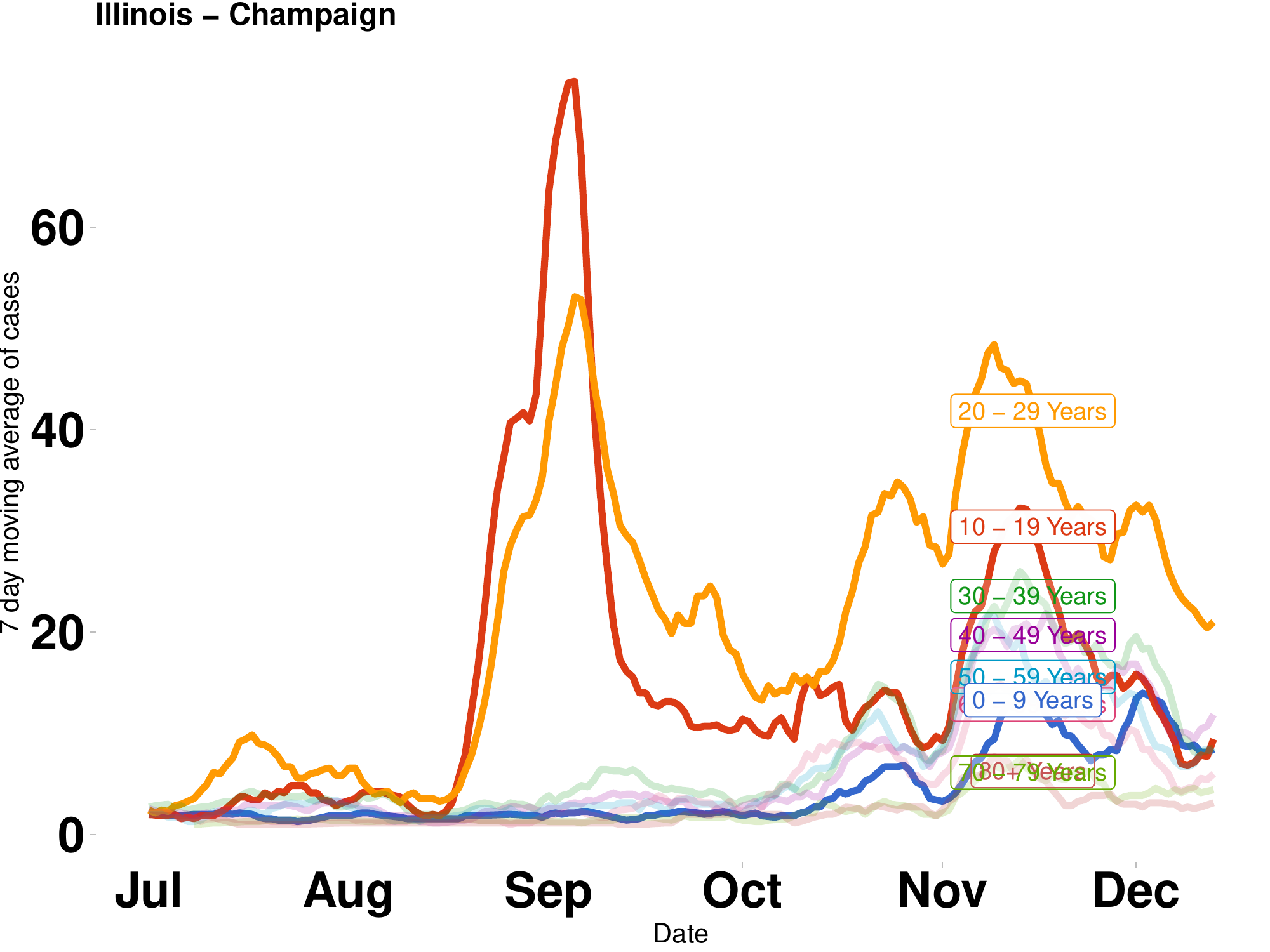}\\  
   College Visits  &  College Visits  & College Visits  &  College Visits  & College Visits  \\   
      \includegraphics[width=0.20\textwidth]{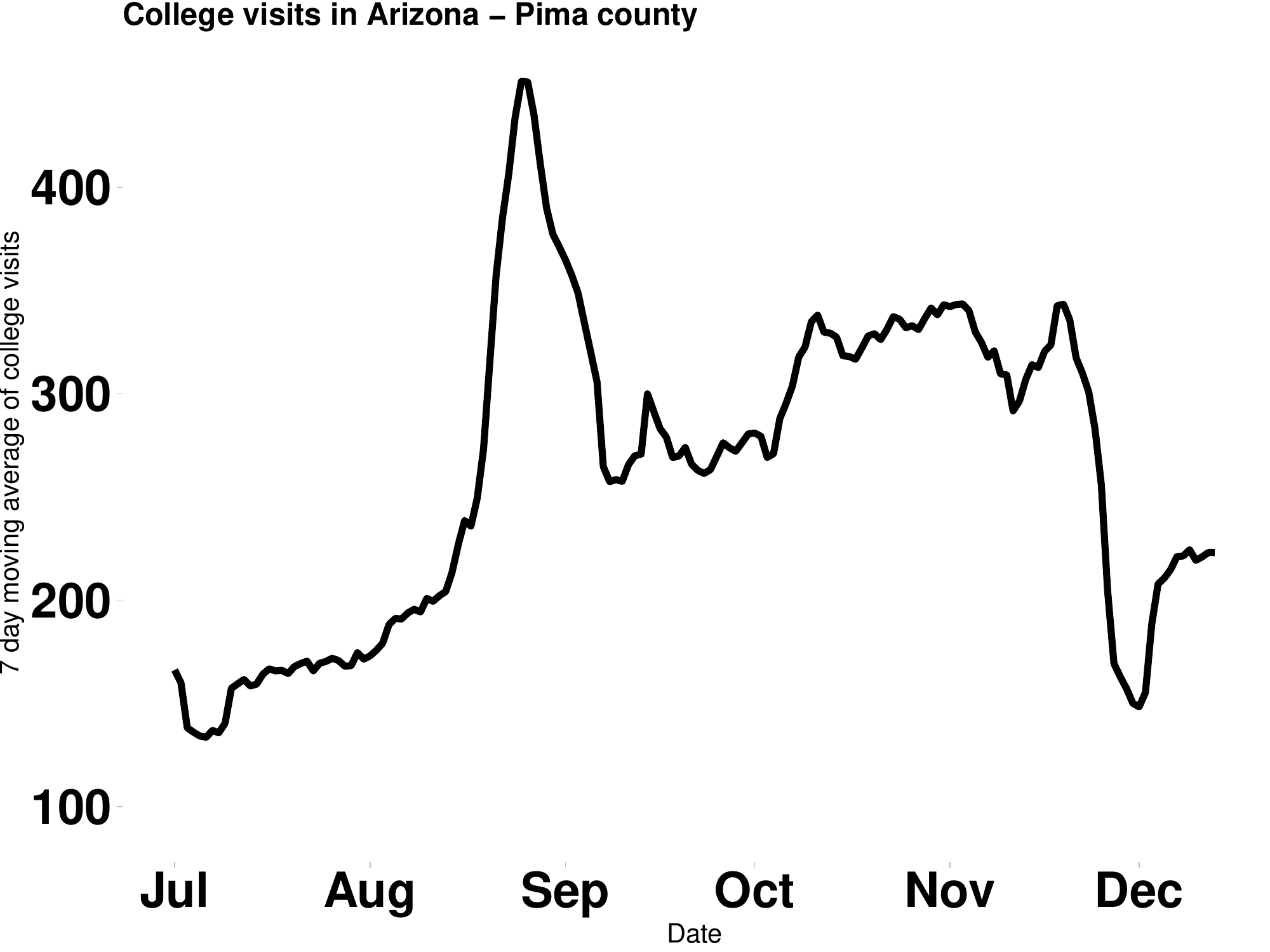}&
      \includegraphics[width=0.20\textwidth]{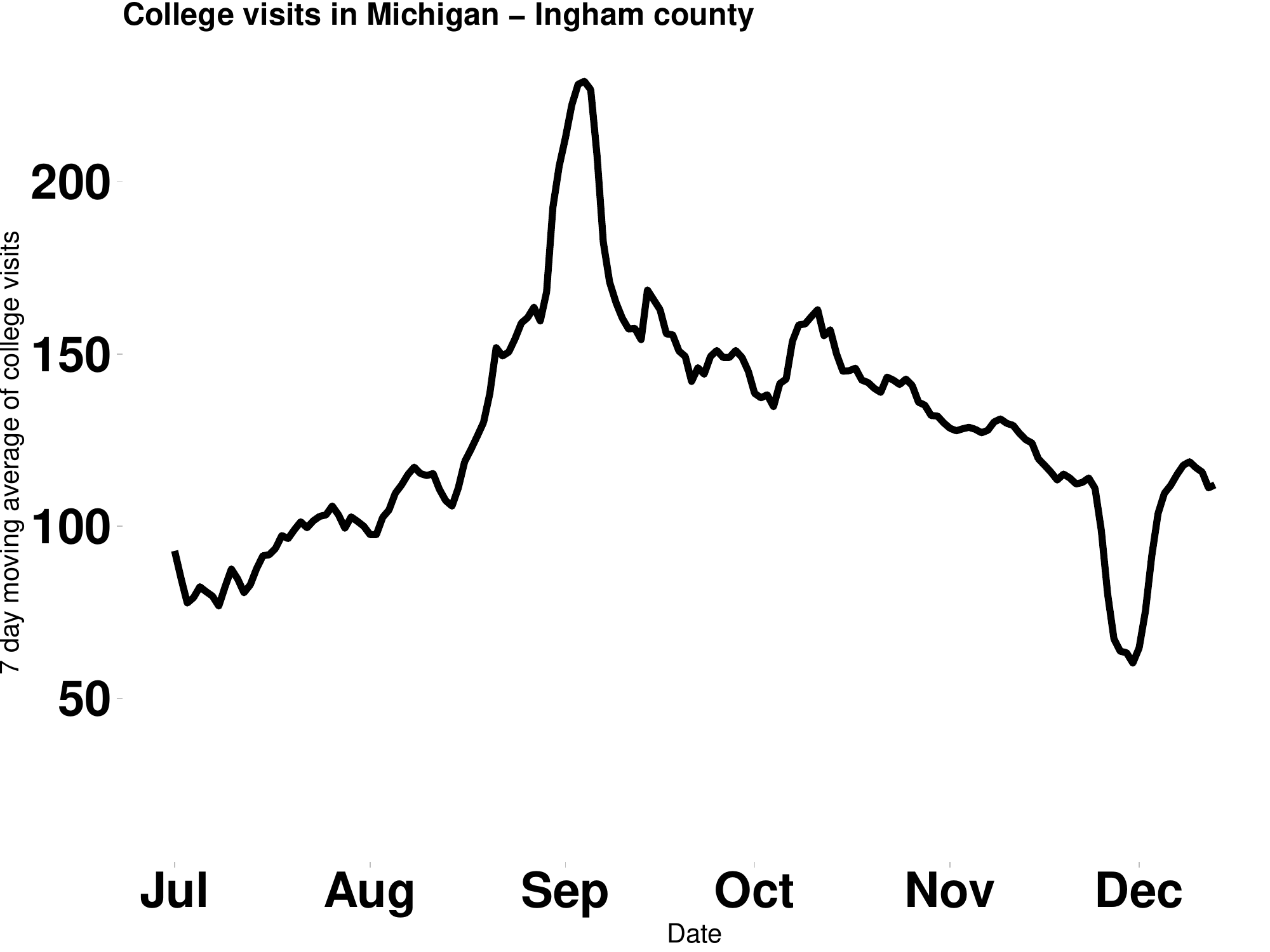}&
      \includegraphics[width=0.20\textwidth]{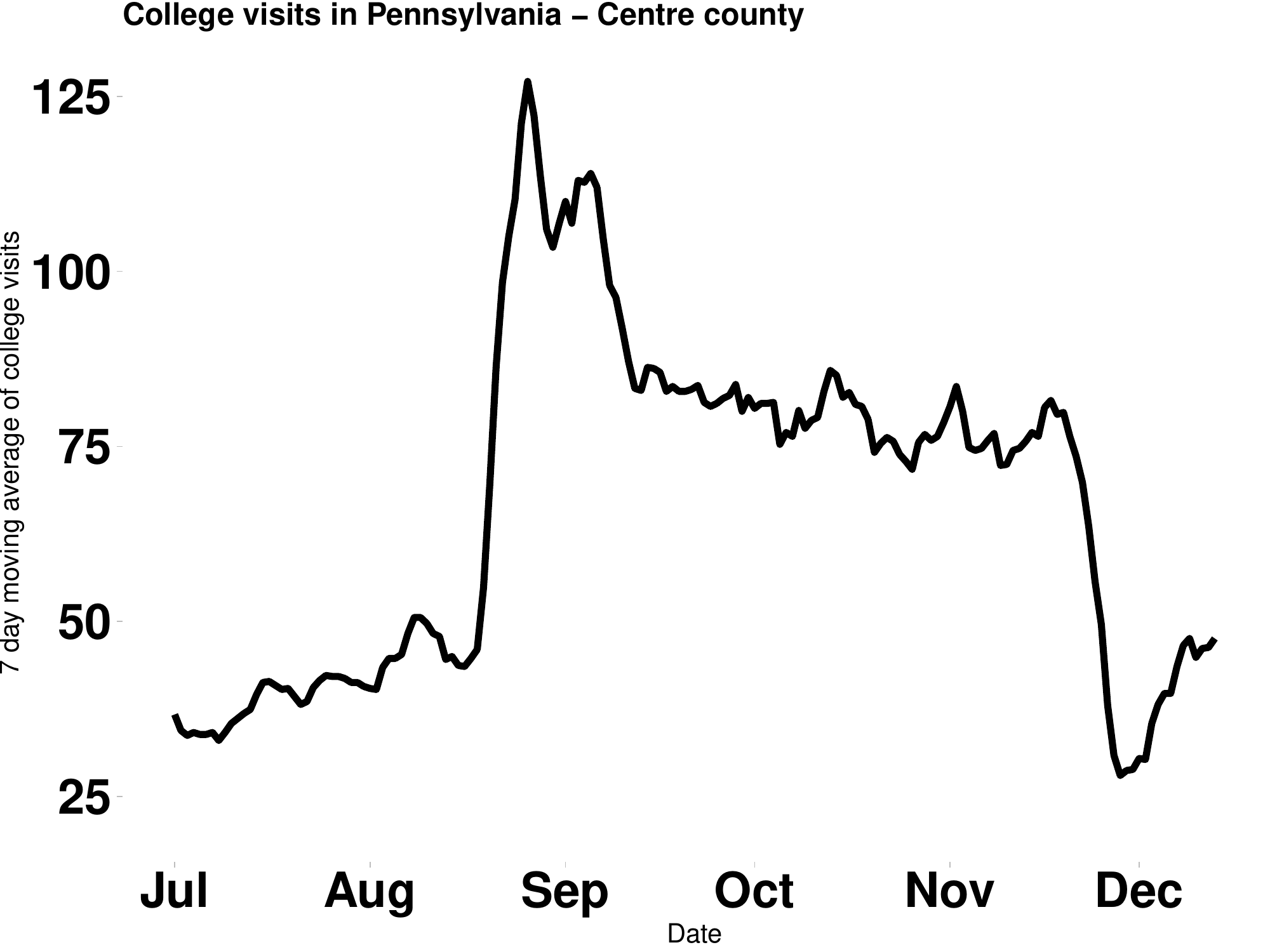}&
      \includegraphics[width=0.20\textwidth]{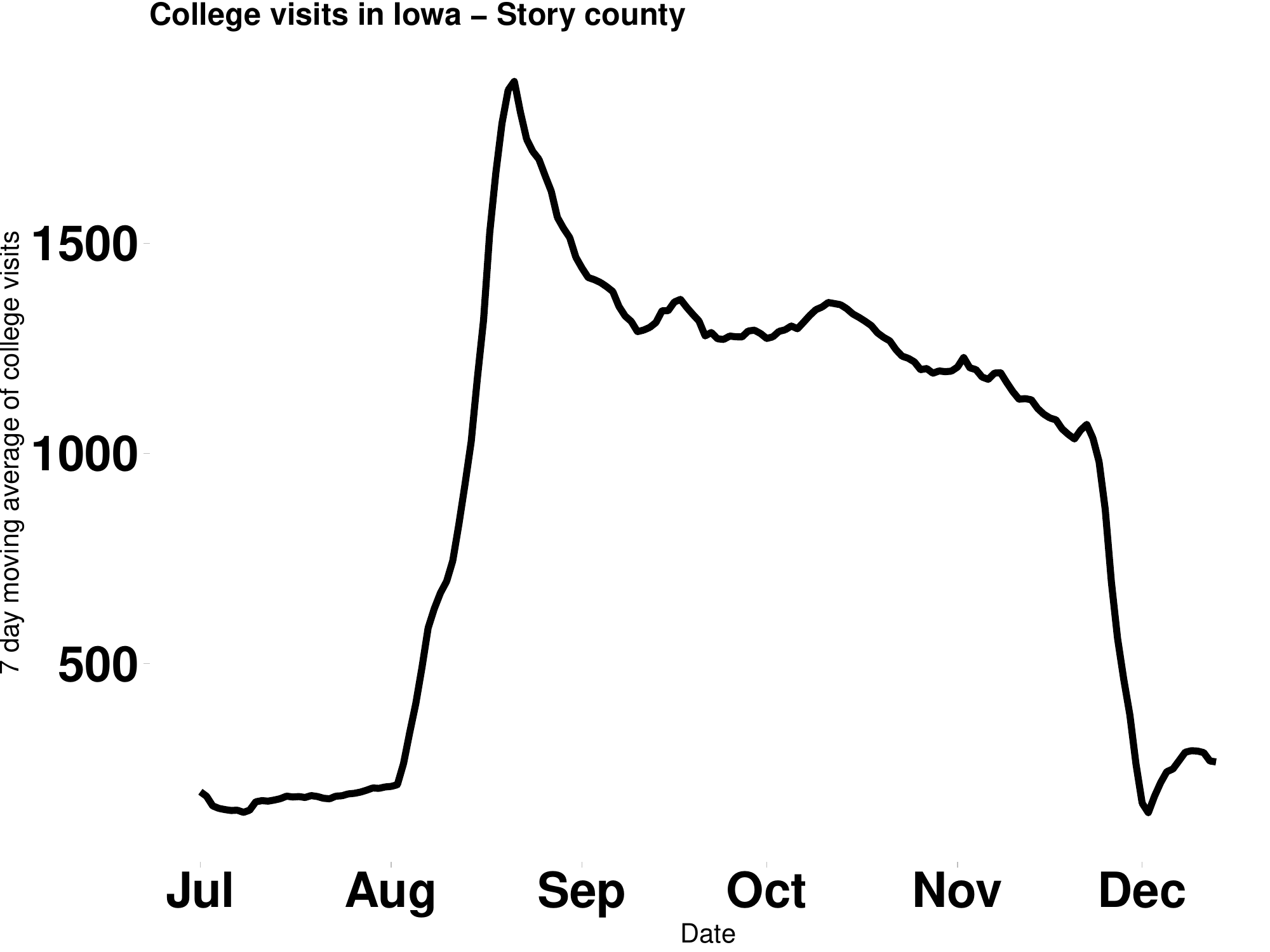}& 
      \includegraphics[width=0.20\textwidth]{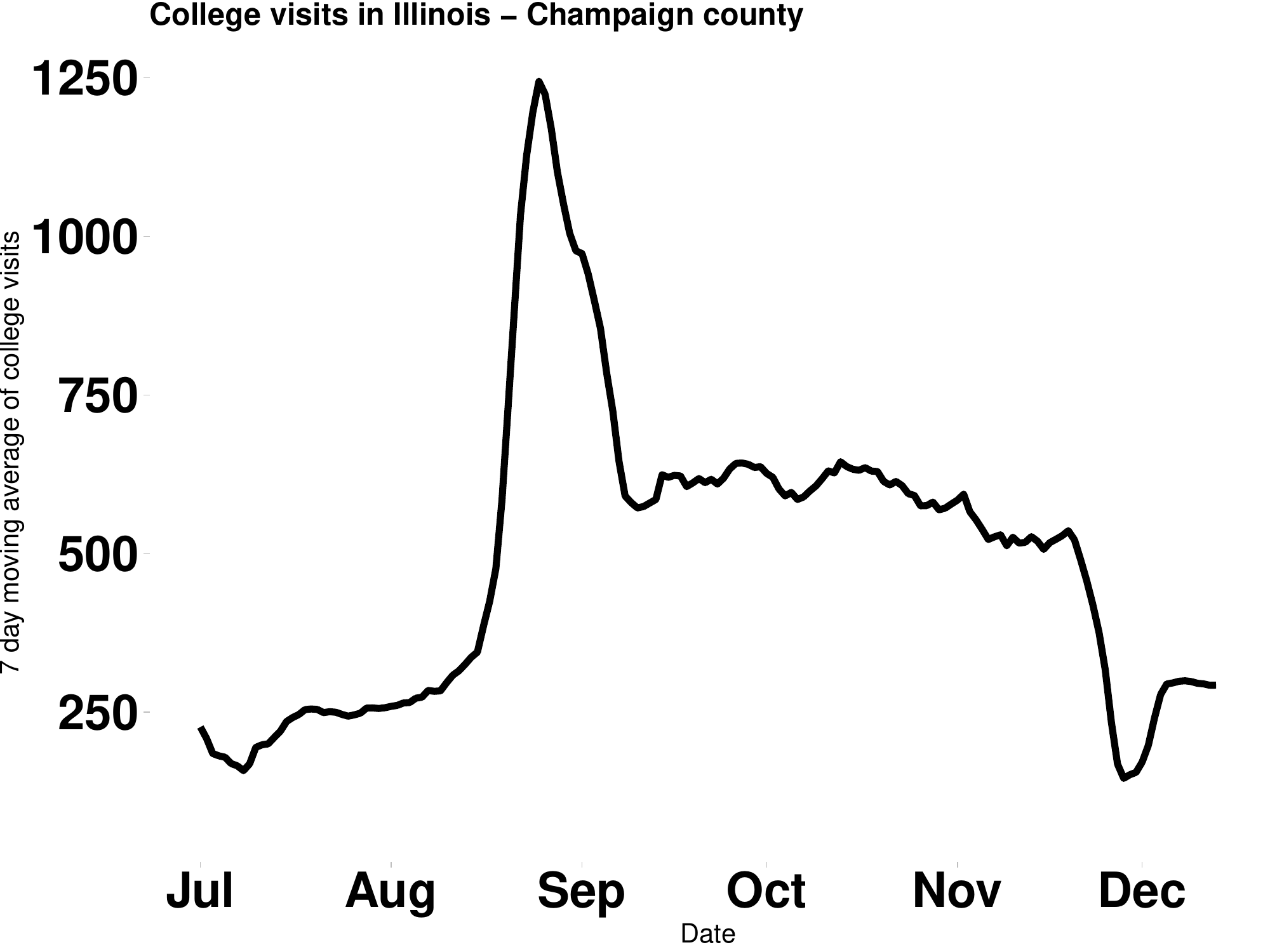}\\  
   Bar Visits  &  Bar Visits & Bar Visits  &  Bar Visits & Bar Visits     \\  
      \includegraphics[width=0.20\textwidth]{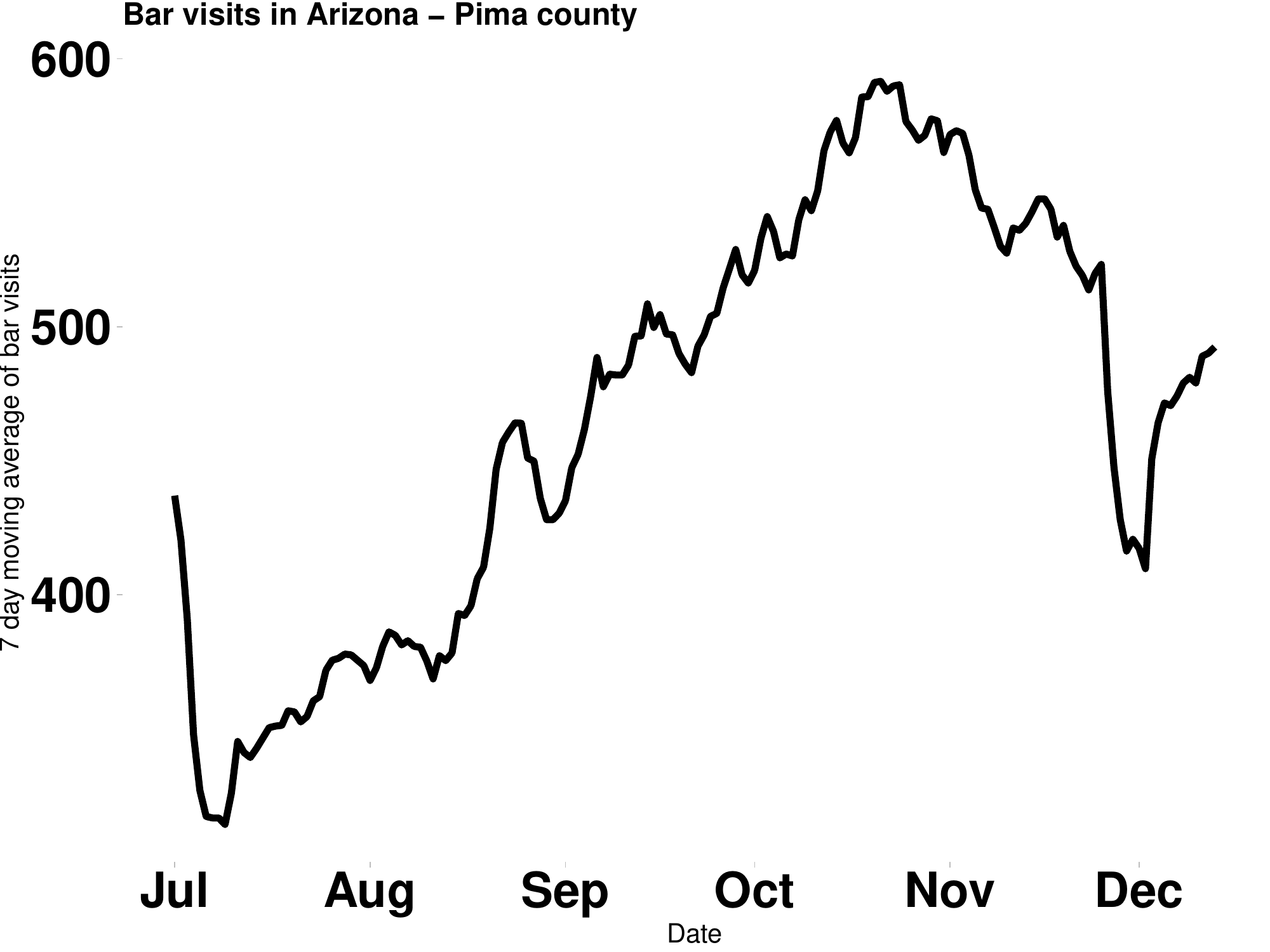}&
      \includegraphics[width=0.20\textwidth]{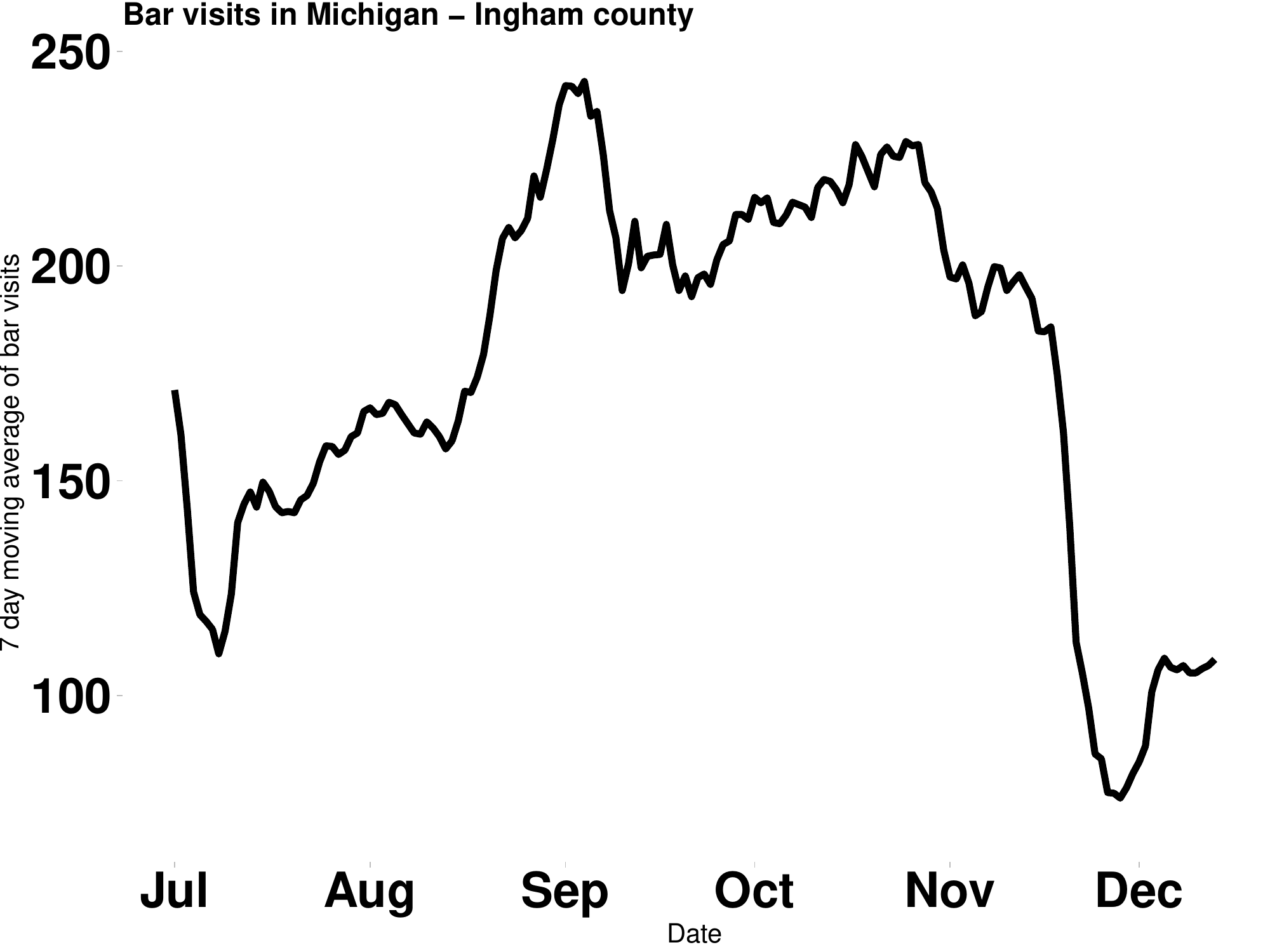}&
      \includegraphics[width=0.20\textwidth]{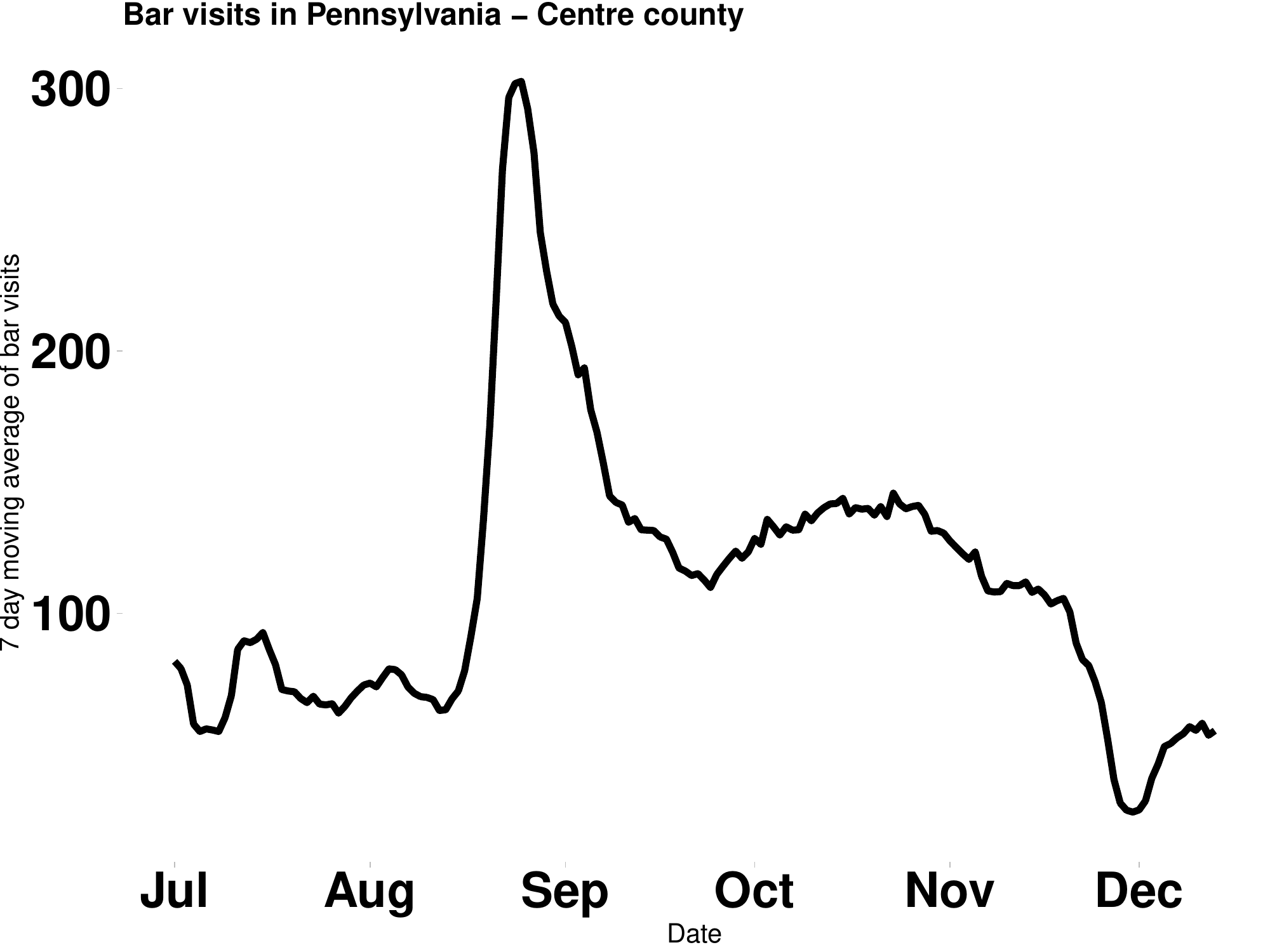}&
      \includegraphics[width=0.20\textwidth]{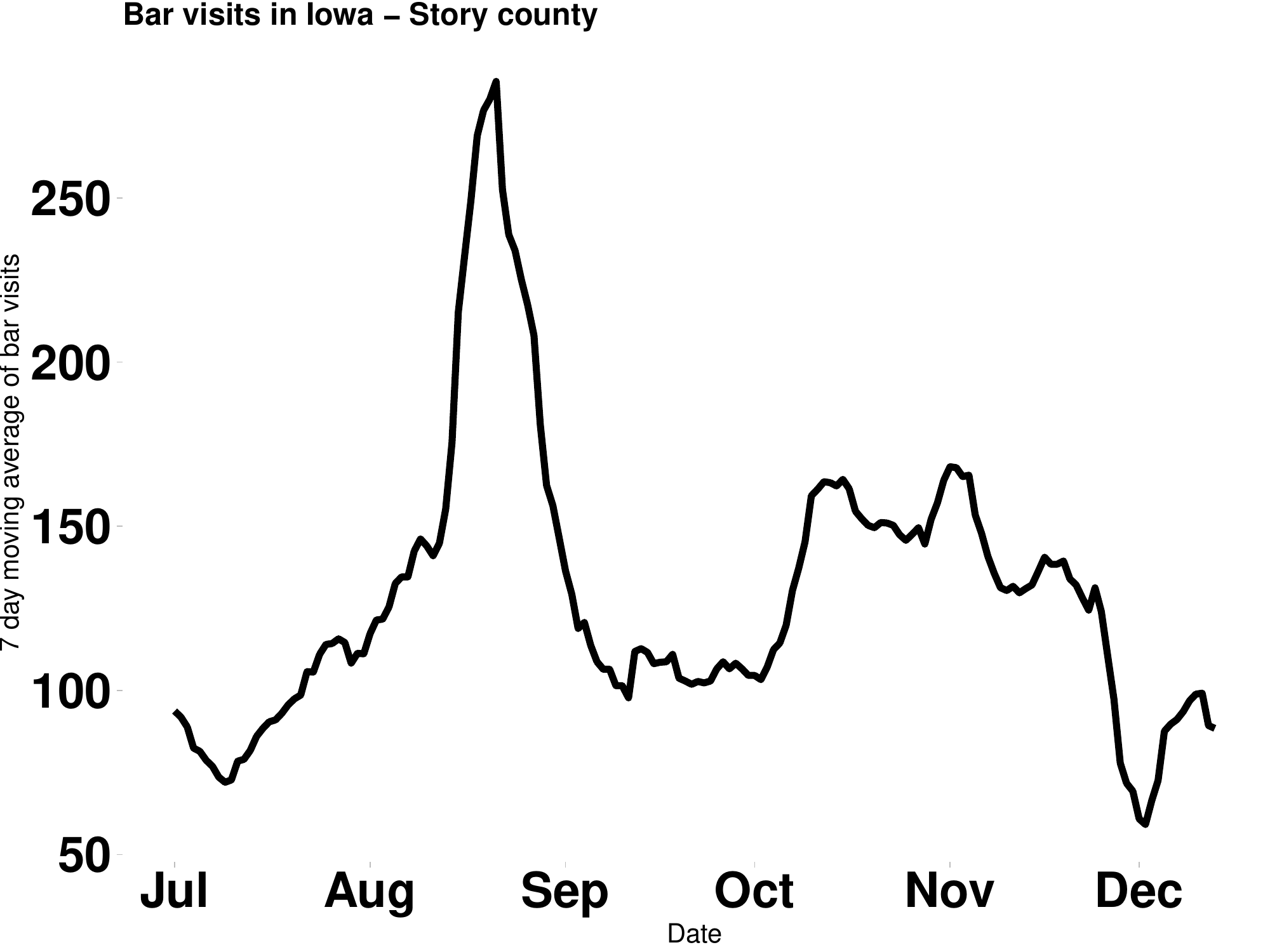}& 
      \includegraphics[width=0.20\textwidth]{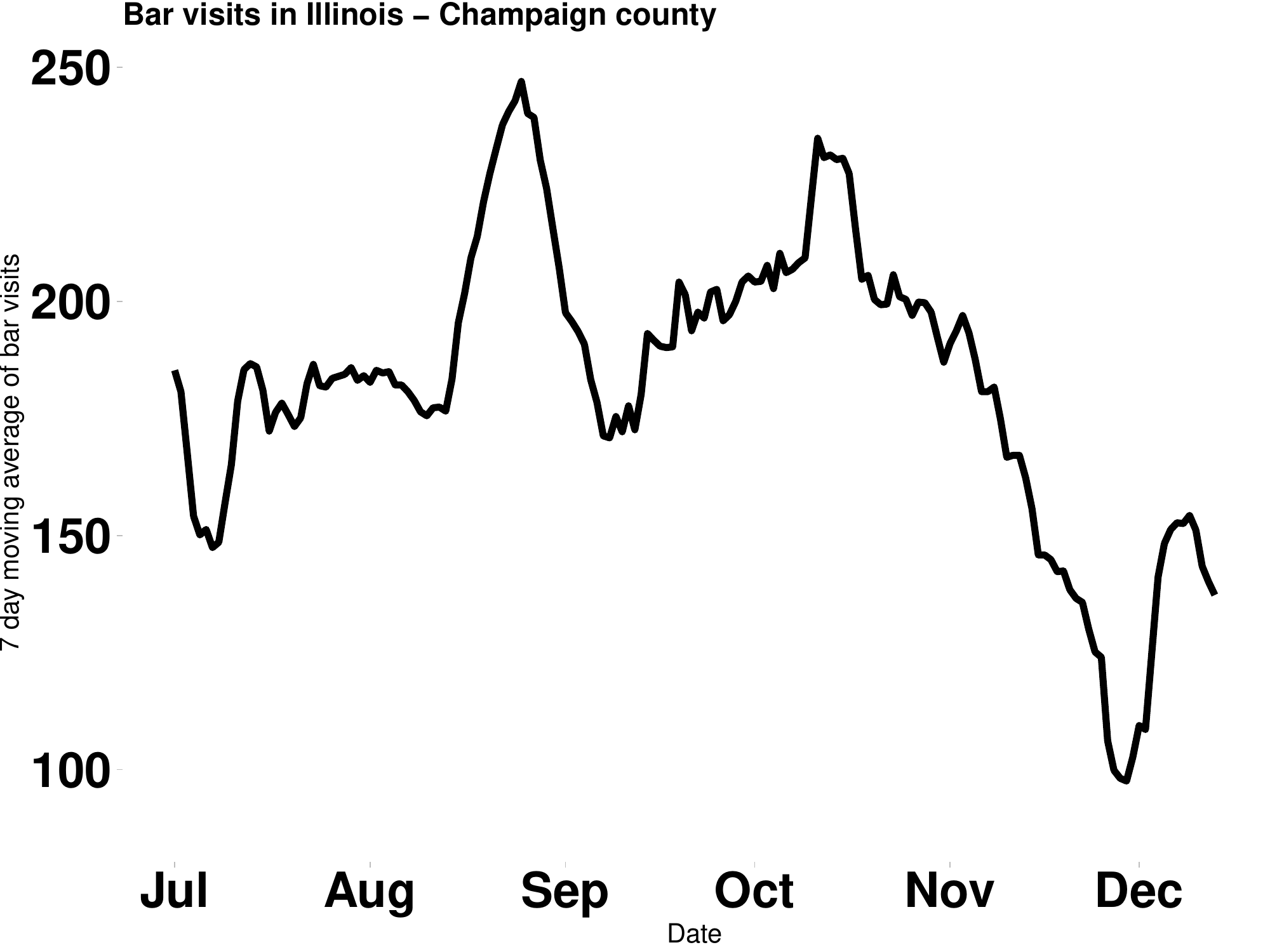}\\     
   Restaurant Visits  &  Restaurant Visits  & Restaurant Visits  &  Restaurant Visits  & Restaurant Visits    \\ 
      \includegraphics[width=0.20\textwidth]{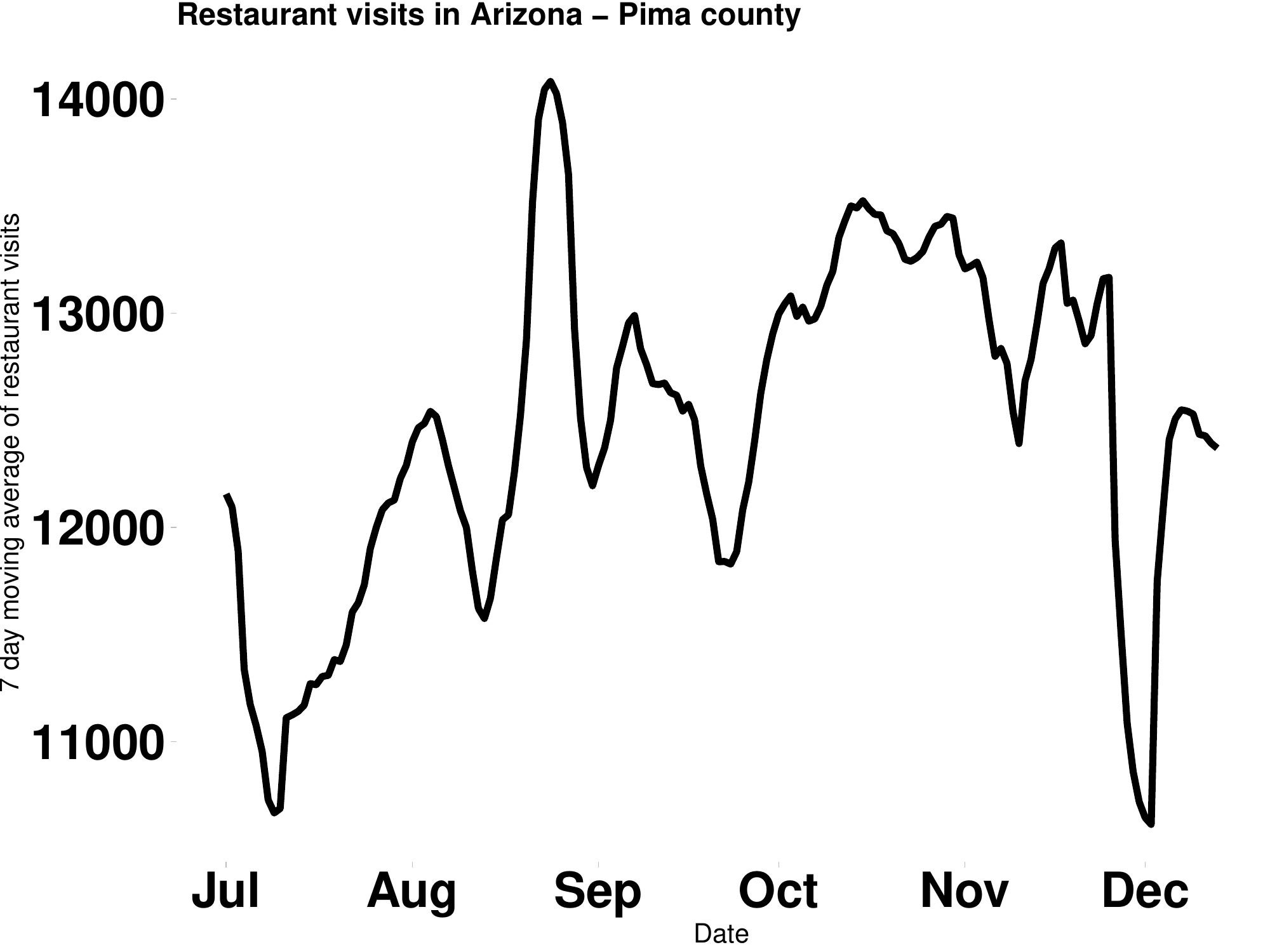}&
      \includegraphics[width=0.20\textwidth]{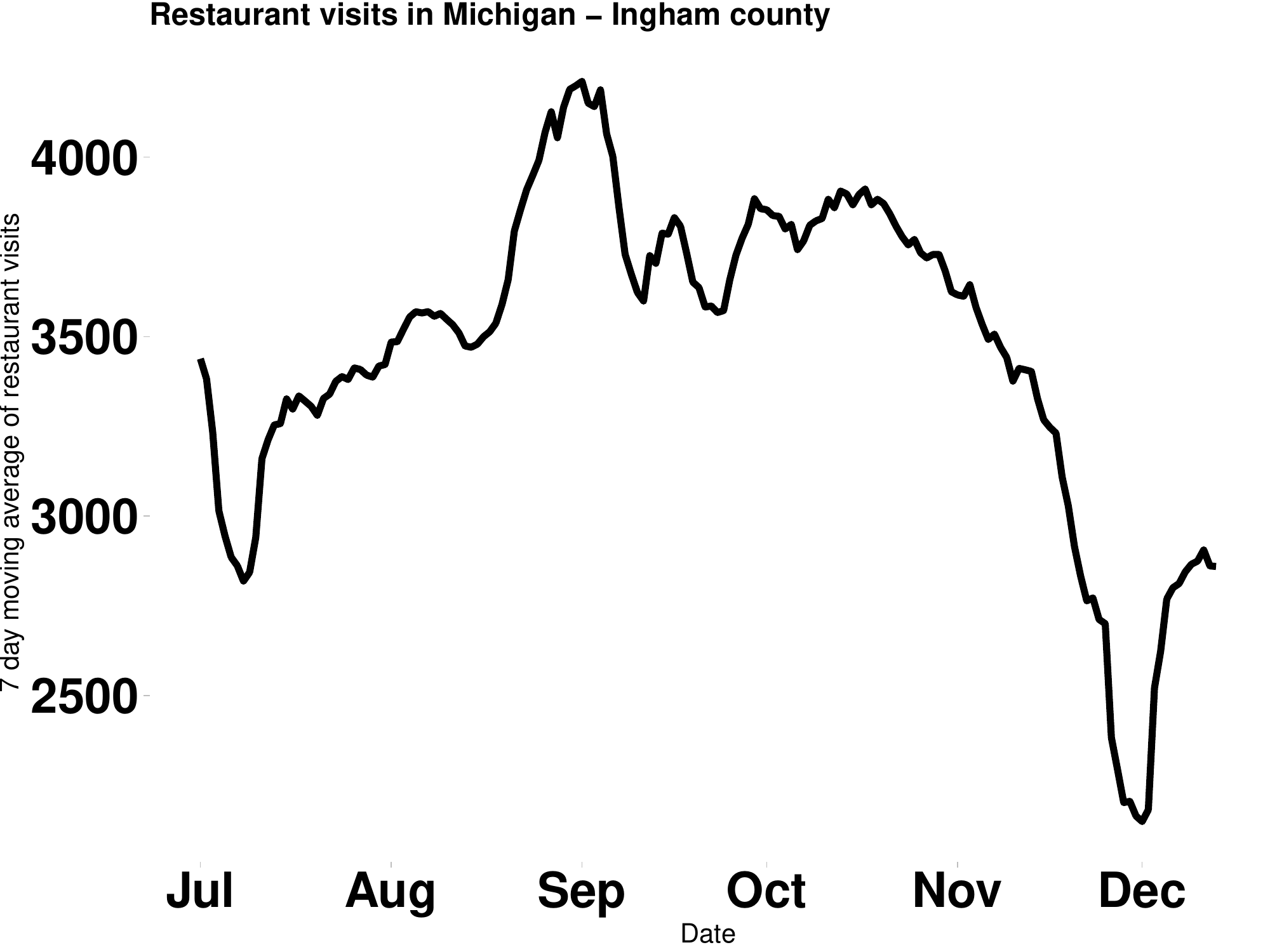}&
      \includegraphics[width=0.20\textwidth]{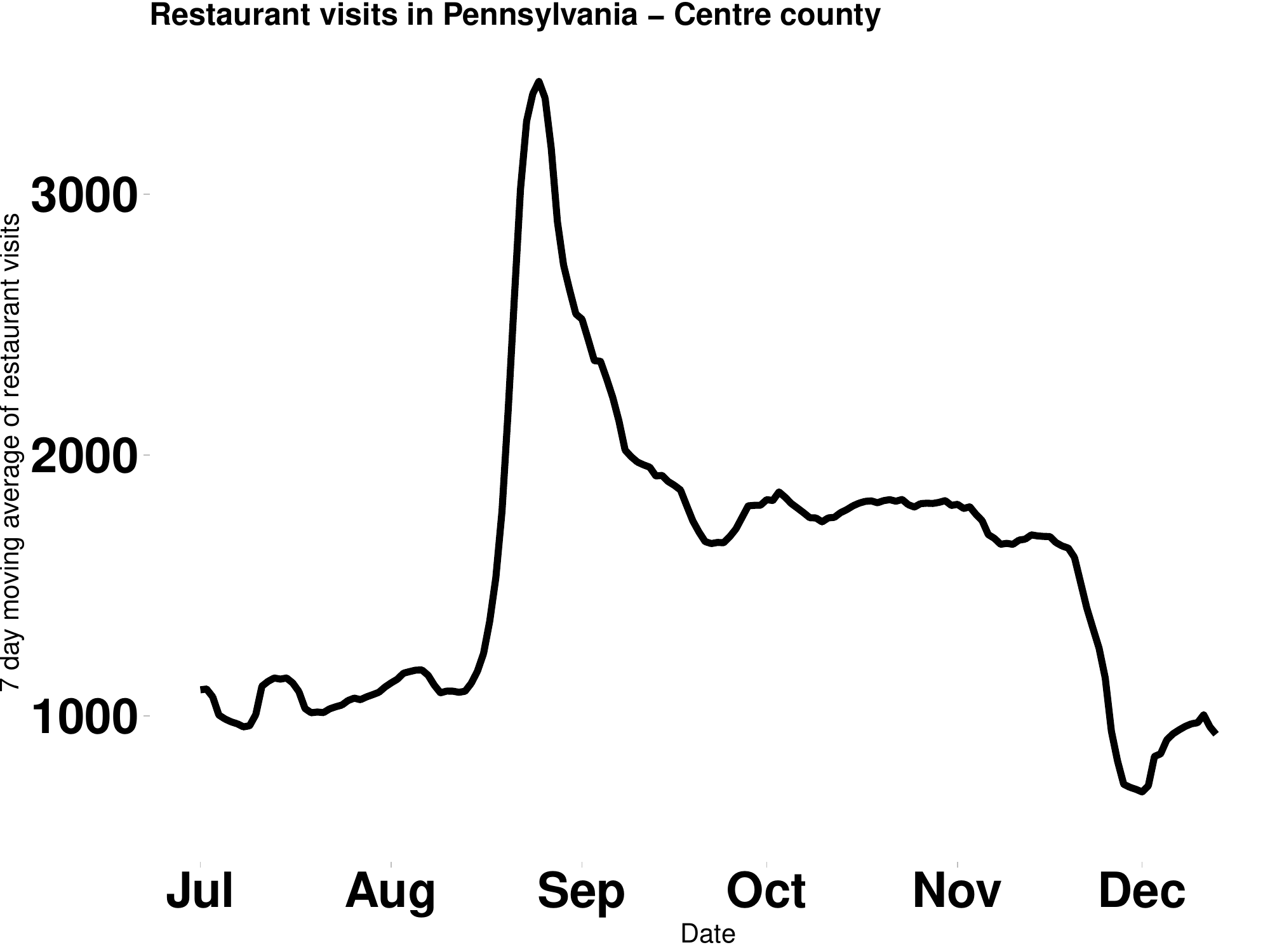}&
      \includegraphics[width=0.20\textwidth]{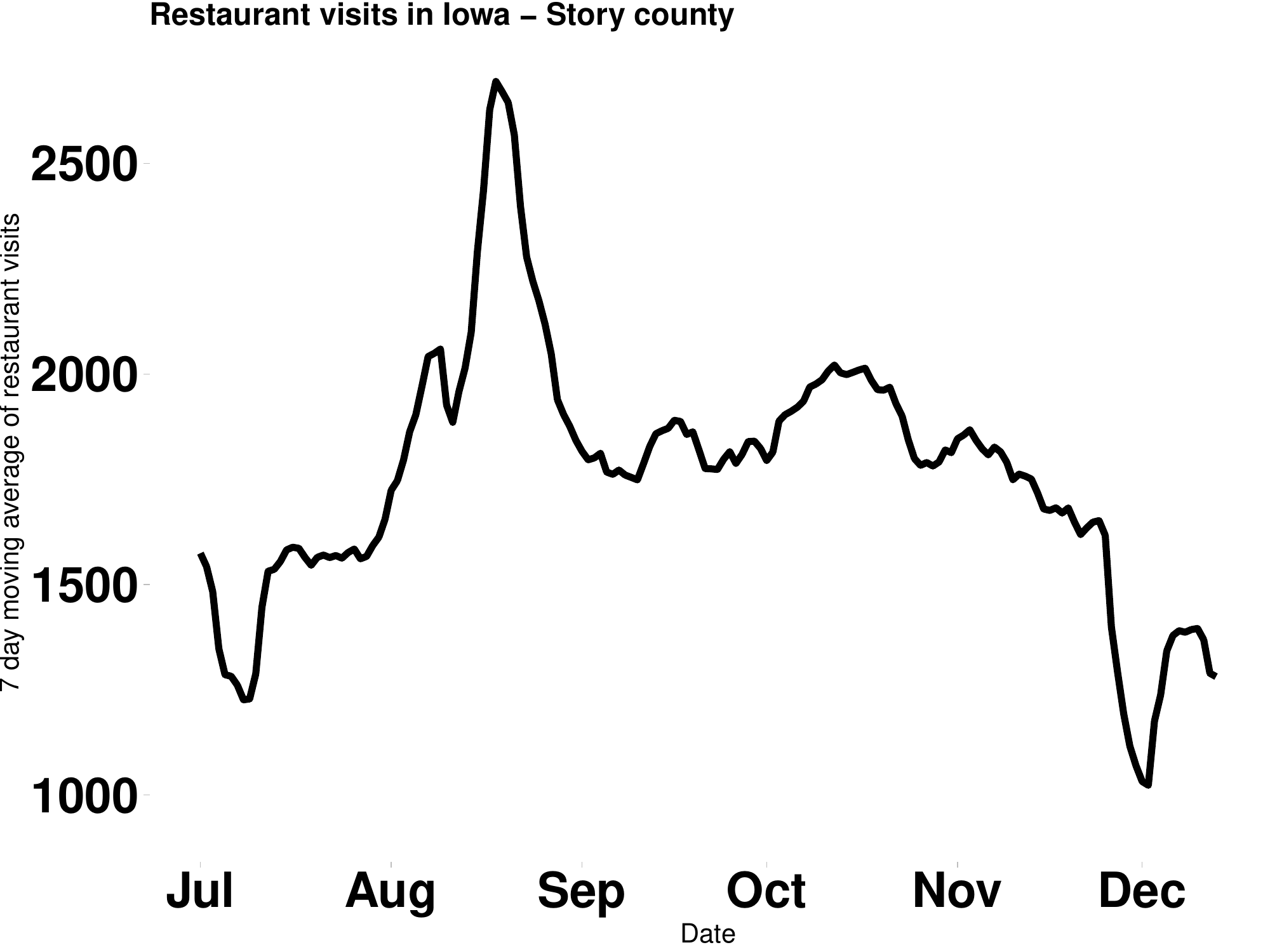}& 
      \includegraphics[width=0.20\textwidth]{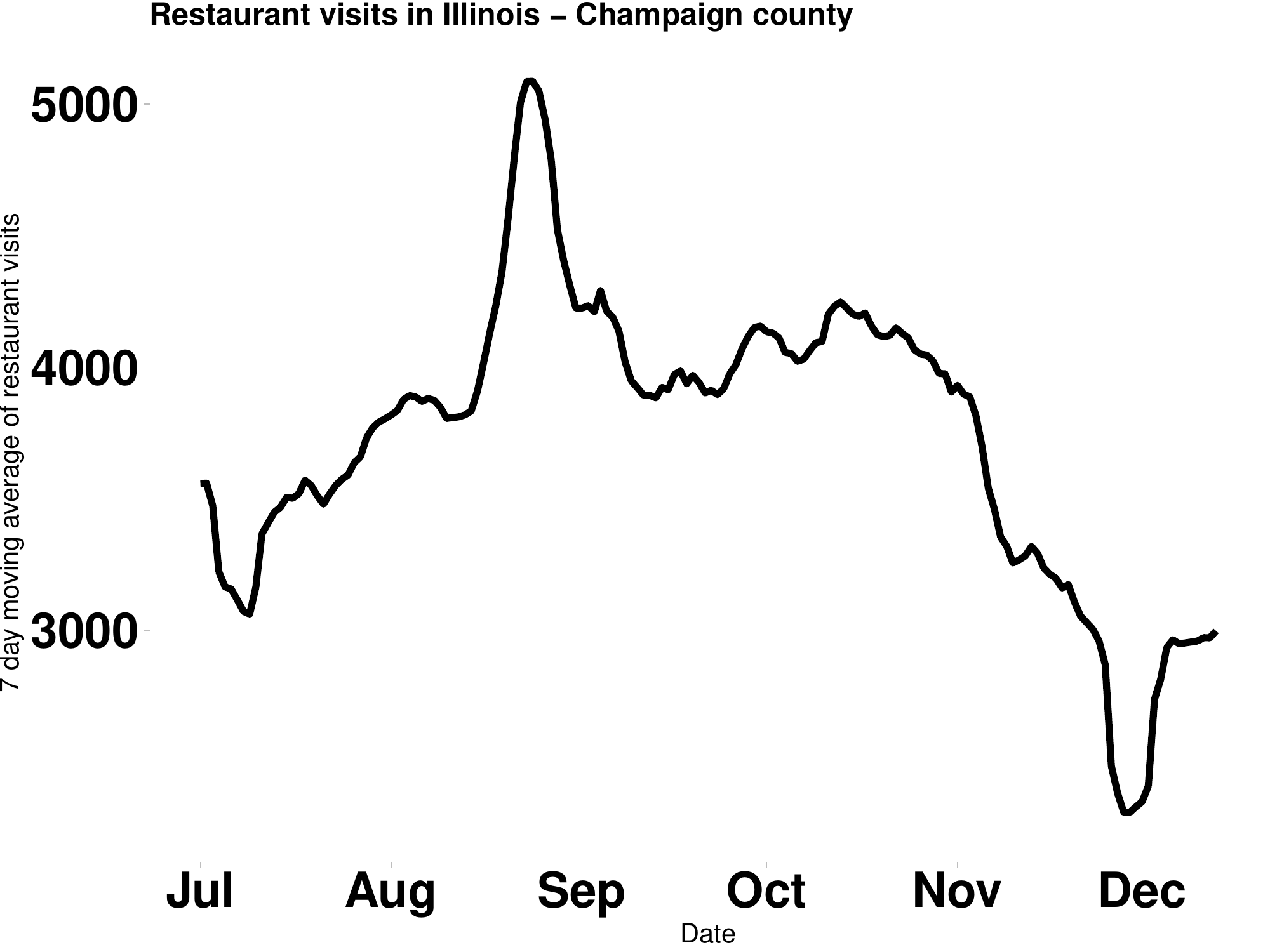}\\   
   Rec. Facilitiy Visits  &  Rec. Facilitiy Visits  & Rec. Facilitiy Visits  &  Rec. Facilitiy Visits& Rec. Facilitiy Visits     \\  
      \includegraphics[width=0.20\textwidth]{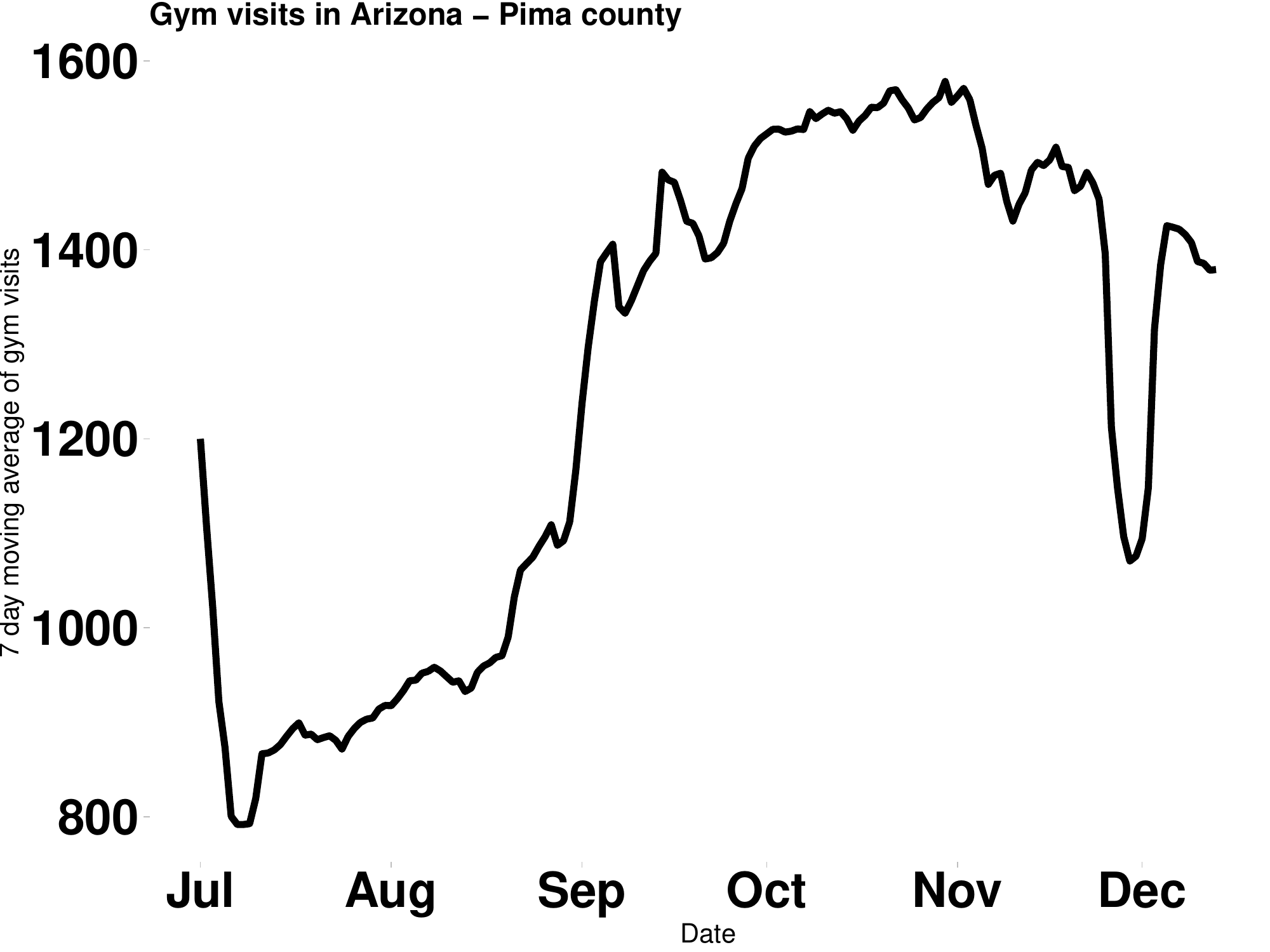}&
      \includegraphics[width=0.20\textwidth]{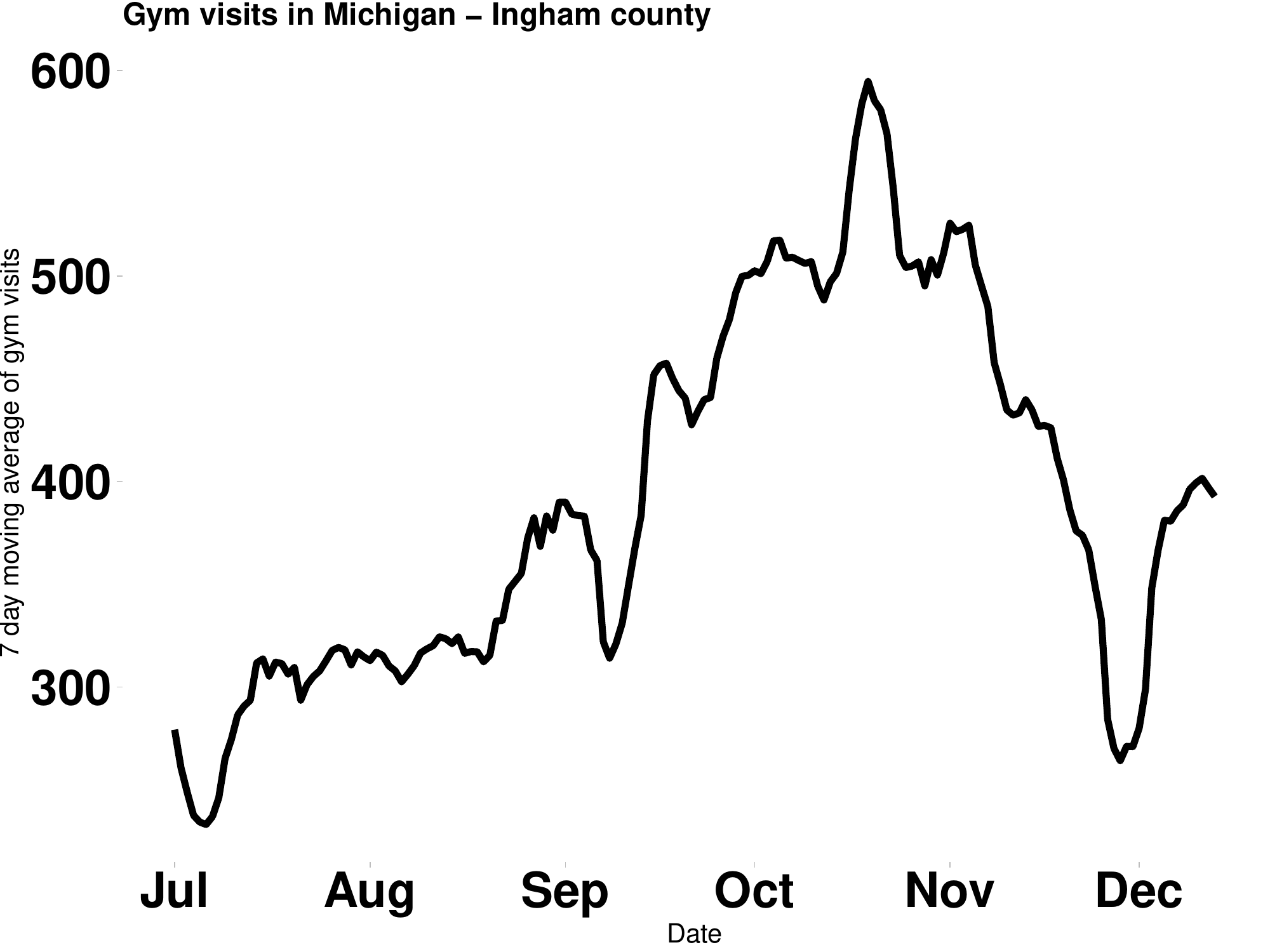}&
      \includegraphics[width=0.20\textwidth]{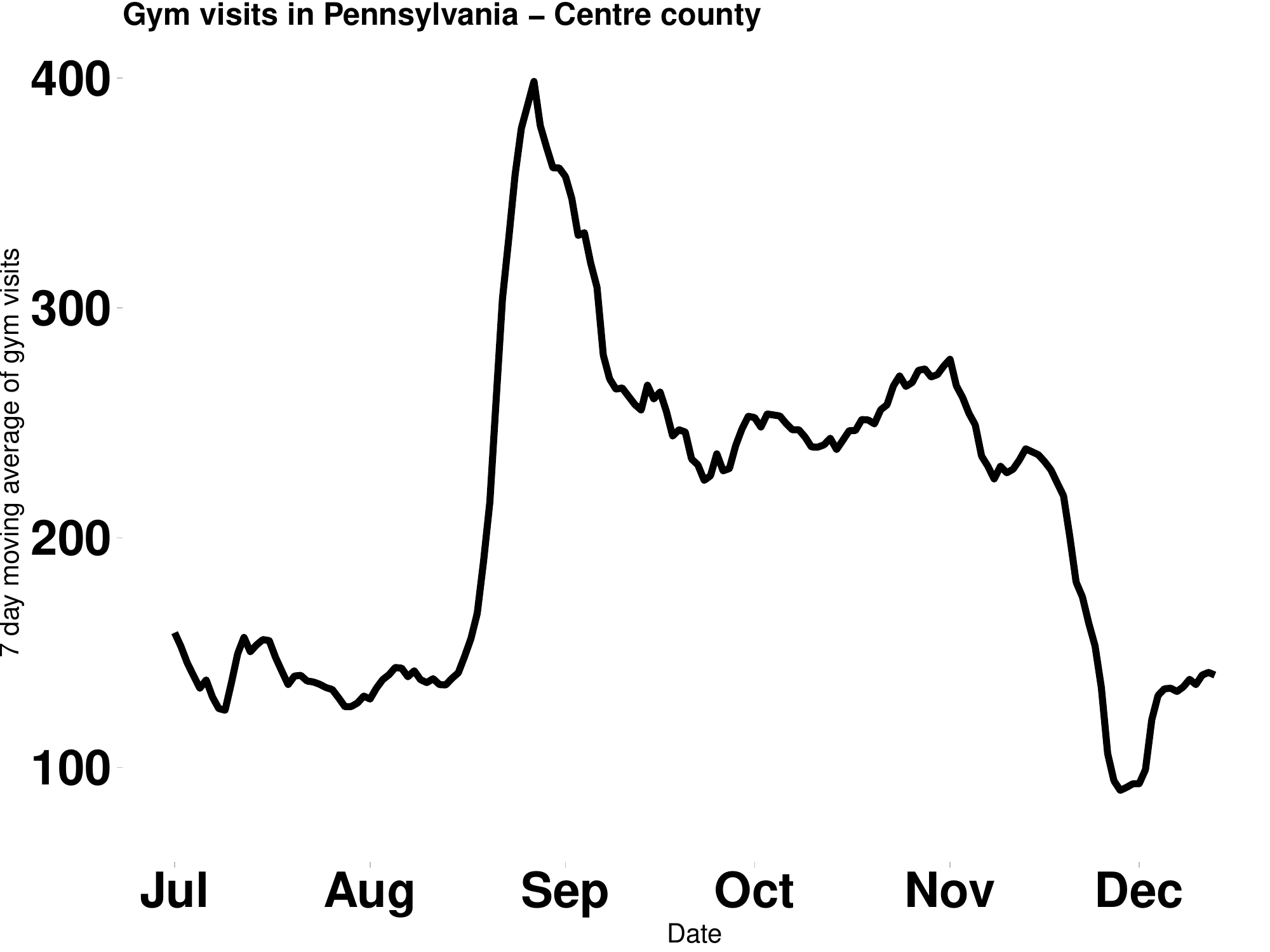}&
      \includegraphics[width=0.20\textwidth]{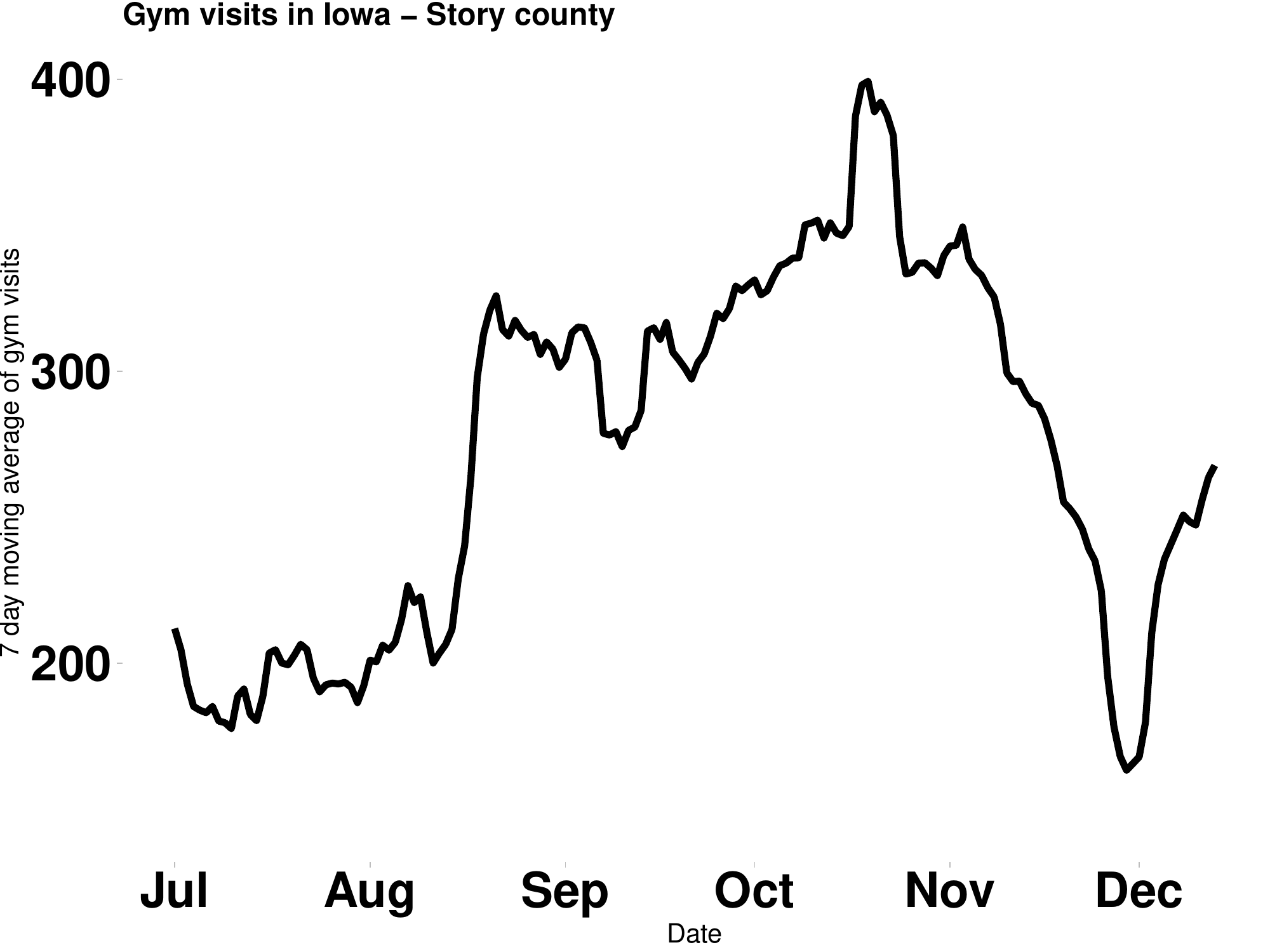}& 
      \includegraphics[width=0.20\textwidth]{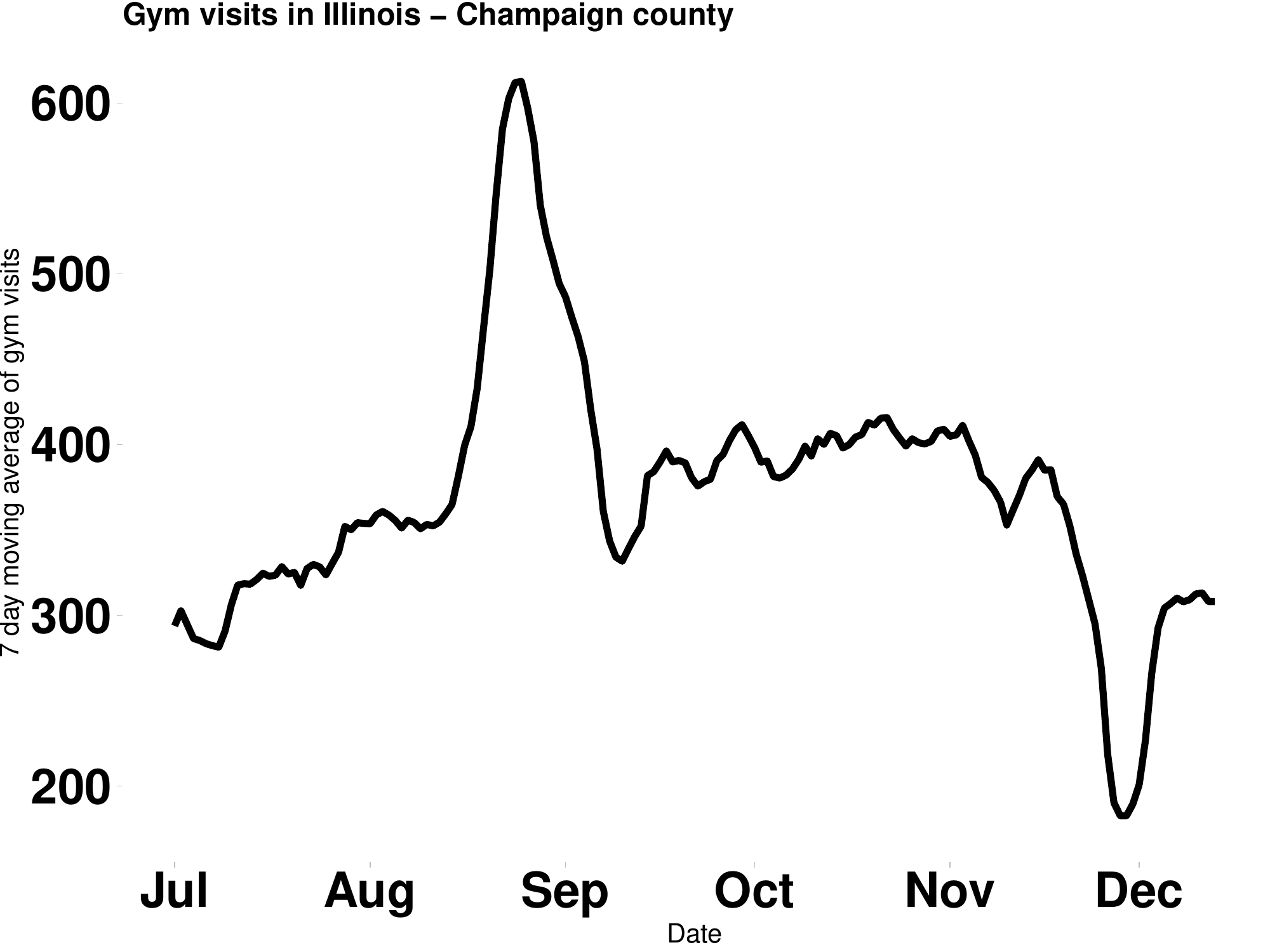}\\   
   School Visits  &  School Visits &School Visits  &  School Visits &School Visits   \\  
      \includegraphics[width=0.20\textwidth]{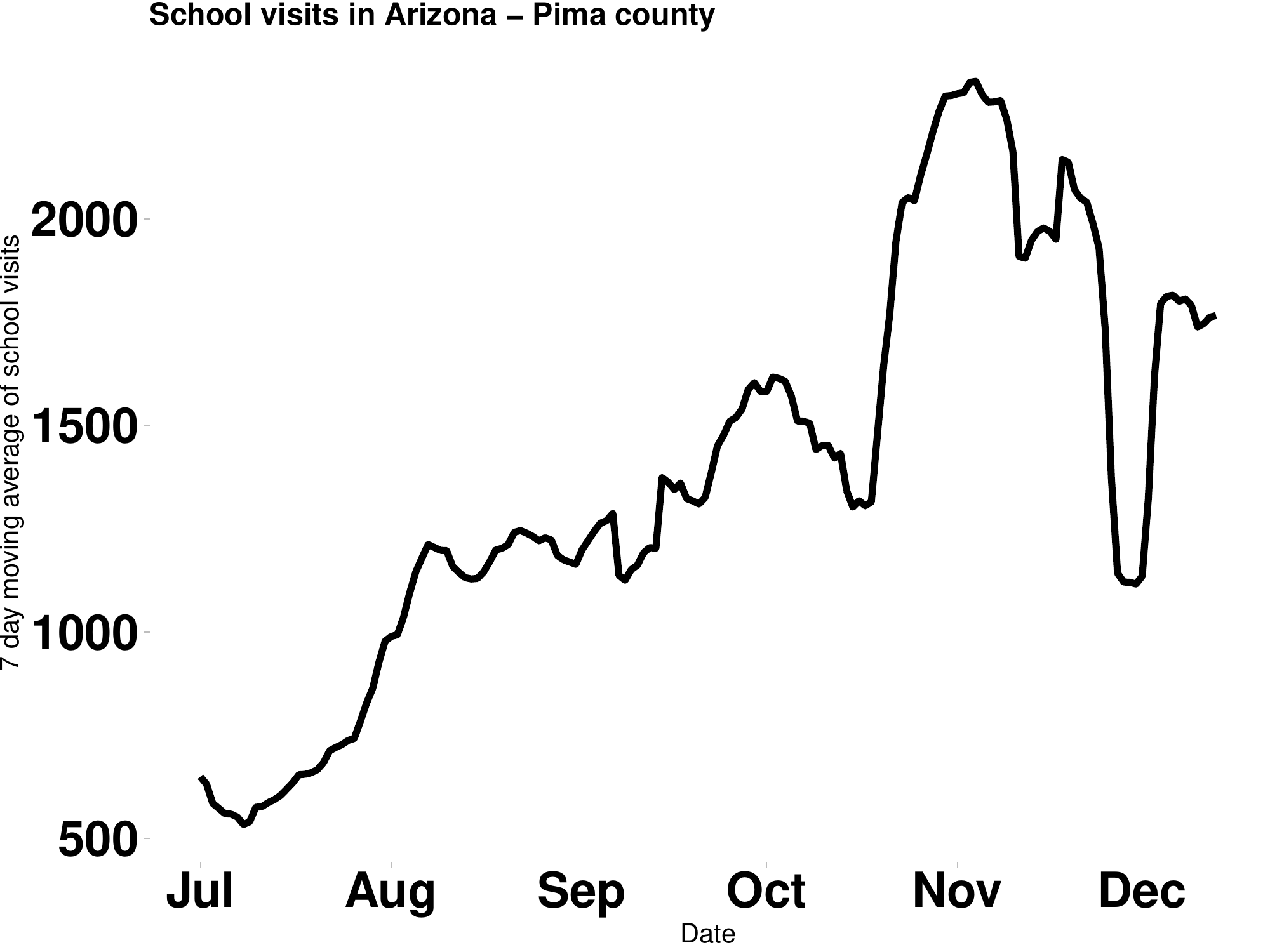}&
      \includegraphics[width=0.20\textwidth]{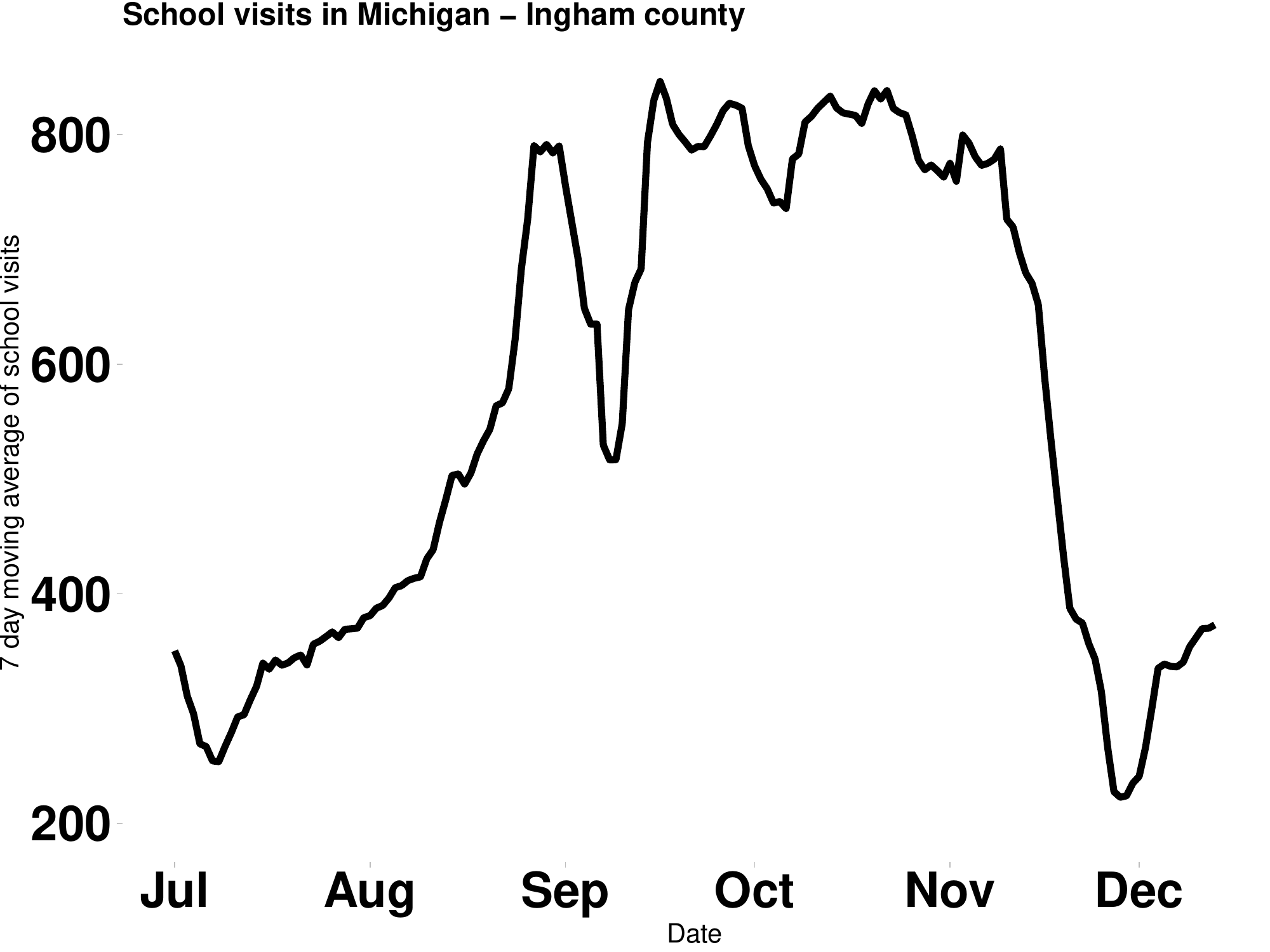}&
      \includegraphics[width=0.20\textwidth]{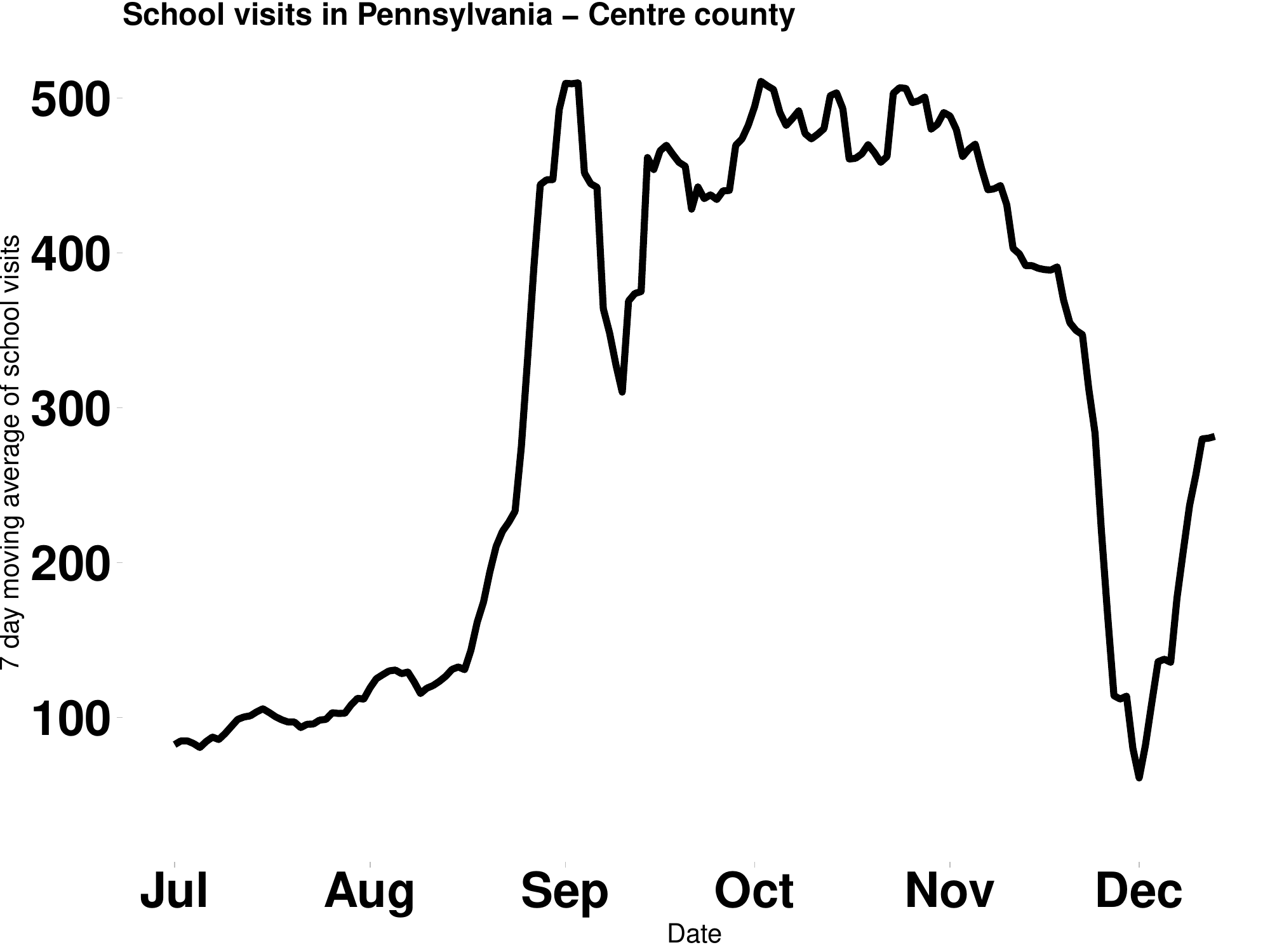}&
      \includegraphics[width=0.20\textwidth]{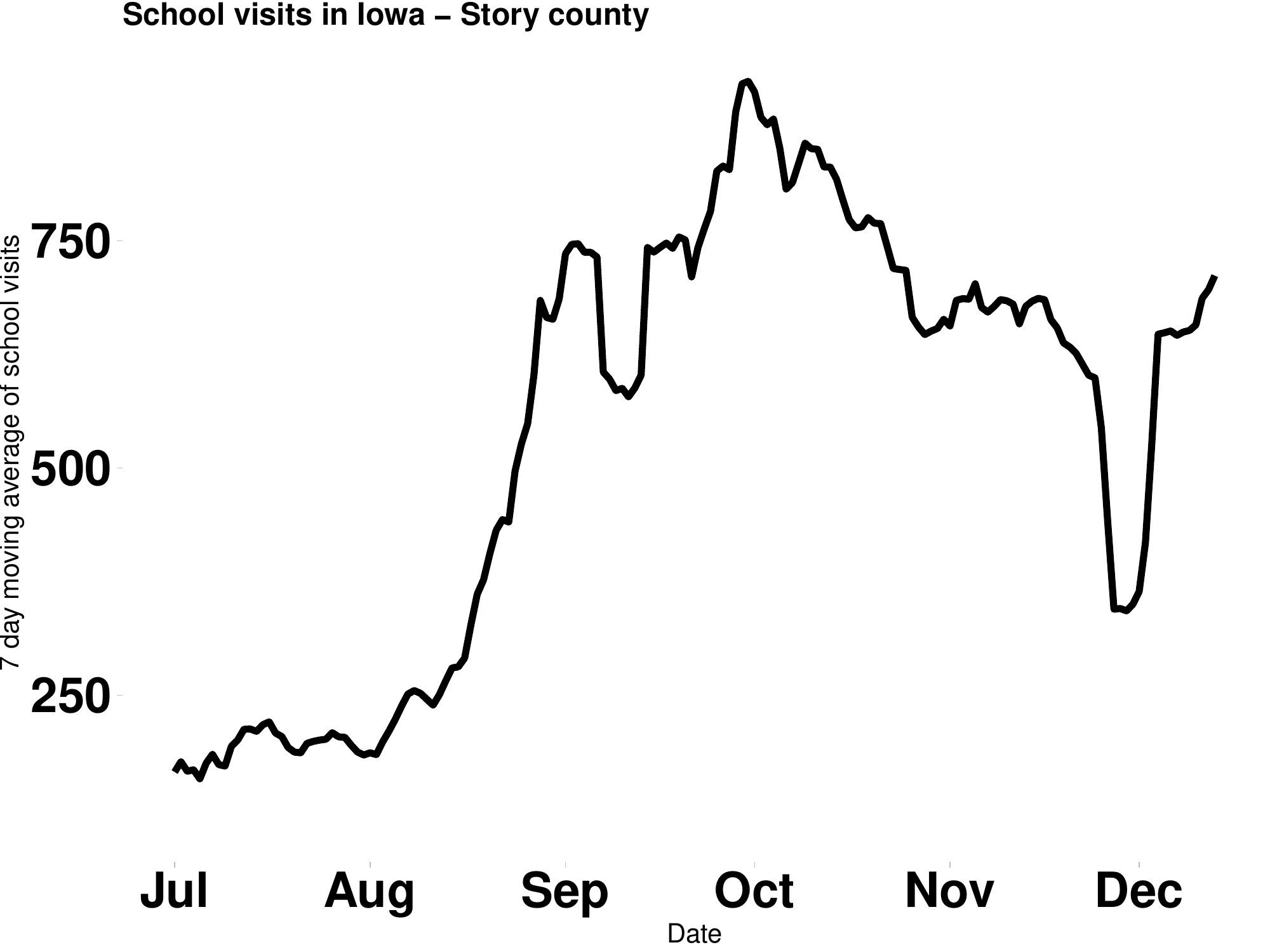}& 
      \includegraphics[width=0.20\textwidth]{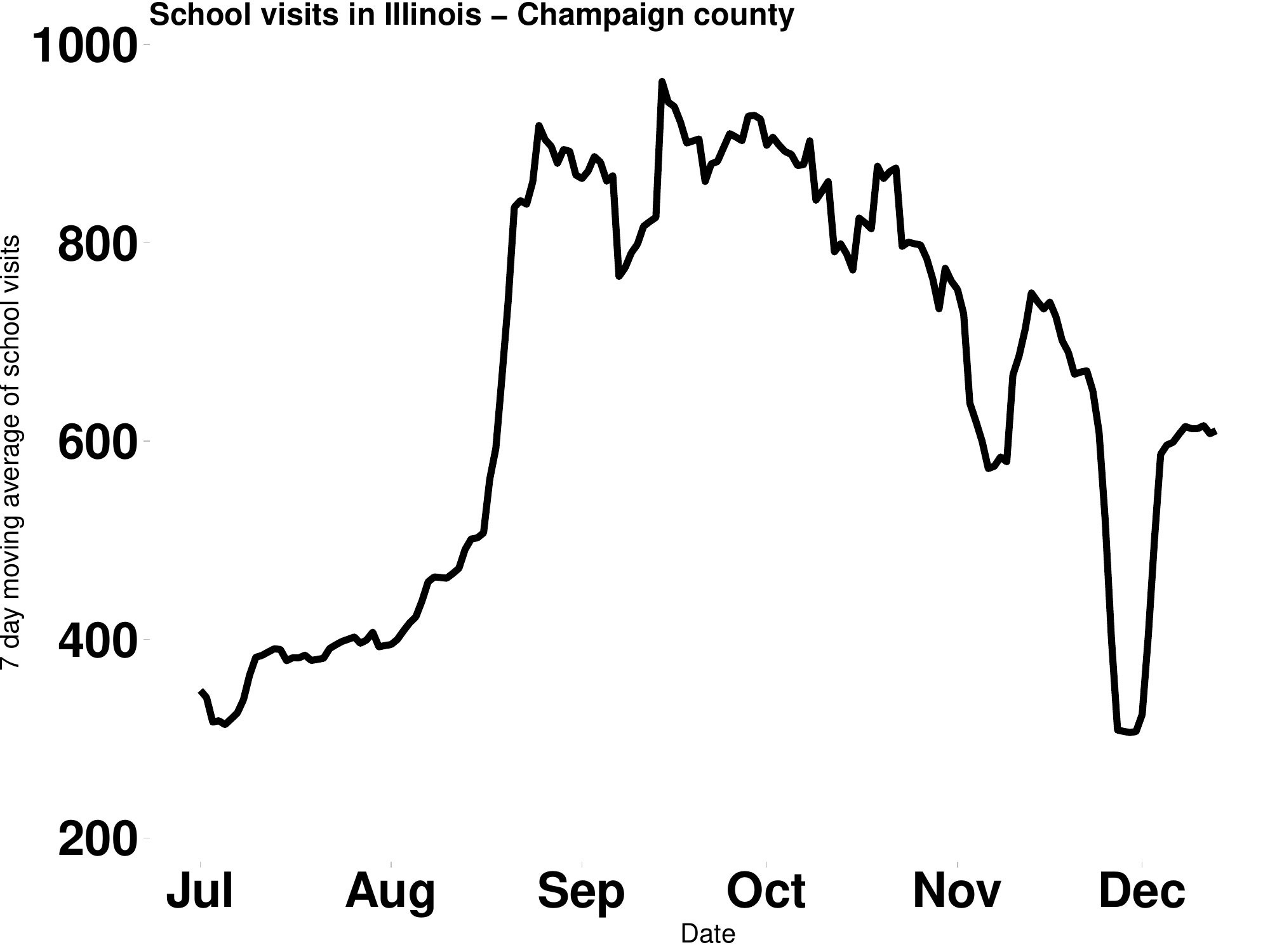}\\    
   CDC vs. NYT  Cases &  CDC vs. NYT  Cases & CDC vs. NYT  Cases &  CDC vs. NYT  Cases& CDC vs. NYT  Cases  \\ 
          \includegraphics[width=0.20\textwidth]{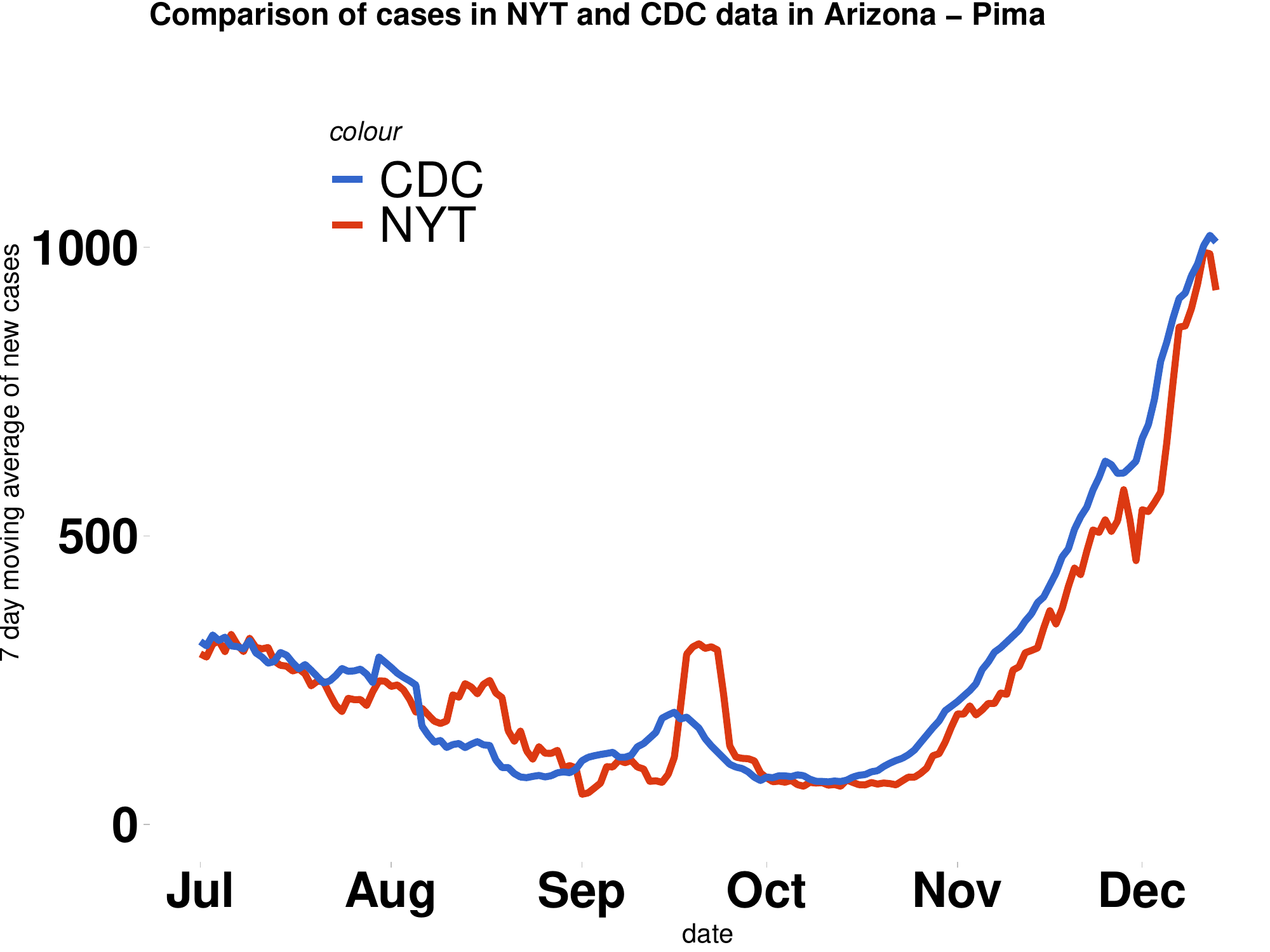}&
      \includegraphics[width=0.20\textwidth]{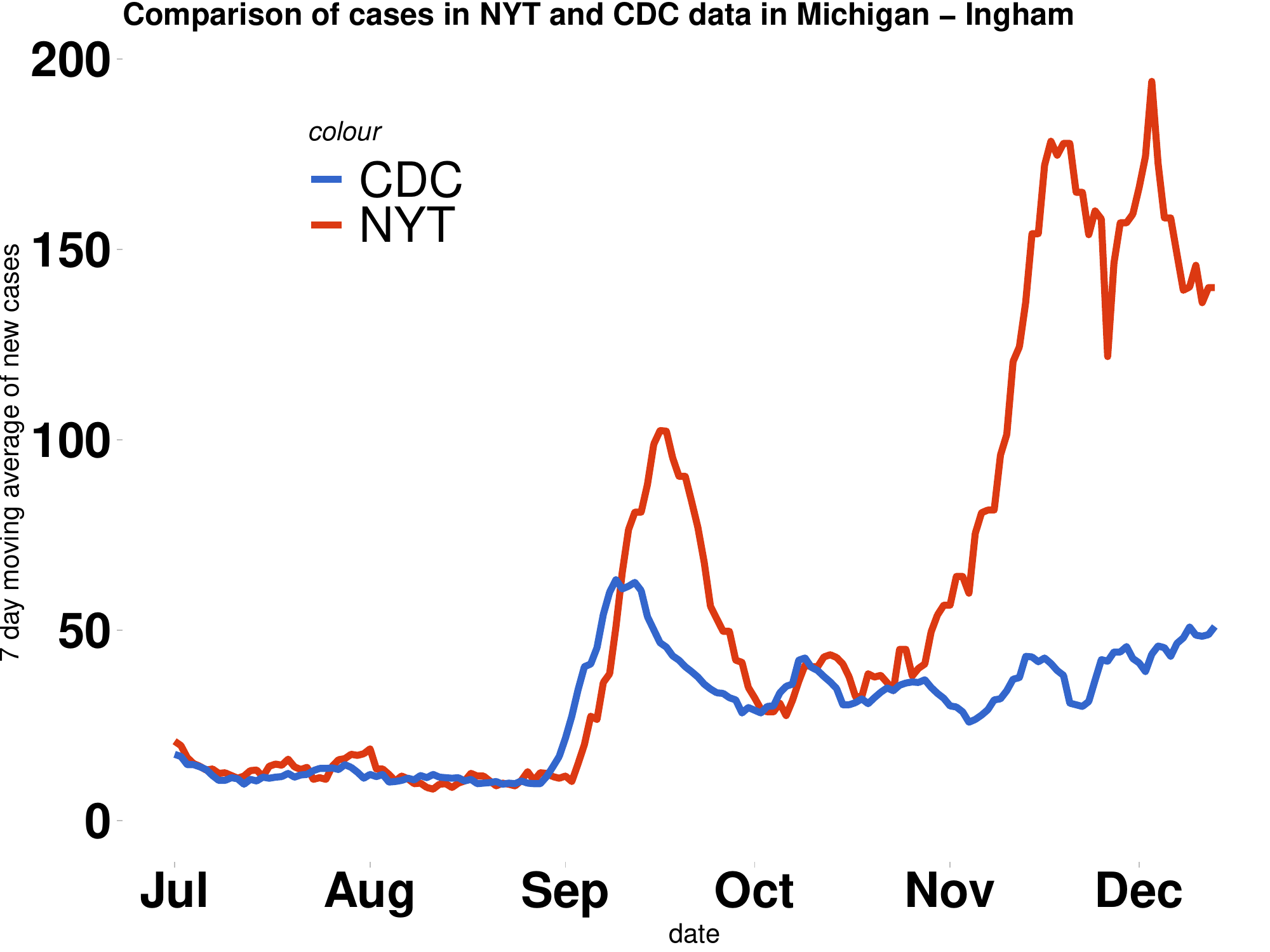}&
      \includegraphics[width=0.20\textwidth]{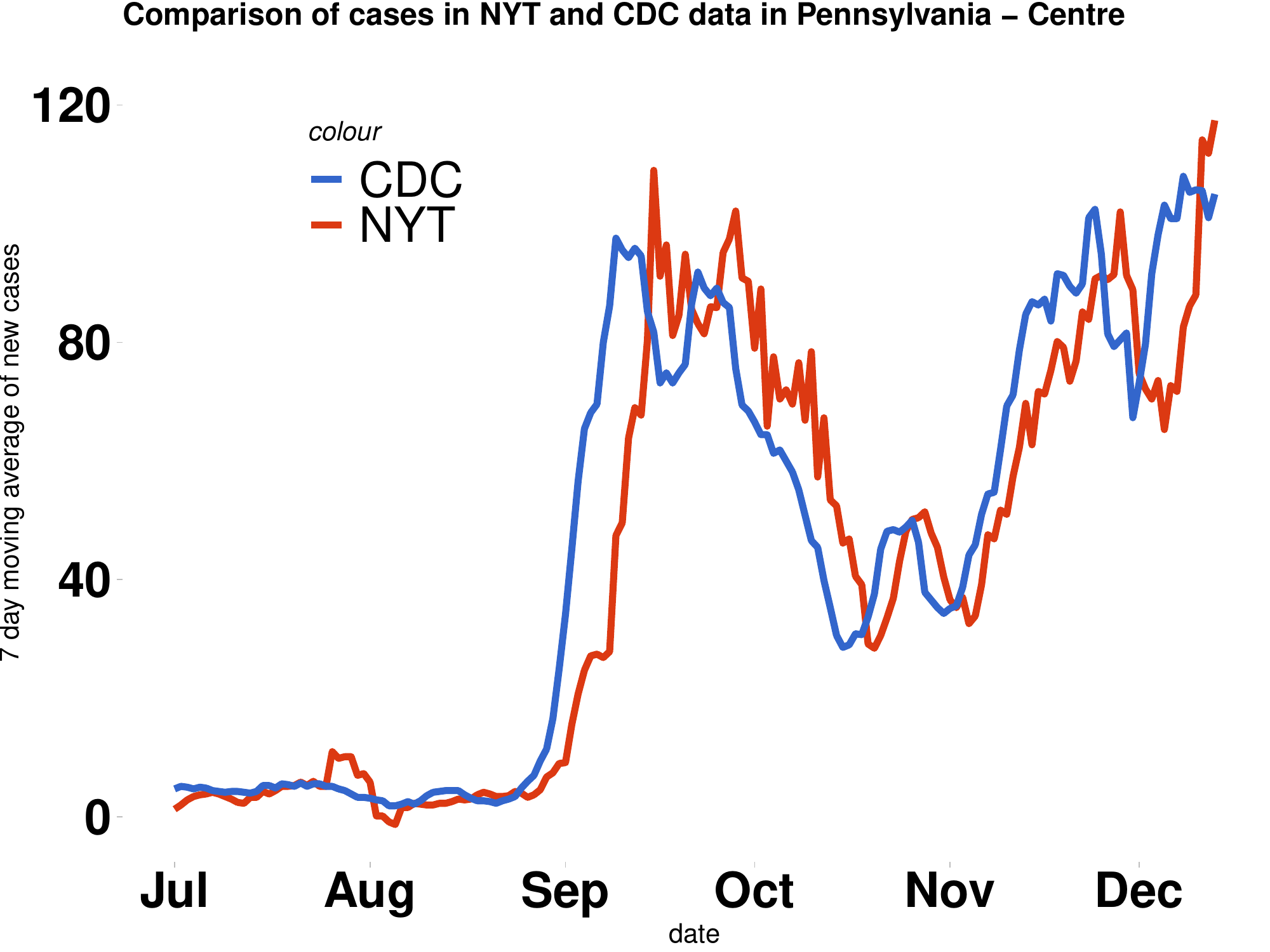}&
      \includegraphics[width=0.20\textwidth]{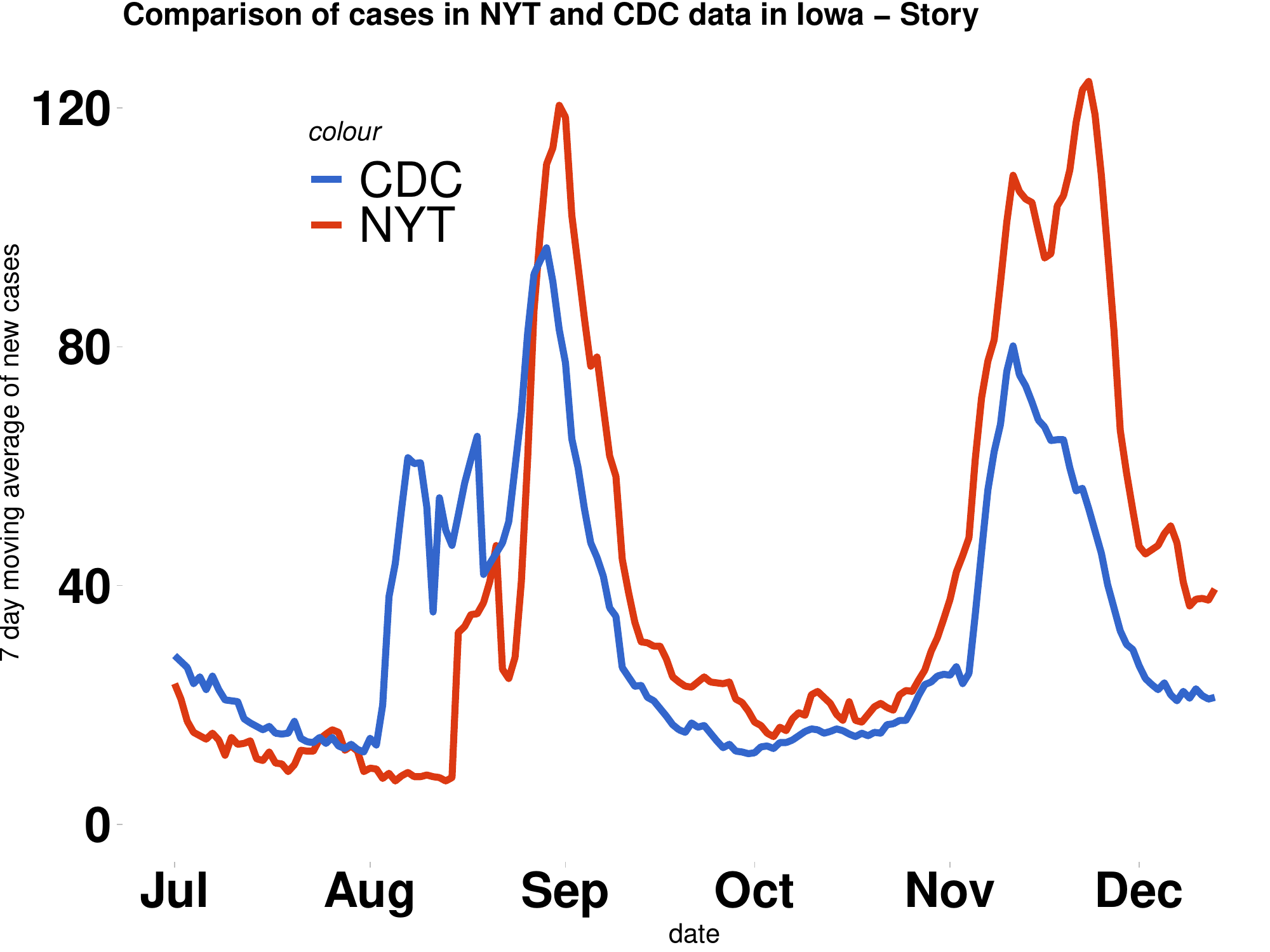}& 
      \includegraphics[width=0.20\textwidth]{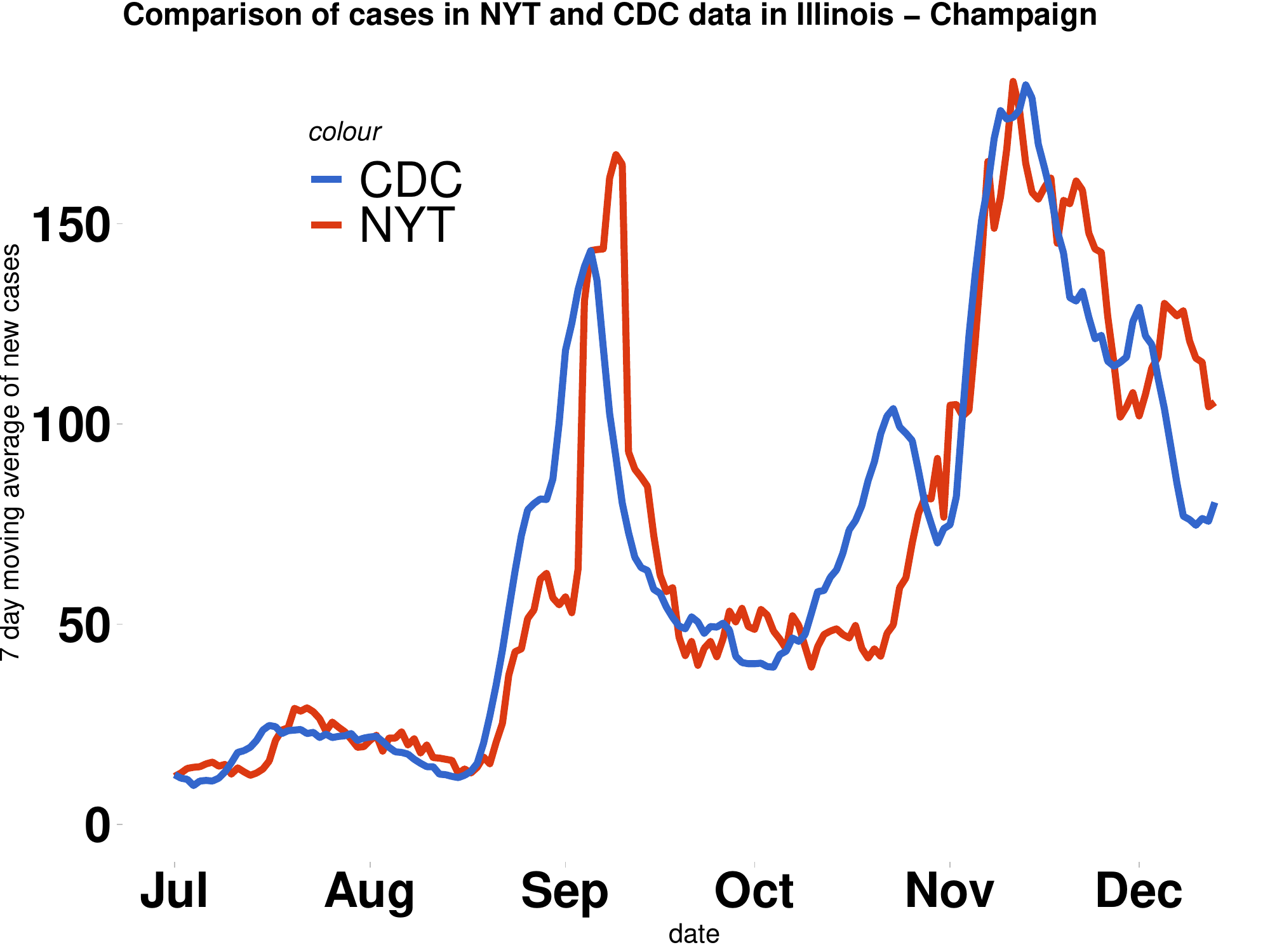}\\     
    \end{tabular} 
  \end{minipage}}  
   {\scriptsize
\begin{flushleft}
Notes: Figure corresponds to Fig. S2 but for Pima, AZ, Ingham, MI, Centre, PA, Story, IA, and Champaign, IL. Across various counties, we also report the evolution of visits to recreation facilities and K-12 school visits. The last panel at the bottom compares the sum of  weekly cases across all age groups reported in CDC dataset with the weekly reported case in NYT dataset.
\end{flushleft}  } 
 \end{figure}

Fig.  \ref{fig:dane} and \ref{fig:dane-SI}  provide descriptive evidence that opening colleges and universities may be associated with the spread of COVID-19 in counties where the University of Wisconsin(UW)-Madison, the University of Oregon, the University of Arizona,  the Michigan State University, the Pennsylvania State University, the Iowa State University, and the University of Illinois-Champaign are located. What happened in Dane county, WI, is also illustrative. The left panel of  Fig.  \ref{fig:dane} presents the evolution of the number of cases by age groups,  the number of visits to colleges and universities, and the number of visits to bars and restaurants in Dane county, WI. The first panel shows that the number of cases for age groups of 10-19  and 20-29 sharply increased in mid-September while few cases were reported for other age groups. The second to the fourth panels suggest that this sharp increase in cases among the 10-29 age cohort in mid-September is associated with an increase in visits to colleges/universities, bars, and restaurants in late August and early September. The fall semester with in-person classes at the UW-Madison began on September 2, 2020, when many undergraduates started living together in residential halls and likely visited bars and restaurants. This resulted in increases in COVID-19 cases on campus; according to the letter from Dane County Executive Joe Parisi to the UW-Madison \cite{parisi2020}, nearly 1,000 positive cases were confirmed on the UW-Madison campus by September 9, 2020, accounting for at least 74 percent of confirmed cases from September 1 to 8, 2020 in Dane county.  The UW-Madison offered no-cost testing to all students and the tests are mandatory upon returning to campus for those who returned to residence halls (\href{https://news.wisc.edu/uw-madison-establishes-free-campus-wide-covid-19-testing-to-support-campus-reopening/}{source 1} and \href{https://chancellor.wisc.edu/blog/uw-in-a-semester-of-covid/}{source 2}). Therefore, an increase in confirmed cases among college-age people  in Dane County is likely to be partly driven by an increase in the number of tests. 

One likely reason why college openings may increase cases is that students go out for bars \citep{Fisher2020,Chang2021}, where properly wearing masks and practicing social distancing are difficult.  Table \ref{tab:PItoB-SI} presents how visits to restaurants and bars are associated with colleges/universities from panel regressions using per-device visits to restaurants and bars as outcome variables. These results indicate that bar visits are positively associated with college visits, consistent with a hypothesis that the transmission of SARS-CoV-2 may be partly driven by an increase in visits to bars by students.

\begin{figure}[!ht] 
\caption{Evolution of Cases/Deaths per 1000 Persons,  Case/Death Growth, Visits to K-12 Schools, Colleges, Restaurants, Bars, Recreational Facilities,   Churches, K-12 School Opening Modes, and NPIs  across U.S. counties \label{fig:evolution-SI}}\medskip
\hspace{-1cm}\resizebox{\columnwidth}{!}{
\begin{minipage}{\linewidth}
    \centering
        \begin{tabular}{ccc} 
   (a) Wkly Case Growth &  (b)  Wkly Cases per 1000&  (c) Visits to K-12 Schools \\
      \includegraphics[width=0.33\textwidth]{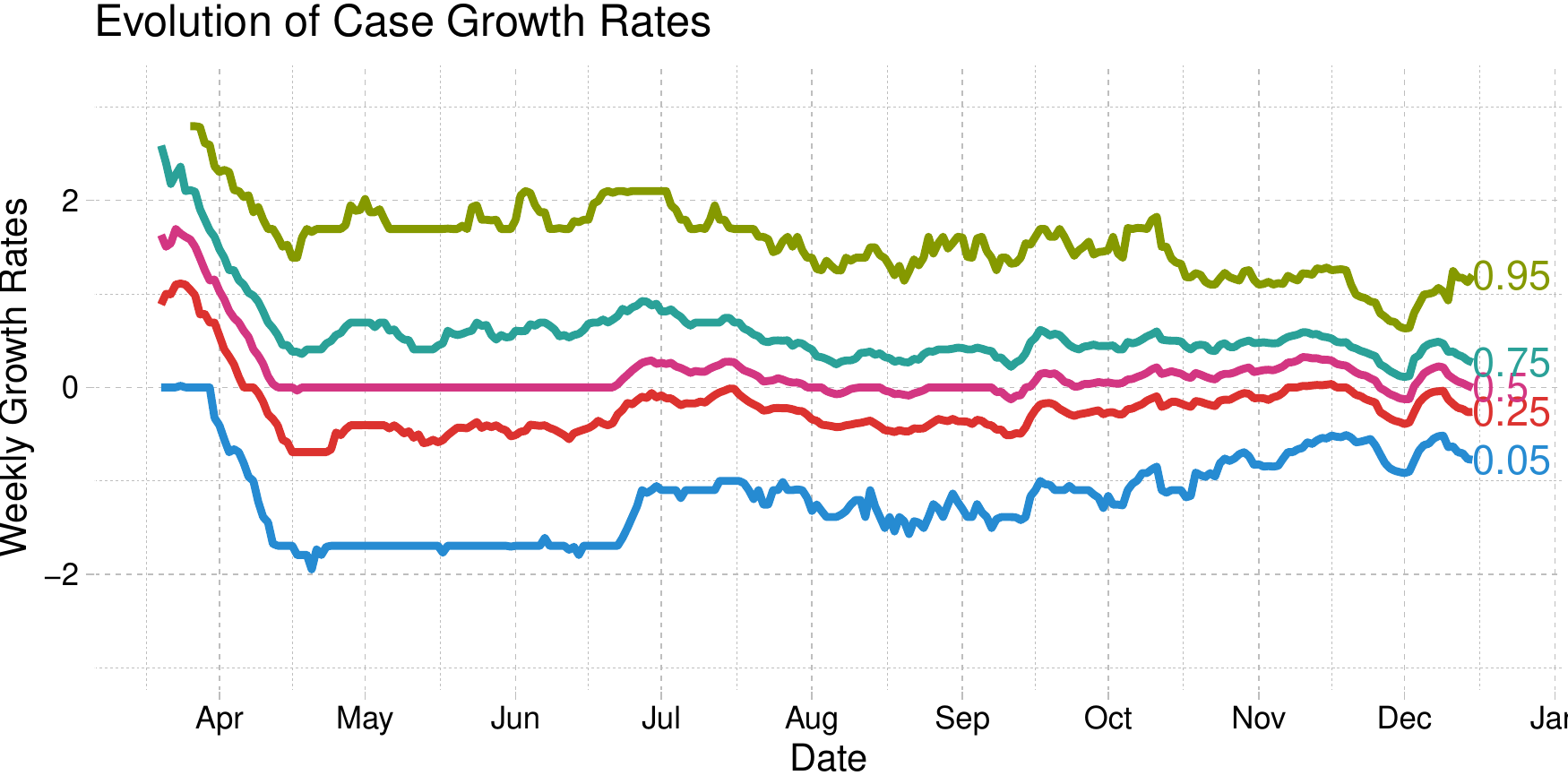}&
      \includegraphics[width=0.33\textwidth]{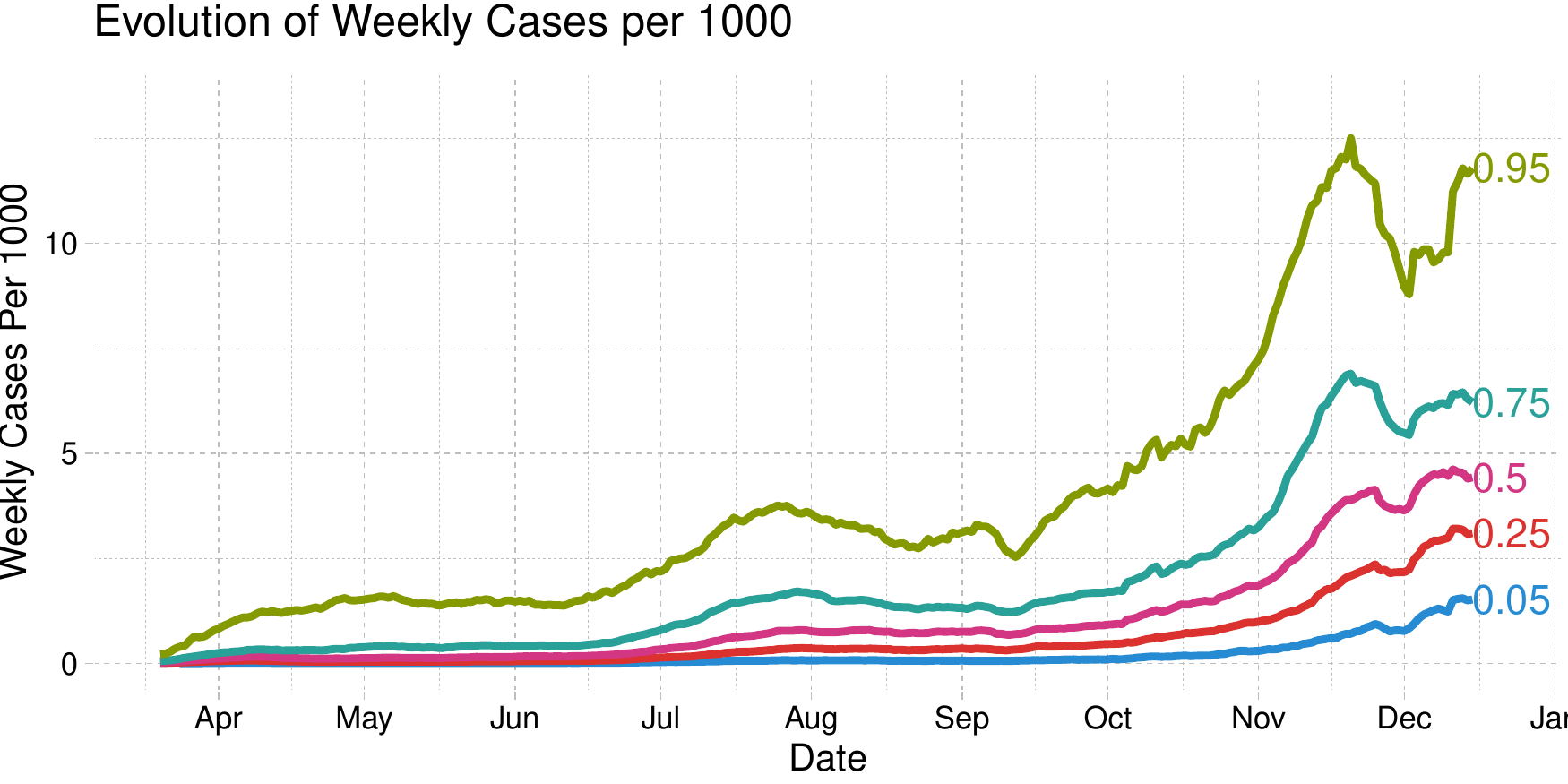}&
  \includegraphics[width=0.33\textwidth]{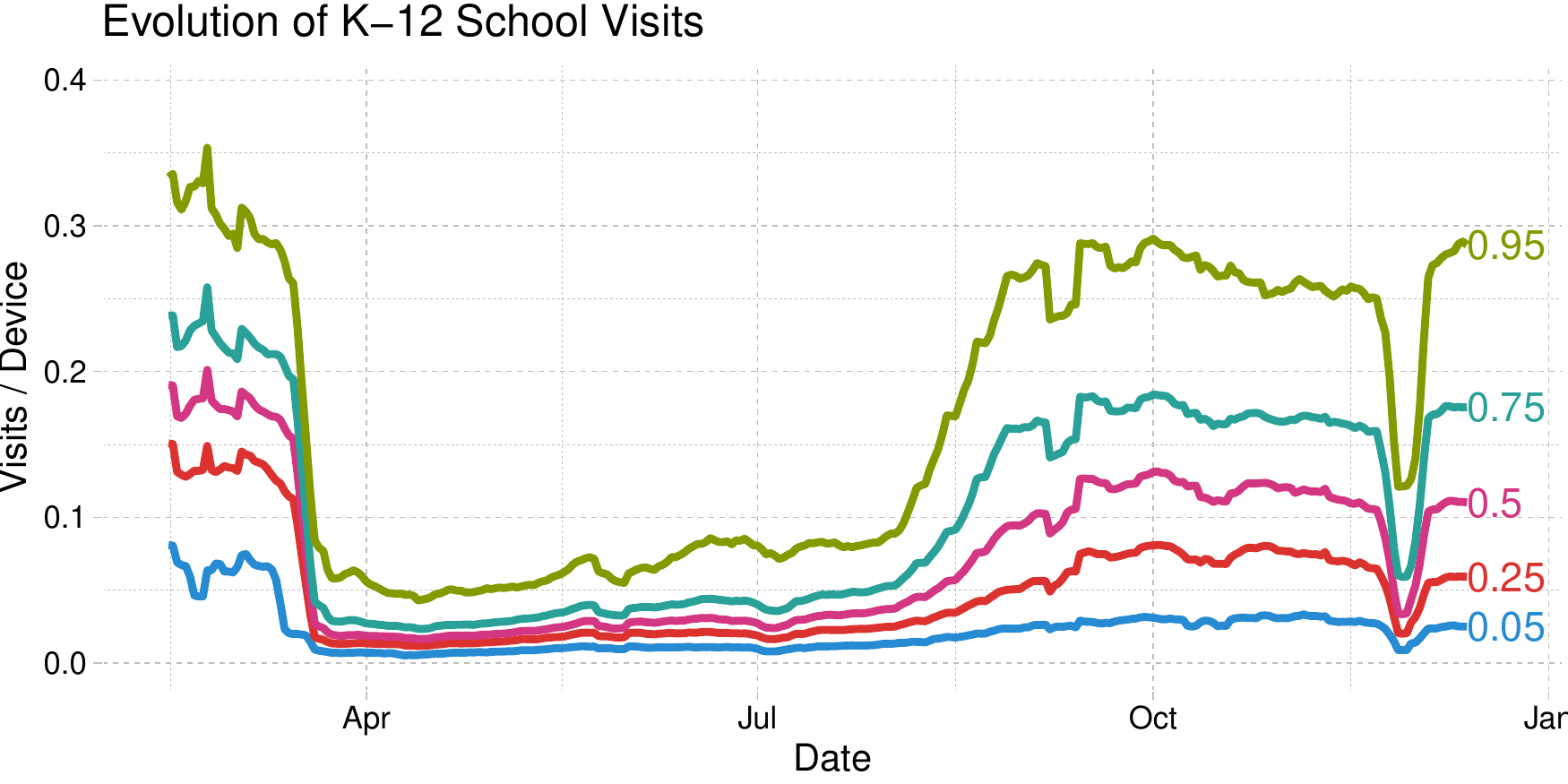}\\ 
   (d) Wkly Death Growth &  (e)  Wkly Deaths per 1000 & (f) Visits to Colleges\\
      \includegraphics[width=0.33\textwidth]{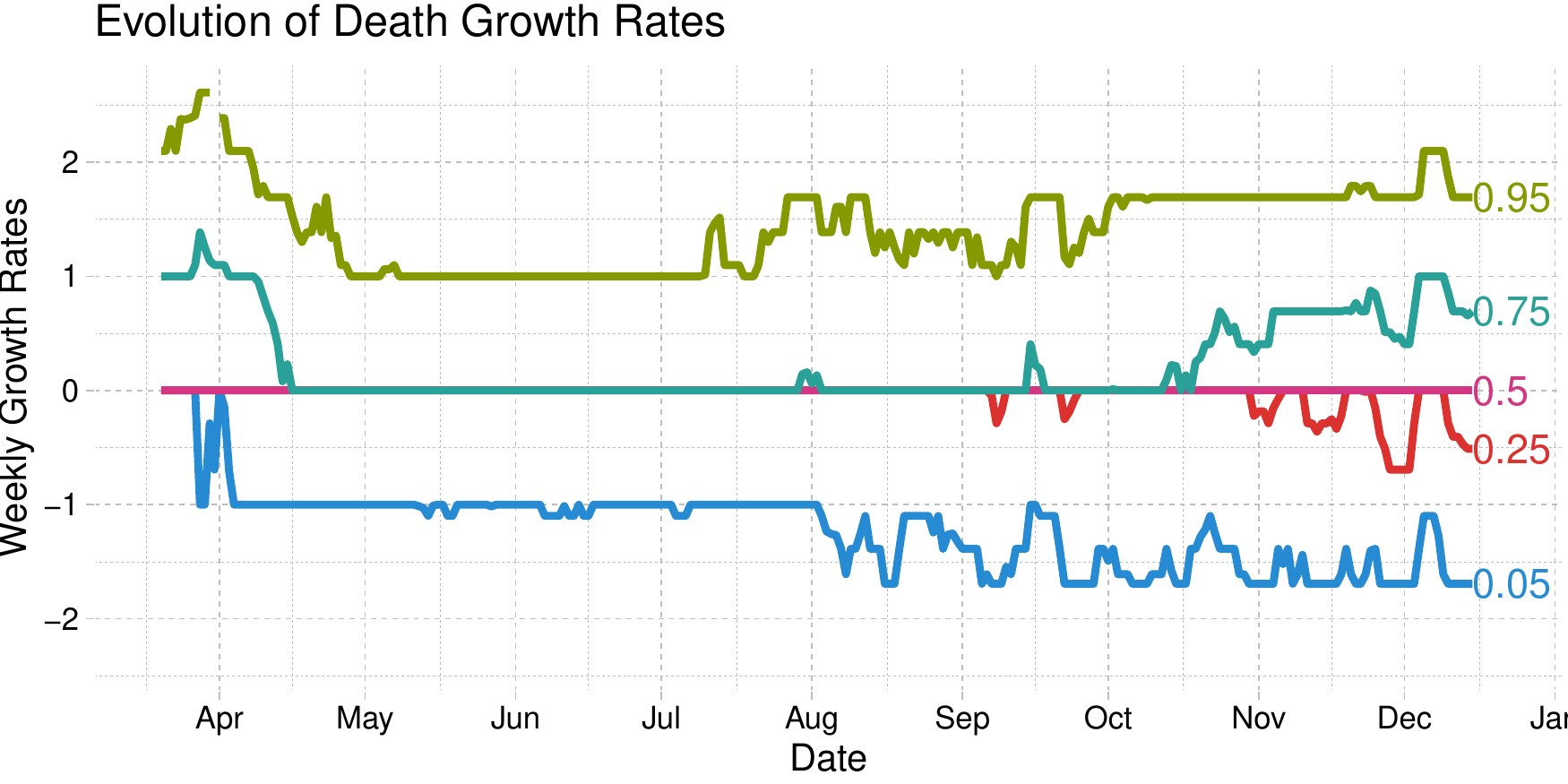}&
      \includegraphics[width=0.33\textwidth]{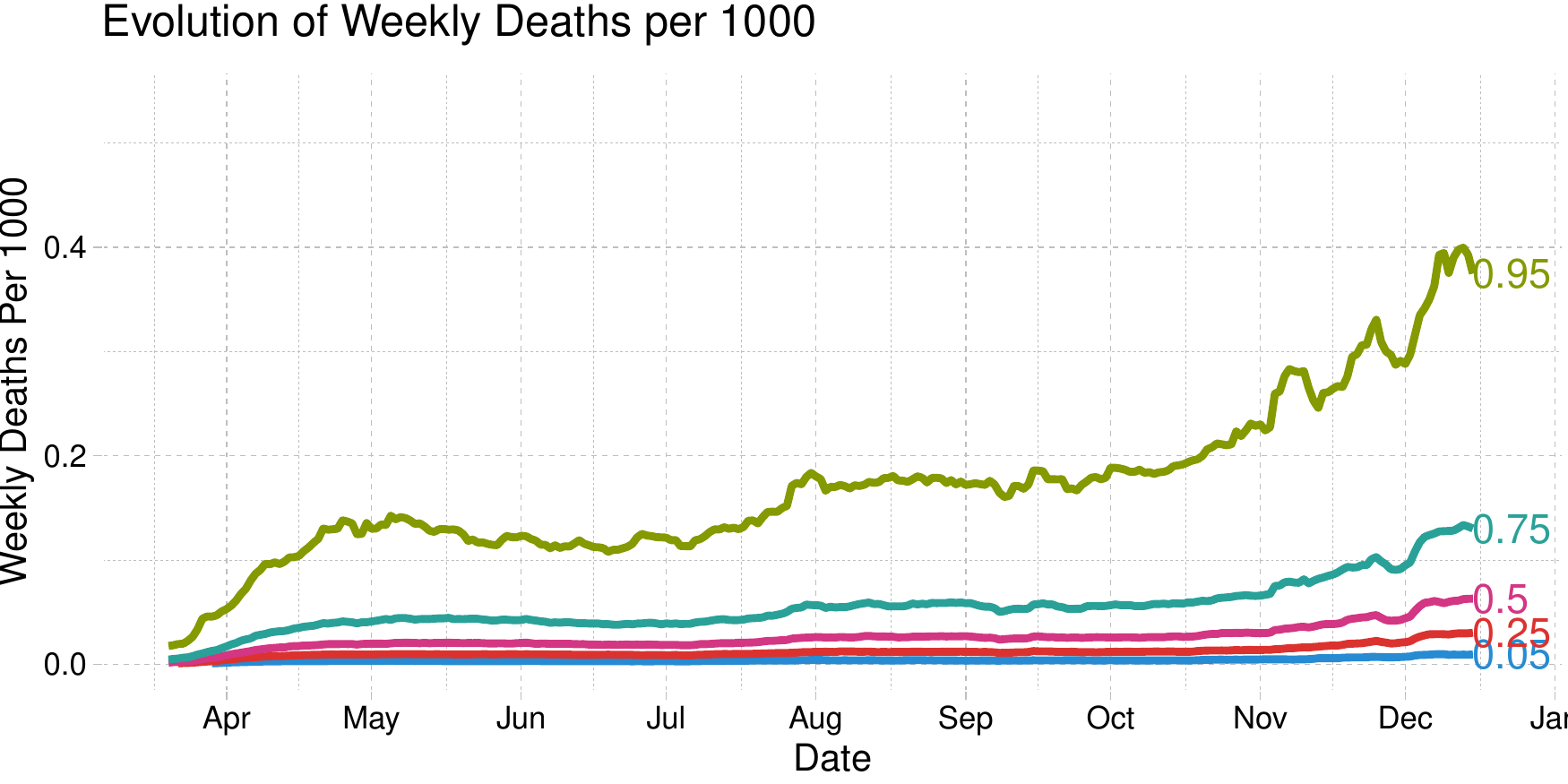}&
  \includegraphics[width=0.33\textwidth]{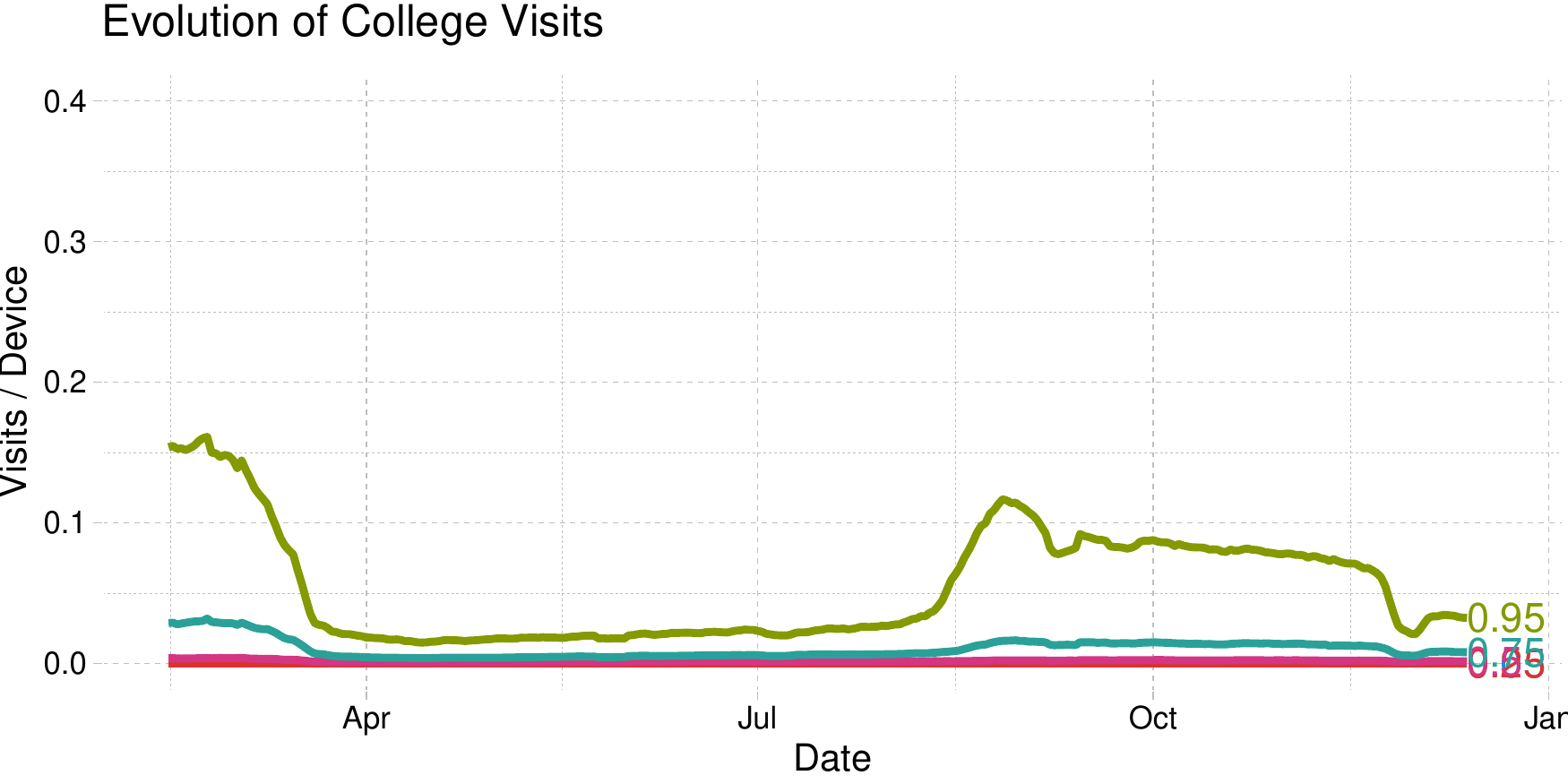}\\  
  (g) Visits to Restaurants & (h) Visits to Bars & (i) Visits to Rec. Facilities\\
  \includegraphics[width=0.33\textwidth]{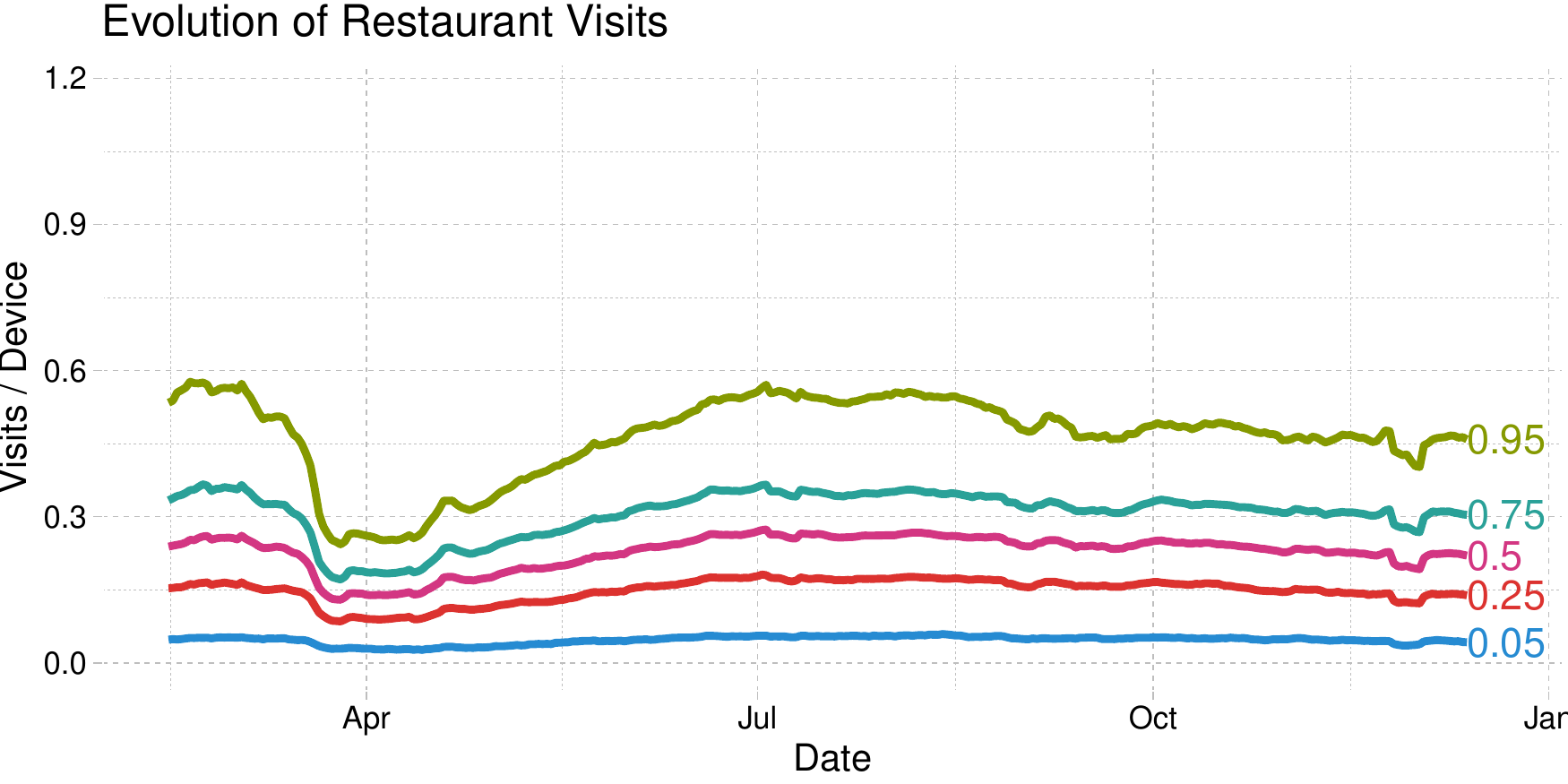}&
  \includegraphics[width=0.33\textwidth]{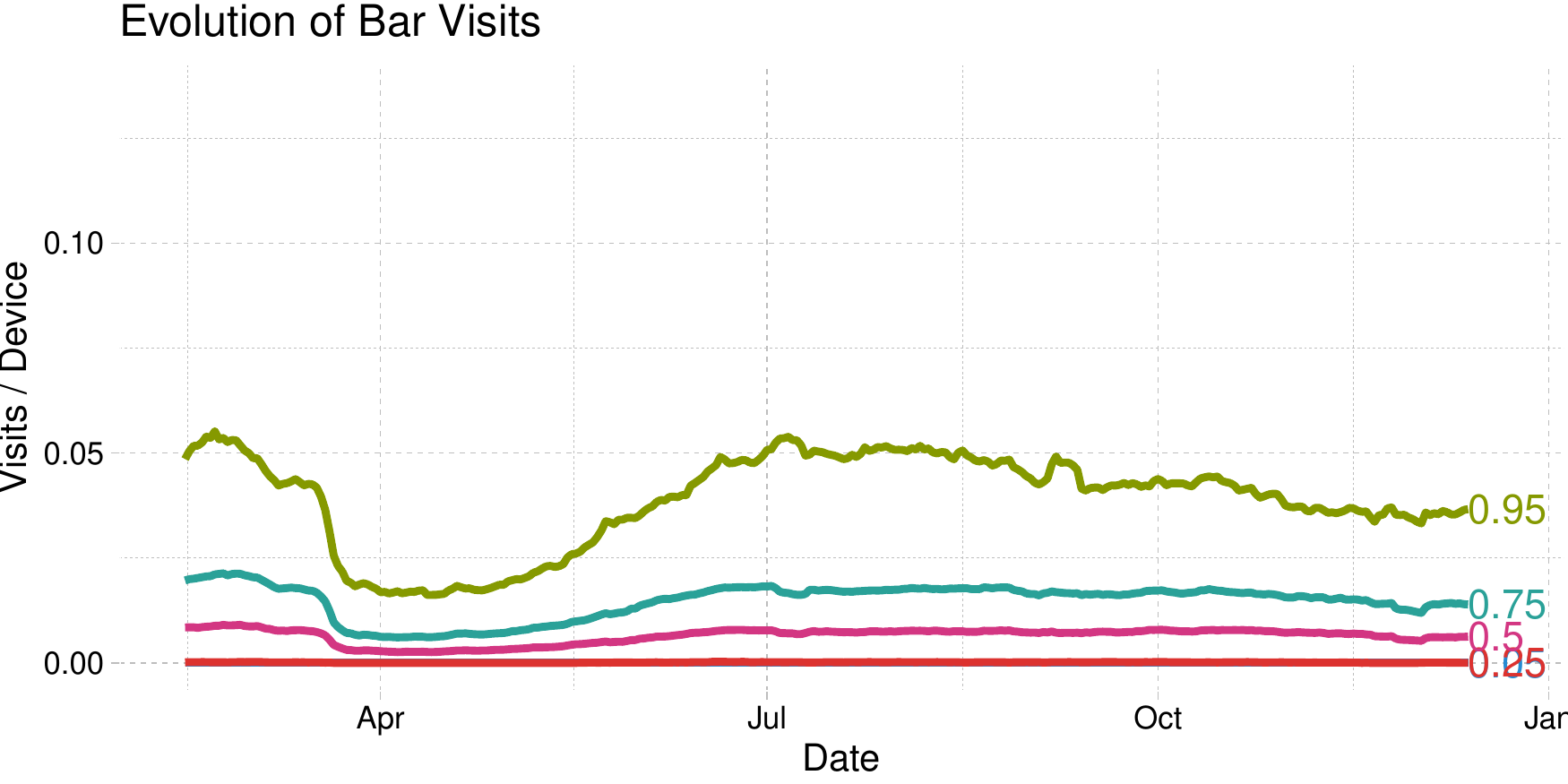}&
  \includegraphics[width=0.33\textwidth]{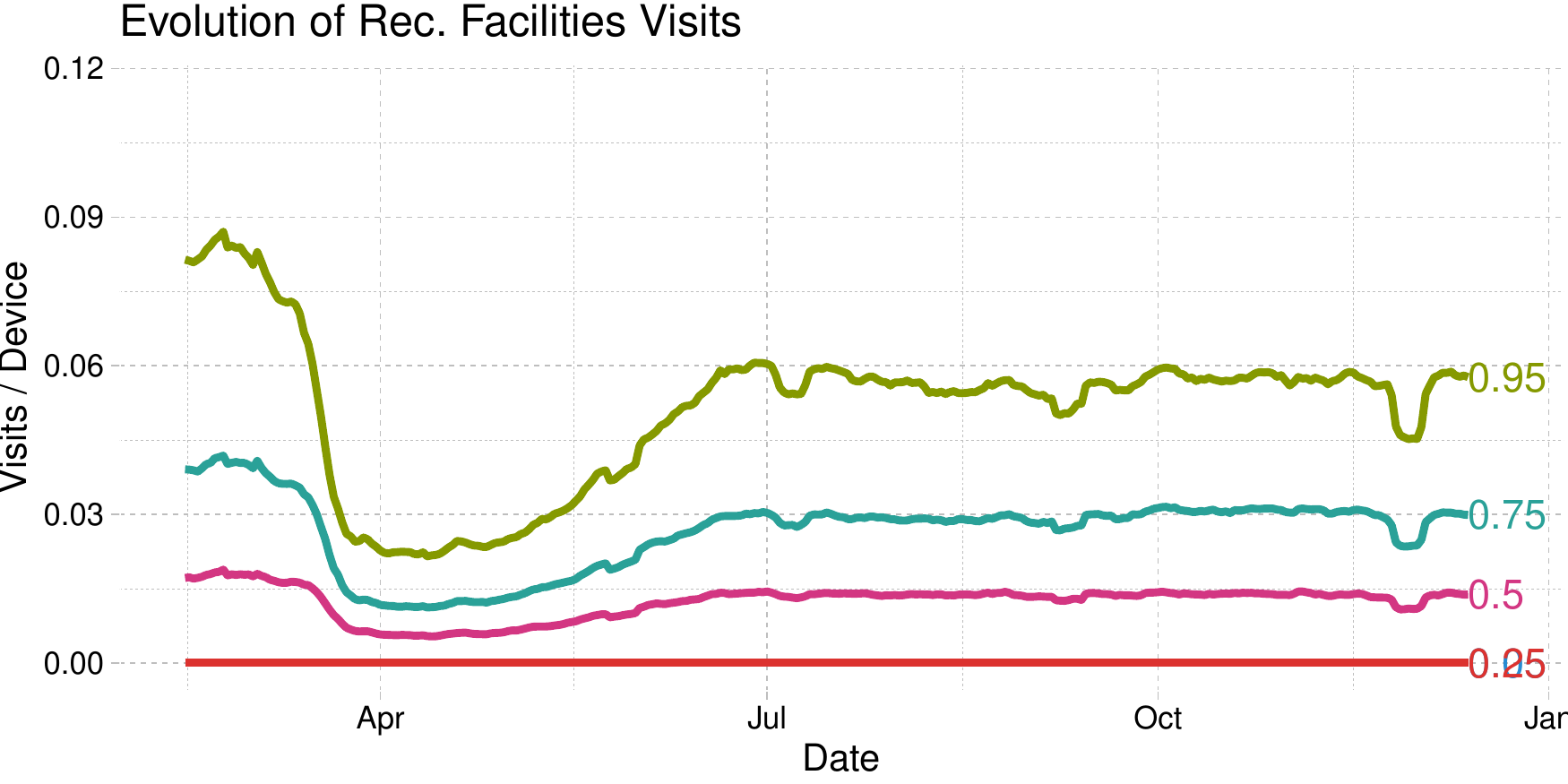}\\
   (j) Visits to Churches&(k) School Opening Modes & (l) NPIs\\ 
  \includegraphics[width=0.33\textwidth]{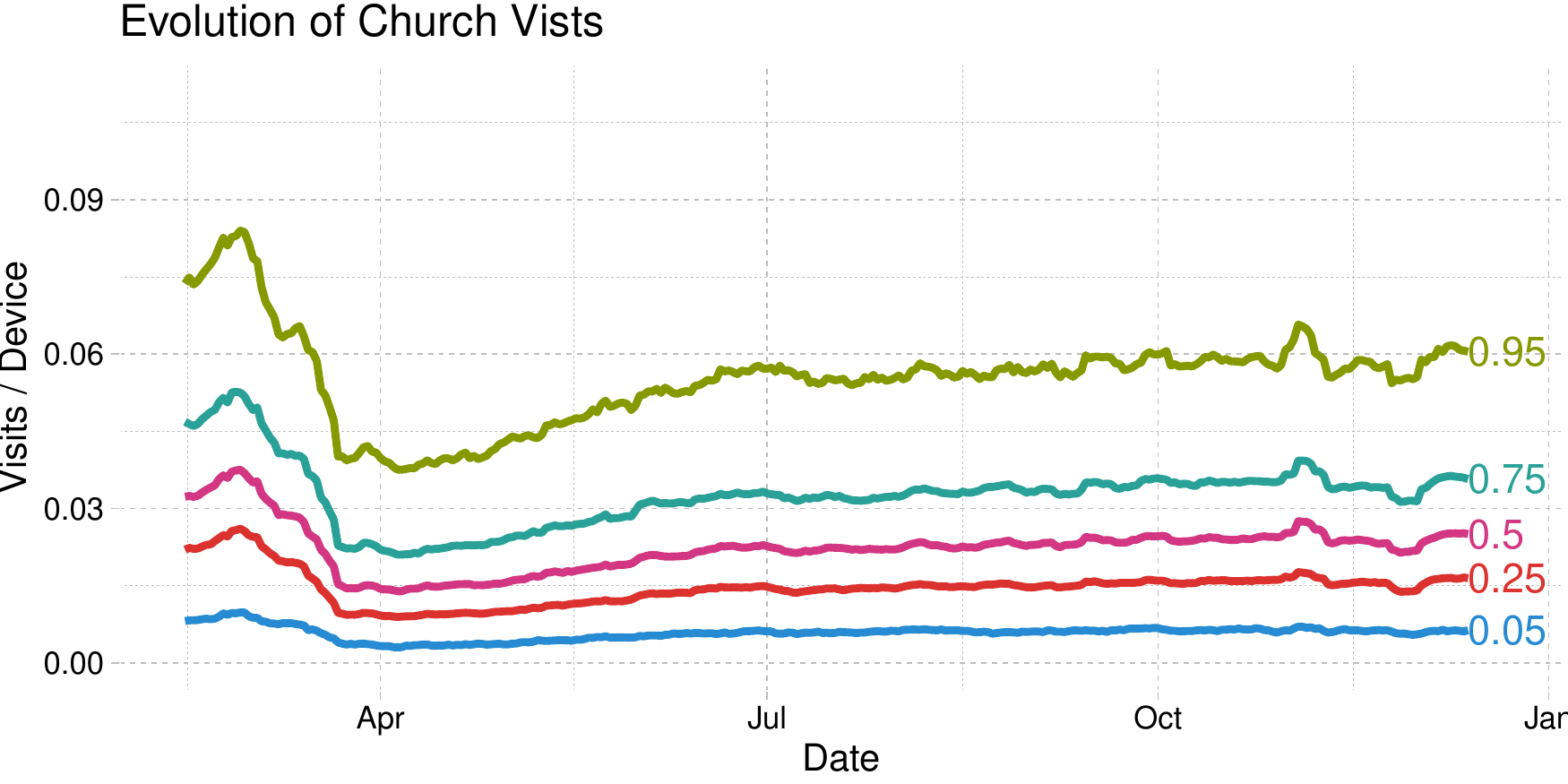}&  \includegraphics[width=0.33\textwidth]{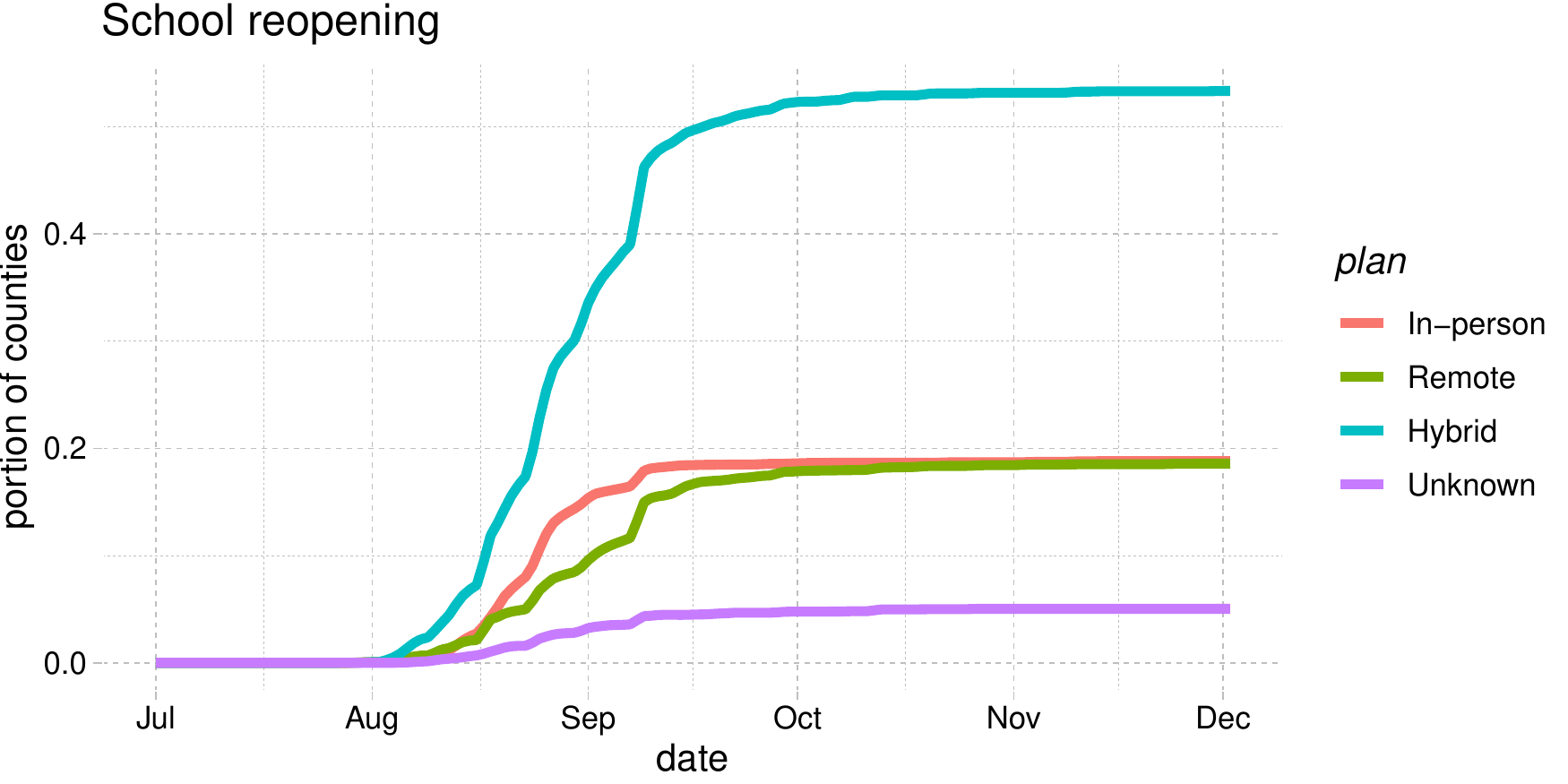}&
  \includegraphics[width=0.33\textwidth]{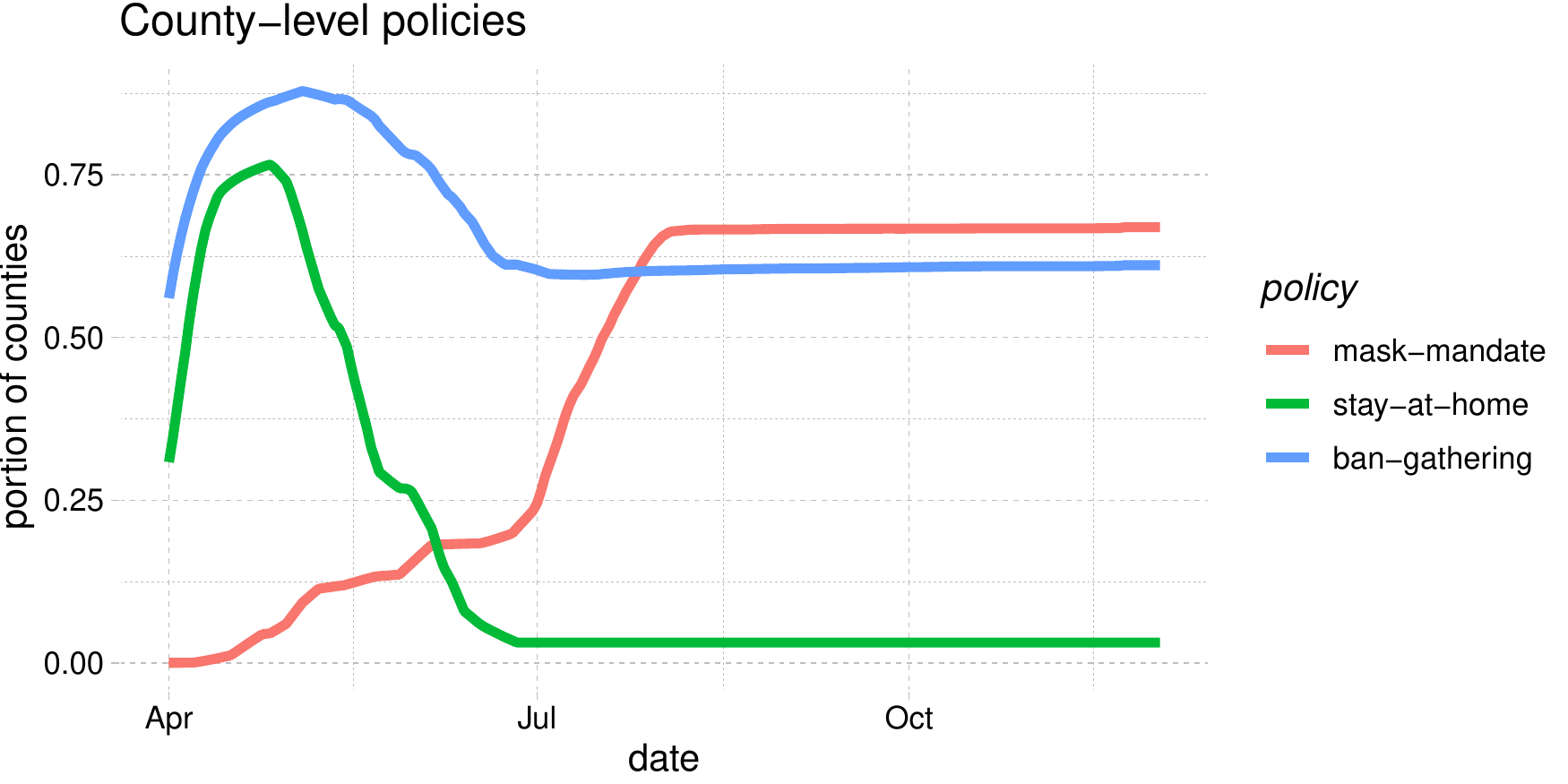}\\
    \end{tabular} 
  \end{minipage}}
  \begin{flushleft}{\ 
Notes:    (a)-(k) report the evolution of various percentiles of corresponding variables in the title over time.  (10)  reports the proportion of counties that open K-12 schools with different teaching methods including ``Unknown'' over time while   (l) reports the proportion of counties that implement three NPIs over time. }
\end{flushleft}   
\end{figure}

\begin{table}[!ht] 
\caption{Summary Statistics \label{fig:summary-SI}}
\hspace{-4cm}\resizebox{0.9\columnwidth}{!}{
\begin{minipage}{\linewidth}
    \centering
\begin{tabular}{@{\extracolsep{5pt}}lccccccc} 
\\[-1.8ex]\hline 
\hline \\[-1.8ex] 
Statistic & \multicolumn{1}{c}{N} & \multicolumn{1}{c}{Mean} & \multicolumn{1}{c}{St. Dev.} & \multicolumn{1}{c}{Min} & \multicolumn{1}{c}{Pctl(25)} & \multicolumn{1}{c}{Pctl(75)} & \multicolumn{1}{c}{Max} \\ 
\hline \\[-1.8ex] 
Wkly Case Growth Rate over 7 days$^{\text{a}}$& 698,278 & 0.099 & 0.901 & $-$8.107 & $-$0.288 & 0.495 & 8.002 \\ 
Wkly Death Growth Rate over 7 days$^{\text{b}}$ & 698,278 & 0.023 & 0.790 & $-$6.170 & 0.000 & 0.000 & 6.170 \\ 
Wkly Death Growth Rate over 21 days$^{\text{c}}$ & 633,617 & 0.115 & 0.988 & $-$5.159 & $-$0.211 & 0.693 & 5.277 \\ 
log(Wkly Cases)$^{\text{d}}$  & 703,702 & 2.829 & 2.140 & $-$1.000 & 1.386 & 4.331 & 10.488 \\ 
log(Wkly Deaths)$^{\text{e}}$& 703,702 & $-$0.269 & 1.147 & $-$1.000 & $-$1.000 & 0.000 & 6.479 \\ 
College Visits  & 728,228 & 0.010 & 0.031 & 0.000 & 0.000 & 0.008 & 1.827 \\ 
K-12 School Visits   & 728,228 & 0.074 & 0.072 & 0.000 & 0.024 & 0.103 & 1.167 \\ 
K-12 opening, in-person& 646,816 & 0.079 & 0.207 & 0.000 & 0.000 & 0.000 & 1.000 \\ 
K-12 opening, Hybrid & 646,816 & 0.224 & 0.357 & 0.000 & 0.000 & 0.424 & 1.000 \\ 
K-12 opening, Remote& 646,816 & 0.078 & 0.227 & 0.000 & 0.000 & 0.000 & 1.000 \\ 
No-Mask for Staffs & 577,680 & 0.293 & 0.455 & 0.000 & 0.000 & 1.000 & 1.000 \\ 
Mandatory Mask & 728,944 & 0.461 & 0.495 & 0 & 0 & 1 & 1 \\ 
Ban Gathering & 728,944 & 0.658 & 0.472 & 0 & 0 & 1 & 1 \\ 
Stay at Home & 728,944 & 0.143 & 0.345 & 0 & 0 & 0 & 1 \\ 
Full Time Workplace Visits  & 728,206 & 0.054 & 0.018 & 0.010 & 0.042 & 0.061 & 0.484 \\ 
Part Time Workplace Visits & 728,206 & 0.101 & 0.025 & 0.023 & 0.084 & 0.113 & 0.567 \\ 
Staying Home Devices & 728,206 & 0.342 & 0.116 & 0.021 & 0.267 & 0.393 & 3.657 \\ 
Recreational Place Visits & 728,228 & 0.017 & 0.022 & 0.000 & 0.000 & 0.026 & 0.786 \\ 
Church Visits & 728,228 & 0.025 & 0.018 & 0.000 & 0.014 & 0.032 & 0.583 \\ 
Drinking Place Visits & 728,228 & 0.012 & 0.024 & 0.000 & 0.0001 & 0.015 & 1.461 \\ 
Restaurant Visits & 728,228 & 0.250 & 0.175 & 0.000 & 0.150 & 0.315 & 4.261 \\ 
Test Growth Rates& 698,278 & 0.067 & 1.099 & $-$13.616 & $-$0.051 & 0.178 & 13.111 \\ 
Population in 2018 (millions) & 706,966 & 0.104 & 0.331 & 0.0002 & 0.012 & 0.071 & 10.106 \\ 
 \hline \\[-1.8ex] 
\end{tabular} 
  {\scriptsize
\begin{flushleft}
Notes:  Based on observations from April 15, 2020 to December 2, 2020 for the maximum of 3142 counties. The growth rates of weekly reported cases/deaths over 7 days are measured by the log-difference over 7 days in weekly cases/deaths in (a) and (b) while the growth rates of weekly deaths over 21 days is measured by the log-difference over 21 days in weekly cases/deaths in (c). For (a)-(e),  the log of weekly cases and deaths is set to be $-1$ when we observe zero weekly cases and deaths. 
\end{flushleft}}   
 \end{minipage}}
\end{table}

\begin{table}[ht] 
\caption{Correlation across variables  \label{fig:corr-SI}}
\hspace{-6cm}\resizebox{0.8\columnwidth}{!}{
\begin{minipage}{\linewidth}
    \centering
\begin{tabular}{lcccccccccccccccc}
\toprule
\rotatebox{90}{ } & \rotatebox{90}{College Visits} & \rotatebox{90}{K-12 School Visits} & \rotatebox{90}{Open K-12 In-person} & \rotatebox{90}{Open K-12 Hybrid} & \rotatebox{90}{Open K-12 Remote} & \rotatebox{90}{No-Mask for Staffs} & \rotatebox{90}{Mandatory mask} & \rotatebox{90}{Ban gatherings} & \rotatebox{90}{Stay at home} & \rotatebox{90}{Full-time Workplace Visits} & \rotatebox{90}{Part-time Workplace Visits} & \rotatebox{90}{Staying Home Devices} & \rotatebox{90}{Bar Visits} & \rotatebox{90}{Restaurant Visits} & \rotatebox{90}{Rec. Facilities Visits} & \rotatebox{90}{Church Vists}\\
\midrule
College Visits & 1.00 &  &  &  &  &  &  &  &  &  &  &  &  &  &  & \\
K-12 School Visits & 0.09 & 1.00 &  &  &  &  &  &  &  &  &  &  &  &  &  & \\
Open K-12 In-person & 0.05 & 0.43 & 1.00 &  &  &  &  &  &  &  &  &  &  &  &  & \\
Open K-12 Hybrid & 0.11 & 0.44 & 0.04 & 1.00 &  &  &  &  &  &  &  &  &  &  &  & \\
Open K-12 Remote & 0.05 & 0.09 & -0.06 & -0.06 & 1.00 &  &  &  &  &  &  &  &  &  &  & \\
\addlinespace
No-Mask for Staffs & 0.01 & 0.16 & 0.15 & -0.02 & -0.10 & 1.00 &  &  &  &  &  &  &  &  &  & \\
Mandatory mask & 0.09 & 0.10 & 0.03 & 0.24 & 0.23 & -0.31 & 1.00 &  &  &  &  &  &  &  &  & \\
Ban gatherings & -0.03 & -0.14 & -0.09 & -0.06 & 0.00 & -0.03 & -0.09 & 1.00 &  &  &  &  &  &  &  & \\
Stay at home & -0.06 & -0.24 & -0.14 & -0.21 & -0.13 & -0.08 & -0.19 & 0.20 & 1.00 &  &  &  &  &  &  & \\
Full-time Workplace Visits & 0.04 & 0.56 & 0.37 & 0.36 & 0.10 & 0.12 & 0.05 & -0.17 & -0.21 & 1.00 &  &  &  &  &  & \\
\addlinespace
Part-time Workplace Visits & 0.06 & 0.60 & 0.32 & 0.34 & 0.04 & 0.20 & -0.01 & -0.12 & -0.31 & 0.71 & 1.00 &  &  &  &  & \\
Staying Home Devices & -0.04 & -0.27 & -0.18 & -0.23 & -0.02 & -0.10 & 0.01 & -0.00 & 0.27 & 0.06 & -0.19 & 1.00 &  &  &  & \\
Bar Visits & 0.05 & 0.08 & 0.02 & -0.02 & 0.01 & 0.08 & 0.01 & -0.06 & -0.08 & 0.12 & 0.11 & 0.15 & 1.00 &  &  & \\
Restaurant Visits & 0.17 & 0.02 & -0.10 & 0.00 & 0.06 & -0.10 & 0.14 & 0.10 & -0.08 & -0.07 & 0.07 & 0.04 & 0.34 & 1.00 &  & \\
Rec. Facilities Visits & 0.15 & -0.00 & -0.07 & 0.03 & 0.11 & -0.08 & 0.18 & 0.03 & -0.08 & -0.05 & -0.03 & 0.09 & 0.26 & 0.52 & 1.00 & \\
\addlinespace
Church Vists & 0.06 & 0.32 & 0.13 & 0.08 & -0.03 & 0.17 & -0.06 & -0.07 & -0.16 & 0.15 & 0.37 & -0.18 & 0.11 & 0.18 & 0.03 & 1.00\\
\bottomrule
\end{tabular}\smallskip {\scriptsize
\begin{flushleft}
Notes:  Based on observations from April 15, 2020 to December 2, 2020 for the maximum of 3142 counties. \end{flushleft}}   
 \end{minipage}}\bigskip
\end{table}
\medskip

\begin{table}[ht] 
\caption{Correlation across mitigation variables  \label{fig:corr-mitigation}}
    \centering
\begin{tabular}{lcccc}
\rotatebox{90}{ } & \rotatebox{90}{\parbox{1.5cm}{No mask\\ for staffs}} & \rotatebox{90}{\parbox{1.5cm}{No mask\\ for students}} & \rotatebox{90}{\parbox{1.5cm}{Yes sports \\for students}} & \rotatebox{90}{\parbox{1.5cm}{No online \\instruction}}\\
\midrule
No mask for staffs & 1.00 &  &  & \\
No mask for students & 0.74$^{***}$ & 1.00 &  & \\
Yes sports for students & 0.14$^{***}$  & 0.07$^{***}$  & 1.00 & \\
No online instruction & 0.13$^{***}$ & 0.12$^{***} $ & 0.06$^{**}$  & 1.00\\
\bottomrule
\end{tabular} {\scriptsize
\begin{flushleft}
Notes:  Based on cross-sectional observations on October 1, 2020.  $^{**}$p$<$0.05; $^{***}$p$<$0.01
 \end{flushleft}}
\end{table}  
 

\begin{figure}[ht]
  \caption{Average weekly cases and deaths are associated with different modes of opening K-12 schools,  visits to K-12 schools, and visits to colleges/universities \label{fig:case-growth-SI}}  
\resizebox{0.35\columnwidth}{!}{
\hspace*{-8cm}\begin{minipage}{\linewidth} 
        \begin{tabular}{cccc}  
  \textbf{(a) K-12 School Visits}   &   \textbf{(b) Restaurant Visits    } &  \textbf{(c)  Recreation Facilitiy Visits  }  &  \textbf{(d) Church Visits}  \\ 
  \includegraphics[width=0.25\textwidth]{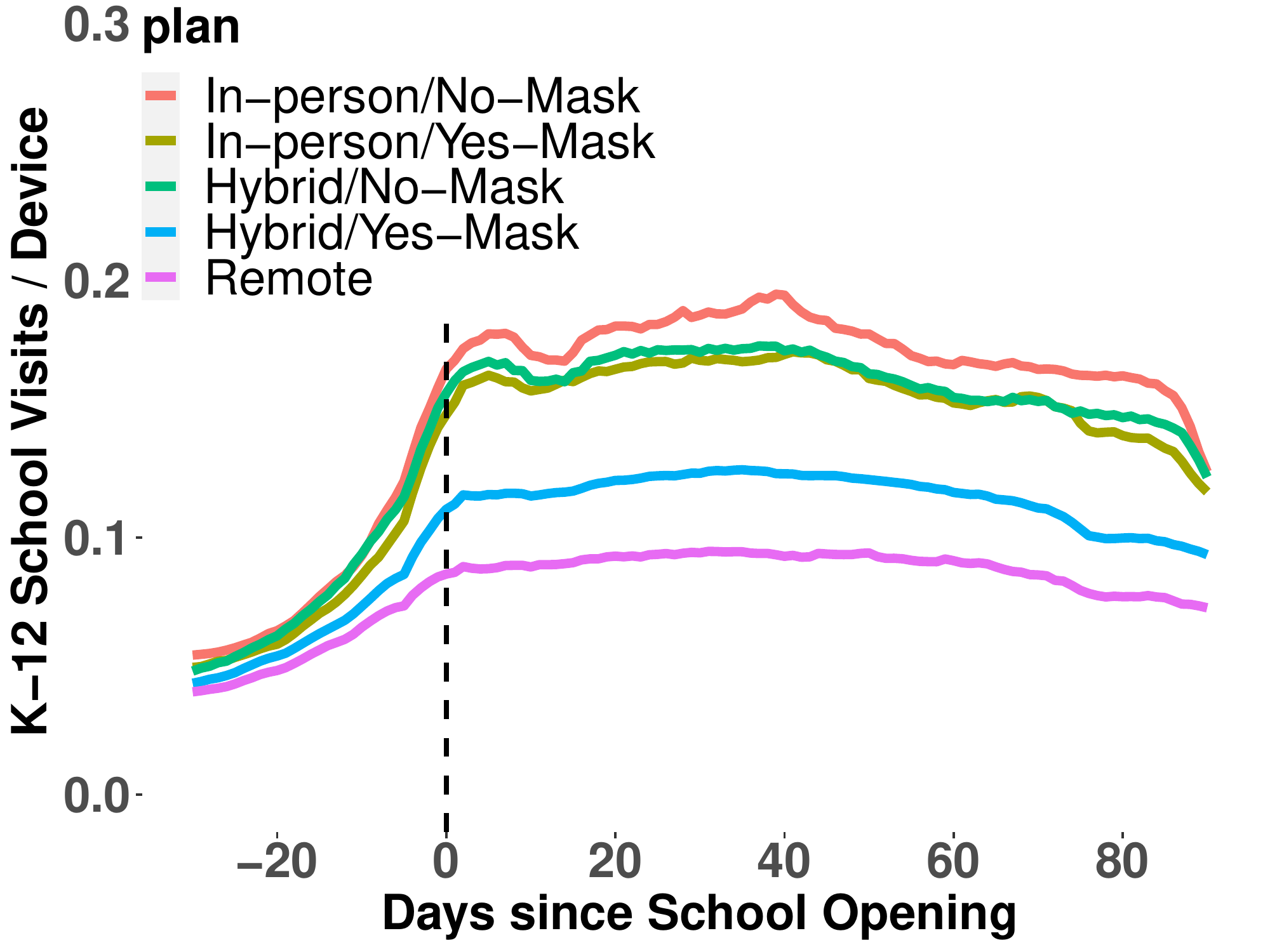}&
  \includegraphics[width=0.25\textwidth]{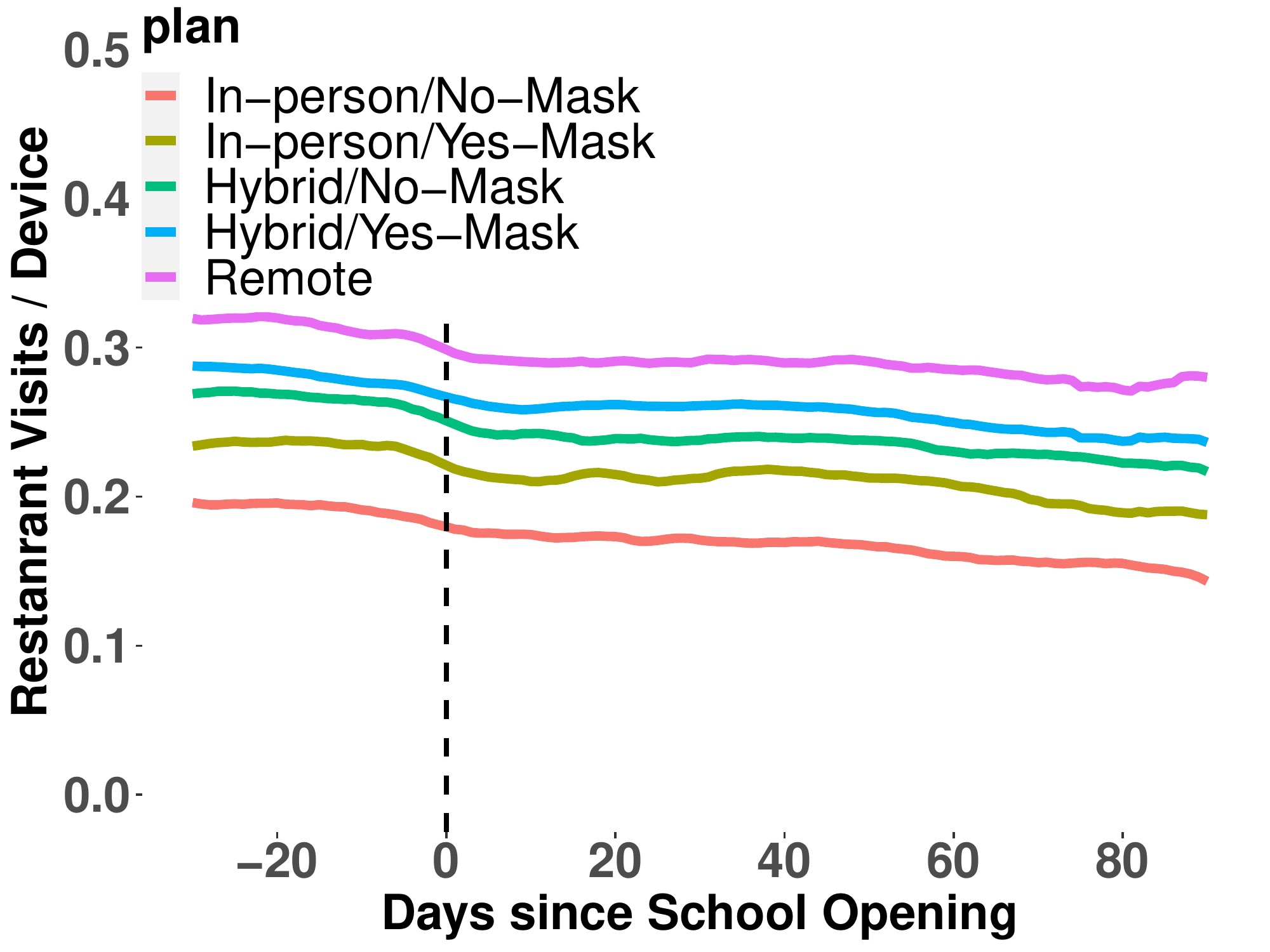}&  
  \includegraphics[width=0.25\textwidth]{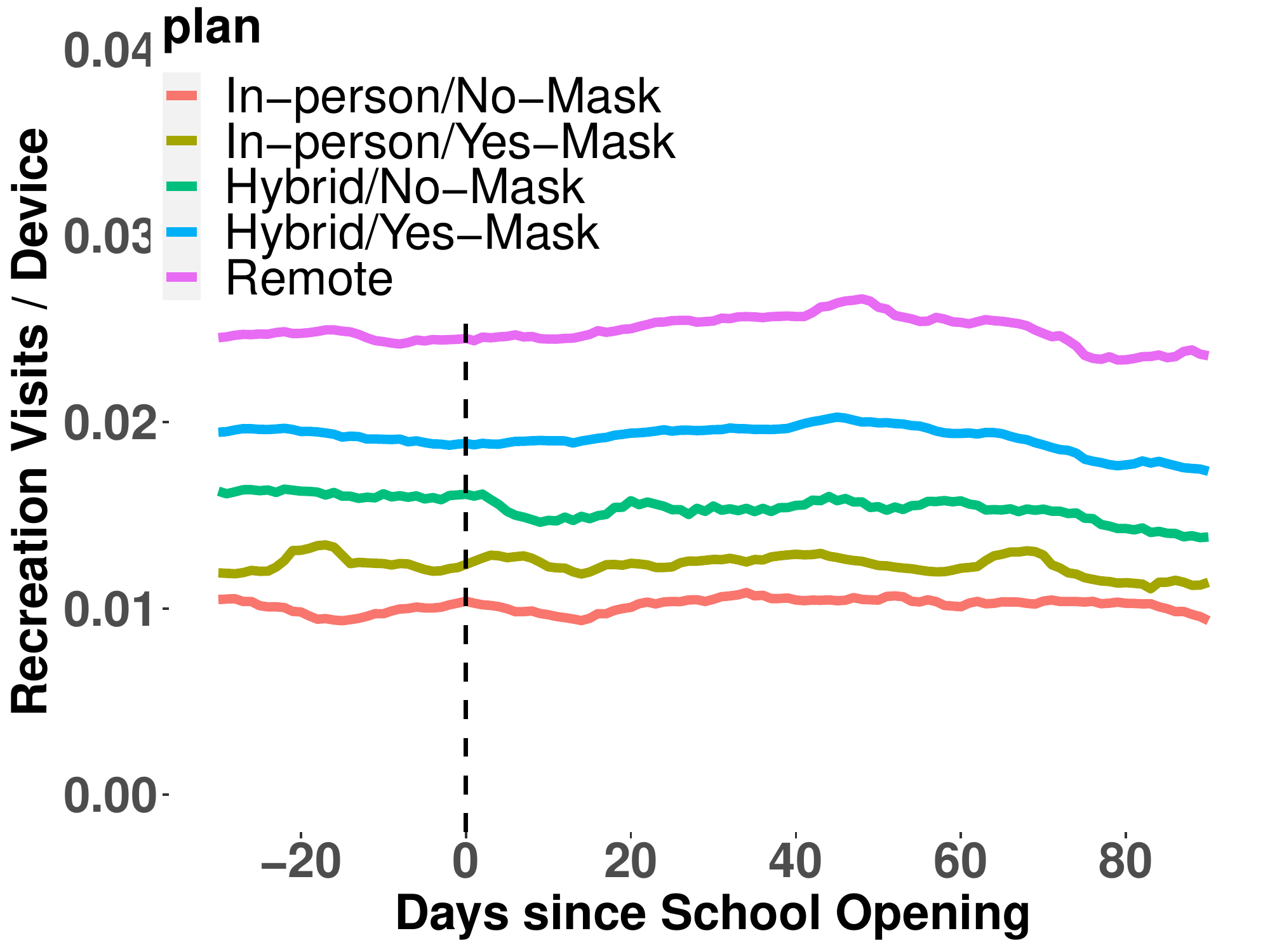}&  
  \includegraphics[width=0.25\textwidth]{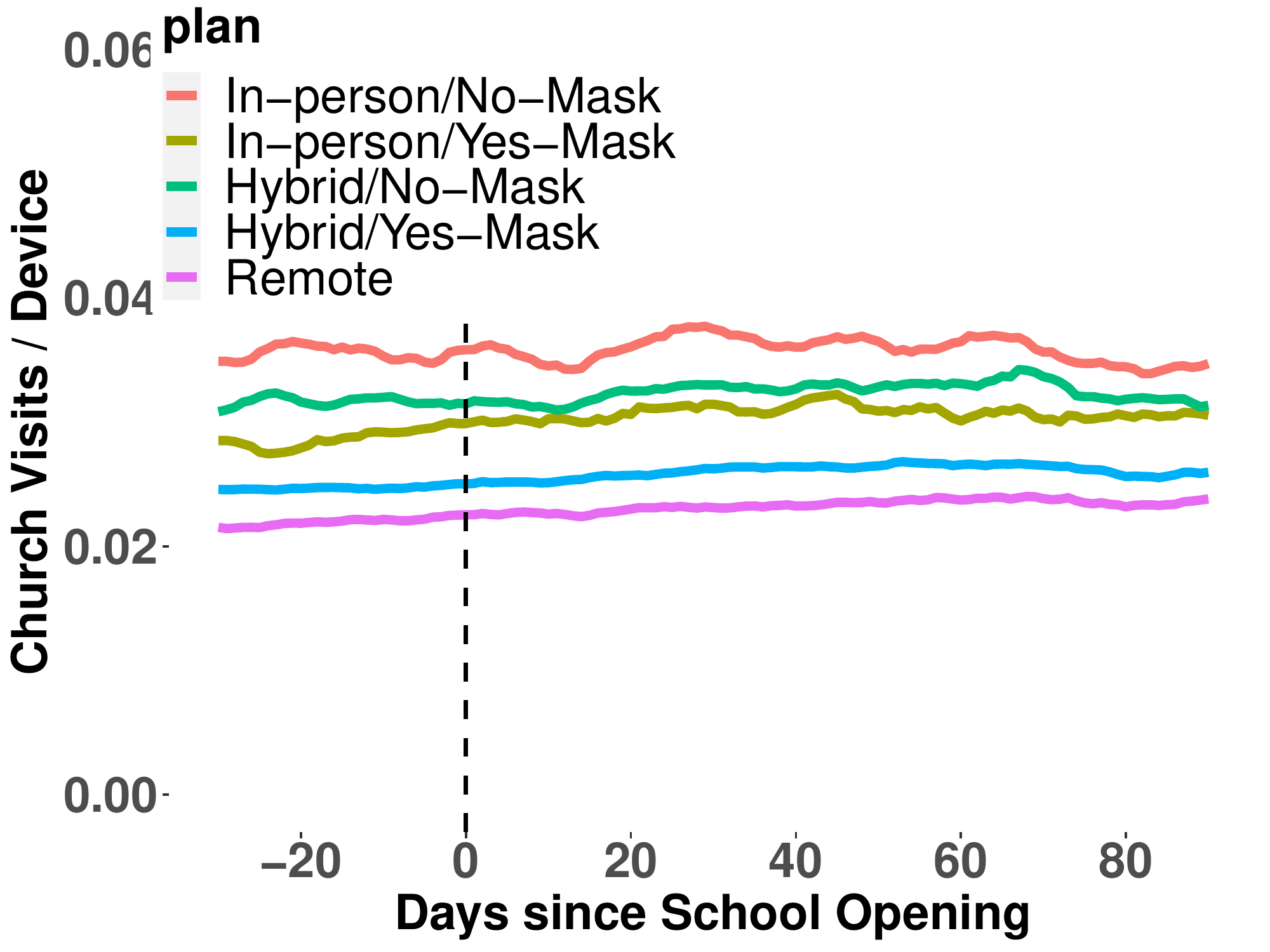}\\  
    \textbf{(e) Full-Time Workplace Visits  }&\textbf{(f) Part-Time Workplace Visits     }&\textbf{(g) Staying Home Devices }&\textbf{(h) Bar Visits } \\ 
  \includegraphics[width=0.25\textwidth]{tables_and_figures/schoolmode-event-fullwork}&
  \includegraphics[width=0.25\textwidth]{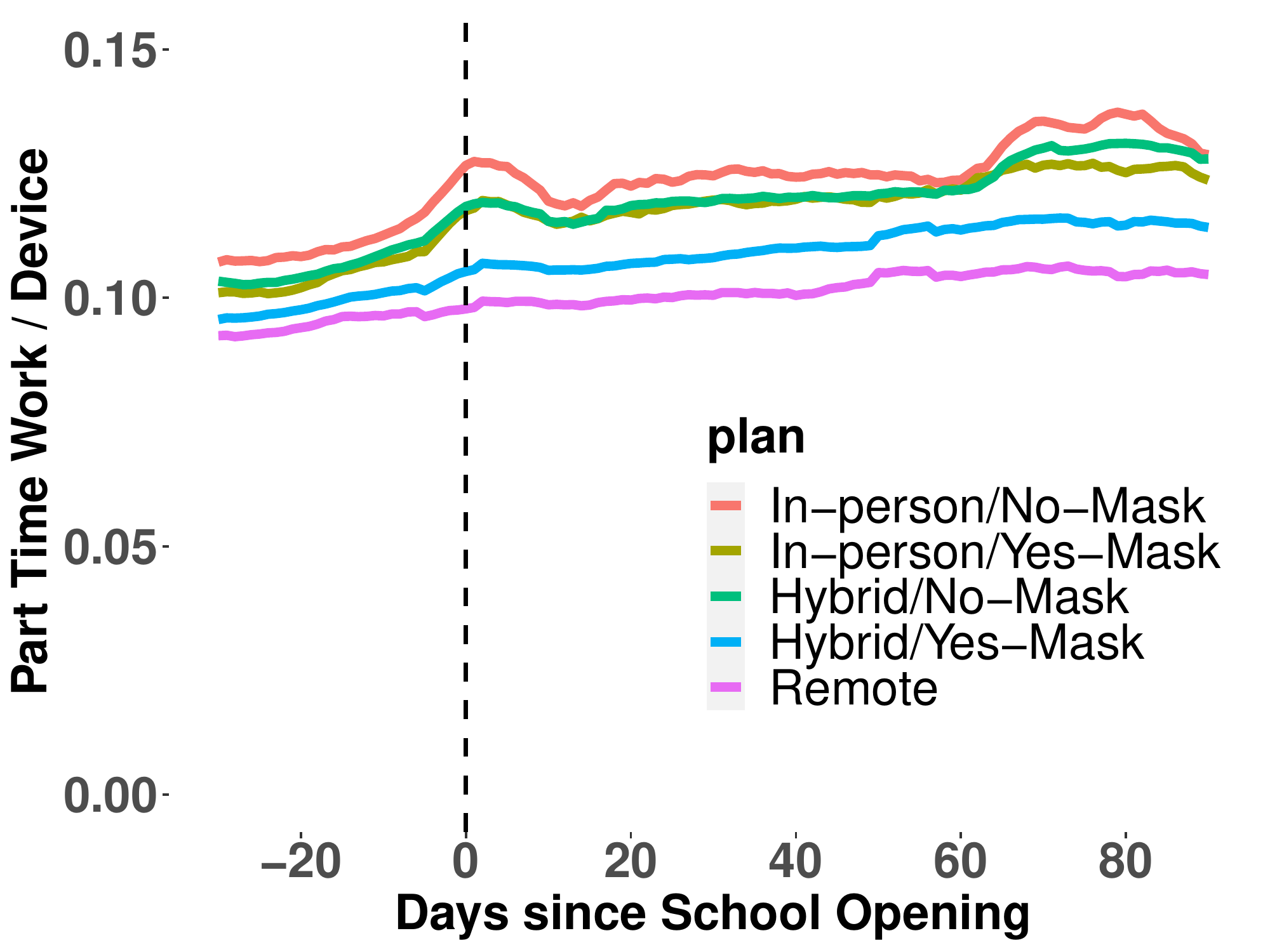}&  
  \includegraphics[width=0.25\textwidth]{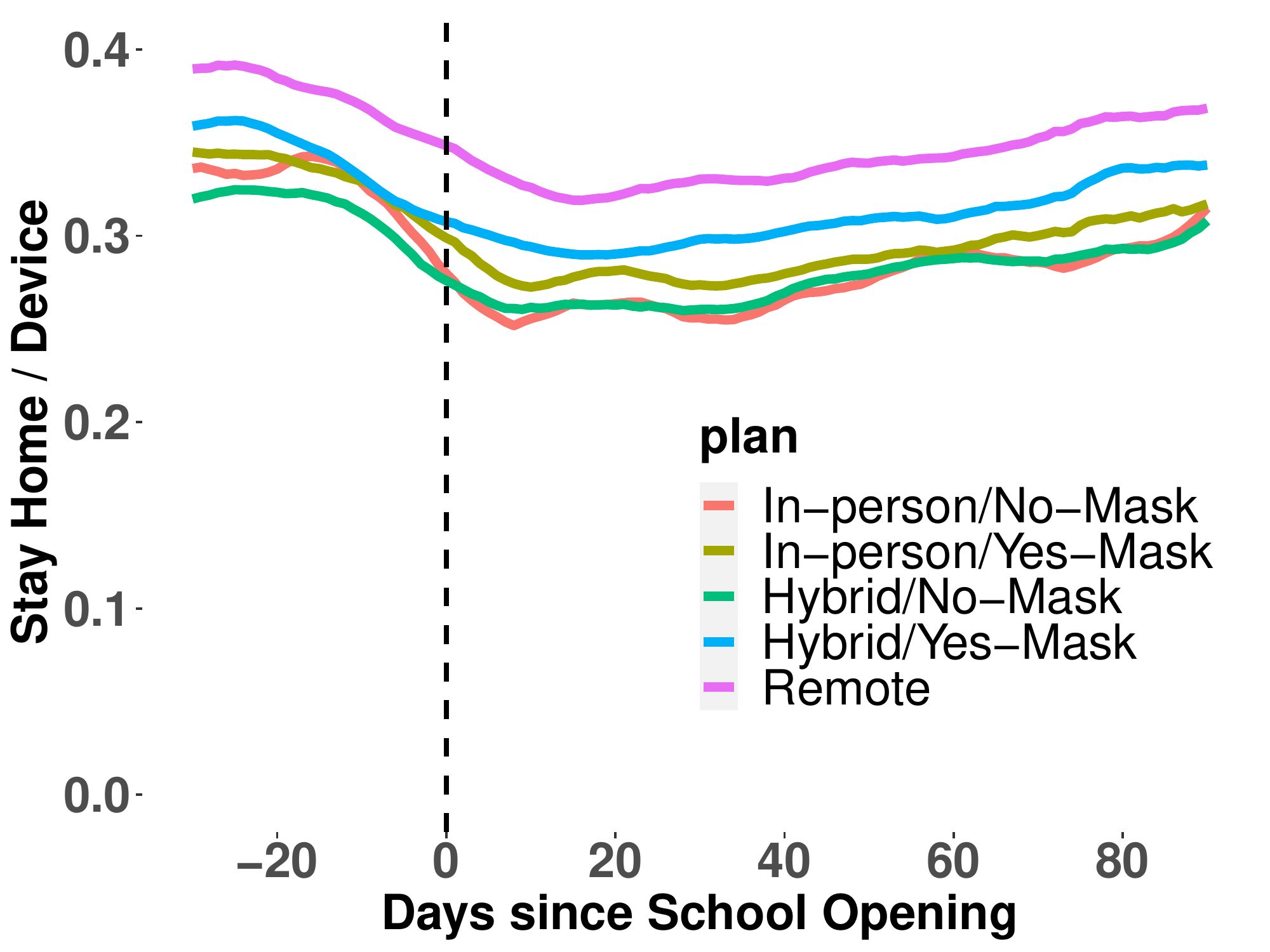}&  
  \includegraphics[width=0.25\textwidth]{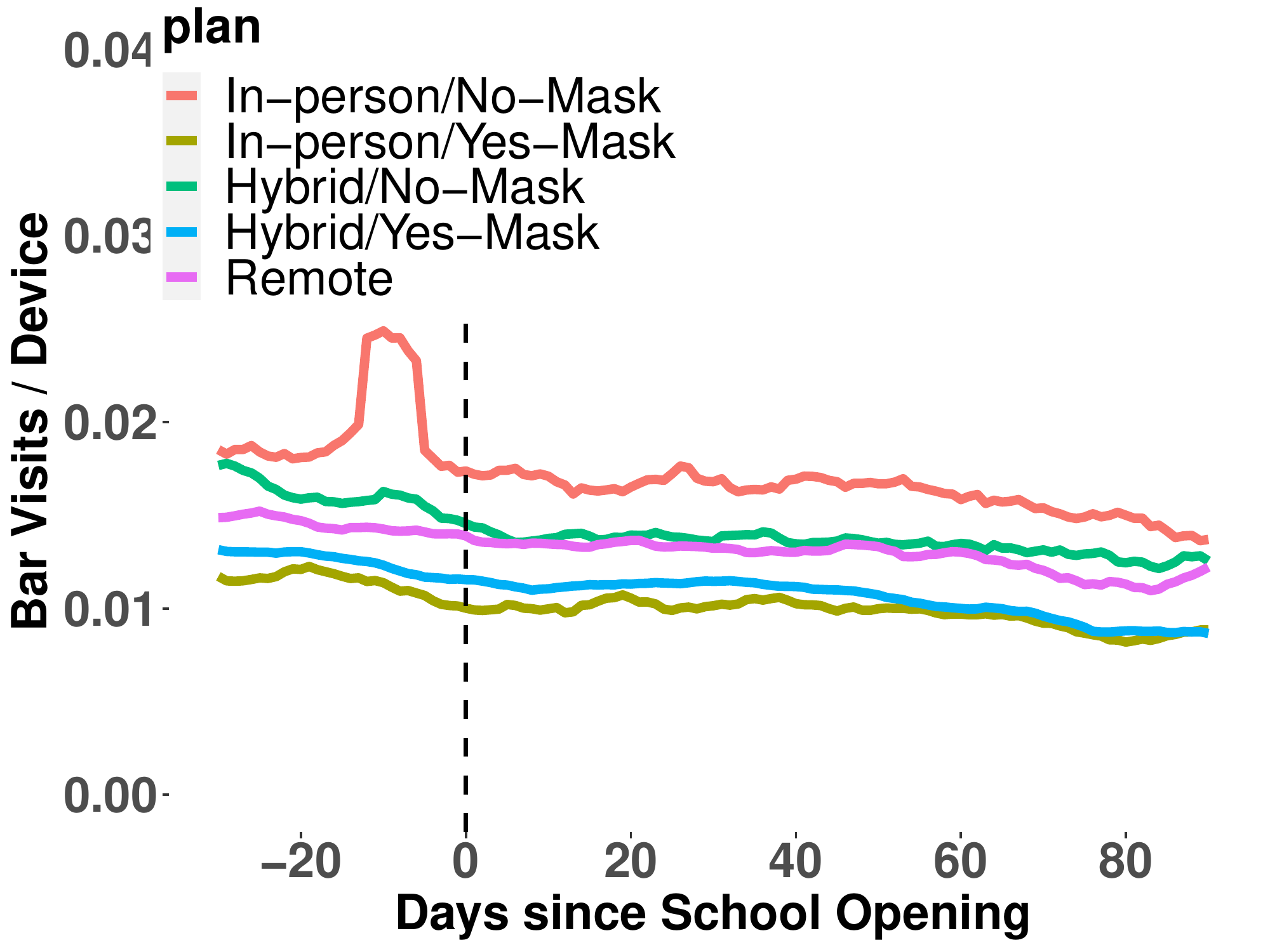}\\  
      \textbf{(i)     Cases by  Opening Modes }&\textbf{(j) Cases by  Opening Modes }&\textbf{(k) Cases by  Opening Modes}&\textbf{(l) Cases by  Opening Modes }\\
    (Student Mask Requirements)&(Staff Mask Requirements)&   (Sports Activities)&   (Online Instruction Increase) \\  
      \includegraphics[width=0.25\textwidth]{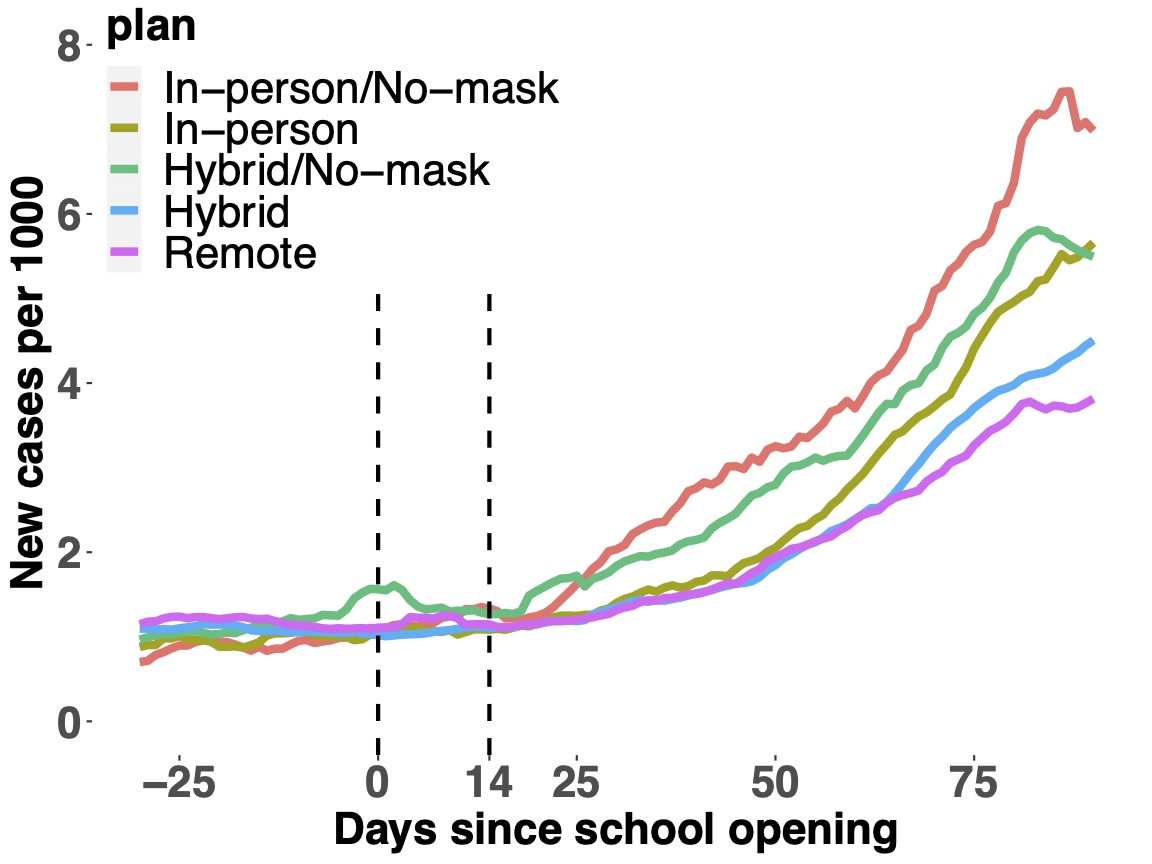}&  
      \includegraphics[width=0.25\textwidth]{tables_and_figures/schoolmode-event-staff-newcases}&  
      \includegraphics[width=0.25\textwidth]{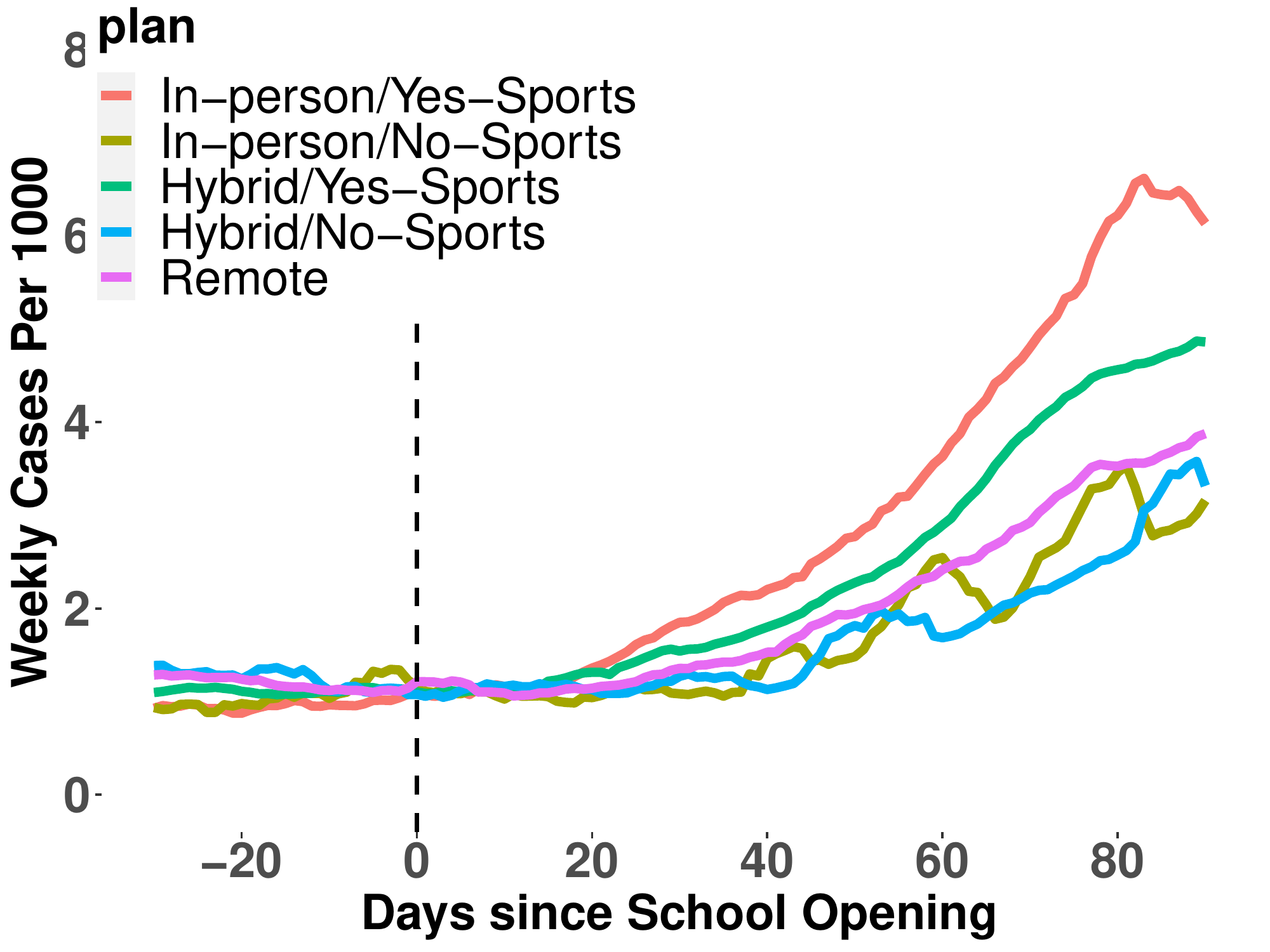}&  
      \includegraphics[width=0.25\textwidth]{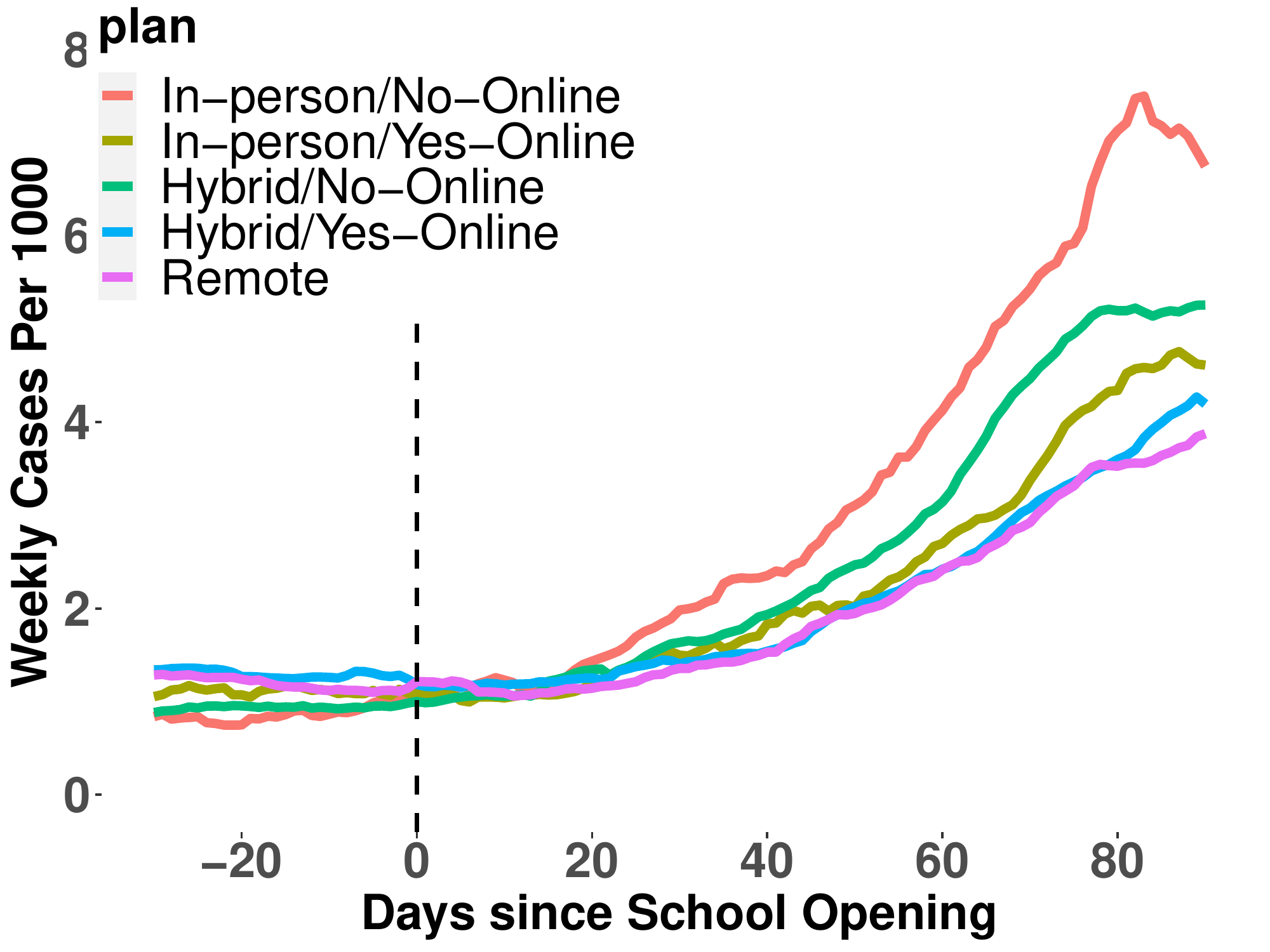}  \\
    \textbf{(m)   Deaths by  Opening Modes }&\textbf{(n) Deaths by  Opening Modes }&\textbf{(o) Deaths by  Opening Modes}&\textbf{(p) Deaths  by  Opening Modes }\\
    (Student Mask Requirements)&(Staff Mask Requirements)&   (Sports Activities)&   (Online Instruction Increase) \\  
          \includegraphics[width=0.25\textwidth]{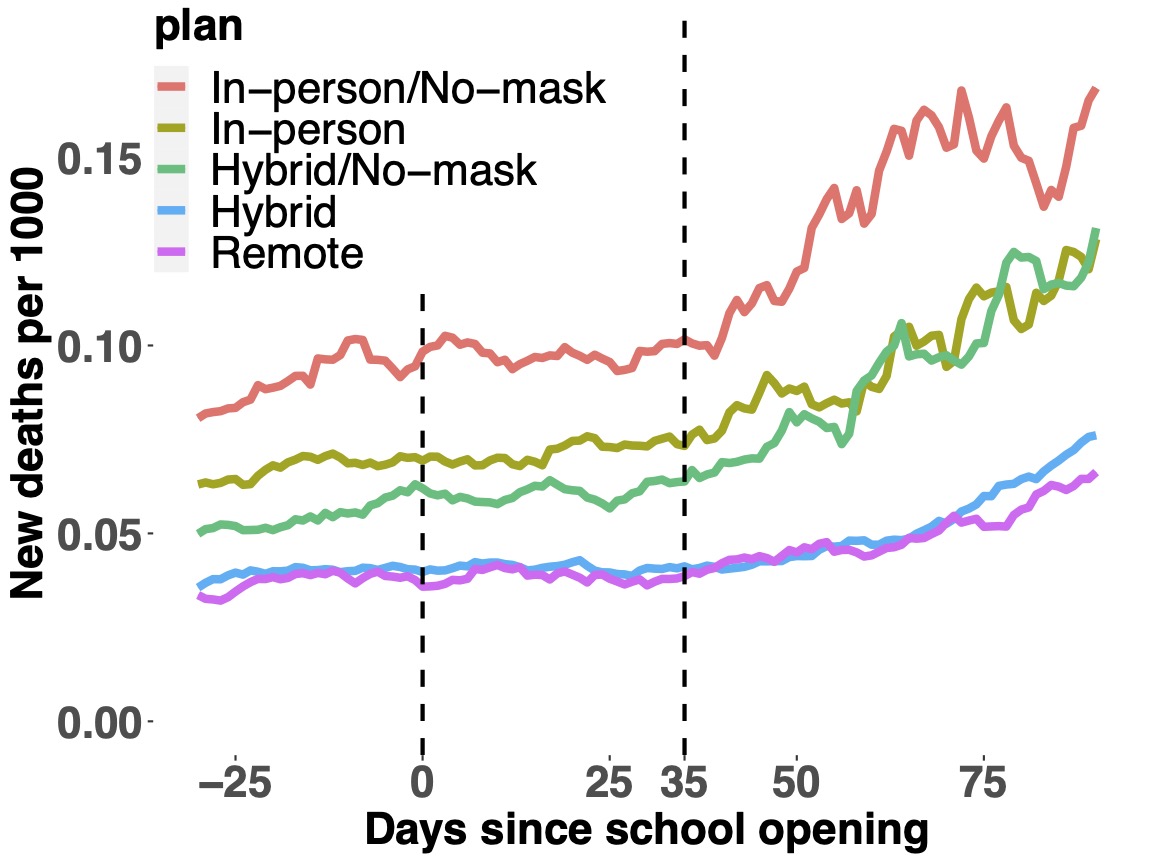}&  
      \includegraphics[width=0.25\textwidth]{tables_and_figures/schoolmode-event-staff-newdeaths}&  
      \includegraphics[width=0.25\textwidth]{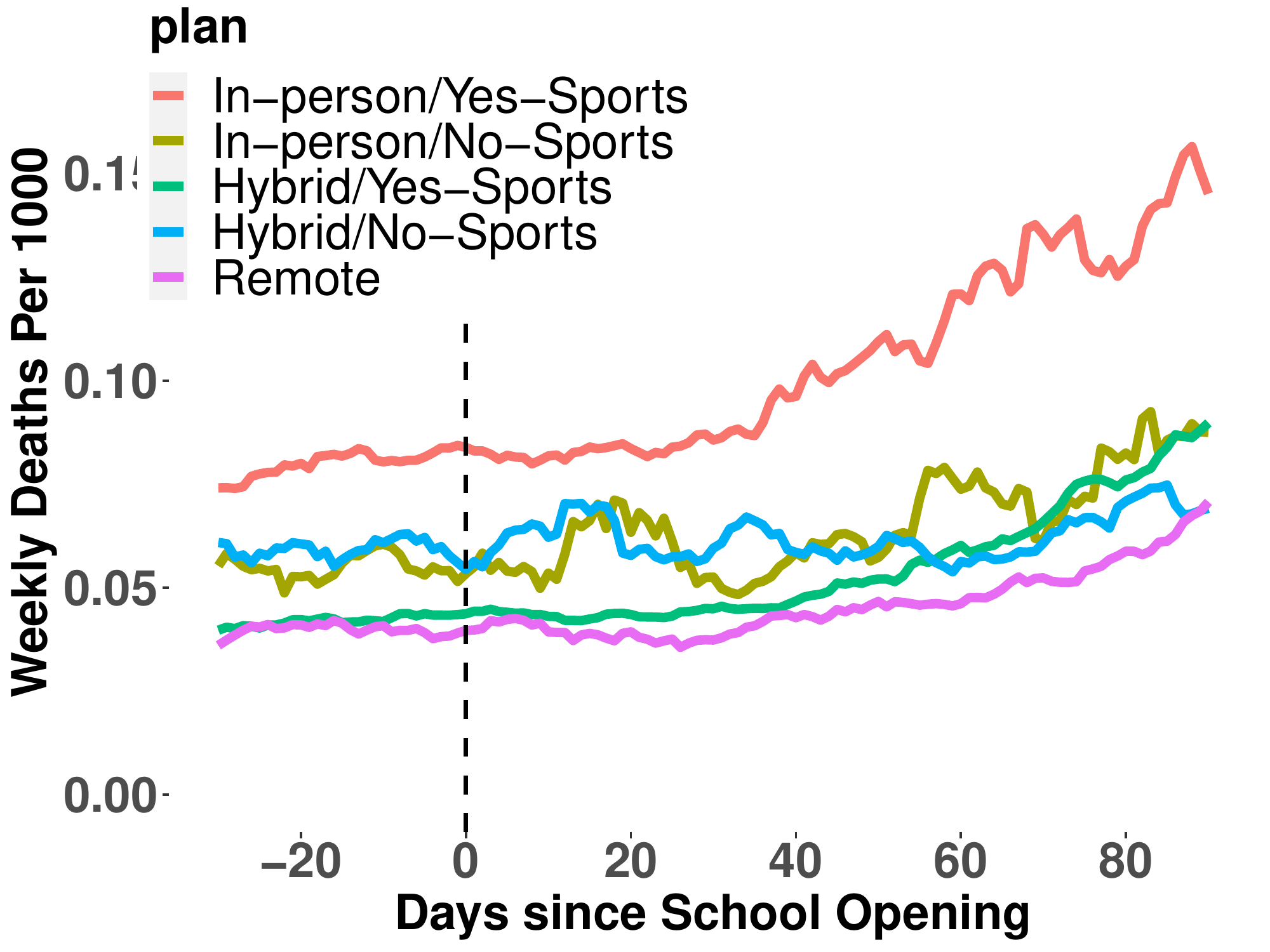}&  
      \includegraphics[width=0.25\textwidth]{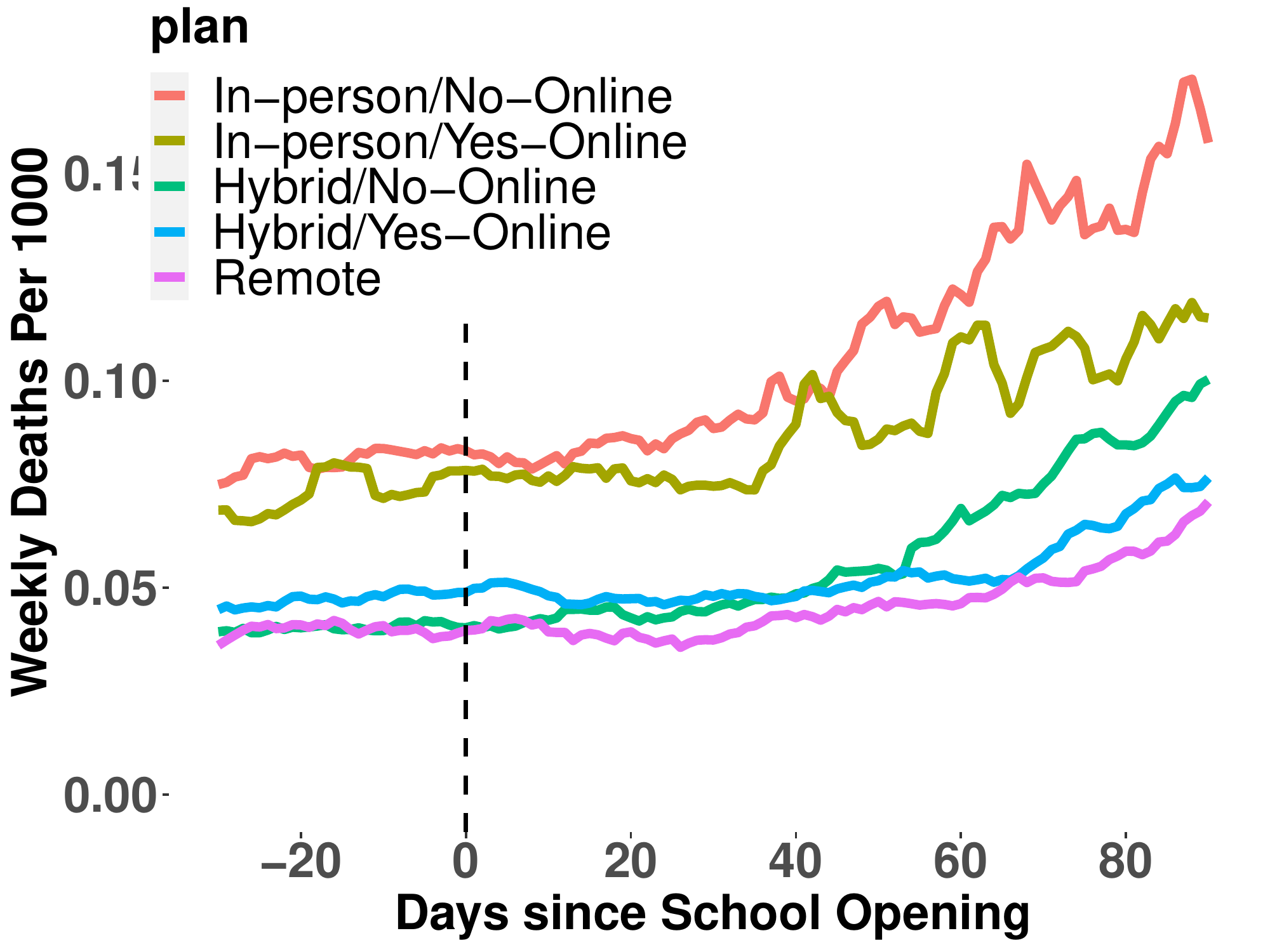}  \\  
      \textbf{(q) Cases by K-12 Visits }&\textbf{(r) Cases by K-12 Visits }&\textbf{(s) Cases by K-12 Visits }&\textbf{(t) Cases by K-12 Visits } \\
    (Student Mask Requirements)&(Staff Mask Requirements)&   (Sports Activities)&   (Online Instruction Increase) \\  
      \includegraphics[width=0.25\textwidth]{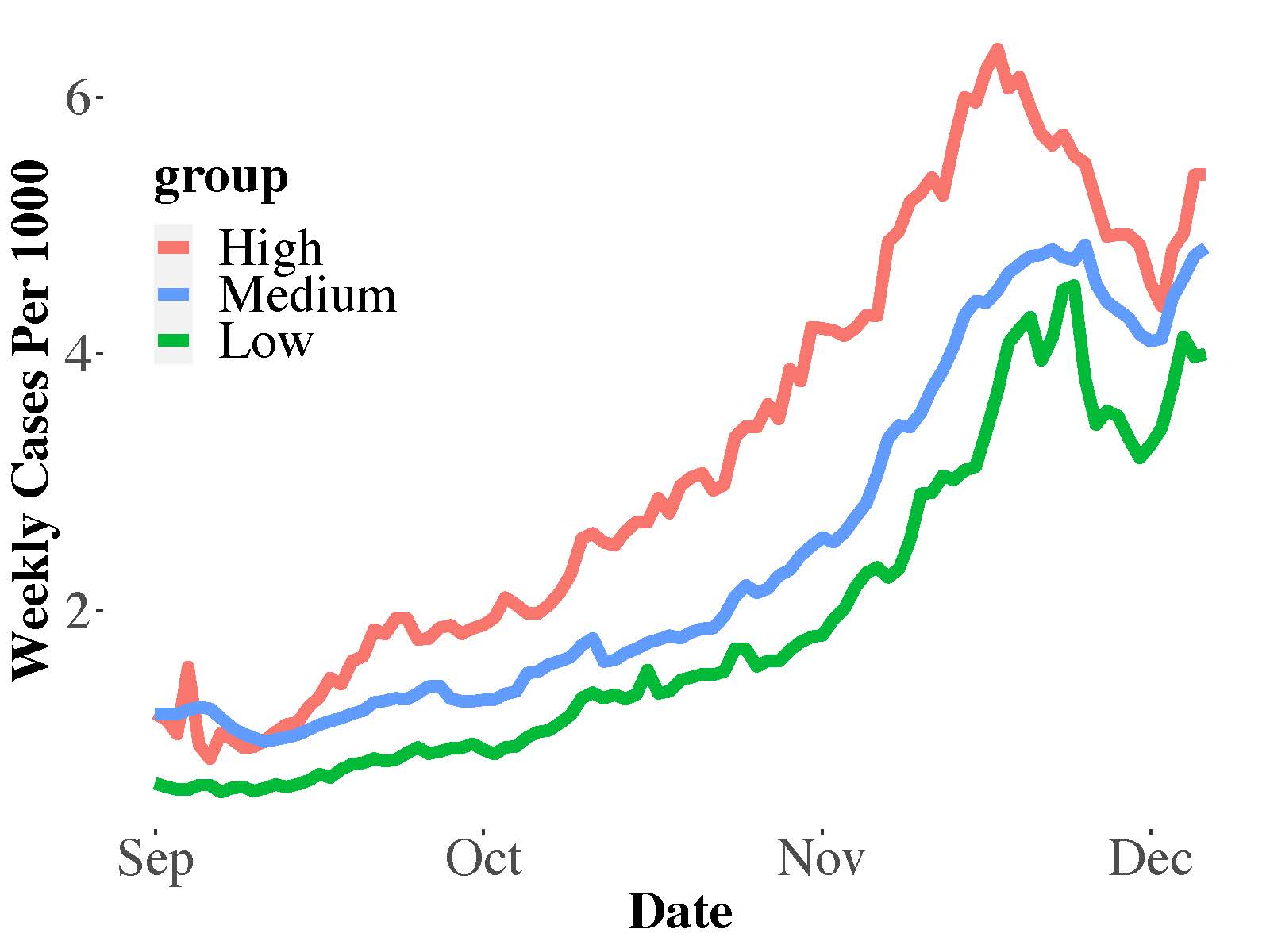}&  
      \includegraphics[width=0.25\textwidth]{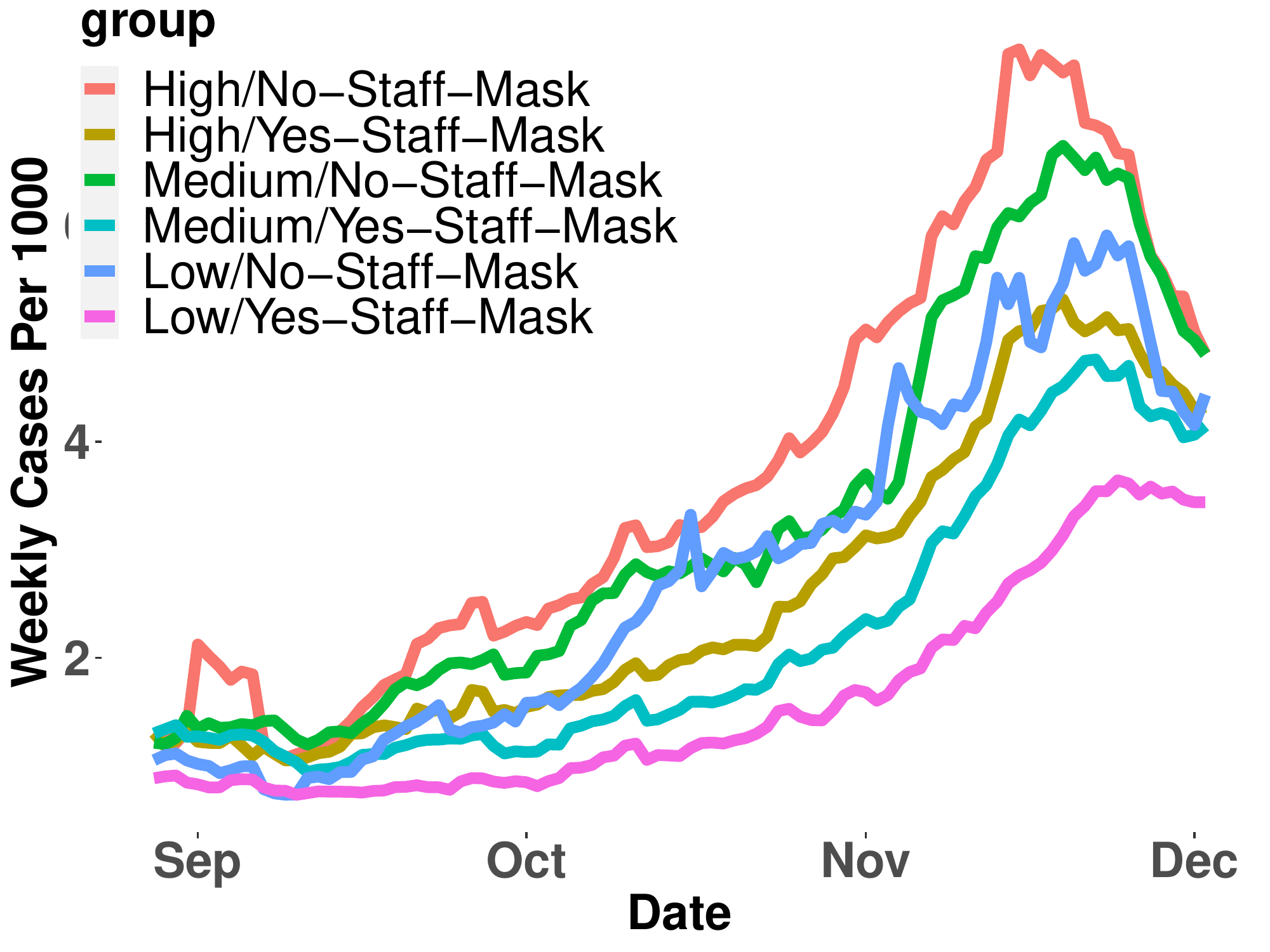}&  
      \includegraphics[width=0.25\textwidth]{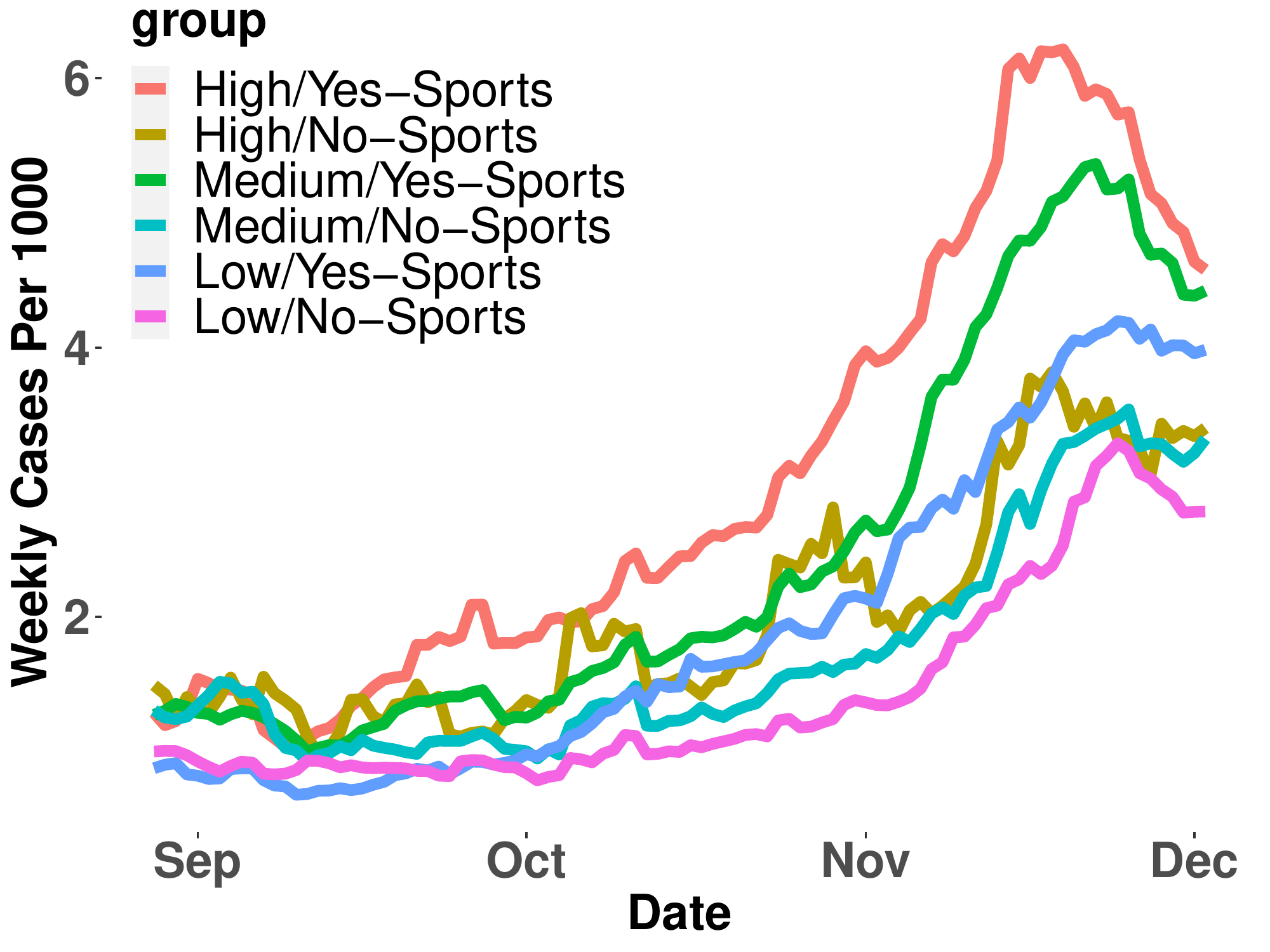}&  
      \includegraphics[width=0.25\textwidth]{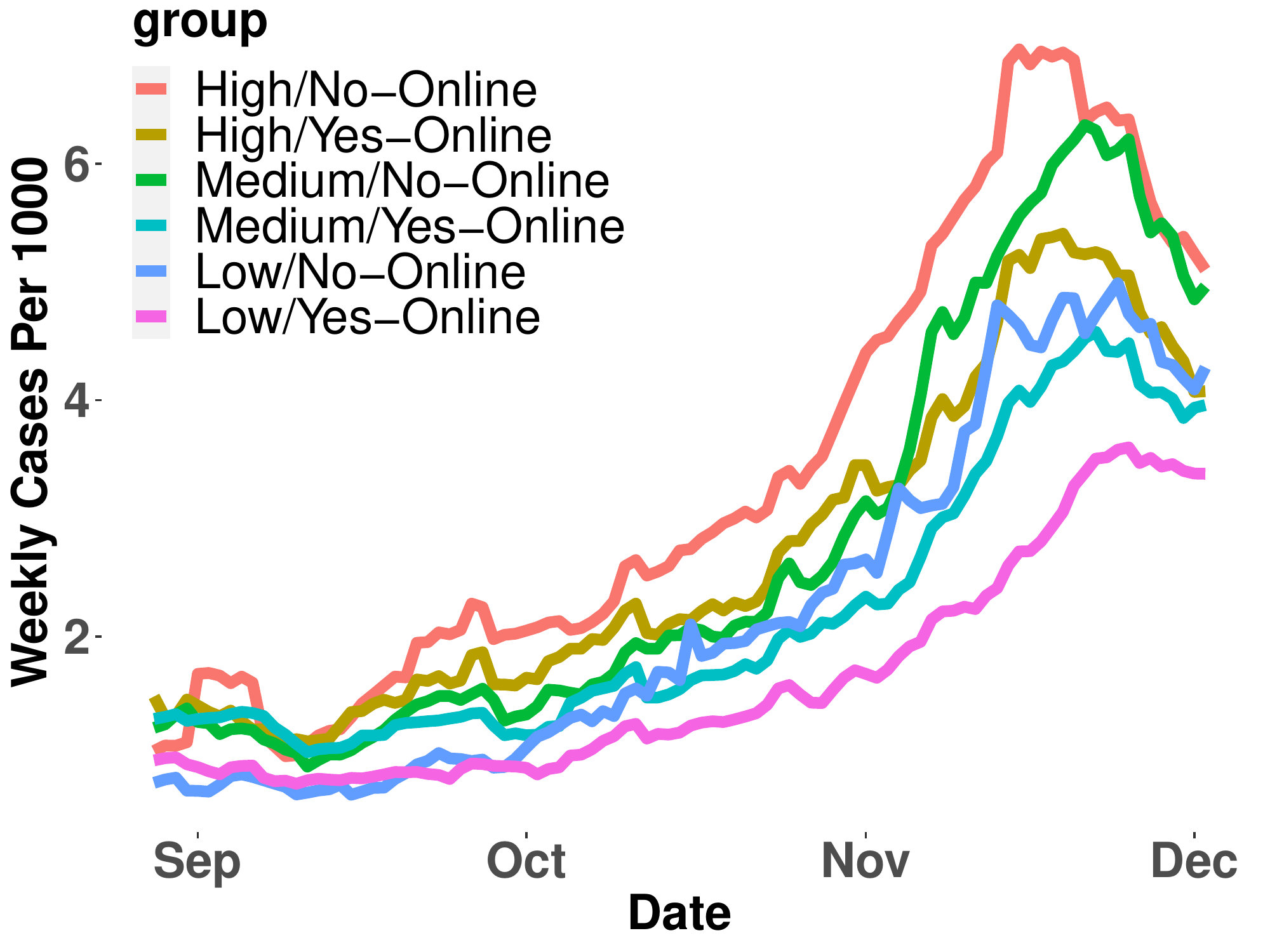}  \\
    \textbf{(u)  Deaths by K-12 Visits}  &\textbf{(v) Deaths by K-12 Visits}&\textbf{(w) Deaths by K-12 Visits }&\textbf{(x) Deaths by K-12 Visits}\\
    (Student Mask Requirements)&(Staff Mask Requirements)&   (Sports Activities)&   (Online Instruction Increase) \\  
      \includegraphics[width=0.25\textwidth]{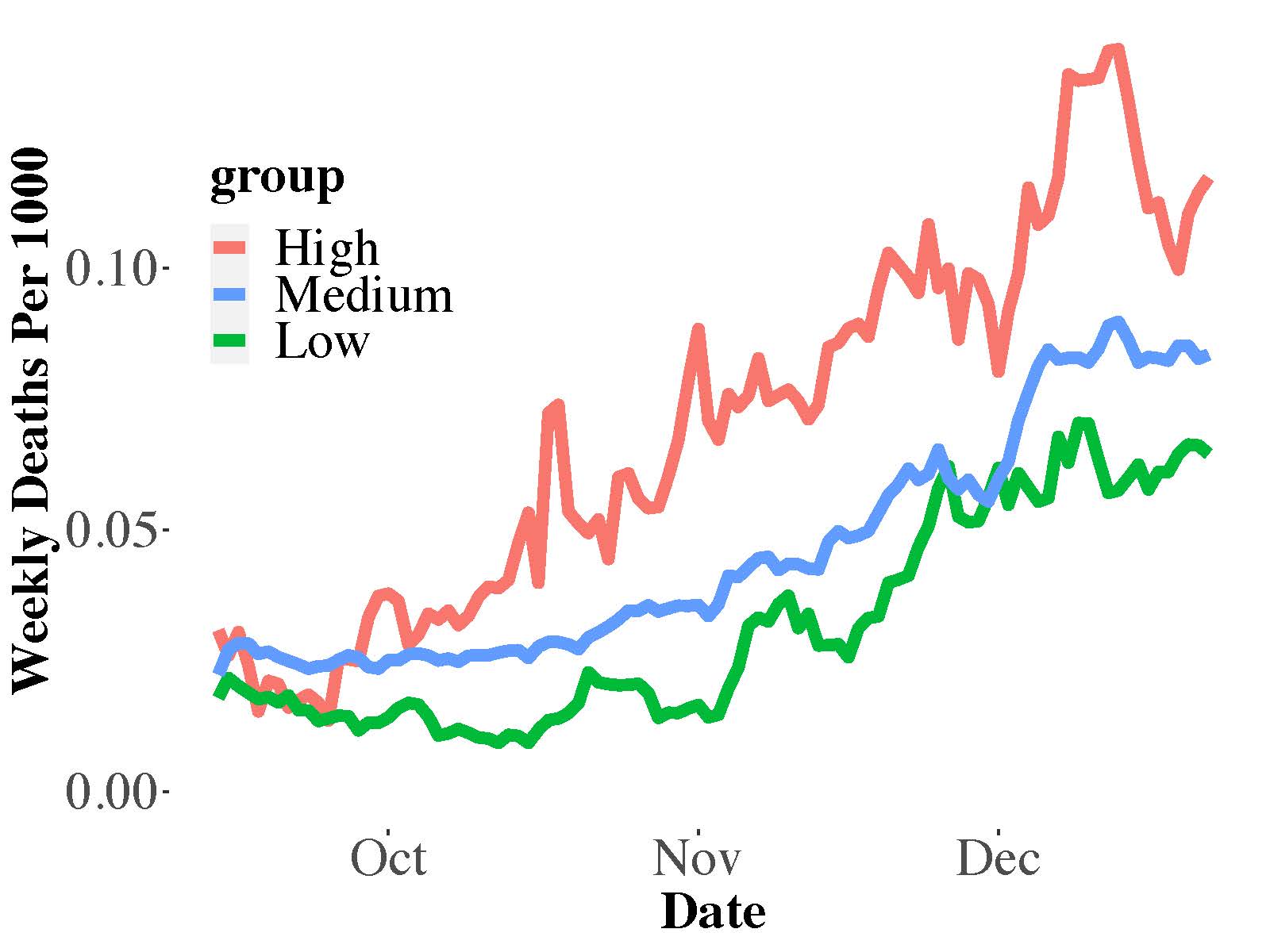}&  
      \includegraphics[width=0.25\textwidth]{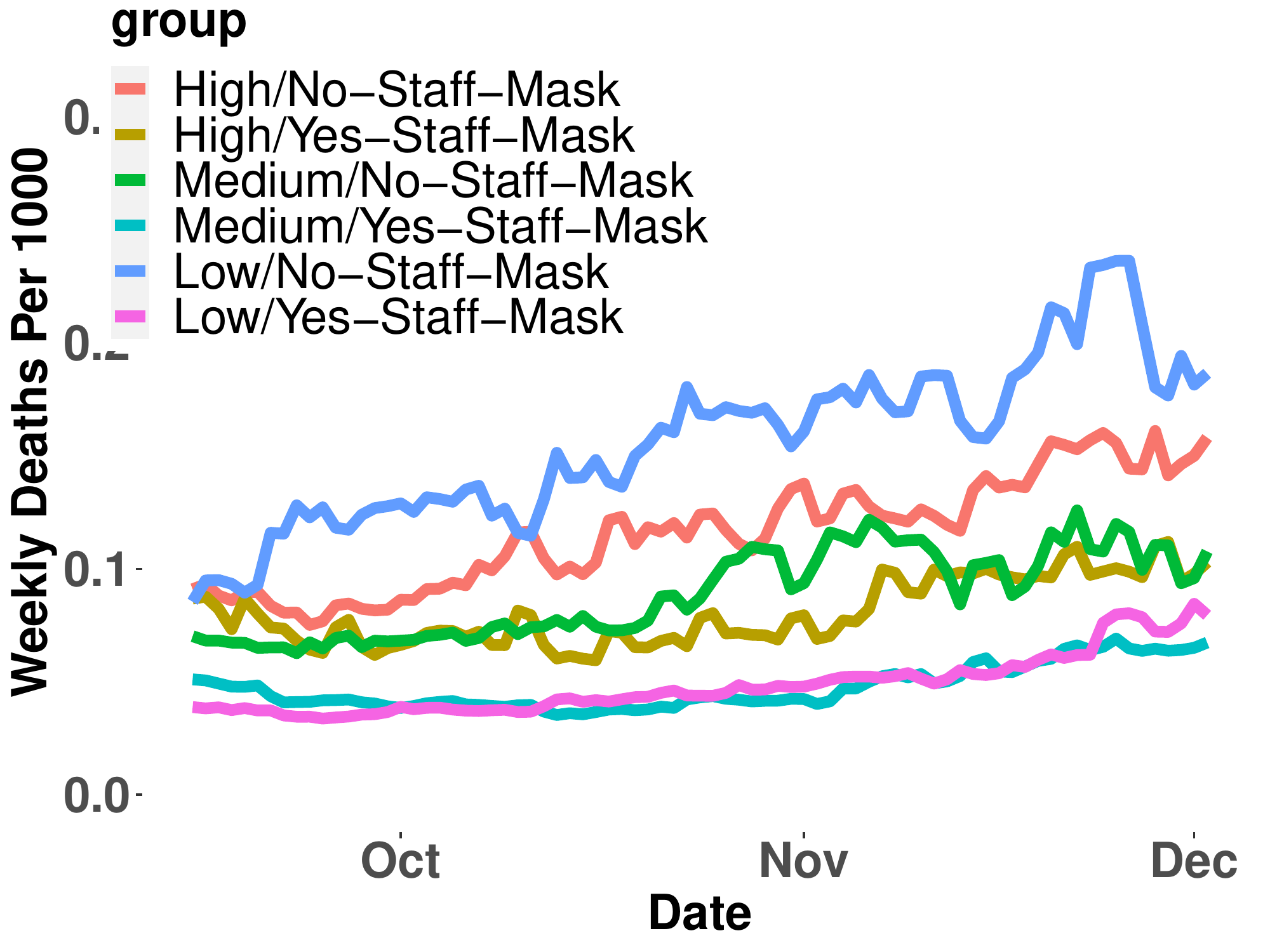}&  
      \includegraphics[width=0.25\textwidth]{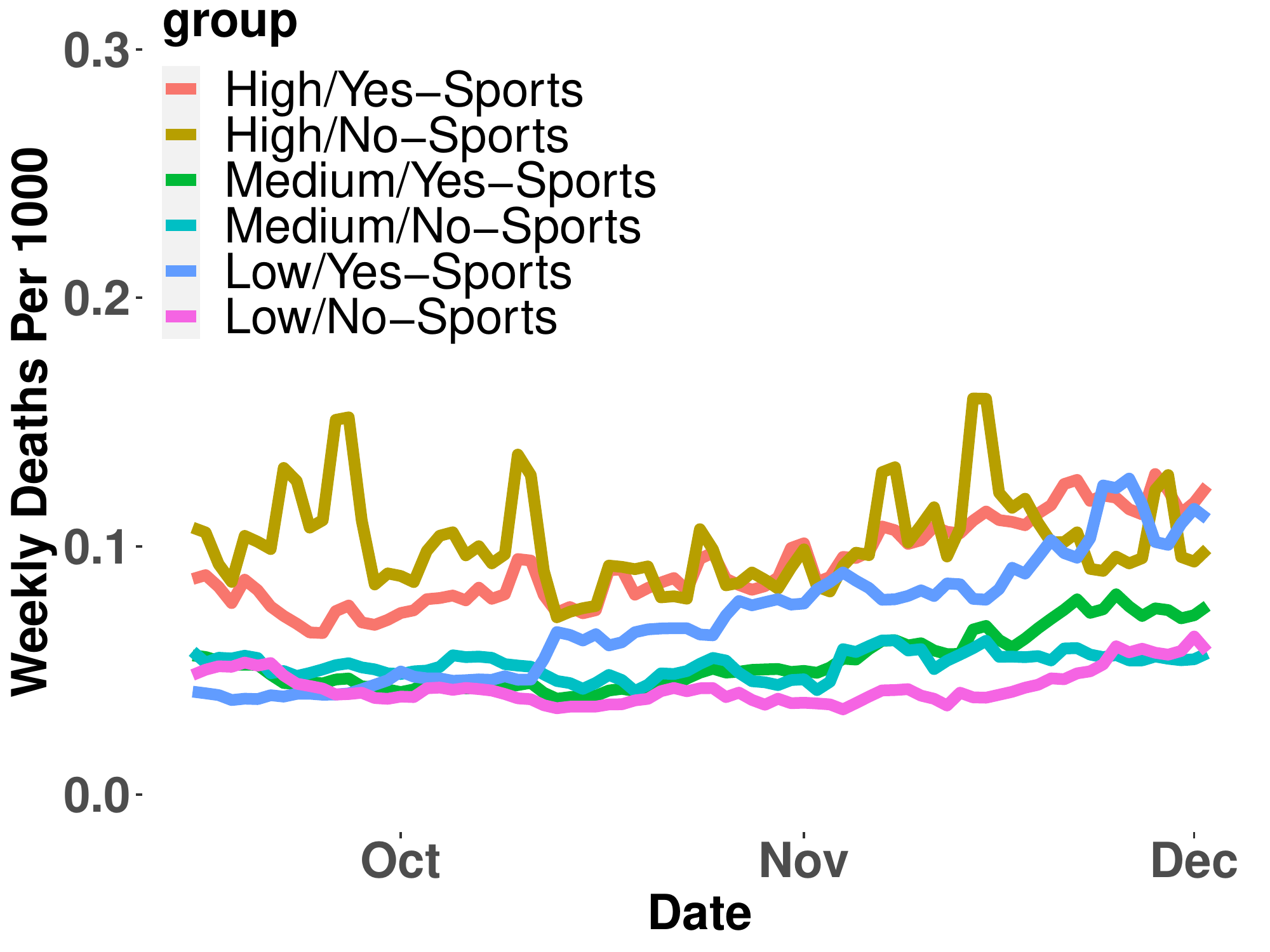} &  
      \includegraphics[width=0.25\textwidth]{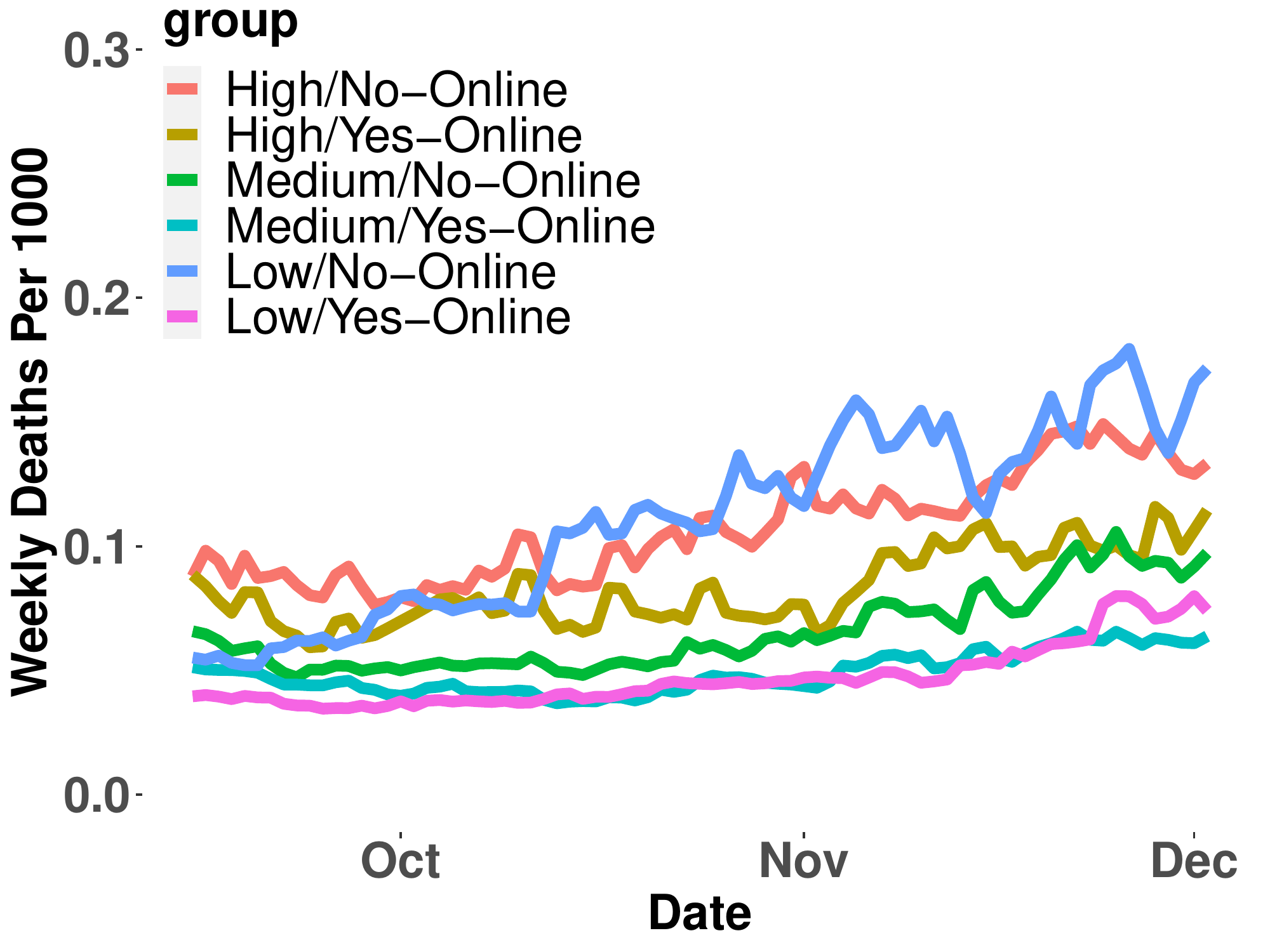}  
    \end{tabular} 
  \end{minipage}}  
\vspace{-0.2cm}  {\scriptsize
\begin{flushleft}
Notes:  (a)-(h)  plot the evolution of corresponding variables in the title before and after the day of school openings and corresponding to figures reported in Fig. 1(c)(d) in the main text. (i)-(p) corresponds to Fig.(a)(b) and  plot the evolution of weekly cases or deaths per 1000 persons averaged across counties within each group of counties classified by K-12 school teaching methods and different mitigation strategies (mask requirements for students, mask requirements for staffs, allowing for sports activities, and increase in online instructions) against the days since K-12 school opening. In (i) and (m), counties that implement in-person teaching  are classified into ``In-person/Yes-Mask" and  ``In-person/No-Mask" based on whether at least one school district requires students to wear masks or not. In (k) and (o),  counties that implement in-person teaching are classified into ``In-person/Yes-Sports" and  ``In-person/No-Sports" based on whether at least one school district requires students to allow sports activities or not. In (l) and (p),  counties that implement in-person teaching   are classified into ``In-person/No-Online" and  ``In-person/Yes-Online" based on whether at least one school district answer that no increase in online instruction. (q)-(x) are similar to (i)-(p) but classify counties by the volume of per-device K-12 school visits and take the calendar dates instead of the days since opening schools as x-axis, where  "Low,"   "Middle," and "High'' are county-day observations of which 14 days lagged per-device K-12 school visits less than the first quartile, between the first and the third quartiles, and larger than the third quartile, respectively. In (q) and (u),   "Low/No-Mask,"   "Middle/No-Mask," and "High/No-Mask'' are a subset of low, middle, and high visits groups of counties for which at least one school district does not require students to wear masks. 
 \end{flushleft}    }
\end{figure}

\begin{figure*}[ht]
  \caption{The evolution of  visits to K-12 schools and full-time workplaces before and after the opening of K-12 schools using the daily frequency data for counties with the same school opening dates for all school districts within a county\label{fig:case-growth-day}}\smallskip
\resizebox{\columnwidth}{!}{
\begin{minipage}{\linewidth}
        \begin{tabular}{ccc}
    \textbf{(c) K-12 School Visits } &     \textbf{(d) Full-Time Workplace Visits}\\
  \qquad \qquad\includegraphics[width=0.4\textwidth]{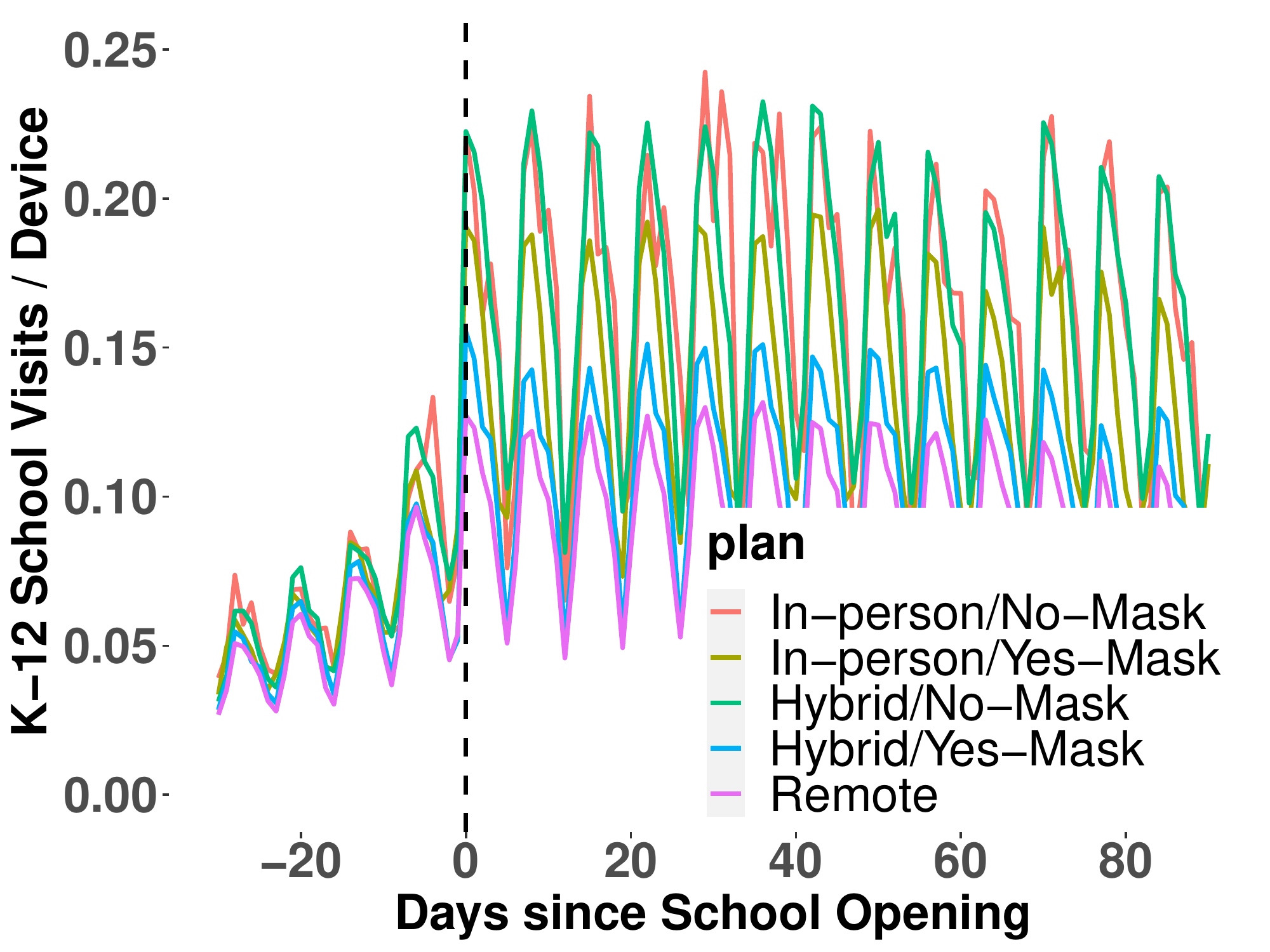}&
  \qquad \qquad\includegraphics[width=0.4\textwidth]{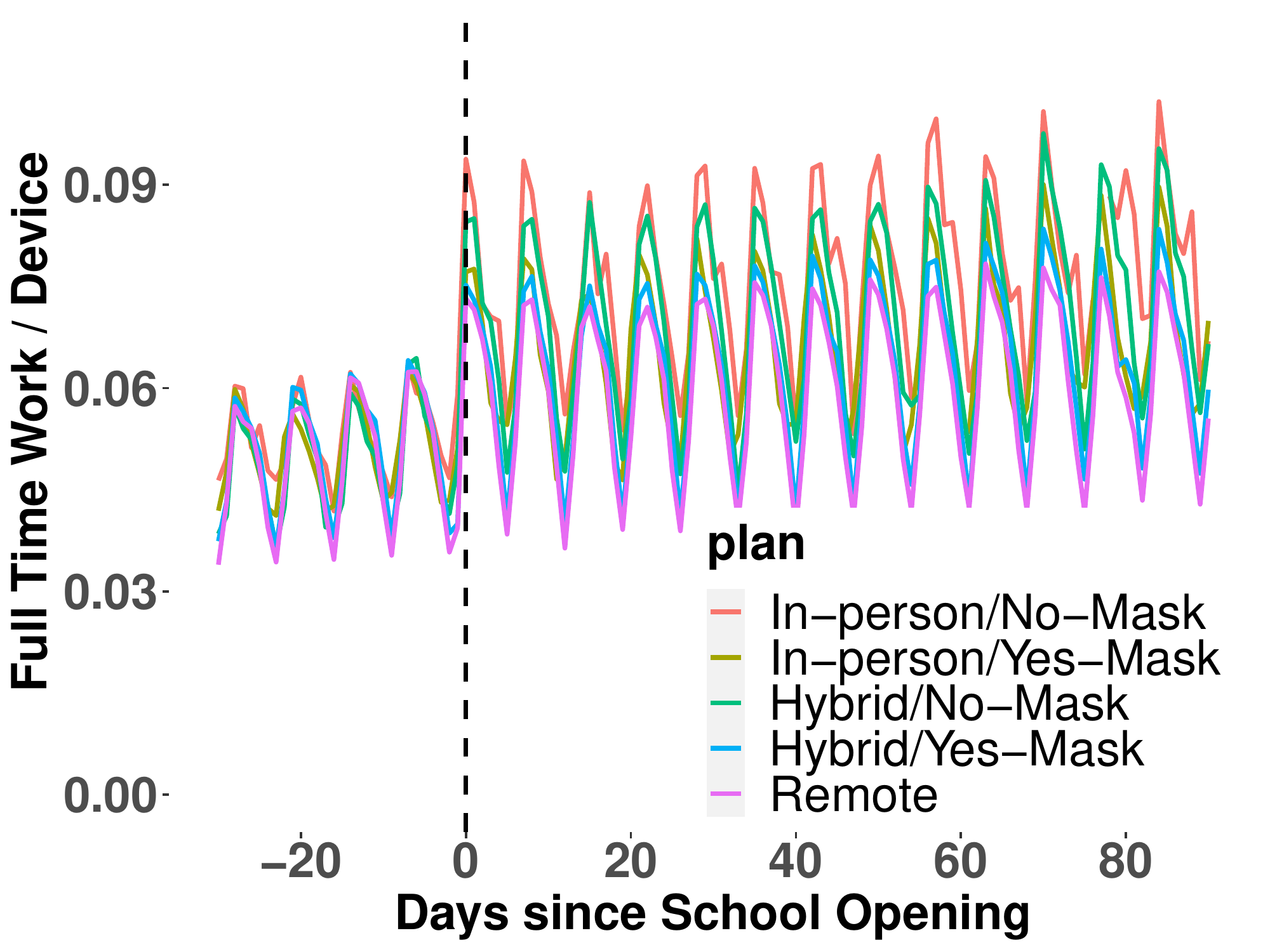} 
    \end{tabular}
  \end{minipage}}
\vspace{-0.2cm}  {\scriptsize
\begin{flushleft}
Notes:   (c) and (d) plot the evolution of per-device visits to K-12 schools and full-time workplaces, respectively, against the days since K-12 school opening.  Here, we restrict our sample to a subset of counties for which the school opening date is the same across all school districts within a county and the measurement of school visits and workplace visits is a daily measure rather than 7-days moving average.
 \end{flushleft}}
\end{figure*}

\begin{figure}[!ht]
  \caption{The event-study regression estimates for behavior variables  obtained using the DID method from Callway and Sant'Anna (2020)\label{fig:ca-behavior}} \medskip
\hspace{-7cm}\resizebox{0.67\columnwidth}{!}{
\begin{minipage}{\linewidth}\medskip
\footnotesize \centering
        \begin{tabular}{cccc}  
 \textbf{(a) Restaurants: In-person  }&\textbf{(b) Restaurants: Hybrid }&\textbf{(c) Restaurants: In-person/No-Mask }&\textbf{(d) Restaurants:  Hybrid/No-Mask }\smallskip\\ 
 \includegraphics[width=0.4\textwidth]{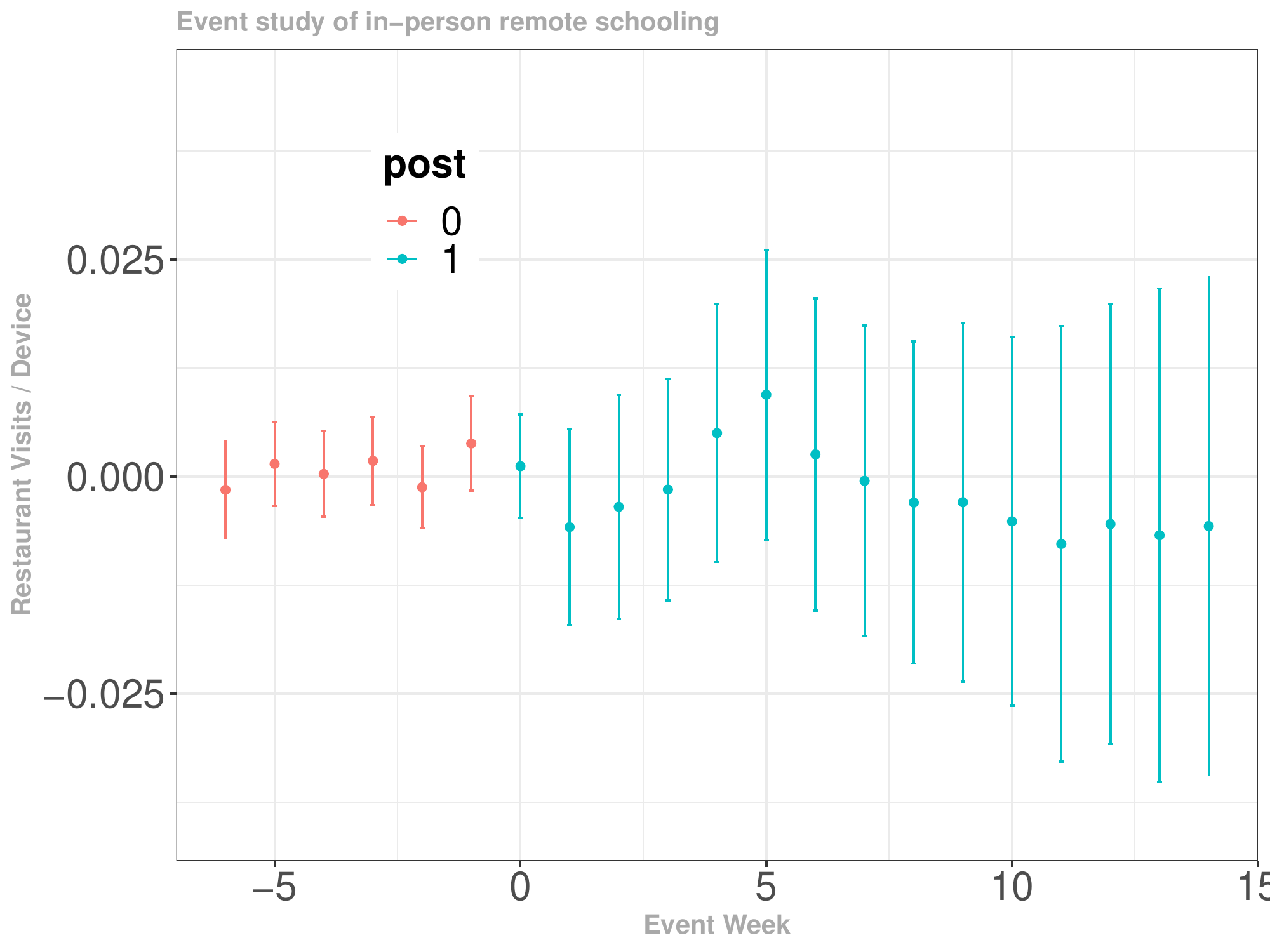}& \includegraphics[width=0.4\textwidth]{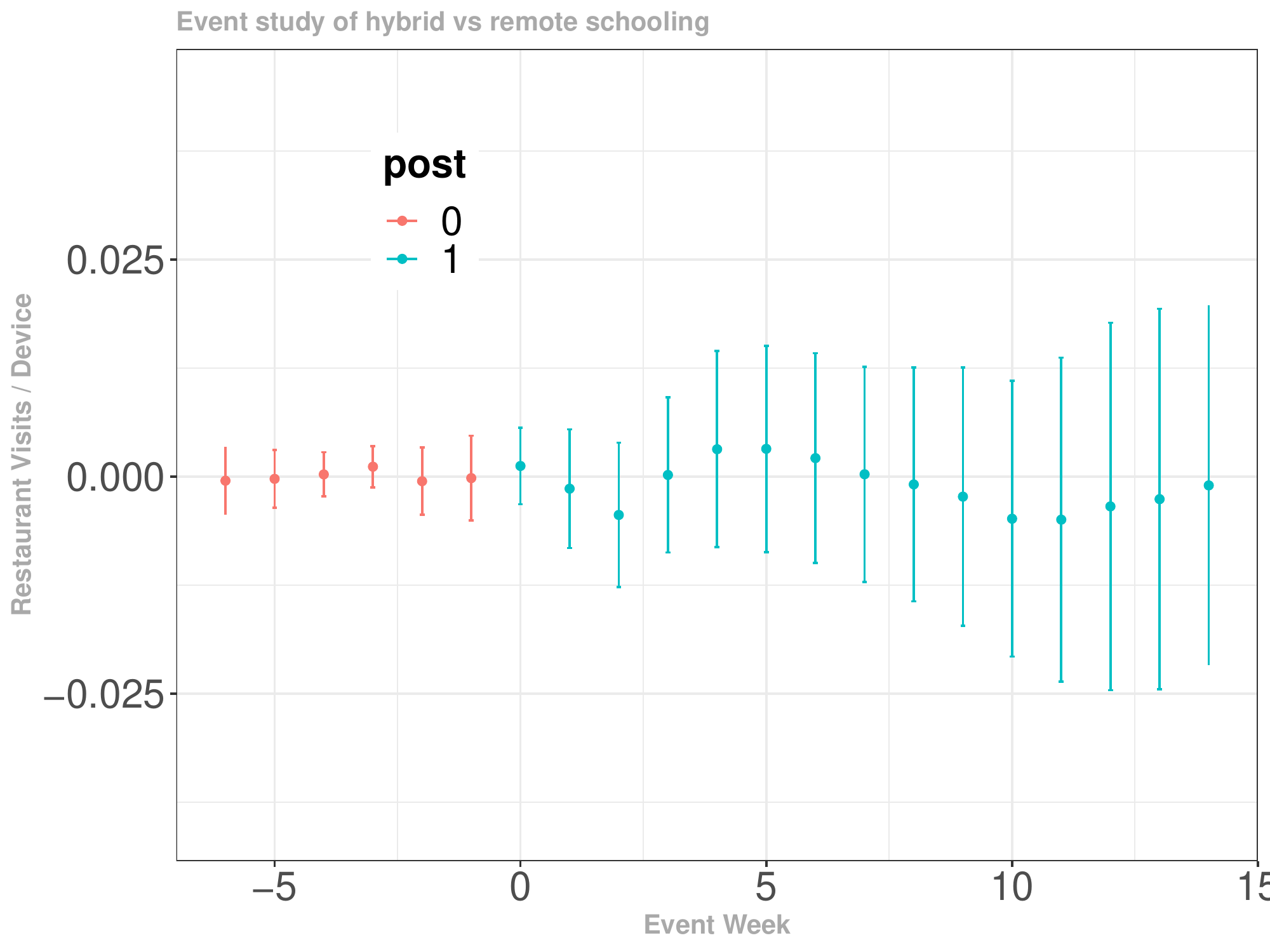} 
 &
 \includegraphics[width=0.4\textwidth]{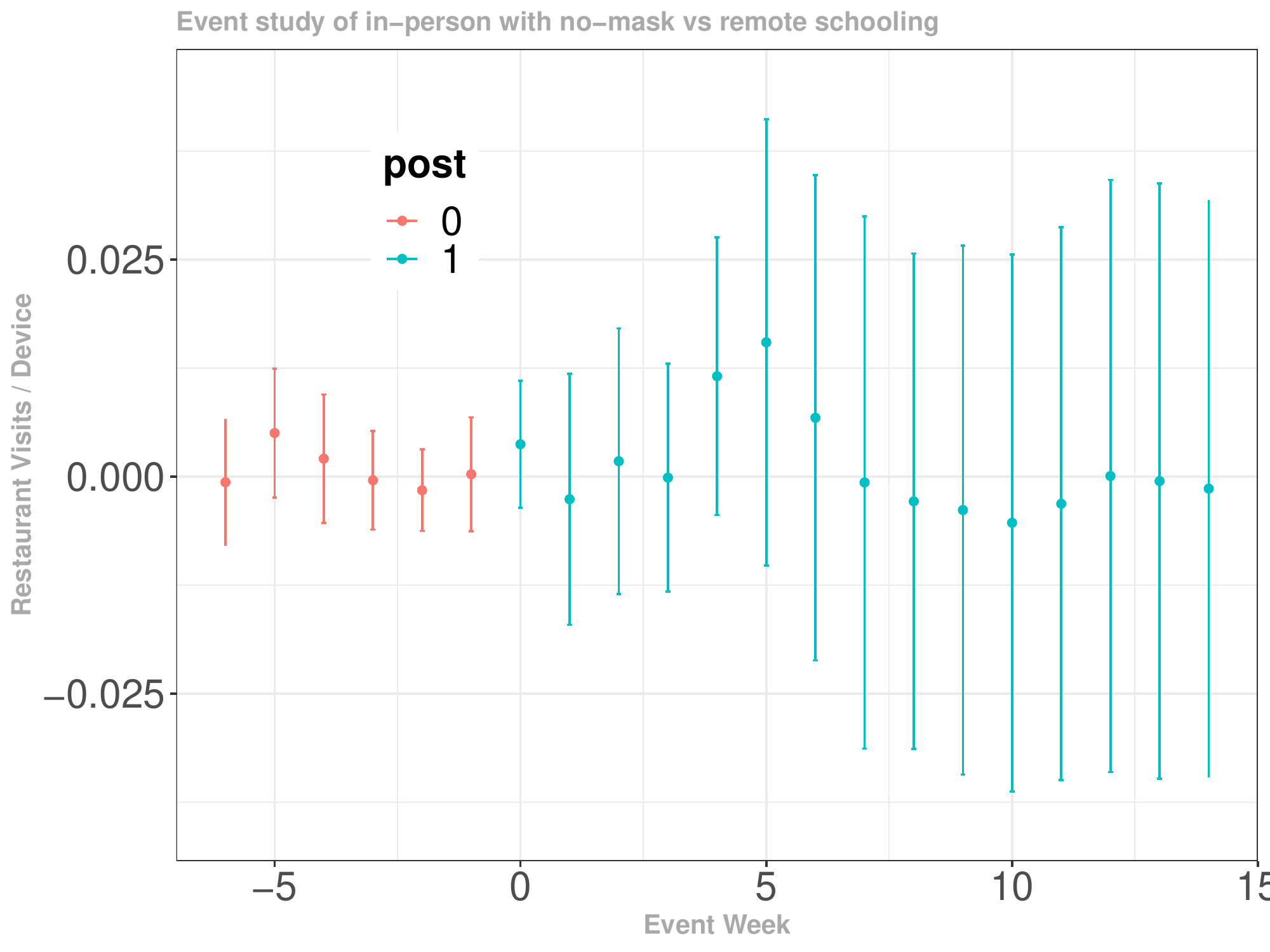}& \includegraphics[width=0.4\textwidth]{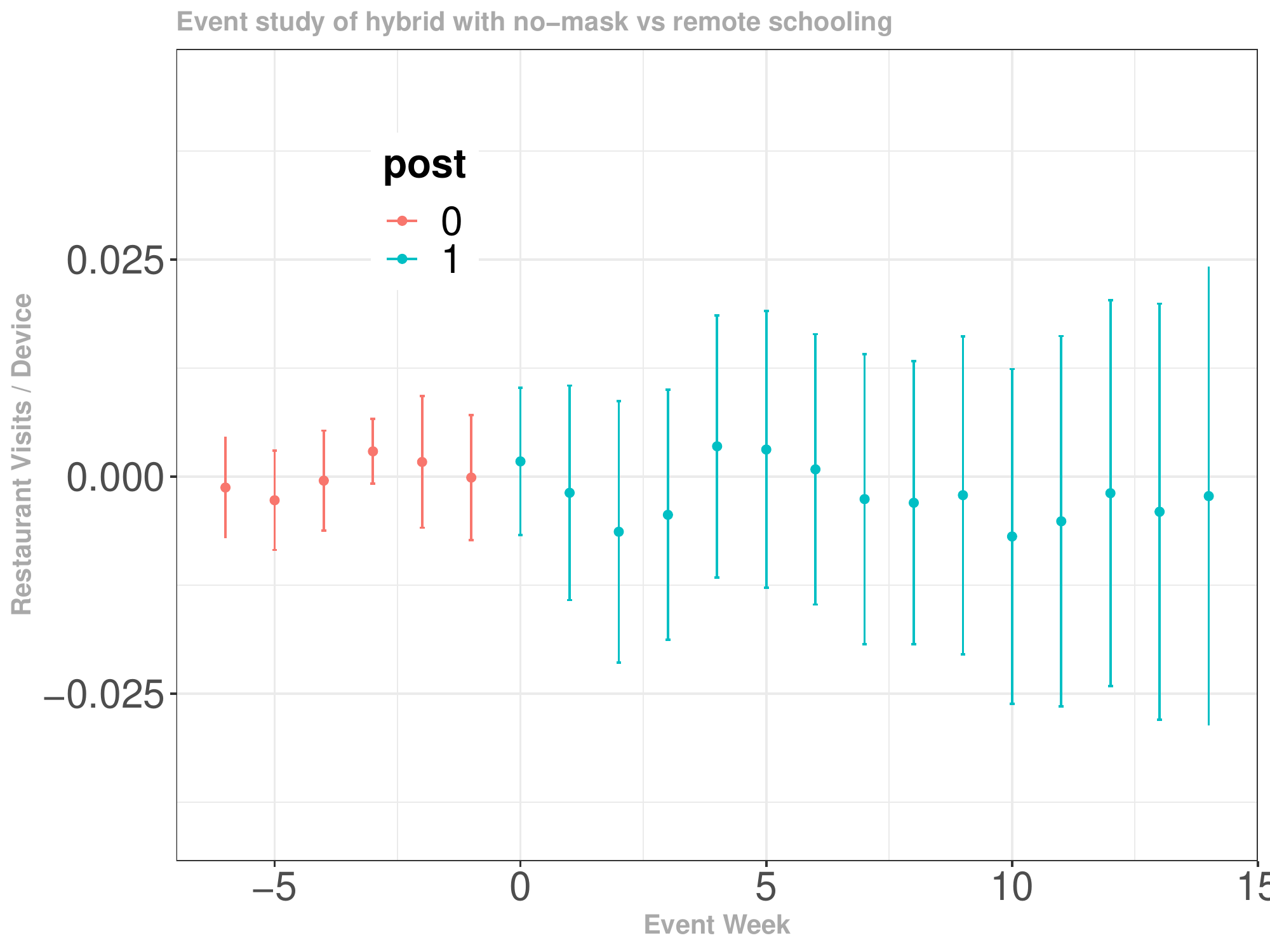}  \smallskip\\ 
 \textbf{(e) Drinking Places: In-person  }&\textbf{(f) Drinking Places: Hybrid }&\textbf{(g) Drinking Places: In-person/No-Mask }&\textbf{(h) Drinking Places:  Hybrid/No-Mask }\smallskip\\ 
 \includegraphics[width=0.4\textwidth]{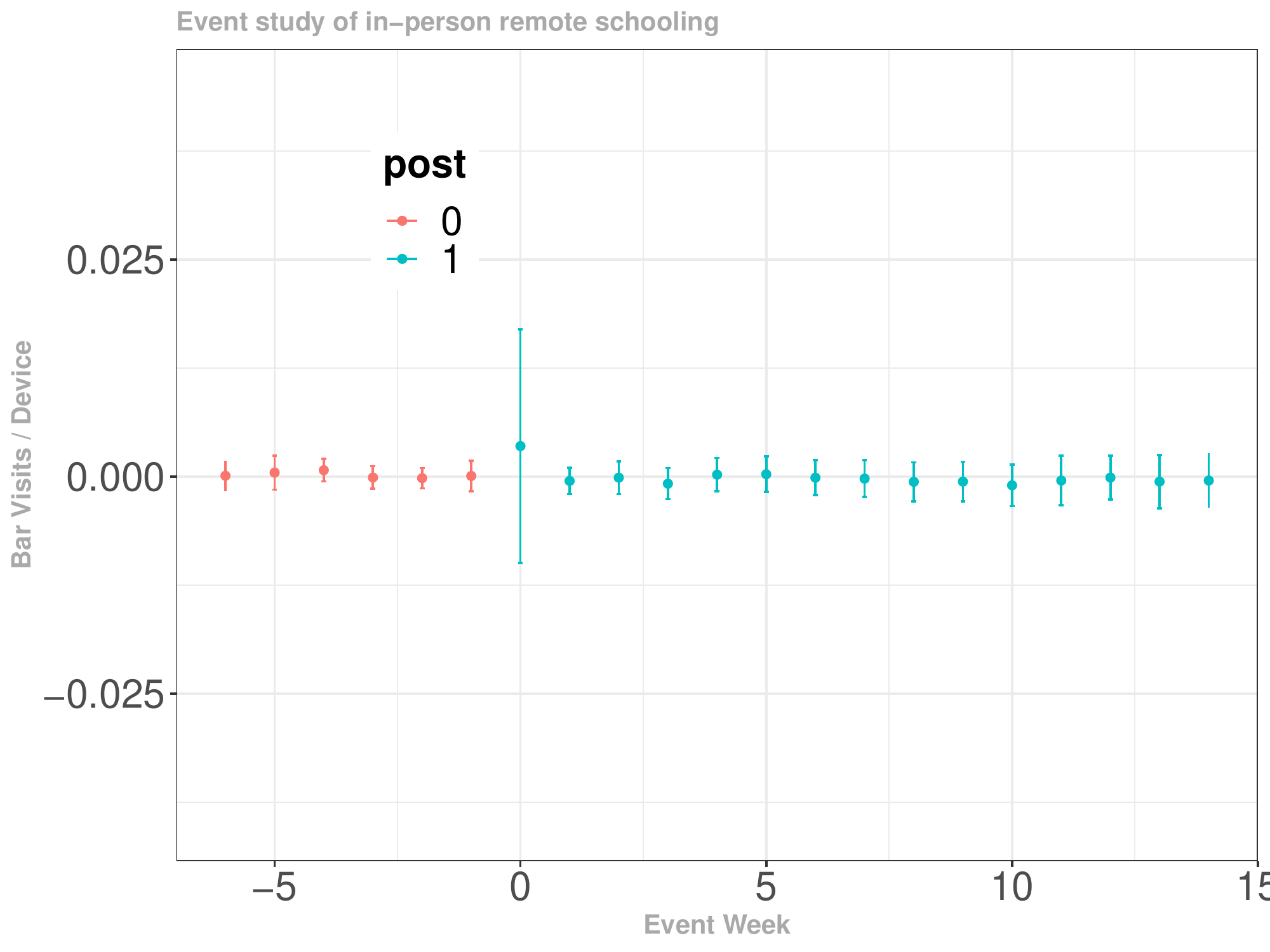}& \includegraphics[width=0.4\textwidth]{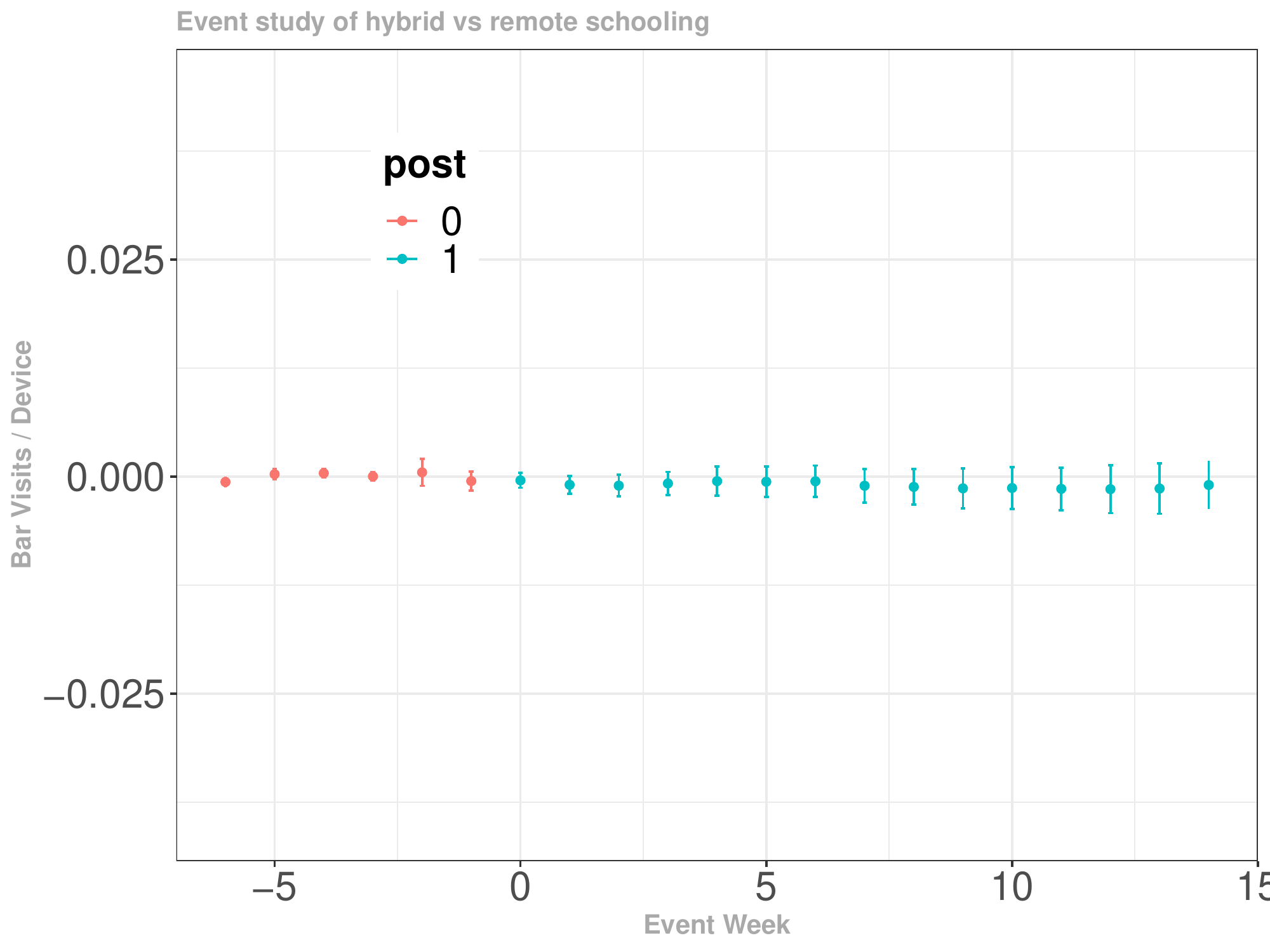} 
 &
 \includegraphics[width=0.4\textwidth]{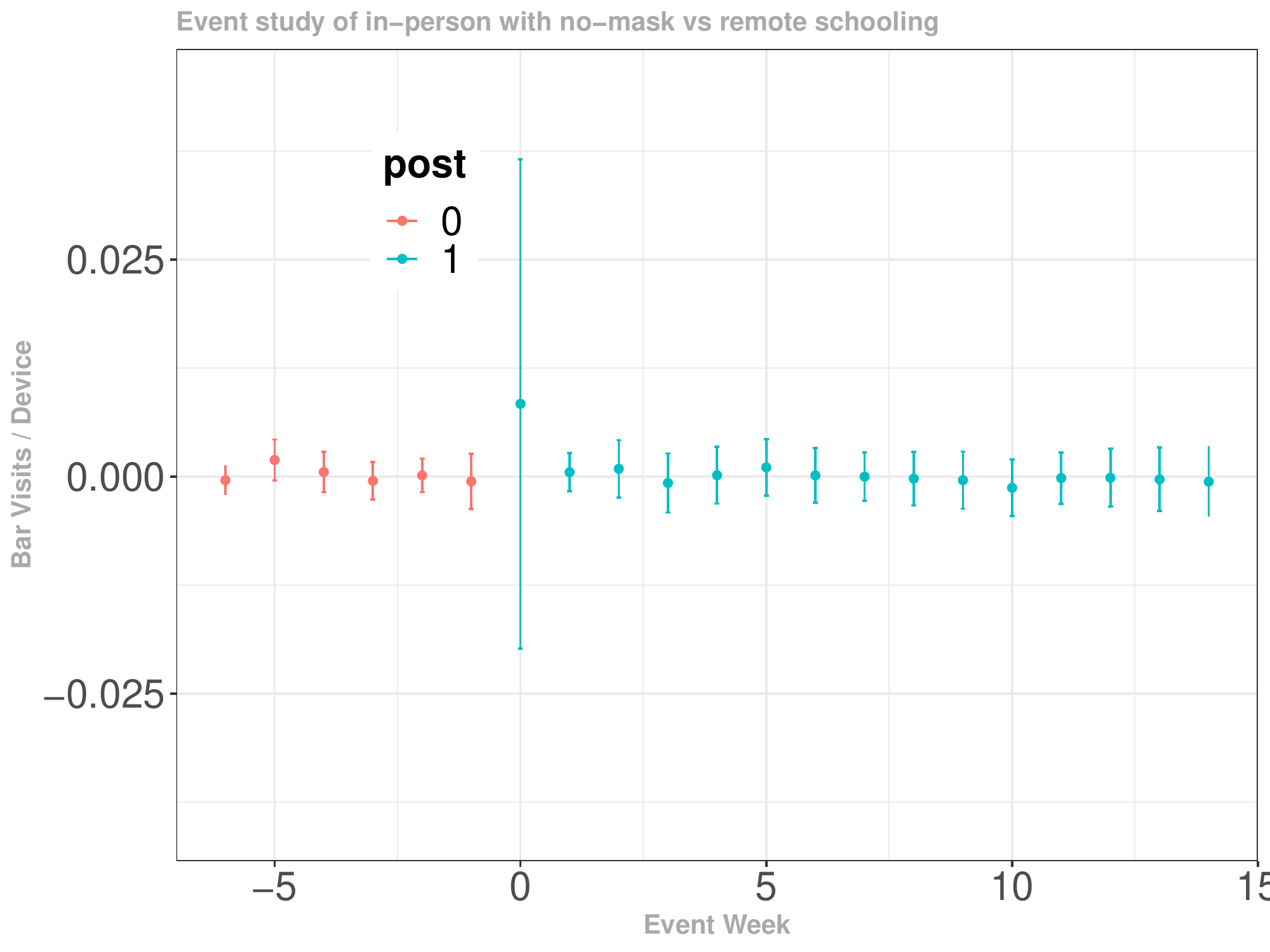}& \includegraphics[width=0.4\textwidth]{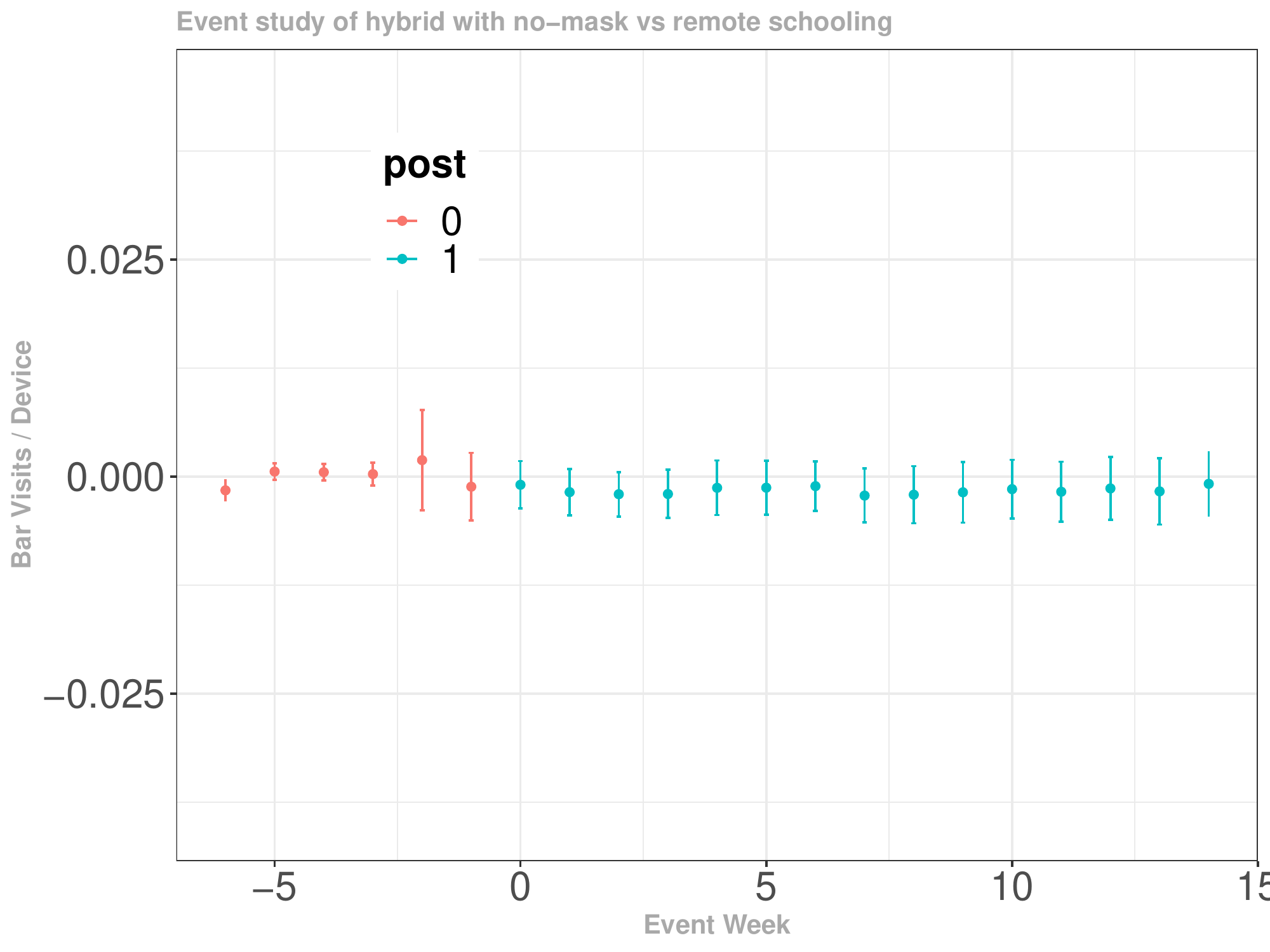}  \smallskip\\ 
  \textbf{(i) Rec. Facilities: In-person  }&\textbf{(j) Rec. Facilities: Hybrid }&\textbf{(k) Rec. Facilities: In-person/No-Mask }&\textbf{(l) Rec. Facilities:  Hybrid/No-Mask }\smallskip\\ 
 \includegraphics[width=0.4\textwidth]{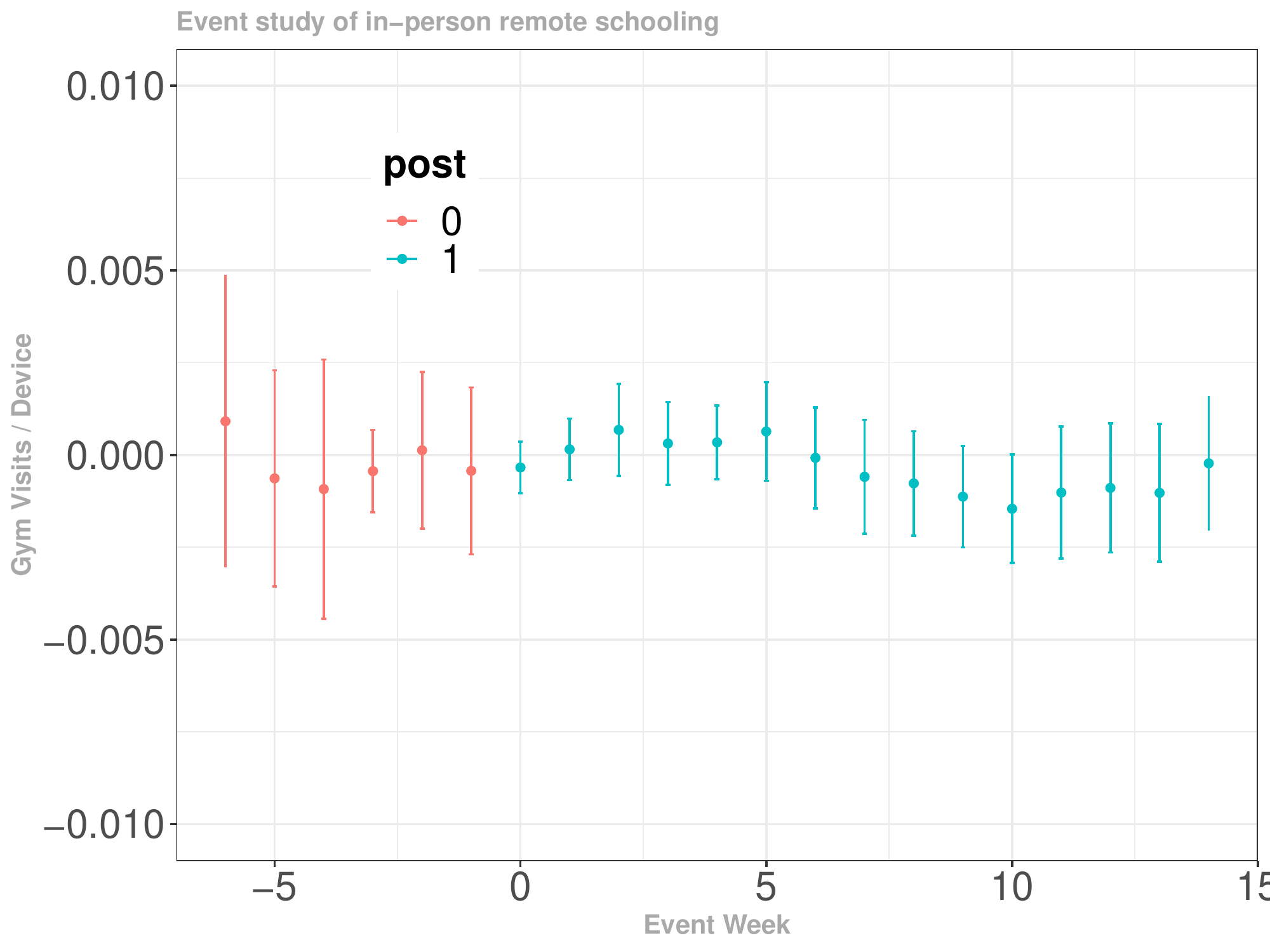}& \includegraphics[width=0.4\textwidth]{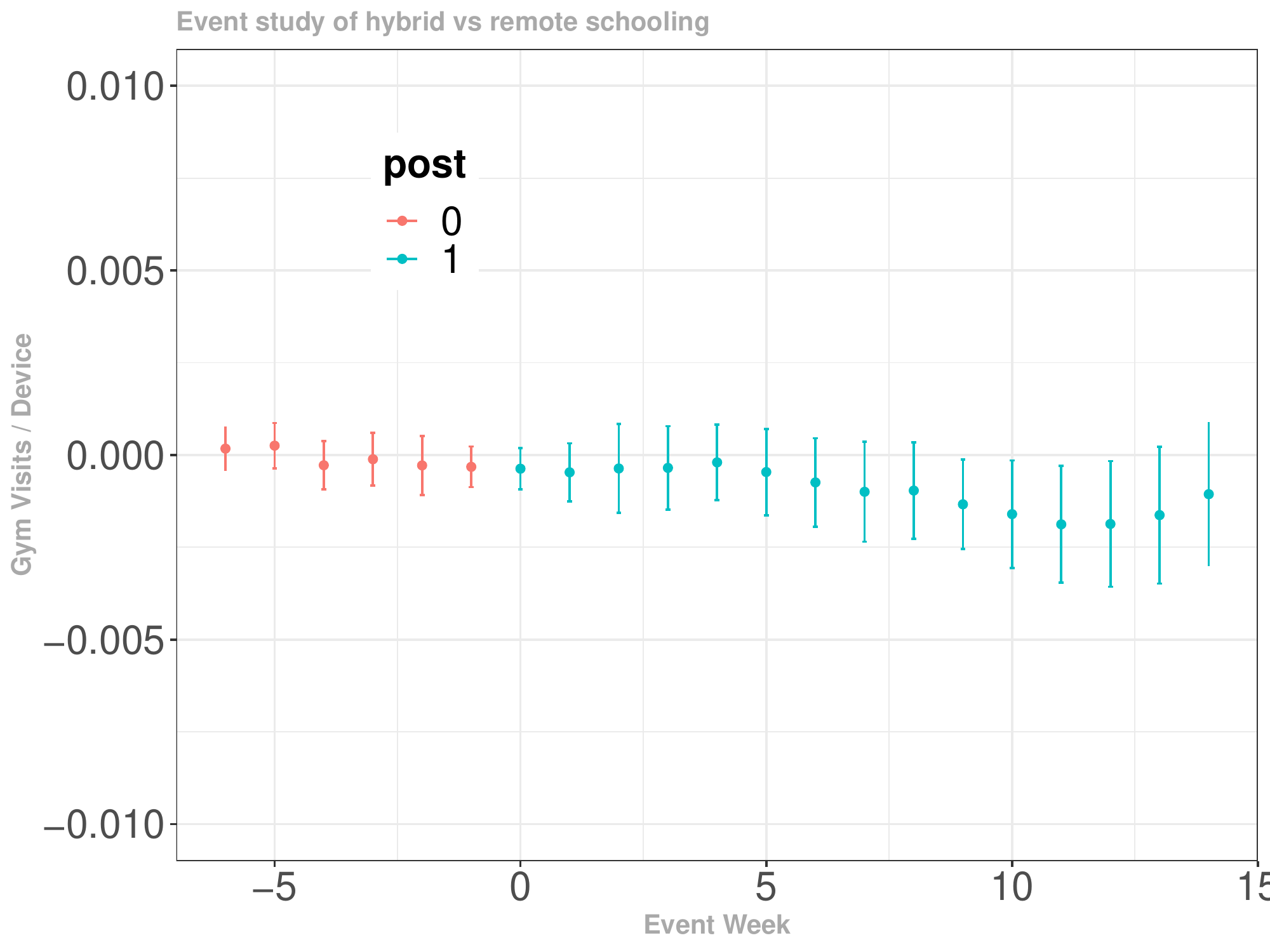} 
 &
 \includegraphics[width=0.4\textwidth]{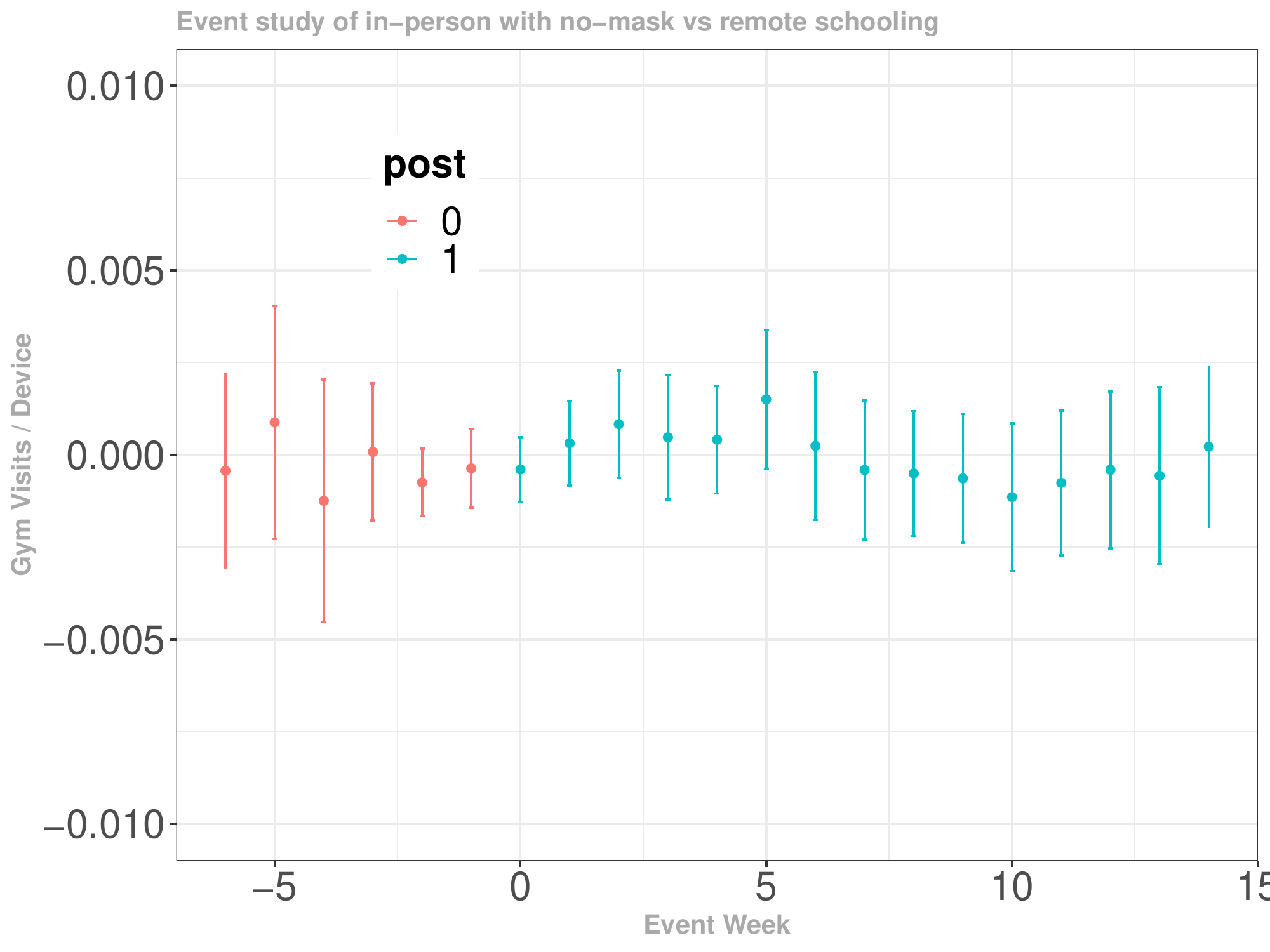}& \includegraphics[width=0.4\textwidth]{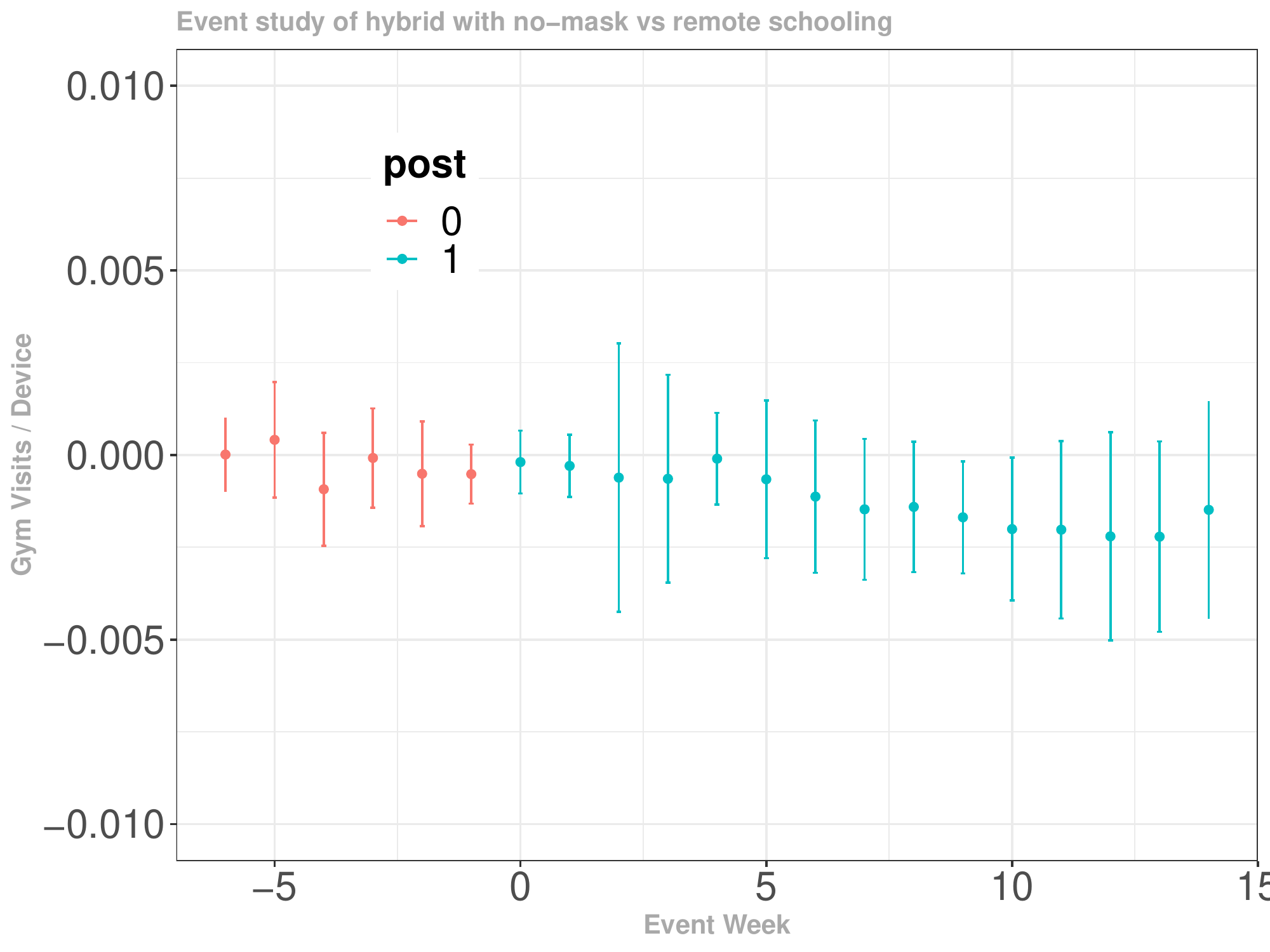}  \smallskip\\ 
  \textbf{(m) Church: In-person  }&\textbf{(n) Church: Hybrid }&\textbf{(o) Church: In-person/No-Mask }&\textbf{(p) Church:  Hybrid/No-Mask }\smallskip\\ 
 \includegraphics[width=0.4\textwidth]{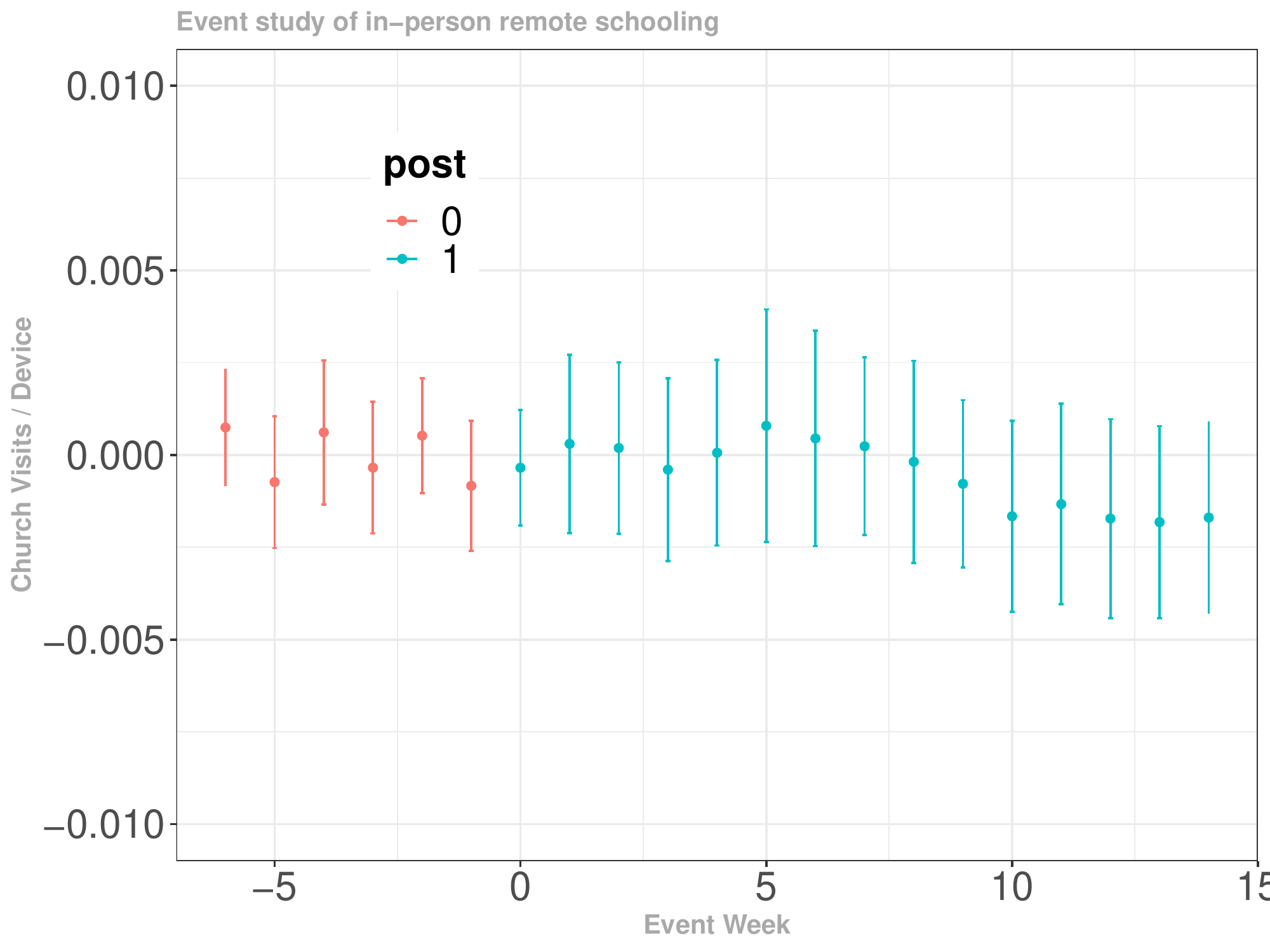}& \includegraphics[width=0.4\textwidth]{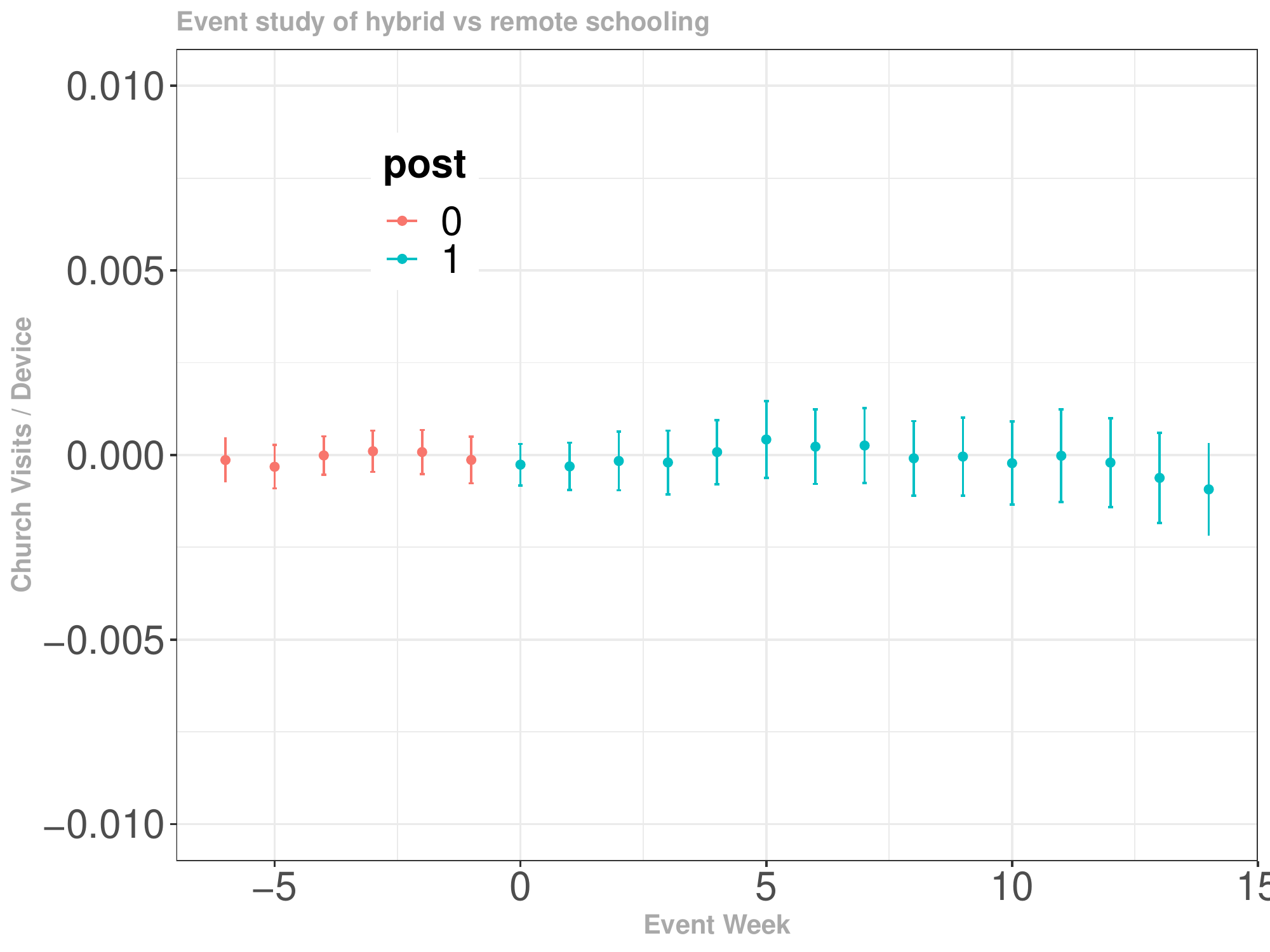} 
 &
 \includegraphics[width=0.4\textwidth]{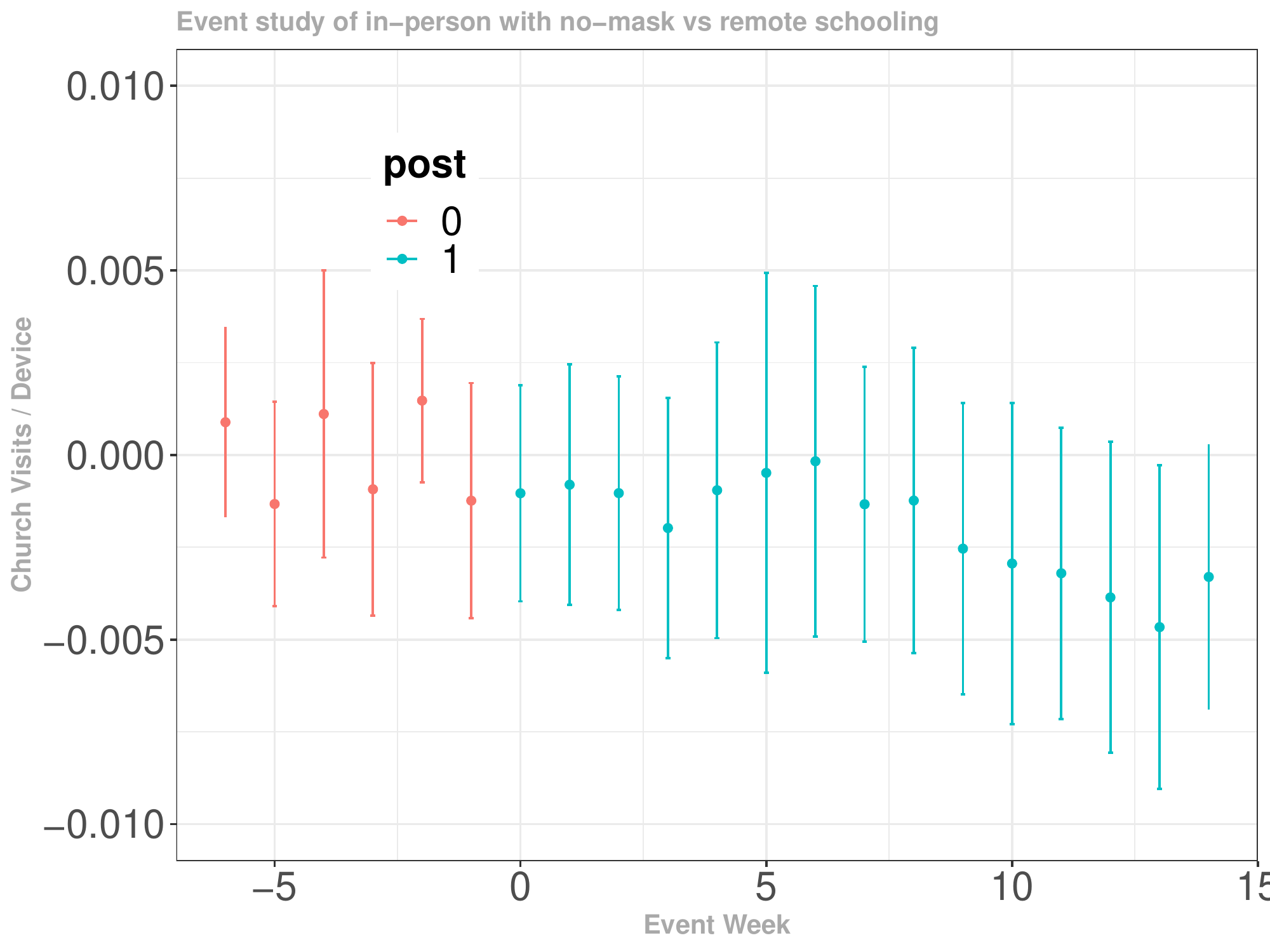}& \includegraphics[width=0.4\textwidth]{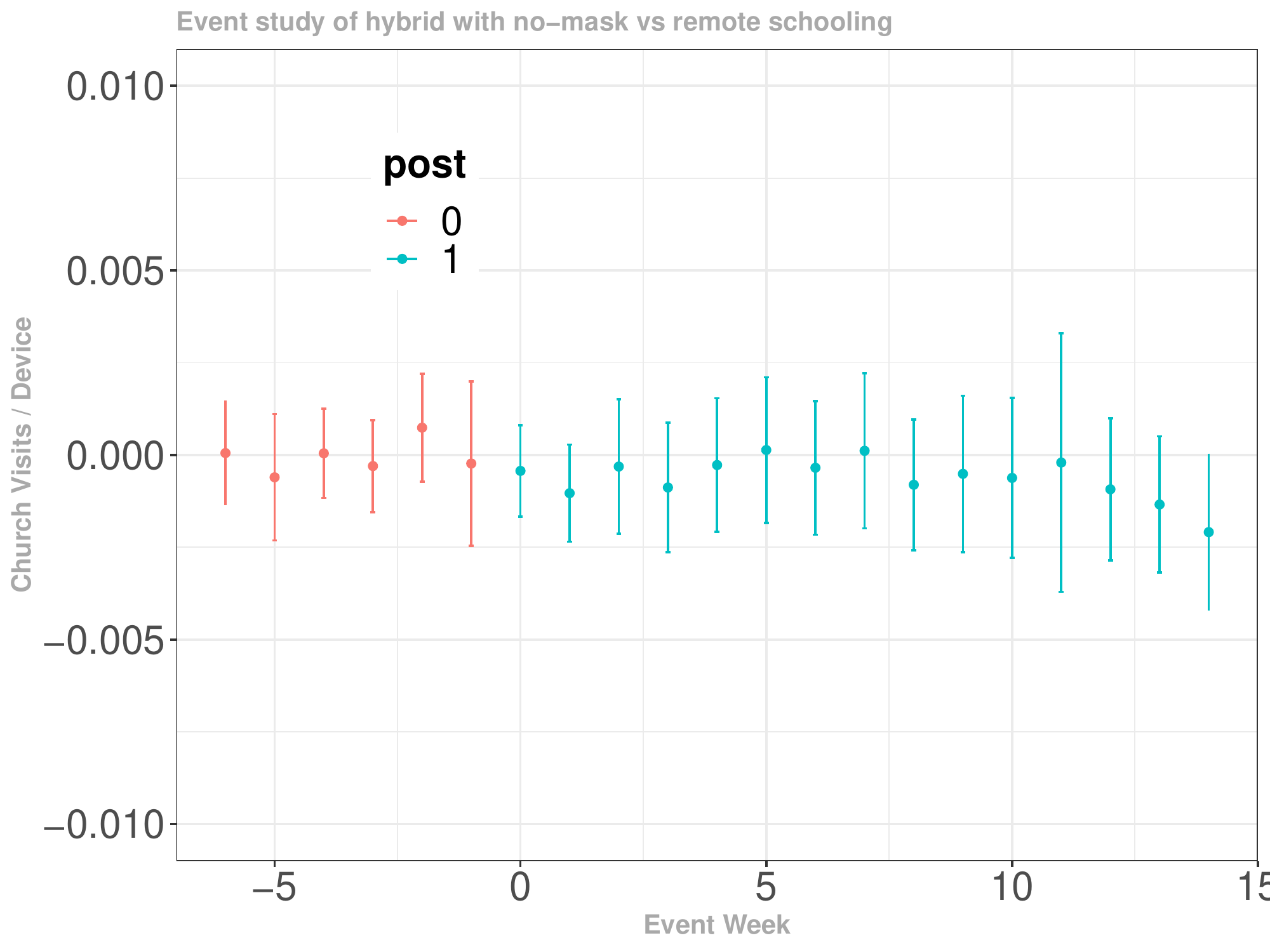}  \smallskip\\ 
  \textbf{(q) College: In-person  }&\textbf{(r) College: Hybrid }&\textbf{(s) College: In-person/No-Mask }&\textbf{(t) College:  Hybrid/No-Mask }\smallskip\\ 
 \includegraphics[width=0.4\textwidth]{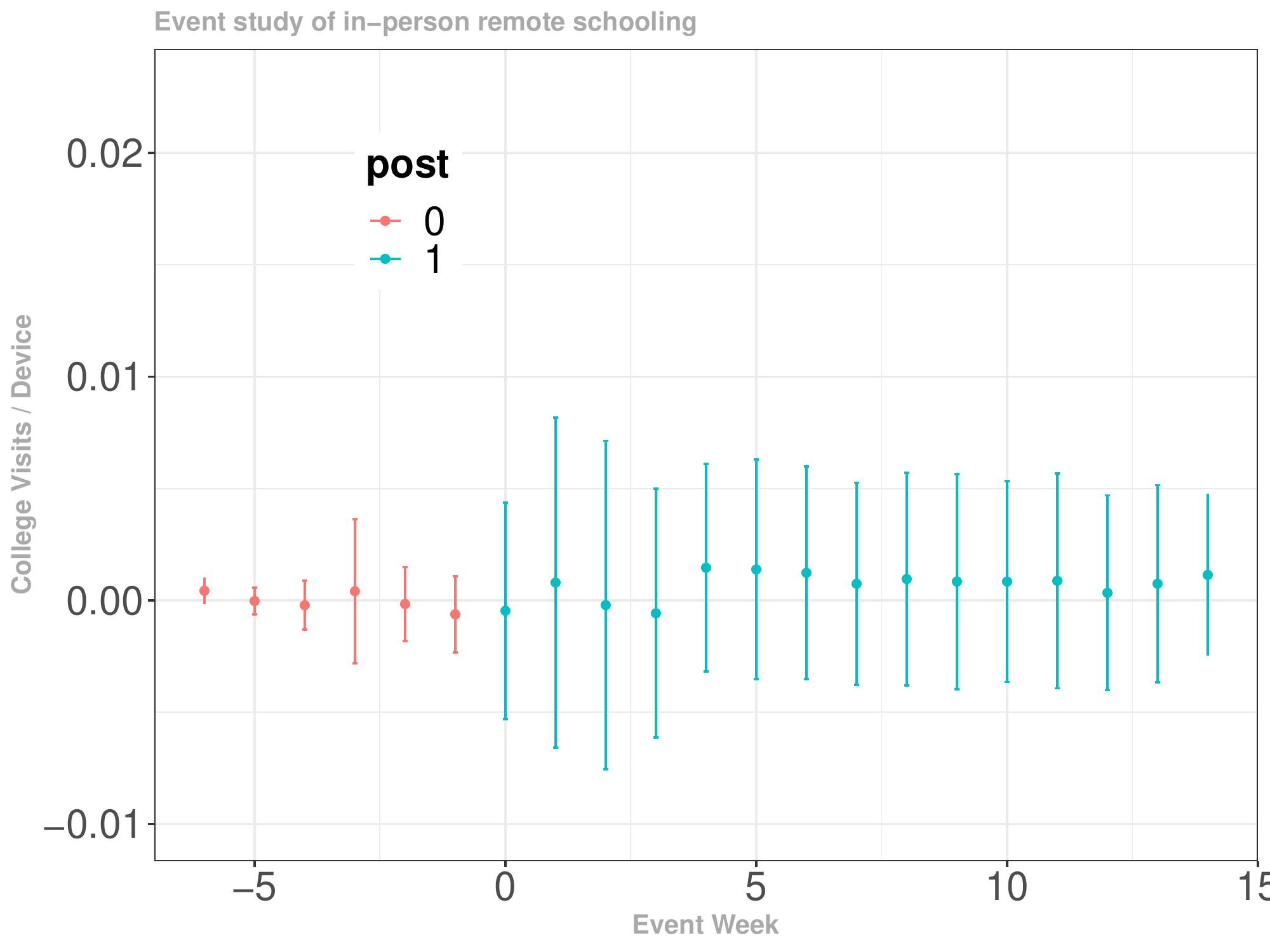}& \includegraphics[width=0.4\textwidth]{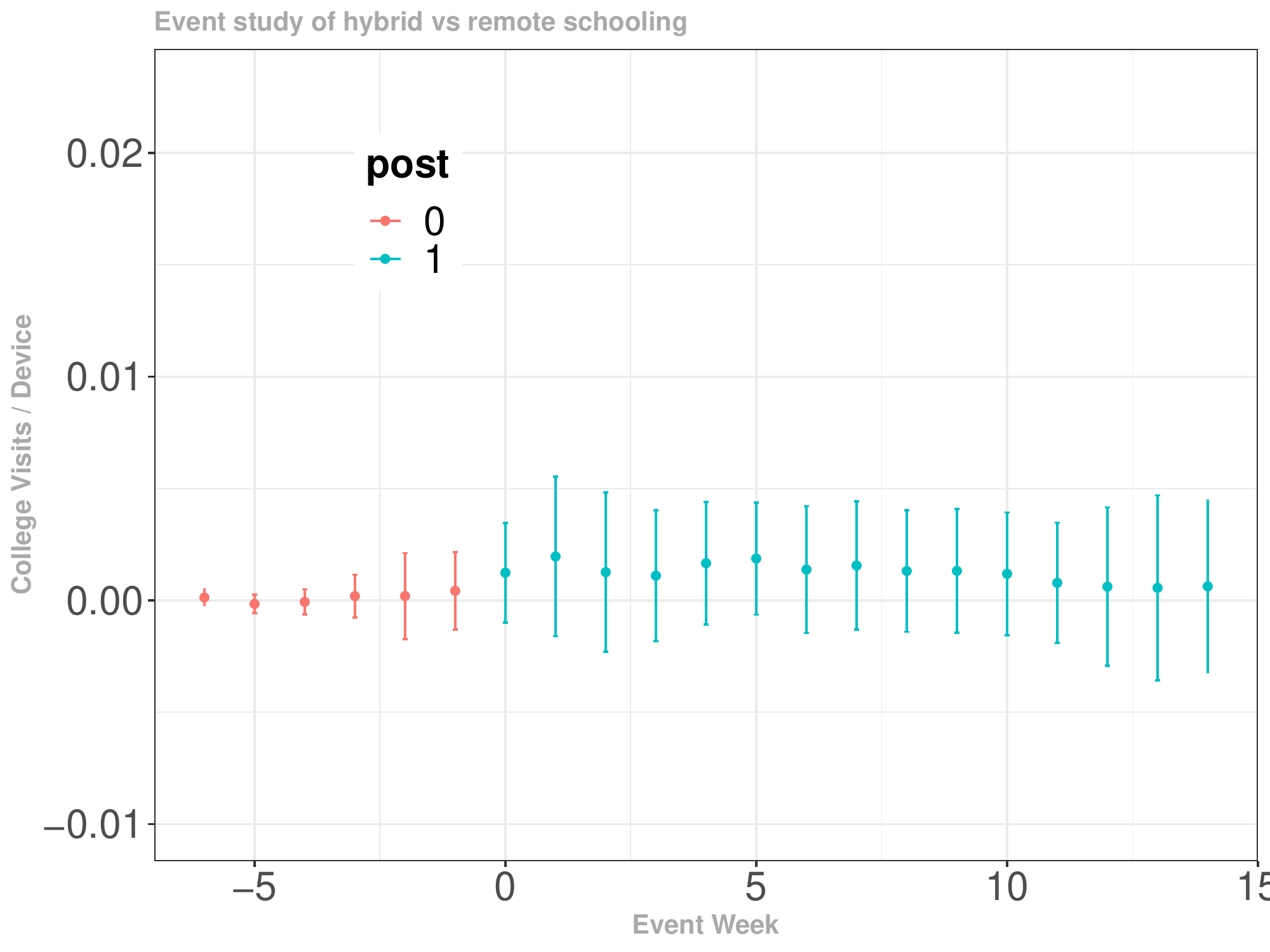} 
 &
 \includegraphics[width=0.4\textwidth]{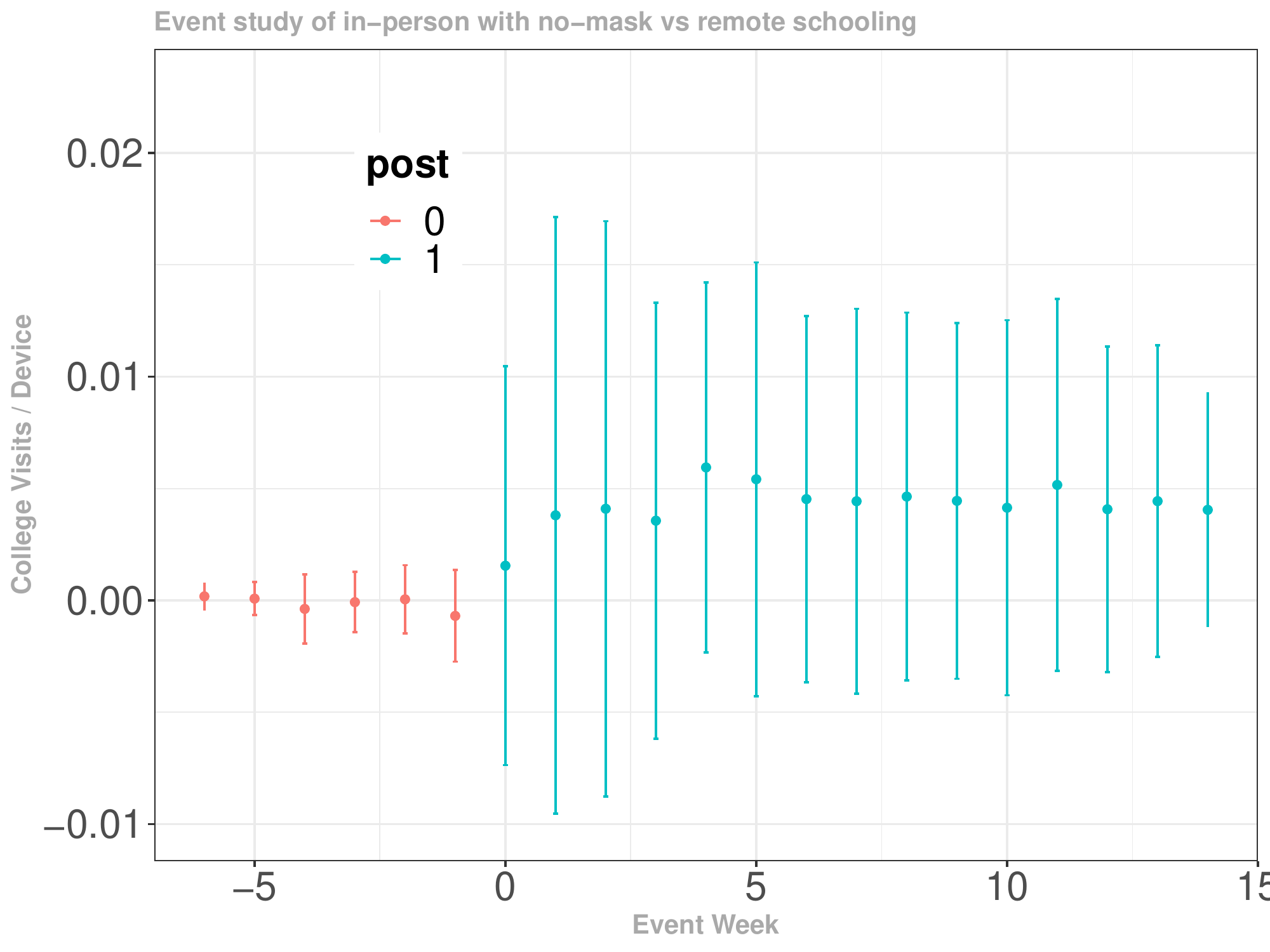}& \includegraphics[width=0.4\textwidth]{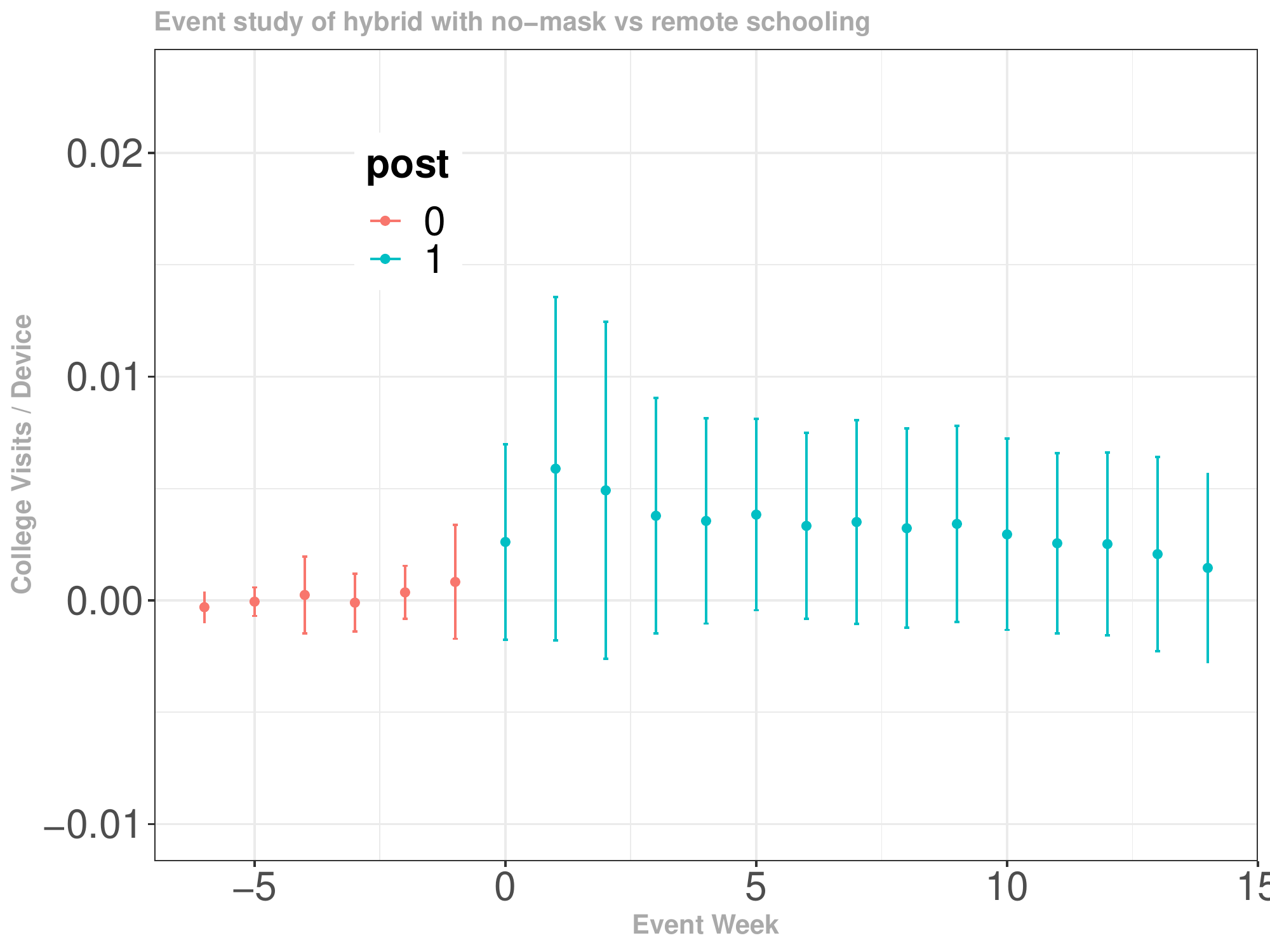}  
\end{tabular}
 \end{minipage}}
  {\scriptsize
\begin{flushleft}
Notes: (a) plots the estimates and 95\% simultaneous confidence intervals of  the average dynamic treatment effect of in-person openings relative to the counties with remote openings as well as the counties that have not opened yet on per-devise restaurant visits using a subset of counties with either in-person opening or remote-opening, where we use the estimation method of \cite{Callaway2020} implemented by their \href{https://cran.r-project.org/web/packages/did/vignettes/did-basics.html}{did R package}.  Similarly, (b), (c), and (d) plots the estimates of  the average dynamic treatment effect of school opening with hybrid, in-persion/mask mandates, and hybrid/mask mandates teaching methods, respectively, using a subset of counties with the corresponding teaching method as well as remote-opening. (e)-(h), (i)-(l), (m)-(p), and (q)-(t)  report the estimates of the average dynamic treatment effect on drinking places, recreational facilities, churches, and colleges,  respectively.  
 \end{flushleft}}
\end{figure}

 \begin{figure}[ht]
  \caption{Sensitivity analysis for the estimated coefficients of K-12 visits and college visits of case growth regressions:  Standard Fixed Effects Estimator without Bias Correction \label{fig:sensitivity-fe-SI}}   
 \resizebox{0.7\columnwidth}{!}{
\centering
        \begin{tabular}{c}  
      (a) Case Growth Estimates   \\
       \includegraphics[width=0.45\textwidth]{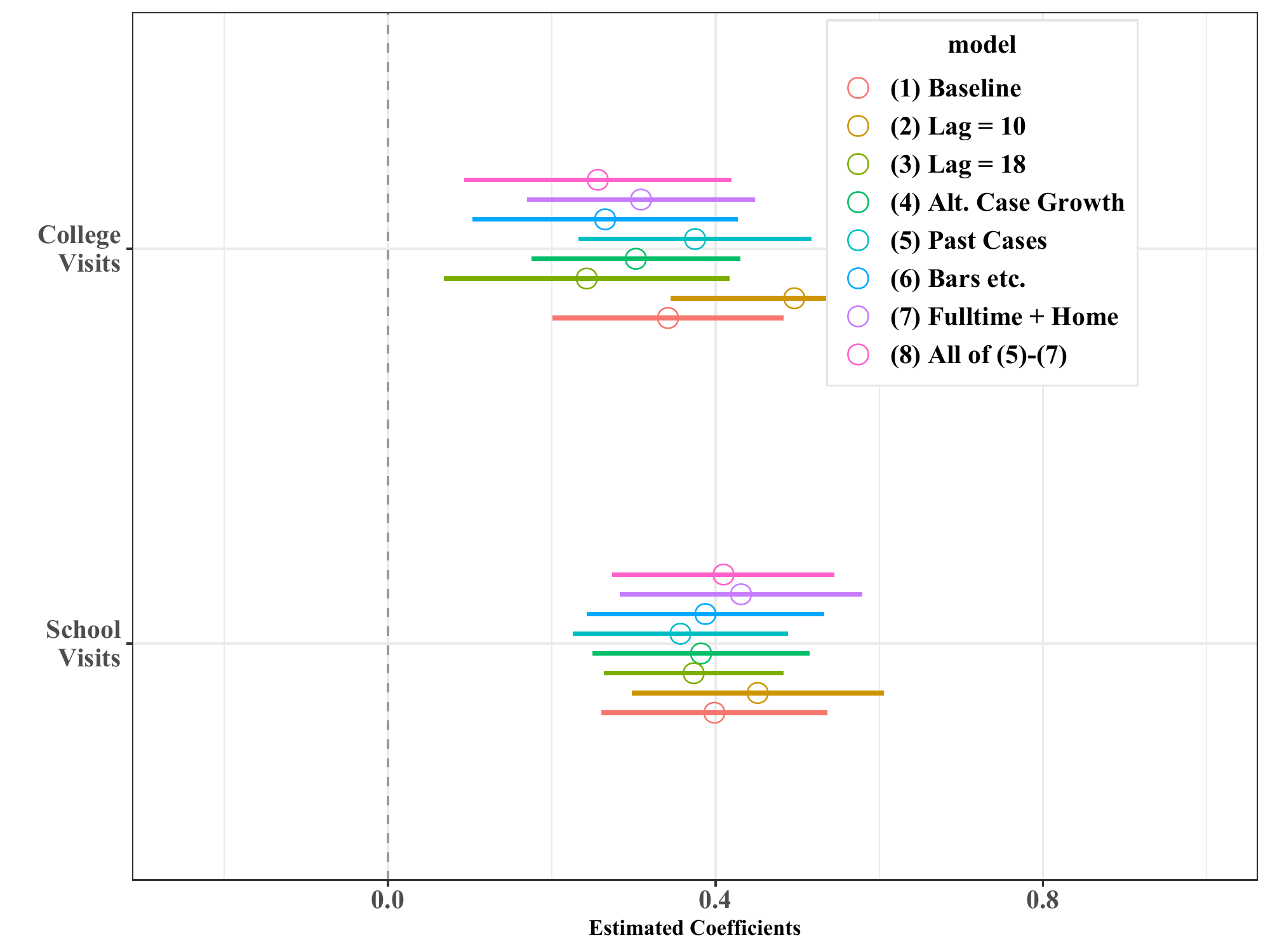} \\
       (b) Case Growth Estimates with School Visits $\times$ No Mask   \\  
       \includegraphics[width=0.45\textwidth]{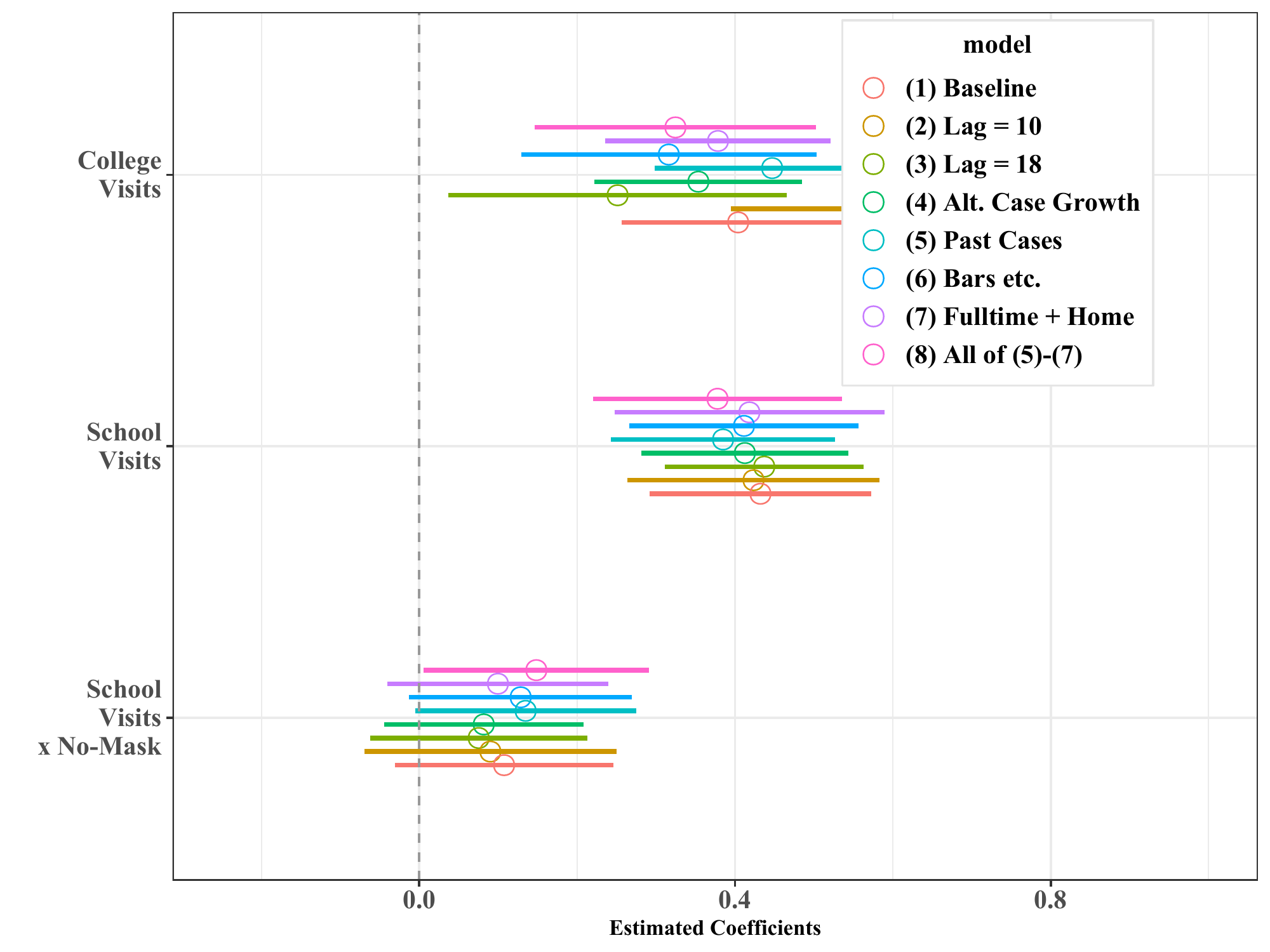}  \\
       \end{tabular}}
\vspace{-0.2cm}   { \scriptsize
\begin{flushleft}
Notes: These figures corresponds to Fig. 5 of the main text but report the result of the (standard) fixed effects estimator without bias correction.  
\end{flushleft}  }
\end{figure} 

\begin{table}[!htbp] \centering
 \caption{The Association of School/College Openings  with Visits to Restaurants and Bars in the United States}\vspace{-0.3cm}
 \label{tab:PItoB-SI}
 \smallskip
\resizebox{0.8\columnwidth}{!}{
 \begin{tabular}{@{\extracolsep{1pt}}lcc|cc}
\\[-1.8ex]\hline
\hline
 & \multicolumn{4}{c}{\textit{Dependent variable}} \\
\cline{2-5}
 & Restaurants & Restaurants & Bars & Bars \\
& (1) & (2) & (3) & (4)\\
\hline
\hline \\[-1.8ex] 
  K-12 School Visits & 0.094$^{**}$ &  & 0.012$^{**}$ &  \\ 
  & (0.046) &  & (0.006) &  \\ 
  Open K-12 In-person &  & $-$0.578 &  & $-$0.059 \\ 
  &  & (0.404) &  & (0.041) \\ 
  Open K-12 Hybrid &  & $-$0.991$^{***}$ &  & $-$0.101$^{***}$ \\ 
  &  & (0.272) &  & (0.038) \\ 
  Open K-12 Remote &  & $-$0.785$^{**}$ &  & $-$0.006 \\ 
  &  & (0.295) &  & (0.056) \\  \hline
  College Visits & 0.288$^{***}$ & 0.270$^{***}$ & 0.022$^{***}$ & 0.020$^{***}$ \\ 
  & (0.053) & (0.051) & (0.006) & (0.005) \\ 
  Mandatory mask & 1.096$^{***}$ & 0.744$^{*}$ & 0.180$^{***}$ & 0.116 \\ 
  & (0.371) & (0.403) & (0.067) & (0.069) \\ 
  Ban gatherings & $-$0.545 & $-$0.477 & $-$0.103 & $-$0.105 \\ 
  & (0.920) & (0.897) & (0.117) & (0.118) \\ 
  Stay at Home & $-$1.885$^{***}$ & $-$1.870$^{***}$ & $-$0.182$^{***}$ & $-$0.167$^{***}$ \\ 
  & (0.203) & (0.242) & (0.025) & (0.025) \\ \hline
  log(Cases) & $-$0.020 & $-$0.019 & $-$0.001 & 0.002 \\ 
  & (0.047) & (0.050) & (0.006) & (0.007) \\ 
 log(Cases), 7d lag& $-$0.026 & $-$0.039 & $-$0.005 & $-$0.003 \\ 
  & (0.032) & (0.032) & (0.005) & (0.005) \\ 
 log(Cases), 14d lag & $-$0.066 & $-$0.078$^{*}$ & $-$0.007 & $-$0.009 \\ 
  & (0.043) & (0.042) & (0.005) & (0.005) \\  
 \hline \\[-1.8ex]
Observations & 670,895 & 595,872 & 670,895 & 595,872 \\ 
R$^{2}$ & 0.881 & 0.883 & 0.807 & 0.807 \\   \hline
\hline \\[-1.8ex]
\end{tabular}
}
{\scriptsize \begin{flushleft}
Notes:  All regression specifications include county fixed effects and state-week fixed effects.   The standard fixed effects estimator without bias correction is used. Clustered standard errors at the state level are reported in the bracket.   {$^{*}$p$<$0.1; $^{**}$p$<$0.05; $^{***}$p$<$0.01}
\end{flushleft}}
\end{table}

\begin{table}[!htbp] \centering
 \caption{The Association of School  Openings and NPI Policies  with Case Growth in the United States: Standard Fixed Effects Estimator without Bias Correction}\vspace{-0.3cm}
 \label{tab:PItoY-fe-SI}  
\resizebox{0.8\columnwidth}{!}{
\begin{tabular}{@{\extracolsep{1pt}}lcc|cc} 
\\[-1.8ex]\hline 
\hline \\ [-1.8ex] 
 & \multicolumn{4}{c}{\textit{Dependent variable:  \textbf{Case Growth Rate}}} \\ 
\cline{2-5} 
& (1) & (2) & (3) & (4)\\ 
\hline 
 K-12 Visits, 14d  lag & 0.393$^{***}$ & 0.429$^{***}$ &  &  \\ 
  & (0.070) & (0.070) &  &  \\ 
  K-12 Visits $\times$ No-Mask &  & 0.100 &  &  \\ 
  &  & (0.070) &  &  \\ 
   K-12  In-person, 14d  lag&  &  & 0.062$^{***}$ & 0.062$^{***}$ \\ 
  &  &  & (0.017) & (0.021) \\ 
  K-12  Hybrid, 14d  lag &  &  & 0.040$^{***}$ & 0.033$^{**}$ \\ 
  &  &  & (0.014) & (0.013) \\ 
 K-12  Remote, 14d  lag &  &  & 0.030$^{*}$ & 0.027$^{*}$ \\ 
  &  &  & (0.016) & (0.015) \\ 
  K-12   In-person $\times$ No-Mask&  &  &  & 0.009 \\ 
  &  &  &  & (0.019) \\ 
 K-12   Hybrid $\times$ No-Mask &  &  &  & 0.032$^{*}$ \\ 
  &  &  &  & (0.017) \\  \hline
 College Visits, 14d  lag & 0.359$^{***}$ & 0.412$^{***}$ & 0.326$^{***}$ & 0.371$^{***}$ \\ 
  & (0.071) & (0.073) & (0.064) & (0.076) \\ 
 Mandatory mask 14d  lag & $-$0.006 & $-$0.006 & $-$0.015 & $-$0.017 \\ 
  & (0.018) & (0.017) & (0.020) & (0.019) \\ 
 Ban gatherings 14d  lag & $-$0.066$^{*}$ & $-$0.068 & $-$0.068$^{**}$ & $-$0.067 \\ 
  & (0.033) & (0.044) & (0.033) & (0.042) \\ 
  Stay at home 14d  lag& $-$0.203$^{***}$ & $-$0.198$^{***}$ & $-$0.200$^{***}$ & $-$0.200$^{***}$ \\ 
  & (0.031) & (0.039) & (0.034) & (0.040) \\  \hline
   log(Cases), 14d  lag & $-$0.088$^{***}$ & $-$0.092$^{***}$ & $-$0.088$^{***}$ & $-$0.092$^{***}$ \\ 
  & (0.009) & (0.010) & (0.010) & (0.010) \\ 
  log(Cases), 21d  lag  & $-$0.042$^{***}$ & $-$0.043$^{***}$ & $-$0.043$^{***}$ & $-$0.043$^{***}$ \\ 
  & (0.005) & (0.005) & (0.005) & (0.005) \\ 
  log(Cases), 28d  lag& $-$0.017$^{***}$ & $-$0.020$^{***}$ & $-$0.018$^{***}$ & $-$0.021$^{***}$ \\ 
  & (0.003) & (0.003) & (0.004) & (0.004) \\ 
  Test Growth Rates & 0.009$^{**}$ & 0.008$^{*}$ & 0.009$^{**}$ & 0.009$^{*}$ \\ 
  & (0.004) & (0.004) & (0.004) & (0.004) \\ 
 \hline 
County Dummies & Yes & Yes &  Yes  &  Yes  \\   
State$\times$ Week Dummies&Yes & Yes &  Yes  &  Yes  \\
\hline 
Observations & 690,297 & 545,131 & 612,963 & 528,941 \\ 
R$^{2}$ & 0.092 & 0.093 & 0.092 & 0.094 \\  \hline 
\hline 
\end{tabular}}
  {\scriptsize
\begin{flushleft}
Notes: Dependent variable is the log difference over 7 days in weekly positive cases. Regressors are 7-day moving averages of corresponding daily variables and lagged by 2 weeks to reflect the time between infection and case reporting except that we don't take any lag for the log difference in test growth rates. All regression specifications include county fixed effects and state-week fixed effects to control for any unobserved county-level factors and time-varying state-level factors such as various state-level policies as well as 2, 3, and 4 weeks lagged log of cases. The standard fixed effects estimator without bias-correction is applied.  Asymptotic clustered standard errors at the state level are reported in the bracket.  {$^{*}$p$<$0.1; $^{**}$p$<$0.05; $^{***}$p$<$0.01}
\end{flushleft}}   
\end{table}

\begin{table}[!htbp] \centering
 \caption{The Association of School  Openings and NPIs with log of Cases in the United States: Debiased Estimator} 
 \label{tab:PItoY-level}
\resizebox{0.8\columnwidth}{!}{
\begin{tabular}{@{\extracolsep{1pt}}lcc|cc}
\\[-1.8ex]\hline
\hline \\ [-1.8ex]
 & \multicolumn{4}{c}{\textit{Dependent variable:  $\log$\textbf{(Weekly Cases)}}} \\
\cline{2-5}
& (1) & (2) & (3) & (4)\\
\hline 
K-12 Visits, 14d  lag & 1.402$^{***}$ & 0.978$^{***}$ &  &  \\ 
  & (0.193) & (0.222) &  &  \\ 
 K-12 Visits $\times$ No-Mask, 14d  lag   &  & 0.792$^{***}$ &  &  \\ 
  &  & (0.194) &  &  \\ 
K-12  In-person, 14d  lag  &  &  & 0.338$^{***}$ & 0.303$^{***}$ \\ 
  &  &  & (0.046) & (0.048) \\ 
K-12 Hybrid, 14d  lag  &  &  & 0.150$^{***}$ & 0.104$^{***}$ \\ 
  &  &  & (0.029) & (0.031) \\ 
K-12  Remote, 14d  lag &  &  & $-$0.013 & 0.009 \\ 
  &  &  & (0.038) & (0.044) \\ 
 K-12   In-person $\times$ No-Mask, 14d  lag  &  &  &  & 0.059 \\ 
  &  &  &  & (0.057) \\ 
  K-12  Hybrid $\times$ No-Mask, 14d  lag&  &  &  & 0.167$^{***}$ \\ 
  &  &  &  & (0.048) \\ \hline
  College Visits, 14d  lag  & 1.927$^{***}$ & 2.016$^{***}$ & 1.945$^{***}$ & 1.879$^{***}$ \\ 
  & (0.253) & (0.264) & (0.233) & (0.254) \\ 
Mandatory mask, 14d  lag & $-$0.278$^{***}$ & $-$0.244$^{***}$ & $-$0.270$^{***}$ & $-$0.256$^{***}$ \\ 
  & (0.048) & (0.051) & (0.050) & (0.052) \\ 
 Ban gatherings, 14d  lag & $-$0.112 & $-$0.067 & $-$0.113 & $-$0.055 \\ 
  & (0.155) & (0.151) & (0.155) & (0.149) \\ 
Stay at home, 14d  lag & 0.412$^{***}$ & 0.435$^{***}$ & 0.464$^{***}$ & 0.469$^{***}$ \\ 
  & (0.066) & (0.073) & (0.071) & (0.079) \\ \hline 
log(Weekly Cases), 14d  lag  & 0.408$^{***}$ & 0.405$^{***}$ & 0.409$^{***}$ & 0.402$^{***}$ \\ 
  & (0.010) & (0.009) & (0.010) & (0.009) \\ 
 log(Weekly Cases), 21d  lag & 0.133$^{***}$ & 0.135$^{***}$ & 0.133$^{***}$ & 0.134$^{***}$ \\ 
  & (0.005) & (0.004) & (0.005) & (0.005) \\ 
   log(Weekly Cases), 28d  lag  & 0.025$^{***}$ & 0.027$^{***}$ & 0.023$^{***}$ & 0.025$^{***}$ \\ 
  & (0.006) & (0.005) & (0.006) & (0.006) \\ 
  Test Growth Rates & 0.005$^{**}$ & 0.004$^{*}$ & 0.005$^{**}$ & 0.004$^{*}$ \\ 
  & (0.002) & (0.002) & (0.002) & (0.002) \\ 
 \hline \\[-1.8ex] 
County Dummies & Yes & Yes &  Yes  &  Yes  \\
State$\times$ Week Dummies&Yes & Yes &  Yes  &  Yes  \\
\hline \\[-1.8ex] 
Observations & 760,422 & 600,958 & 675,405 & 583,119 \\ 
R$^{2}$ & 0.867 & 0.862 & 0.866 & 0.861 \\   \hline
\hline 
\end{tabular}}
  {\scriptsize
\begin{flushleft}
Notes: Dependent variable is the log of weekly positive cases. Regressors are 7-days moving averages of corresponding daily variables  and lagged by 2 weeks to reflect the time between infection and case reporting except that we don't take any lag for the log difference in test growth rates.   Because the 2 weeks lagged log cases variable is in the control, the estimated coefficients can be interpreted as the covariate's association with case growth over 2 weeks.   All regression specifications include county fixed effects and state-week fixed effects to control for any unobserved county-level factors and time-varying state-level factors such as  various state-level policies.
The debiased fixed effects estimator is applied.  The results from the estimator without bias correction is presented in  SI Appendix, Table S1.
Asymptotic clustered standard errors at the state level are reported in bracket.   {$^{*}$p$<$0.1; $^{**}$p$<$0.05; $^{***}$p$<$0.01}
\end{flushleft}}
\end{table}

\begin{table}[!htbp] \centering
 \caption{The Association of School  Openings and NPIs with log of Cases in the United States: Standard Fixed Effects Estimator without Bias Correction} 
 \label{tab:PItoY-level-nolag}
\resizebox{0.8\columnwidth}{!}{
\begin{tabular}{@{\extracolsep{1pt}}lcc|cc}
\\[-1.8ex]\hline
\hline \\ [-1.8ex]
 & \multicolumn{4}{c}{\textit{Dependent variable:  $\log$\textbf{(Weekly Cases)}}} \\
\cline{2-5}
& (1) & (2) & (3) & (4)\\
\hline 
K-12 Visits, 14d  lag  & 0.955$^{***}$ & 0.640$^{**}$ &  &  \\ 
  & (0.232) & (0.297) &  &  \\ 
 K-12 Visits $\times$ No-Mask, 14d  lag  &  & 0.616$^{**}$ &  &  \\ 
  &  & (0.290) &  &  \\ 
K-12  In-person, 14d  lag  &  &  & 0.368$^{***}$ & 0.332$^{***}$ \\ 
  &  &  & (0.068) & (0.079) \\ 
K-12 Hybrid, 14d  lag  &  &  & 0.184$^{***}$ & 0.138$^{***}$ \\ 
  &  &  & (0.047) & (0.046) \\ 
K-12  Remote, 14d  lag &  &  & $-$0.007 & 0.009 \\ 
  &  &  & (0.053) & (0.059) \\ 
 K-12   In-person $\times$ No-Mask, 14d  lag  &  &  &  & 0.052 \\ 
  &  &  &  & (0.102) \\ 
  K-12  Hybrid $\times$ No-Mask, 14d  lag&  &  &  & 0.186$^{***}$ \\ 
  &  &  &  & (0.069) \\   \hline
   College Visits, 14d  lag  & 1.871$^{***}$ & 1.964$^{***}$ & 1.972$^{***}$ & 1.916$^{***}$ \\ 
  & (0.398) & (0.407) & (0.376) & (0.398) \\ 
Stay at home, 14d  lag & $-$0.093$^{*}$ & $-$0.068 & $-$0.097$^{*}$ & $-$0.072 \\ 
  & (0.055) & (0.055) & (0.056) & (0.055) \\ 
 Ban gatherings, 14d  lag  & $-$0.167$^{*}$ & $-$0.181$^{*}$ & $-$0.193$^{**}$ & $-$0.186$^{*}$ \\ 
  & (0.094) & (0.106) & (0.093) & (0.103) \\ 
Stay at home, 14d  lag& 0.038 & 0.054 & 0.060 & 0.081 \\ 
  & (0.085) & (0.097) & (0.092) & (0.103) \\  
  Test Growth Rates & 0.006$^{**}$ & 0.005$^{**}$ & 0.006$^{***}$ & 0.005$^{**}$ \\ 
  & (0.002) & (0.002) & (0.002) & (0.002) \\ 
 \hline \\[-1.8ex]  
County Dummies & Yes & Yes &  Yes  &  Yes  \\
State$\times$ Week Dummies&Yes & Yes &  Yes  &  Yes  \\
\hline \\[-1.8ex] 
Observations & 760,422 & 600,958 & 675,405 & 583,119 \\ 
R$^{2}$ & 0.867 & 0.862 & 0.866 & 0.861 \\   \hline
\hline 
\end{tabular}}
 {\scriptsize
\begin{flushleft}
Notes: Dependent variable is the log of weekly positive cases. Regressors are 7-days moving averages of corresponding daily variables  and lagged by 2 weeks to reflect the time between infection and case reporting except that we don't take any lag for the log difference in test growth rates.  All regression specifications include county fixed effects and state-week fixed effects to control for any unobserved county-level factors and time-varying state-level factors such as  various state-level policies.
Asymptotic clustered standard errors at the state level are reported in bracket.   {$^{*}$p$<$0.1; $^{**}$p$<$0.05; $^{***}$p$<$0.01}
\end{flushleft}}
\end{table}

\begin{table}[!htbp] \centering
 \caption{The Association of School  Openings, NPI Policies, Full-time/Part-time Work, and Staying Home Devices with Case Growth in the United States: Debiased Fixed Effects Estimator}\vspace{-0.3cm}
 \label{tab:PBItoY-SI}  
\resizebox{0.8\columnwidth}{!}{
\begin{tabular}{@{\extracolsep{1pt}}lcc|cc} 
\\[-1.8ex]\hline 
\hline \\ [-1.8ex] 
 & \multicolumn{4}{c}{\textit{Dependent variable:  \textbf{Case Growth Rate}}} \\ 
\cline{2-5} 
& (1) & (2) & (3) & (4)\\ 
\hline 
   K-12 Visits, 14d  lag & 0.393$^{***}$ & 0.283$^{***}$ &  &  \\ 
  & (0.075) & (0.087) &  &  \\ 
   K-12 Visits $\times$ No-Mask &  & 0.287$^{***}$ &  &  \\ 
  &  & (0.071) &  &  \\ 
  K-12  In-person, 14d  lag&  &  & 0.015 & $-$0.007 \\ 
  &  &  & (0.016) & (0.020) \\ 
  K-12  Hybrid, 14d  lag &  &  & $-$0.028$^{**}$ & $-$0.055$^{***}$ \\ 
  &  &  & (0.013) & (0.013) \\ 
  K-12  Remote, 14d  lag &  &  & $-$0.094$^{***}$ & $-$0.115$^{***}$ \\ 
  &  &  & (0.015) & (0.014) \\ 
  K-12   In-person $\times$ No-Mask &  &  &  & 0.034$^{*}$ \\ 
  &  &  &  & (0.020) \\ 
  K-12   Hybrid $\times$ No-Mask &  &  &  & 0.043$^{***}$ \\ 
  &  &  &  & (0.017) \\ \hline
  Full-time Work Device, 14d  lag& $-$0.117 & 0.186 & 0.956$^{**}$ & 0.967$^{**}$ \\ 
  & (0.417) & (0.490) & (0.384) & (0.436) \\ 
 Part-time Work Device, 14d  lag & 0.262 & 0.466 & 0.820$^{***}$ & 0.915$^{***}$ \\ 
  & (0.259) & (0.305) & (0.276) & (0.309) \\ 
 Staying Home Device, 14d  lag& $-$0.290$^{***}$ & $-$0.283$^{***}$ & $-$0.352$^{***}$ & $-$0.332$^{***}$ \\ 
  & (0.057) & (0.069) & (0.061) & (0.067) \\ \hline
 College Visits, 14d  lag  & 0.060 & 0.012 & 0.114$^{*}$ & 0.010 \\ 
  & (0.071) & (0.072) & (0.065) & (0.075) \\ 
  Mandatory mask 14d  lag & $-$0.114$^{***}$ & $-$0.124$^{***}$ & $-$0.128$^{***}$ & $-$0.128$^{***}$ \\ 
  & (0.018) & (0.017) & (0.019) & (0.019) \\ 
 Ban gatherings 14d  lag & $-$0.120$^{***}$ & $-$0.127$^{***}$ & $-$0.125$^{***}$ & $-$0.126$^{***}$ \\ 
  & (0.034) & (0.044) & (0.034) & (0.043) \\ 
  Stay at home 14d  lag & $-$0.246$^{***}$ & $-$0.241$^{***}$ & $-$0.232$^{***}$ & $-$0.239$^{***}$ \\ 
  & (0.033) & (0.040) & (0.034) & (0.040) \\  \hline
   log(Cases), 14d  lag & $-$0.100$^{***}$ & $-$0.101$^{***}$ & $-$0.096$^{***}$ & $-$0.098$^{***}$ \\ 
  & (0.009) & (0.010) & (0.010) & (0.010) \\ 
   log(Cases), 21d  lag & $-$0.060$^{***}$ & $-$0.059$^{***}$ & $-$0.059$^{***}$ & $-$0.058$^{***}$ \\ 
  & (0.004) & (0.005) & (0.005) & (0.005) \\ 
   log(Cases), 28d  lag& $-$0.030$^{***}$ & $-$0.033$^{***}$ & $-$0.030$^{***}$ & $-$0.033$^{***}$ \\ 
  & (0.003) & (0.003) & (0.004) & (0.003) \\ 
  Test Growth Rates & 0.009$^{**}$ & 0.008$^{*}$ & 0.009$^{**}$ & 0.009$^{**}$ \\ 
  & (0.004) & (0.004) & (0.004) & (0.004) \\ 
 \hline 
County Dummies & Yes & Yes &  Yes  &  Yes  \\   
State$\times$ Week Dummies&Yes & Yes &  Yes  &  Yes  \\
\hline 
Observations & 690,297 & 545,131 & 612,963 & 528,941 \\ 
R$^{2}$ & 0.092 & 0.093 & 0.092 & 0.094 \\  
\hline 
\hline 
\end{tabular}}
  {\scriptsize
\begin{flushleft}
Notes: Dependent variable is the log difference over 7 days in weekly positive cases.  All regression specifications include county fixed effects and state-week fixed effects to control for any unobserved county-level factors and time-varying state-level factors such as  various state-level policies.
The debiased fixed effects estimator is applied.  Asymptotic clustered standard errors at the state level are reported in the bracket.  {$^{*}$p$<$0.1; $^{**}$p$<$0.05; $^{***}$p$<$0.01}
\end{flushleft}}   
\end{table}

\begin{table}[!htbp] \centering
 \caption{The Association of School  Openings and NPI Policies with Death Growth in the United States: Debiased Fixed Effects Estimator}\vspace{-0.3cm}
 \label{tab:PItoY-death-SI}  
\resizebox{0.8\columnwidth}{!}{
\begin{tabular}{@{\extracolsep{1pt}}lcc|cc} 
\\[-1.8ex]\hline 
\hline \\ [-1.8ex] 
 & \multicolumn{4}{c}{\textit{Dep. variable: \textbf{Death Growth Rate over 3 weeks}  }} \\ 
\cline{2-5} 
& (1) & (2) & (3) & (4)\\ 
\hline 
  & (0.113) & (0.165) &  &  \\ 
 K-12 Visits $\times$ No-Mask,  35d lag &  & 0.617$^{***}$ &  &  \\ 
  &  & (0.222) &  &  \\ 
 K-12 In-person,  35d lag  &  &  & 0.015 & 0.040 \\ 
  &  &  & (0.052) & (0.063) \\ 
   K-12 Hybrid,  35d lag &  &  & 0.004 & $-$0.041 \\ 
  &  &  & (0.027) & (0.028) \\ 
  K-12 Remote,  35d lag&  &  & $-$0.048 & $-$0.109$^{***}$ \\ 
  &  &  & (0.030) & (0.040) \\ 
  K-12 In-person $\times$ No-Mask,  35d lag&  &  &  & 0.050 \\ 
  &  &  &  & (0.068) \\ 
K-12 Hybrid $\times$ No-Mask,  35d lag &  &  &  & 0.141$^{***}$ \\ 
  &  &  &  & (0.050) \\ \hline
College Visits, 35d lag  & 0.313 & 0.530$^{***}$ & 0.507$^{***}$ & 0.561$^{***}$ \\ 
  & (0.199) & (0.203) & (0.196) & (0.193) \\  
Mandatory mask,   35d lag & $-$0.127$^{***}$ & $-$0.125$^{***}$ & $-$0.142$^{***}$ & $-$0.123$^{***}$ \\ 
  & (0.027) & (0.024) & (0.027) & (0.025) \\ 
  Ban gatherings,  35d lag & $-$0.212$^{***}$ & $-$0.355$^{***}$ & $-$0.292$^{***}$ & $-$0.360$^{***}$ \\ 
  & (0.077) & (0.094) & (0.080) & (0.095) \\ 
  Stay at home,   35d lag  & $-$0.198$^{***}$ & $-$0.127$^{**}$ & $-$0.156$^{**}$ & $-$0.116$^{*}$ \\ 
  & (0.057) & (0.064) & (0.062) & (0.069) \\  \hline
  log(Deaths), 35d lag& $-$0.666$^{***}$ & $-$0.675$^{***}$ & $-$0.662$^{***}$ & $-$0.672$^{***}$ \\ 
  & (0.013) & (0.013) & (0.013) & (0.013) \\ 
 log(Deaths), 42d lag & $-$0.047$^{***}$ & $-$0.050$^{***}$ & $-$0.045$^{***}$ & $-$0.049$^{***}$ \\ 
  & (0.006) & (0.007) & (0.007) & (0.007) \\ 
  log(Deaths), 49d lag & $-$0.062$^{***}$ & $-$0.064$^{***}$ & $-$0.059$^{***}$ & $-$0.063$^{***}$ \\ 
  & (0.005) & (0.006) & (0.005) & (0.006) \\ 
 \hline \\[-1.8ex] 
County Dummies & Yes & Yes &  Yes  &  Yes  \\   
State$\times$ Week Dummies&Yes & Yes &  Yes  &  Yes  \\ 
\hline \\[-1.8ex] 
Observations & 615,523 & 485,776 & 546,519 & 471,386 \\ 
R$^{2}$ & 0.268 & 0.274 & 0.273 & 0.276 \\    \hline 
\hline 
\end{tabular}}	
  {\scriptsize
\begin{flushleft}
Notes: Dependent variable is the log difference over 21 days in weekly reported deaths. Regressors are 7-day moving averages of corresponding daily variables and lagged by 35 days to reflect the time between infection and  death reporting. All regression specifications include county fixed effects and state-week fixed effects to control for any unobserved county-level factors and time-varying state-level factors such as various state-level policies. 
The debiased fixed effects estimator is applied.  Asymptotic clustered standard errors at the state level are reported in the bracket.  {$^{*}$p$<$0.1; $^{**}$p$<$0.05; $^{***}$p$<$0.01}
\end{flushleft}}   
\end{table}

\begin{table}[!htbp] \centering
 \caption{The Association of School  Openings, NPI Policies, Full-time/Part-time Work, and Staying Home Devices  with Death Growth in the United States: Debiased Fixed Effects Estimator}\vspace{-0.3cm}
 \label{tab:PItoY-death-fulltime-SI}  
\resizebox{0.8\columnwidth}{!}{
\begin{tabular}{@{\extracolsep{1pt}}lcc|cc} 
\\[-1.8ex]\hline 
\hline \\ [-1.8ex] 
 & \multicolumn{4}{c}{\textit{Dep. variable: \textbf{Death Growth Rate over 3 weeks}  }} \\ 
\cline{2-5} 
& (1) & (2) & (3) & (4)\\ 
\hline

 K-12 Visits,  35d lag  & 0.494$^{***}$ & 0.270 &  &  \\ 
  & (0.121) & (0.168) &  &  \\ 
 K-12 Visits $\times$ No-Mask,  35d lag  &  & 0.622$^{***}$ &  &  \\ 
  &  & (0.221) &  &  \\ 
 K-12 In-person,  35d lag  &  &  & $-$0.018 & 0.024 \\ 
  &  &  & (0.050) & (0.060) \\ 
K-12 Hybrid,  35d lag &  &  & $-$0.019 & $-$0.052$^{*}$ \\ 
  &  &  & (0.027) & (0.028) \\ 
 K-12 Remote,  35d lag&  &  & $-$0.055$^{*}$ & $-$0.105$^{***}$ \\ 
  &  &  & (0.030) & (0.039) \\ 
  K-12 In-person $\times$ No-Mask,  35d lag&  &  &  & 0.056 \\ 
  &  &  &  & (0.068) \\ 
 K-12 Hybrid $\times$ No-Mask,  35d lag &  &  &  & 0.131$^{**}$ \\ 
  &  &  &  & (0.051) \\ \hline
College Visits, 35d lag  & 0.331$^{*}$ & 0.590$^{***}$ & 0.625$^{***}$ & 0.706$^{***}$ \\ 
  & (0.196) & (0.199) & (0.199) & (0.192) \\ 
Mandatory mask,   35d lag  & $-$0.125$^{***}$ & $-$0.123$^{***}$ & $-$0.139$^{***}$ & $-$0.124$^{***}$ \\ 
  & (0.027) & (0.025) & (0.027) & (0.026) \\ 
 Ban gatherings,  35d lag & $-$0.218$^{***}$ & $-$0.356$^{***}$ & $-$0.291$^{***}$ & $-$0.358$^{***}$ \\ 
  & (0.079) & (0.096) & (0.081) & (0.097) \\ 
  Stay at home,   35d lag & $-$0.207$^{***}$ & $-$0.143$^{**}$ & $-$0.155$^{**}$ & $-$0.125$^{*}$ \\ 
  & (0.058) & (0.064) & (0.062) & (0.069) \\ \hline
  log(Deaths), 35d lag& $-$0.665$^{***}$ & $-$0.675$^{***}$ & $-$0.661$^{***}$ & $-$0.672$^{***}$ \\ 
  & (0.013) & (0.013) & (0.013) & (0.013) \\ 
 log(Deaths), 42d lag& $-$0.046$^{***}$ & $-$0.050$^{***}$ & $-$0.044$^{***}$ & $-$0.049$^{***}$ \\ 
  & (0.006) & (0.006) & (0.007) & (0.007) \\ 
   log(Deaths), 49d lag  & $-$0.061$^{***}$ & $-$0.064$^{***}$ & $-$0.058$^{***}$ & $-$0.063$^{***}$ \\ 
  & (0.005) & (0.006) & (0.005) & (0.006) \\ 
    Full-time Work Device, 14d  lag& $-$1.803$^{***}$ & $-$2.080$^{***}$ & 0.033 & $-$0.711 \\ 
  & (0.668) & (0.642) & (0.700) & (0.690) \\ 
  Part-time Work Device, 14d  lag& 2.041$^{***}$ & 1.920$^{***}$ & 2.162$^{***}$ & 2.325$^{***}$ \\ 
  & (0.487) & (0.552) & (0.499) & (0.492) \\ 
  lStaying Home Devices, 35d lag& 0.412$^{***}$ & 0.531$^{***}$ & 0.383$^{***}$ & 0.559$^{***}$ \\ 
  & (0.109) & (0.099) & (0.116) & (0.097) \\ 
 \hline \\[-1.8ex]  
County Dummies & Yes & Yes &  Yes  &  Yes  \\   
State$\times$ Week Dummies&Yes & Yes &  Yes  &  Yes  \\ 
\hline \\[-1.8ex] 
Observations & 615,523 & 485,776 & 546,519 & 471,386 \\ 
R$^{2}$ & 0.268 & 0.274 & 0.273 & 0.276 \\  \hline 
\hline 
\end{tabular}}	
  {\scriptsize
\begin{flushleft}
Notes: Dependent variable is the log difference over 21 days in weekly reported deaths. Regressors are 7-day moving averages of corresponding daily variables and lagged by 35 days to reflect the time between infection and  death reporting. All regression specifications include county fixed effects and state-week fixed effects to control for any unobserved county-level factors and time-varying state-level factors such as various state-level policies. 
The debiased fixed effects estimator is applied.  Asymptotic clustered standard errors at the state level are reported in the bracket.  {$^{*}$p$<$0.1; $^{**}$p$<$0.05; $^{***}$p$<$0.01}
\end{flushleft}}   
\end{table}

\end{document}